\documentclass{aa}
\usepackage{etex}
\usepackage{natbib}
\usepackage{tablefootnote}
\usepackage{hyperref}
\usepackage[dvipsnames]{xcolor}
\usepackage{subcaption}
\usepackage[flushleft]{threeparttable}

\newcommand{\fig}[1]{Fig.~\ref{#1}}

\newcommand{\tab}[1]{Table~\ref{#1}}
\newcommand{\eq}[1]{Eq.~\ref{#1}}
\newcommand{\sect}[1]{Sect.~\ref{#1}}

\def\mdot{\dot{\rm M}}
\def\vturb{v_{\rm turb}}
\def\sigmaturb{\sigma_{\rm disp}}
\def\vdisp{\sigma_{\rm disp}^\star}
\def\vtherm{c_{\rm s}}
\def\vturbb{v_{\rm turb,3D}}
\def\ldriv{l_{\rm driv}}
\def\msunpc2{\rm M_\odot/pc^2}
\def\msunpc3{\rm M_\odot/pc^3}
\def\sfr{\rm \dot{\Sigma}_\star}
\def\rhosfr{\dot{\rho}_\star}
\def\phiv{\phi_{\rm{v}}}

\def\tauffcl{t_{\rm ff,cl}}
\def\tauvis{t_{\rm vis}}
\def\taudep{t_{\rm dep}}
\def\tauturbcl{t_{\rm turb,cl}}
\def\sigmastar{\dot{\Sigma}_\star}
\def\hi{\rm H\,{\textsc{I}}}
\def\ion{\zeta_{\rm CR}}
\def\chico{\chi^2_{\rm CO}}
\def\chicoa{\chi^2_{\rm CO(1-0)}}
\def\chicob{\chi^2_{\rm CO(2-1)}}
\def\chihi{\chi^2_{\rm HI}}
\def\chisfr{\chi^2_{\rm SFR}}
\def\chihcn{\chi^2_{\rm HCN}}
\def\chihco{\chi^2_{\rm HCO^+}}

\def\chitot{\chi^2_{\rm tot}}
\def\aco{\alpha_{\rm CO}}
\def\acomw{\alpha_{\rm CO}^{\rm MW}}
\def\ahcn{\alpha_{\rm HCN}}

\def\aunit{\rm M_\odot\,pc^{-2}\,(K\,km/s)^{-1}}

\begin{document}

\title{Predicting HCN, HCO$^+$, multi-transition CO, and dust emission of star-forming galaxies}
\subtitle{Constraining the properties of resolved gas and dust disks of local spiral galaxies}

   \author{T.~Liz\'ee\inst{1}, B.~Vollmer\inst{1},  J.~Braine\inst{2}, P.~Gratier\inst{2}, F.~Bigiel\inst{3}}

   \institute{Universit\'e de Strasbourg, CNRS, Observatoire astronomique de Strasbourg, UMR 7550, F-67000 Strasbourg, France \and 
          Laboratoire d'Astrophysique de Bordeaux, Univ. Bordeaux, CNRS, B18N, all\'ee Geoffroy Saint-Hilaire, 33615 Pessac, France \and
          Argelander-Institut fur Astronomie, Universit\"at Bonn, Auf dem H\"ugel 71, D-53121 Bonn, Germany}

 \abstract{ The interstellar medium (ISM) is a turbulent, multi-phase, and multi-scale medium following scaling relations linking the surface density, volume density, and velocity dispersion with the cloud size. Galactic clouds range from  below 1 pc to about 100 pc in size.  Extragalactic clouds appear to follow the same range although they are only now becoming observable in atomic and molecular lines. Analytical models of galactic gaseous disks need to take into account the multi-scale and multi-phase nature of the interstellar medium. They can be described as clumpy star-forming accretion disks in vertical hydrostatic equilibrium, with the mid-plane pressure balancing the gravity of the gaseous and stellar disk. ISM turbulence is taken into account by applying Galactic scaling relations to the cold atomic and molecular gas phases. Turbulence is maintained through energy injection by supernovae. With the determination of the gas mass fraction at a given spatial scale, the equilibrium gas temperature between turbulent heating and line cooling, the molecular abundances, and the molecular line emission can be calculated. The resulting model radial profiles of IR, H{\sc i}, CO, HCN, and HCO$^+$ emission are compared to THINGS, HERACLES, EMPIRE, SINGS, and GALEX observations of 17 local spiral galaxies. The model free parameters were constrained for each galactic radius independently. The Toomre parameter, which measures the stability against star formation (cloud collapse), exceeds unity in the inner disk of a significant number of galaxies. In two galaxies it also exceeds unity in the outer disk. Therefore, in spiral galaxies $Q_{\rm tot}=1$ is not ubiquitous. The model gas velocity dispersion is consistent with the observed H{\sc i} velocity dispersion where available. Within our model HCN and HCO$^+$ is already detectable in relatively low-density gas ($\sim 1000$~cm$^{-3}$). CO and HCN conversion factors and molecular gas depletion time were derived. Both conversion factors are consistent with values found in the literature. Whereas in the massive galaxies the viscous timescale greatly exceeds the star formation timescale, the viscous timescale is smaller than the star formation timescale within $\rm{R}~\sim~2~\rm{R}_{\rm d}$, the disk scale length, in the low-mass galaxies. We suggest that massive spiral galaxies undergo starvation in the absence of gas accretion from the halo, whereas in low-mass galaxies the fuel for star-formation reaches $\rm{R}~\sim~2~\rm{R}_{\rm d}$ from outside via a thick gas disk component with a high radial infall velocity observable in the H{\sc i} line.
}


\authorrunning{Liz\'ee et al.}
\titlerunning{Predicting HCN, HCO$^+$, multi-transition CO, and dust emission of star-forming galaxies}
\maketitle

%
\newpage

\section{Introduction}
 
The ISM of spiral galaxies {involves} multiple scales, multiple densities and temperatures, and multiple phases (ionized, atomic, molecular).
The observed gas velocity dispersions ($\sim 10$~km\,s$^{-1}$ in the {atomic} gas and several km\,s$^{-1}$ in the molecular gas) indicate 
that the ISM is supersonically turbulent. Within the star-forming disk, turbulence is triggered by thermal instabilities and maintained by the energy 
injection through stellar feedback.
Scaling relations for giant molecular clouds (GMCs) were first established by \citet{1981MNRAS.194..809L}: power-law relationships between the 
velocity dispersion and the gas density {on} the one hand, and the size of the emitting regions on the other hand.
The power-law indices were refined by, e.g., \citet{2009ApJ...699.1092H} and \citet{2010A&A...519L...7L}. Galactic H{\sc i} scaling relations
were established by \citet{1979MNRAS.186..479L} and \citet{1983Ap&SS..93...37Q} and used in the two-phase H{\sc i} model of the Galaxy by \citet{2003ApJ...587..278W}.

The molecular gas depletion time, $t_{\rm dep,H2}=M_{\rm H2}/\rm{SFR}$, measured at kpc-scales in large spiral galaxies is about constant with $t_{\rm dep,H2} \sim 2~{\rm Gyr}$ (e.g., \citealt{1998ApJ...498..541K}, \citealt{1998ARA&A..36..189K},\citealt{2008AJ....136.2846B}, \citealt{2008AJ....136.2782L},\citealt{2021MNRAS.501.4777E}). This relation between the molecular gas mass and the star-formation rate breaks down at scales smaller than a few hundred parsecs (\citealt{2011IAUS..270..327B},\citealt{2014MNRAS.439.3239K}, \citealt{2019ApJ...887...49S}) because of the life cycle of the star-formation process. At the scale of individual giant molecular clouds (GMCs), regions of massive star-formation and cold molecular gas are not correlated because subsequent phases of the star-formation process are observed. At these scales, a gas depletion time up to 17 times shorter than that of nearby galaxies (\citealt{2010ApJ...723.1019H}) or broken power laws (\citealt{2017ApJ...839..113R}) were found. 

The star-formation efficiency, which is the inverse of the gas depletion time, depends on the determination of the molecular gas mass, which
is hampered by the uncertainty of the conversion between the
observed line emission (mostly CO, but also HCN and HCO$^{+}$ for the dense gas) and the mass of molecular hydrogen.
Whereas the Galactic CO conversion factor is well established ($\alpha_{\rm CO}=\Sigma_{\rm H_2}/I_{\rm CO}=4.3~\aunit$ {including He}; \citealt{2013ARA&A..51..207B}) and can be regarded as canonical in star-forming galaxies of about solar metallicity, the widely used HCN conversion factor of $\alpha_{\rm HCN} = 10~ \aunit$ was derived by \citet{2004ApJ...606..271G} assuming virialized (self-gravitating) optically thick dense gas cores with a gas density of $n \sim 3 \times 10^4$~cm$^{-3}$ and constant brightness temperatures of 35~K.
\citet{2010ApJS..188..313W} found a twice higher HCN conversion factor of $\alpha_{\rm HCN} \sim 20~ \aunit$.

The  theory  of  clumpy  gas  disks  of \citet{2003A&A...404...21V}  provides  analytic expressions for large-scale and small-scale properties of galactic gas disks. 
The large-scale properties considered are the gas surface density, volume density,  disk  height,  turbulent  driving  length  scale, velocity  dispersion,  gas  viscosity,  
volume  filling  factor, and molecular  fraction.  Small-scale  properties  are  the  mass,  size, density, turbulent crossing time, free-fall time, and molecular formation timescale of the most massive self-gravitating gas clouds. These quantities depend on the stellar surface density, angular velocity, disk radius $R$, and three free 
parameters, which are the Toomre parameter $Q$ of the gas, the mass accretion rate $\dot{M}$, and the ratio $\delta$ between  the driving length scale of 
turbulence  and the cloud size. \citet{2011AJ....141...24V} determined these free parameters using three independent measurements of the radial profiles of the atomic
gas (H{\sc i}), molecular gas (CO), and SFR (FUV + 24~$\mu$m) for a sample of 18 mostly spiral galaxies from \citet{2008AJ....136.2782L}.
The fits of radial profiles were acceptable with reduced $\chi^2$ minimizations (defined as 
the normalized sum of the squared differences divided by the observational 
uncertainties) generally smaller than 2 for all galaxies except 
NGC 5194 (M 51):
\begin{equation} \label{first}
  \chi^2_{\rm tot}  = \sum_{i} \chi^2_{i} = \sum_{i}  \frac{\left(\mathcal{O}_i - \mathcal{M}_i\right)^2}{\sigma^2_{i}}\ ,
\end{equation}
where $\mathcal{O}_i$ corresponds to a single observational measurement, $\mathcal{M}_i$ is the model value, and $\sigma_i$ the observational error.
It was found that the model star formation efficiency is very sensitive to the description of local pressure equilibrium in the disk midplane. The  model-derived  free-fall timescales of self-gravitating clouds were
in good agreement with expectations from observations.  Only low-mass galaxies ($M_* < 10^{10}$~M$_{\odot}$) can balance the gas loss due to star formation by 
radial gas inflow within the galactic disk. \citet{2016MNRAS.458.1671K} elaborated a similar model including radial gas transport and stellar feedback.
They concluded that in spiral galaxies at low redshift turbulence is driven by star formation.

\citet{2017A&A...602A..51V} significantly extended the \citet{2011AJ....141...24V} model by introducing ISM scaling relations into the model.
The extended model simultaneously calculates the total gas mass, HI/H$_2$ mass ratio, the gas velocity dispersion, IR luminosity, IR spectral energy distribution, 
CO spectral line energy distribution (SLED), HCN(1-0) and HCO$^+$(1-0) emission of a galaxy given its size, integrated star formation rate, stellar mass radial 
profile, rotation curve, and Toomre $Q$ parameter. The model was applied to the integrated properties of local spiral galaxies, ultra-luminous
infrared galaxies (ULIRGs), high-z starforming galaxies, and submillimeter galaxies. The model reproduced the observed CO luminosities and SLEDs of all sample 
galaxies within the model uncertainties ($\sim 0.3$~dex). The model CO and HCN conversion factors had uncertainties of a factor of two. Both the HCN and HCO$^+$ 
emissions trace the dense molecular gas to a factor of approximately two for the local spiral galaxies, ULIRGs and smm-galaxies.

In the present article we used the \citet{2017A&A...602A..51V} model to calculate the IR, H{\sc i}, CO, HCN, and HCO$^+$ emission radial profiles 
and compare them to observations of $17$ local spiral galaxies from \citet{2008AJ....136.2782L}. Compared to \citet{2011AJ....141...24V}, the
assumption of a radially constant mass accretion rate $\dot{M}$ was dropped and the Toomre parameter $Q$ was self-consistently determined at each galactic radius.
The new model allowed us to compare the star formation rate based on the FUV and 24~$\mu$m emission to the IR radial profile and to
calculate the radial variations of (i) the CO and HCN conversion factors and (ii) the star formation efficiency.
 
The structure of this article is the following: the observations from the literature are described in Sect.~\ref{sec:observations}.
Our model of a turbulent clumpy accretion disk is outlined{, the search for the best-fit model is explained,
and the detected model degeneracies are presented in Sect.~\ref{sec:model}. The results} are given in 
Sect.~\ref{sec:results}, followed by the discussion (Sect.~\ref{sec:discussion}) and
our conclusions (Sect.~\ref{sec:conclusions}).

\section{Observations \label{sec:observations}}

In this section we {describe the observational data, which is used for the comparison with} the quantities obtained from our models. All multi-wavelength data were convolved to the same spatial resolution and were converted to the same unit, i.e. K~km\,s$^{-1}$ for the atomic and molecular line data and MJy/sr for the infrared data.

\subsection{Atomic hydrogen $\hi$}

The 21-cm $\hi$ radial profiles used in this paper were taken from the VLA THINGS data survey (CDS VizieR table J/AJ/136/2782) presented in \citet{2008AJ....136.2563W}. The spatial resolution is about $7"$ with a spectral resolution of 5~km\,s$^{-1}$ per channel. With an average of 7~hr of observations on source, a typical rms noise of $0.4$~mJy/beam (which corresponds to $5$~K) was reached. 

\subsection{Carbon monoxyde CO(2-1)}

The HERACLES survey is presented in detail in (\citealt{2009AJ....137.4670L}). The data were collected using the HERA multi-pixel receiver from the single dish IRAM 30m telescope (Pico Veleta, Spain). HERA was tuned near 230~GHz to observe the CO(2-1) rotational transition. OTF maps were produced, giving a spatial resolution of $13"$, a spectral resolution of 2.6~km\,s$^{-1}$ and a rms noise about 20~mK. The HERACLES survey focused on targets that were part of the THINGS galaxy sample. We used the radial profiles presented in \citet{2008AJ....136.2782L} (CDS VizieR table J/AJ/136/2782) except for NGC~3627, NGC~5194, {and NGC~7793}, for which we used the CO(2-1) profile presented in \citet{2021MNRAS.504.3221D} derived from PHANGS data \citep{2021ApJS..255...19L}. 

\subsection{Carbon monoxyde CO(1-0), Hydrogen cyanide HCN(1-0), and Formylium HCO$^+$(1-0)}

The CO(1-0), HCN(1-0), and HCO$^+$(1-0) radial profiles were taken from the EMPIRE survey (\citealt{2019ApJ...880..127J}; CDS VizieR table J/ApJ/880/127). The EMPIRE observations were carried out at the IRAM 30m telescope using the dual-polarization EMIR receiver. The spectral resolution is about 4~km\,s$^{-1}$ per channel, with an rms noise about 2-3~mK for the dense gas {($n_{\rm H_2} \sim 10^4$~cm$^{-3}$) HCN and HCO$^+$} observations. Spatial resolutions are 26", 33" and 33" for the CO(1-0), HCN(1-0) and HCO$^+$(1-0) data, respectively. In contrast to the other data presented in this paper {for which radial profiles with $10''$ wide rings were extracted from the moment~0 maps}, \citet{2019ApJ...880..127J} used a stacking method to recover the emission from the regions with low signal-to-noise ratio. 

\subsection{Star-Formation Rate \label{sec:sfr1}}

Star-formation rate profiles were computed by \citet{2009AJ....137.4670L} using the FUV GALEX data (\citealt{2007ApJS..173..185G}) with the 24$\mu$m infrared data from the SPITZER SINGS survey (\citealt{2003PASP..115..928K}). Their spatial resolutions are $6"$ and $5"$, respectively . The following linear combination was  used to compute the SFR:
\begin{equation}
    \dot{\Sigma}_{\star} = (8.1 \times 10^{-2} I_{\rm FUV} + 3.2 \times 10^{-3} I_{\rm 24\mu m}) \times \cos{} i ,
\end{equation}
where $I_{\rm FUV}$ and $I_{\rm 24\mu m}$ are the ultraviolet and infrared fluxes in MJy$\,\rm{sr^{-1}}$ and $i$ is the inclination angle of the galaxy. The star-formation rate surface density $\sfr$ has units of $\rm M_\odot kpc^{-2} yr^{-1}$. 
 
\subsection{Far-infrared data profiles and temperature maps}

We worked with four far-infrared bands from the PACS (100$\mu$m) and SPIRE (250, 350, and 500$\mu$m) instrument on the Herschel satellite. Their spatial resolutions are 7, 18, 25, and $35"$, respectively. The far-infrared profiles were taken from \citet{2015A&A...576A..33H} (CDS VizieR table J/A+A/576/A33).

\subsection{Sample of galaxies}

Our sample comprises {17} star-forming galaxies that the surveys presented in the previous section have in common. This galaxy sample is composed of {five} low-mass ($M_* < 10^{10}$~M$_{\odot}$) galaxies and 12 nearby large spiral galaxies. The general properties of these galaxies are presented in \tab{galaxy_sample}. The rotation curves of the model galaxies were computed by \citet{2008AJ....136.2782L} using the following expression (\citealt{2003MNRAS.346.1215B}): 
   \begin{equation}
      v_{\rm{rot}}=v_{\rm{flat}} \left( 1 - \rm{exp}\left(-\frac{ R }{ l_{\rm{flat}} }\right) \right)\ ,
   \end{equation}
where $v_{\rm{flat}}$ and $l_{\rm{flat}}$ were determined by performing a polynomial fit on the observational curves.
{The common spatial resolution is $400$~pc for the low-mass galaxies and $800$~pc for the large spiral galaxies.}

\begin{table*}[!ht]
   \caption{Galaxy properties}
   \label{galaxy_sample}
   \begin{center}
    \renewcommand{\arraystretch}{1.2}
   \begin{tabular}{cccccccccccc} 
      \hline
      \hline
         Galaxy$^1$ & Type & RA J2000 & DEC J2000 & R$_{25}$ &D & $i$ & PA & log$\, M_{\star}$ & R$_{\rm{d}}$ & $v_{flat}$ & $l_{flat}$ \\
            &  &   &   & (kpc) & (Mpc) & (deg) & (deg) & (10$^8$ M$_\odot$) & (kpc) & (km\,s$^{-1}$) & (kpc) \\
        \hline
        NGC~628 & Sc & 01 36 41.772 & +15 47 0.46 &10.4 &7.3 & 7& 20& 10.1& 2.3& 217& 0.8\\
        NGC~3184 & SBc &10 18 16.985 & +41 25 27.77 &12.0 &11.1 & 16& 179& 10.3& 2.4& 210 & 2.8 \\
        NGC~3627 & SBb &11 20 15.026 & +12 59 28.64 &13.8 &9.3 & 62 & 173&10.6 &2.8 &192 &1.2 \\
        NGC~5055 & Sbc &13 15 49.274 & +42 01 45.73 &17.3 & 10.1& 59& 102& 10.8& 3.2& 192&0.7 \\
        NGC~5194 & SBc &13 29 52.698 & +47 11 42.93 &9.0 & 8.0& 20& 172& 10.6& 2.8& 219& 0.8\\
        NGC~6946 & SBc &20 34 52.332 & +60 09 13.24 &9.9 & 5.9& 33& 243& 10.5& 2.5&186 &1.4 \\
        \hline
        NGC~2841 & Sb	&09 22 02.655 & +50 58 35.32  &14.2 &14.1 & 74& 153& 10.8& 4.0& 302& 0.6\\
        NGC~3198 & SBc &10 19 54.990 & +45 32 58.88 & 13.0 &13.8& 72& 215& 10.1& 3.2& 150&2.8\\
        NGC~3351 & SBb &10 43 57.733 & +11 42 13.00 & 10.6 &10.1& 41& 192& 10.4& 2.2&196 &0.7 \\
        NGC~3521 & SBbc &11 05 48.568 & -00 02 9.23 & 13.0 &10.7& 73& 340& 10.7& 2.9& 227 & 1.4\\      
        NGC~4736 & Sab &12 50 53.148 & +41 07 12.55 & 5.3 &4.7& 41& 296&10.3 & 1.1& 156&0.2 \\
        NGC~7331 & SAb &22 37 04.102 & +34 24 57.31 &19.5 & 14.7& 76& 168&10.9 & 3.3& 244&1.3 \\
        \hline
        NGC~925 & SBcd &02 27 16.913 & +33 34 43.97 & 14.3 & 9.2 & 66& 287& 9.9& 4.1& 136& 6.5\\
        NGC~2403 & SBc &07 36 51.396 & +65 36 09.17 & 7.4 &3.2& 63& 124& 9.7& 1.6& 134& 1.7\\
        NGC~2976 & Sc &09 47 15.458 & +67 54 58.97 & 3.8 &3.6& 65& 335& 9.1& 0.9& 92& 1.2\\        
        NGC~4214 & Irr. &12 15 39.174 & +36 19 36.80 & 2.9 &2.9& 44& 65& 8.8& 0.7& 57& 0.9\\ 
        NGC~7793 & Scd & 23 57 49.754 & -32 35 27.70 & 6.0 & 3.9 & 50 & 290 & 9.5 & 1.3 & 115 & 1.5 \\
      \hline
   \end{tabular}

      \end{center}
      \footnote{a}{In order of appearance: massive galaxies (log$\, M_{\star} > 10$) with EMPIRE data, massive galaxies without EMPIRE data, low mass galaxies (M$_* < 10^{10}$~M$_{\odot}$) without EMPIRE data.}
\end{table*} 

\section{Model fitting \label{sec:model}} 

In this section we describe our analytical model and introduce its free parameters. The comparison with the data was performed via a reduced $\chi^2$ minimization (Eq.\ref{first}). We investigated the degeneracies between the free model parameters.

\subsection{The model}
{Our model is a slightly modified version of the analytical model presented by \citet{2017A&A...602A..51V}, itself derived from previous versions presented in \citet{2003A&A...404...21V} and \citet{2011AJ....141...24V}. A detailed description of the model is given in Appendix~\ref{app:model}. 
The analytical model describes galaxies as star-forming, clumpy, and turbulent accretion disks. The interstellar medium (ISM) is considered as a turbulent multi-phase gas. This gas is assumed to be in vertical hydrostatic equilibrium, with the mid-plane pressure balancing the weight of the gaseous and the stellar disk (\citet{1989ApJ...338..178E}; Eq.~\ref{pressure}). The model gas is described as "clumpy", so that the local density can be enhanced relative to the average density of the disk. The local free-fall time of an individual gas clump is taken as the governing timescale for star formation.
The SFR is used to calculate the rate of energy injection by supernova explosions. This rate is related to the turbulent velocity dispersion and the driving 
scale of turbulence. These quantities in turn provide estimates of the clumpiness of gas in the disk (the contrast between local and average density) and the rate 
at which viscosity moves matter inward.
The model relies on several empirical calibrations: the relation between the stellar velocity dispersion and the stellar disk scale length (Eq.~\ref{eq:vdisp}),
the relationship between the SFR and the energy injected into the ISM by supernovae (Eq.~\ref{sfr3}), and the characteristic time of H$_2$ formation, which is
related to the gas metallicity $Z$ (Eqs.~\ref{taumol} to \ref{metallicity}) {and the gas density}.
The equilibrium between the different phases of the ISM and the equilibrium between turbulence and star formation depends on three local timescales: the turbulent crossing time $t^l_{\rm turb}$, the molecule formation timescale $t^l_{\rm mol}$, and the local free-fall timescale $t^l_{\rm ff}$ of a cloud. In addition, photo-dissociation of molecules is taken into account. 

 The model has a large-scale and a small-scale part. The large-scale part gives the surface density, turbulent velocity, disk height, and gas viscosity. The small-scale part begins at densities where gas clouds become self-gravitating ($t^l_{\rm ff}=t^l_{\rm turb}$). The non-self-gravitating and self-gravitating clouds obey different scaling relations, which are set by observations (Appendix~\ref{sec:scaling}). For each gas density, the mass { fraction is} characterized by a lognormal probability distribution function and the Mach number (\citealt{1997MNRAS.288..145P}). { The temperatures of the
gas clouds} are calculated via the equilibrium between turbulent mechanical and cosmic ray heating and gas cooling via CO and H$_2$ line emission 
(Appendix~\ref{sec:heatcool}). The abundances of the different molecules are determined using the gas-grain code NAUTILUS \citep{2009A&A...493L..49H}.
The dust temperatures are calculated via the equilibrium between heating by the interstellar UV and optical radiation field and cooling via infrared emission.

The model simultaneously calculates (i) for the large-scale part: the total gas profile, the gas velocity dispersion, the star formation rate, and the volume filling factor; (ii) for the small-scale part: the molecular fraction, infrared profiles and SED, and molecular line profiles of different molecules such as CO(1-0), CO(2-1), HCN(1-0) and HCO$^+$(1-0). The molecular line emission calculation is based on a two-level approximation and the escape probability formalism.
This approximation greatly decreases the computation time.
An important ingredient for the line emission is the area-filling factor of the gas clouds, which is a result of the small-scale part of the analytic disk model.
The new code was modified to enable an independent treatment of each of the radial data points. The input of a given radial profile of the Toomre parameter $Q$ and the assumption of a constant mass accretion rate $\dot{M}$, which were required to produce the molecular line profiles in \citet{2011AJ....141...24V}, are no longer necessary. 
The radial profiles of the stellar surface density and the rotation curve (both taken from \citealt{2008AJ....136.2782L}) are used as inputs of the model. 
The model yields the radial profiles of $\dot{M}$ and $Q$ for a given $\delta$, which best fit the available observations (SFR, H{\sc i}, CO(1-0), CO(2-1)).
With these profiles the physical properties of the ISM such as the molecular surface density, the gas velocity dispersion, or the
CO and HCN conversion factors can be calculated.
}

\subsection{Model parameters}

{
The model inputs are the rotation curve and the stellar surface density profile.
The model contains three main free parameters: (i) the Toomre parameter $Q$ of the gas, which indicates the gravitational stability ($Q = 1$) of a gas cloud, (ii) the mass accretion rate $\mdot$, which sets the gas turbulent velocity, and (iii) $\delta$, the ratio between the turbulent driving length scale and the size of the largest self-gravitating clouds. At a given disk radius $R$ the model yields the star formation rate per unit area and the H{\sc i} and molecular line emission for a given set of $Q$, $\mdot$, and $\delta$. 

Unlike {\citet{2017A&A...602A..51V}}, we decided to vary other parameters which were previously constant. These parameters present large uncertainties and were identified as having a significant impact on the results. 
The first parameter, $\xi$, relates the SFR to the energy injected into the ISM by supernovae (Eq.~\ref{sfr3}). The second parameter, $\alpha_0$, relates the characteristic time to form H$_2$ out of H to the local free-fall time (Eq.~\ref{molecule}). This parameter depends on to the gas metallicity of the galaxy (Eq.~\ref{metallicity}). { The last parameter, $\gamma$, controls the vertical component of the stellar velocity dispersion} and thus the gravitational restoring force of the stellar disk (Eq.~\ref{pressure}). This is motivated by a possible heating of the stellar disk following a gravitational interaction. Because of the large computation time of the models 
we limited ourselves to {four values for $\delta$, five values of $\xi$ and $\alpha_0$, and three values for $\gamma$: $\delta=[3, 5, 7, 9]$, $\xi=(0.25\,,0.5\,,1\,,2\,,4) \times 4.6 \times 10^{-8}$~(pc/yr)$^2$, $\alpha_0 = (0.25\,,0.5\,,1\,,2\,,4) \times 2.2 \times 10^7~\rm{yr~M_\odot~pc^{-3}}$, and $\gamma=0.5\,,1\,,2$.
We individually varied each of these parameters, with default values corresponding to unity for $\xi$, $\alpha_0$, and $\gamma$. Because of the large computation times we did not consider combinations of these variations. All these free parameters are summarized in Table \ref{free}.}

\begin{table*}[!h]
  \caption{Model free parameters}
  \label{free}
  \begin{center}
   \renewcommand{\arraystretch}{1.5}
   \addtolength{\tabcolsep}{-2.pt}
  \begin{tabular}{ccc} 
     \hline
     \hline
    $Q$ & - & Toomre parameter, gravitational stability of gas clouds (Eq.~\ref{toomre}) \\
    $\dot{M}$ & M$_\odot$ yr$^{-1}$ & radial mass accretion rate within the disk (Eq.~\ref{mdot}) \\
    $\delta$ & - & scaling factor between the diriving length scale and the size of self-gravitating structures (Eq.~\ref{larson})\\ 
    $\xi$ & pc$^2$ yr$^{-1}$ & constant relating supernovae energy input to star-formation (Eq.~\ref{sfr3})\\
    $\alpha_0$ & $\rm{yr~M_\odot~pc^{-3}}$ & constant of molecule formation timescale, inverse of the effective stellar yield (Eq.~\ref{metallicity}) \\
    $\gamma$ & - & scaling factor of the vertical stellar velocity dispersion (Eq.~\ref{pressure}) \\
    \hline
  \end{tabular}

     \end{center}
\end{table*}

The IR emission also depends on the gas-to-dust ratio and exponent $\beta$ of the wavelength dependence of the dust absorption coefficient.
For the search of $[Q(R), \dot{M}(R), \delta, \alpha_0, \xi, \gamma]$  we used a gas-to-dust ratio GDR$=100$ and $\beta=1.5$. 
As a last step, we kept best-fit $[Q(R), \dot{M}(R), \delta, \alpha_0, \xi, \gamma]$ constant and varied GDR and $\beta$ to determine their values by searching for the best fit of the observed infrared profiles.}

Finally, different values of the cosmic ray ionization rate were applied to the best-fit models. The value that was assumed in previous versions of the model was $\ion = 10^{-17}\ \rm s^{-1}$, which corresponds to the standard value for GMCs (\citealt{2006PNAS..10312269D}). Additionally, we used  $\ion = 3 \times 10^{-18}\ \rm s^{-1}$ and $10^{-18}\ \rm s^{-1}$.

\subsection{Determination of the best-fit model using $\chi^2$ minimizations \label{deter}}

{
For each set of $[\delta, \alpha_0, \xi, \gamma]$ we performed a $\chi^2$ minimization at each radius independently to find the best-fit $Q$ and $\dot{M}$. The contributions from the different radial profiles (SFR, H{\sc i}, CO(1-0), CO(2-1))  were summed to obtain a total $\chi^2$:
\begin{equation}
    \chi^2_{\rm tot} = \sum_i \frac{(\mathcal{O}_i-\mathcal{M}_i)^2}{\sigma_i^2} = \chi^2_{\rm HI} + \chi^2_{\rm SFR} + \chi^2_{\rm CO}\ .
\end{equation}
where $\mathcal{O}_i$ and $\mathcal{M}_i$ are the observed and model data points and $\sigma_i$ the uncertainties of the observations.
For the majority of galaxies, we set $\chi^2_{\rm CO} = \chi^2_{\rm CO(2-1)}$ with the CO(2-1) data from the HERACLES survey.
We increased the uncertainties of the PHANGS radial profiles of NGC~3627 and NGC~5194 to those typically found 
for the HERACLES profiles. Without this increase of the uncertainties, the PHANGS profiles would dominate the total $\chi^2$.
For galaxies observed by the EMPIRE survey, we set $\chi^2_{\rm CO} = \chi^2_{\rm CO(2-1)}/2~+~\chi^2_{\rm CO(1-0)}/2$ to take the CO(1-0) emission in the determination of the best-fit model into account. { It turned out that the inclusion of the CO(1-0) emission led to somewhat different fitting results. For consistency between the results of all galaxies, we also performed $\chi^2$ minimizations including only CO(2-1) emission for the galaxies observed by EMPIRE.}
The comparison of the line brightness temperatures obtained with our model and those obtained with RADEX \citep{2007A&A...468..627V} showed good agreement (within $\sim 50$\,\%)
for the CO lines but much less good agreement for the HCN and HCO$^+$ lines (up to a factor of two to three). 
Therefore, $\chi^2_{\rm HCN}$ and $\chi^2_{\rm HCO^+}$ based on the line brightness temperatures obtained with our model were calculated but not included in the $\chi^2_{\rm tot}$ calculation. 
}
We worked within a parameter grid of twenty values of Toomre parameter $Q = [1:10]$ and twenty values of the accretion rate $\mdot = [10^{-3}:1]$ for each radius. {These intervals were chosen according to the results of
\citet{2011AJ....141...24V}. We did not consider gas disks that are unstable to fragmentation ($Q<1$).}
The outer radius of the model is set by the detection limit of the CO observations. 

The best-fit models for each galaxy are listed in \tab{bestfit1} for galaxies with EMPIRE data and \tab{bestfit2} for the rest of the sample. 
{The corresponding radial profiles of $Q$ and $\mdot$ are shown in Figs.~\ref{ngc6946b} and \ref{fig:n628} to \ref{fig:7793} and discussed
in Sect.~\ref{QMV}.} The models presented in the two tables correspond to the models that present a $\chi^2_{\rm tot}~\in~[\chi^2_{\rm tot,min}$,$\chi^2_{\rm tot,min}$+0.1$\chi^2_{\rm tot,min}$]. In \tab{bestfit1}, the {values between parentheses correspond to the $\chi^2_{\rm CO(1-0)}$ that are not included in the $\chi^2_{\rm tot}$ calculations.} 
 
{Because the determination of the best-fit basic models were very time consuming, we decided to vary GDR, $\beta$, and $\ion$ separately:
a second, independent $\chi^2$ minimization was performed on the basic model to determine the value of $\beta$ and GDR to fit the 100, 250, and 500$\mu$m infrared radial profiles. This minimization generally led to values of GDR and $\beta$, which are higher than the default values, decreasing the dust optical depth and IR emission. The corresponding change of the molecular line emission due to a lower background temperature is less than one percent.
The values of $\beta$ and GDR for each galaxy are presented in \tab{betagdr}. We found a mean $\beta=2$ and a mean gas-to-dust ratio GDR $\sim 250$, which
is more than twice the value found by \citet{2013ApJ...777....5S}. The difference probably lies in the dust illumination of our model and that
of \citet{2007ApJ...657..810D}. In our model the radiation field is proportional to the star formation rate, which is constant at a given galactic radius.
In the Draine \& Li model the dust mass is exposed to a power-law distribution of starlight intensities between $U_{\rm min}$ and $U_{\rm max}$ with 
$dM/dU \propto U^{-1}$. This leads to higher dust temperatures and lower dust masses in our model compared to those derived by the Draine \& Li model. 

A third $\chi^2$ minimization is finally performed on the basic model to find which cosmic ray ionization rate reproduces best the HCN and HCO$^+$ radial profiles, with three different value of $\ion = [10^{-18},3\times10^{-18},10^{-17}]$~sec$^{-1}$. These values correspond to the observed range of the
cosmic ray ionization rate (see Sect.~\ref{ioniz}). The best-fit values of $\ion$ are presented in \tab{bestcr}.
}
\begin{table*} [!ht]
  \caption{Best-fit models of the EMPIRE galaxies} 
  \label{bestfit1}
  \begin{center}
   \renewcommand{\arraystretch}{1.2}
  \begin{tabular}{c|cccc|cccc|c}
     \hline
     \hline
         Galaxy & $\delta$ &  $\xi$ & $\alpha_0$ & $\gamma$ & $\chi_{\rm HI}^2$ & $\chi_{\rm SFR}^2$ & $\chi_{\rm CO(2-1)}^2$ & $\chi_{\rm CO(1-0)}^2$ & $\chi_{\rm tot}^2$\\
         \hline
         \hline
NGC~628 
& 5 & - & $0.5\times$ & - & 32 & 22 & 79 & 401 & 294 \\
& 3 & - & $0.5\times$ & - & 31 & 19 & 62 & 468 & 315 \\
& 9 & $2\times$ & - & - & 26 & 18 & 65 & 1074 & 614 \\ \cline{2-10}
& 5 & - & - & $0.5\times$ & 6 & 10 & 7 & (2285) & 23 \\
& 7 & - & - & $0.5\times$ & 6 & 5 & 12 & (2147) & 23 \\
& 5 & - & - & - & 4 & 8 & 13 & (2714) & 24 \\
& 9 & - & - & - & 5 & 6 & 12 & (2132) & 24 \\
\hline
NGC 3184   
& 9 & - & $0.5\times$ & - & 34 & 108 & 27 & 107 & 209 \\
& 9 & $2\times$ & - & - & 24 & 49 & 30 & 254 & 214 \\
& 5 & $2\times$ & - & - & 14 & 74 & 26 & 234 & 218 \\
& 7 & - & $0.5\times$ & - & 27 & 129 & 29 & 105 & 224 \\
& 7 & $2\times$ & - & - & 52 & 57 & 31 & 200 & 224 \\ \cline{2-10}
& 5 & $2\times$ & - & - & 5 & 7 & 9 & (1261) & 20 \\
& 9 & $2\times$ & - & - & 5 & 7 & 8 & (1258) & 20 \\
& 3 & $2\times$ & - & - & 12 & 5 & 9 & (1738) & 26 \\
\hline
NGC~3627 & 5 & - & $0.5\times$ & - & 209 & 225 & 31 & 1208 & 1053 \\
 & 9 & - & $0.5\times$ & - & 235 & 275 & 31 & 1117 & 1083 \\
 & 7 & - & $0.5\times$ & - & 273 & 323 & 30 & 1059 & 1141 \\ \cline{2-10}
& 3 & - & $0.5\times$ & - & 3 & 1 & 40 & (12612) & 44 \\
 & 9 & - & $0.5\times$ & - & 2 & 4 & 39 & (33875) & 44 \\
 & 5 & - & $0.5\times$ & - & 2 & 2 & 40 & (27516) & 45 \\
\hline
NGC 5055 
& 3 & $2\times$ & - & - & 42 & 50 & 211 & 353 & 374 \\
& 5 & $2\times$ & - & - & 25 & 75 & 220 & 368 & 394 \\
& 7 & $2\times$ & - & - & 39 & 59 & 202 & 410 & 404 \\ \cline{2-10}
& 7 & $2\times$ & - & - & 41 & 45 & 91 & (1450) & 177 \\
& 9 & $2\times$ & - & - & 36 & 52 & 91 & (1443) & 179 \\
& 3 & $2\times$ & - & - & 42 & 39 & 99 & (1206) & 181 \\
& 5 & $2\times$ & - & - & 35 & 59 & 89 & (1756) & 182 \\
\hline
NGC~5194  & 3 & $2\times$ & - & - & 90 & 73 & 29 & 259 & 307 \\
 & 5 & - & $0.5\times$ & - & 143 & 113 & 27 & 107 & 322 \\
 & 3 & $2\times$ & - & - & 90 & 73 & 29 & 259 & 307 \\ \cline{2-10}
 & 9 & - & $0.5\times$ & - & 10 & 4 & 16 & (20647) & 30 \\
 & 3 & $2\times$ & - & - & 11 & 8 & 16 & (24194) & 35 \\
 & 7 & - & $0.5\times$ & - & 12 & 5 & 19 & (29896) & 36 \\
\hline
NGC 6946 
& 9 & - & $0.5\times$ & - & 182 & 1355 & 97 & 1033 & 2102 \\
& 7 & - & $0.5\times$ & - & 281 & 1523 & 94 & 893 & 2297 \\
& 5 & - & $0.5\times$ & - & 820 & 1572 & 97 & 438 & 2659 \\ \cline{2-10}
& 7 & - & $0.5\times$ & - & 7 & 7 & 25 & (224629) & 40 \\
& 9 & - & $0.5\times$ & - & 6 & 10 & 26 & (209394) & 42 \\
& 5 & - & $0.5\times$ & - & 7 & 10 & 26 & (248890) & 43 \\
\hline
  \end{tabular}
     \end{center}
     \begin{tablenotes}
      \centering
      \item {Values between parentheses are not included into the total $\chi^2$ calculation,}
      \item {All the parameters are described in Table \ref{free}. }
      \item {The best models are for most galaxies those with $\alpha_0$/2 or $\xi \times 2$ (see Sect.\ref{degeneracies}). }
  \end{tablenotes}
\end{table*}

\begin{table*}[!ht]
  \caption{Best-fit models of the THINGS galaxies without EMPIRE data}
  \label{bestfit2}
  \begin{center}
   \renewcommand{\arraystretch}{1.2}

  \begin{tabular}{c|cccc|ccc|c}
     \hline
     \hline
Galaxy & $\delta$ &  $\xi$ & $\alpha_0$ & $\gamma$ & $\chi_{\rm HI}^2$ & $\chi_{\rm SFR}^2$ & $\chi_{\rm CO(2-1)}^2$ & $\chi_{\rm tot}^2$\\
         \hline
         \hline
NGC~2841  
& 7 & - & - & $0.5\times$ & 5 & 2 & 2 & 9 \\
& 3 & - & $0.5\times$ & - & 3 & 3 & 4 & 10 \\
& 5 & - & $0.5\times$ & - & 2 & 4 & 5 & 11 \\
\hline
NGC~3198
& 9 & $0.5\times$ & - & - & 2 & 8 & 16 & 26 \\
& 7 & $0.5\times$ & - & - & 4 & 8 & 14 & 26 \\
& 9 & - & - & - & 8 & 9 & 10 & 27 \\
\hline
NGC~3351
& 5 & $0.5\times$ & - & - & 13 & 45 & 39 & 96 \\
& 9 & $0.5\times$ & - & - & 22 & 27 & 48 & 97 \\
& 7 & $0.5\times$ & - & - & 29 & 43 & 36 & 108 \\
\hline
NGC~3521
& 9 & $2\times$ & - & - & 83 & 135 & 52 & 270 \\
& 7 & $2\times$ & - & - & 97 & 111 & 67 & 275 \\
& 5 & $2\times$ & - & - & 108 & 124 & 69 & 300 \\
\hline
NGC~4736 
& 3 & - & $2\times$ & - & 4 & 8 & 6 & 18 \\
& 5 & - & $2\times$ & - & 5 & 6 & 10 & 21 \\
& 9 & - & $2\times$ & - & 2 & 9 & 12 & 24 \\
\hline
NGC~7331 
& 9 & - & - & $2\times$ & 31 & 11 & 23 & 65 \\
& 5 & - & - & $2\times$ & 32 & 16 & 34 & 81 \\
& 7 & - & - & $2\times$ & 37 & 19 & 24 & 79 \\
\hline
\hline
NGC~925 & 3 & - & $0.5\times$ & - & 59 & 192 & 216 & 468 \\
 & 5 & - & $0.5\times$ & - & 56 & 167 & 307 & 531 \\
 & 7 & - & $0.5\times$ & - & 69 & 135 & 358 & 562 \\
 \hline
NGC~2403 & 3 & $2\times$ & - & - & 31 & 29 & 28 & 88 \\
 & 5 & $2\times$ & - & - & 29 & 39 & 25 & 93 \\
 & 7 & $2\times$ & - & - & 34 & 36 & 32 & 102 \\
 \hline
NGC~2976 & 7 & - & $0.5\times$ & - & 108 & 86 & 89 & 283 \\
 & 9 & - & $0.5\times$ & - & 131 & 90 & 54 & 275 \\
 & 3 & - & $0.5\times$ & - & 116 & 85 & 90 & 291 \\
 & 5 & - & $0.5\times$ & - & 120 & 83 & 80 & 283 \\
 \hline
NGC~4214 & 3 & $2\times$ & - & - & 1 & 14 & 2 & 17 \\
 & 3 & - & - & - & 2 & 13 & 3 & 18 \\
 & 5 & $2\times$ & - & - & 1 & 13 & 8 & 22 \\
\hline
NGC~7793
& 5 & - & $2\times$ & - & 37 & 10 & 68 & 115 \\
& 7 & - & $2\times$ & - & 38 & 11 & 70 & 118 \\
& 9 & - & $2\times$ & - & 42 & 13 & 64 & 119 \\
& 3 & - & $2\times$ & - & 41 & 13 & 76 & 130 \\
\hline 

  \end{tabular}

 \end{center} 
 \begin{tablenotes}
  \centering
  \item {All the parameters are described in Table \ref{free}. }
  \item {The best models are for most galaxies those with $\alpha_0$/2 or $\xi \times 2$ (see Sect.\ref{degeneracies}).} 
\end{tablenotes}
\end{table*}

\begin{table}[!ht]
  \caption{Infrared profile parameters}
  \label{betagdr}
  \begin{center}
   \renewcommand{\arraystretch}{1.2}
   \addtolength{\tabcolsep}{8.pt}
  \begin{tabular}{ccc}
     \hline
      Galaxy & $\beta^a$ & GDR$^a$ \\
      \hline
      \hline
      NGC~628 & 2 & 300 \\
        & 2 & 300  \\
      \hline        
      NGC~3184 & 2 & 500 \\
      & 2 & 400  \\
      \hline
      NGC~3627 & 2 & 300 \\
      & 2 & 200  \\
      \hline
      NGC~5055 & 2 & 100 \\
      & 2 & 200  \\
      \hline
      NGC~5194 & 1.5 & 400 \\ 
      & 1.5 & 200  \\
      \hline
      NGC~6946 & 2.5 & 200 \\
      & 2.5 & 200  \\
      \hline
      NGC~2841 & 2 & 200 \\
      \hline
      NGC~3198 & 2.5 & 100 \\
      \hline
      NGC~3351 & 2 & 500 \\
      \hline
      NGC~3521 & 2 & 200 \\
      \hline
      NGC~4736 & 2 & 600 \\
      \hline
      NGC~7331 & 2 & 200 \\ 
      \hline
      NGC~925 & 2.5 & 100 \\ 
      \hline
      NGC~2403 & 1.5 & 200 \\ 
      \hline
      NGC~2976 & 2 & 100 \\ 
      \hline
      NGC~4214 & 0.5 & 500 \\ 
      \hline
      NGC~7793 & 2 & 100 \\
      \hline
      \hline
      $\bar{x}$ & $1.9 \pm 0.5$ & $294 \pm 201$ \\
      \hline
  \end{tabular}

     \end{center}

     \begin{tablenotes}
      \centering
      \item $a$: upper line: with $\chicoa$ in the $\chi^2$ calculations. 
      \item Lower line: without $\chicoa$.
      \item $\beta$: exponent of the wavelength dependence of the dust absorption coefficient.
      GDR: gas-to-dust ratio.
      \end{tablenotes}
     
\end{table}

\begin{table*}[!ht]
  \caption{Determination of the HCN and HCO$^+$ best-fit models for different CR ionization rates $\zeta_{\rm CR}$}
  \label{bestcr}
  \begin{center}
   \renewcommand{\arraystretch}{1.5}
  \begin{tabular}{c|ccc|ccc}
     \hline
     \hline
          Galaxy & \begin{tabular}{c}  \\ $10^{-17}$~s$^{-1}$ \end{tabular} & \begin{tabular}{c} with $\chi^2_{CO(1-0)}$ \\ \hline $3 \times 10^{-18}$~s$^{-1}$ \end{tabular} & \begin{tabular}{c} \\ $10^{-18}$~s$^{-1}$ \end{tabular} & \begin{tabular}{c}  \\ $10^{-17}$~s$^{-1}$ \end{tabular} & \begin{tabular}{c} without $\chi^2_{CO(1-0)}$ \\ \hline $3 \times 10^{-18}$~s$^{-1}$ \end{tabular} & \begin{tabular}{c} \\ $10^{-18}$~s$^{-1}$ \end{tabular} \\

          \hline
        NGC~628 & 710 & 121 & \textbf{74} & 2485 & 490 & \textbf{202} \\ 
        NGC~3184 & 16781 & 3594 & \textbf{3348} & 10373 & \textbf{2032} & {2217} \\
        NGC3627  &    826  & \textbf{422} & 502 & 4209  & 594 & \textbf{349} \\
        NGC~5055 & 8468 & 1264 & \textbf{257} & 74241 & 16433 & \textbf{6849} \\
        NGC5194   & 13747   & 5370   & \textbf{4316} & 8130 & 6658  & \textbf{6026} \\
        NGC~6946 & 175271 & \textbf{5524} & 6751 & \textbf{20430} & 37300 & 41532 \\  
        \hline
  \end{tabular}

     \end{center}
      
\end{table*}

\subsection{Degeneracies between free parameters} \label{degeneracies} 

The study of the models revealed several degeneracies between the input parameters. Most of the results stated here have been discussed in our previous work (\citealt{2021A&A...645A.111L}).
\begin{itemize}
  \item The exact value of $\delta$ never appears {to be important for the determination of the best-fit model. A constant value of $\delta=5$ as used in
  \citet{2011AJ....141...24V} and \citet{2017A&A...602A..51V} is thus justified.}  
  \item Increasing the value of $\xi$ by a factor of two leads to similar model {$\chi^2$, $Q$, and $\mdot$} than dividing the value of $\alpha_0$ by a factor of two. 
  \item Increasing the value of $\alpha_0$ by a factor of two leads to similar model {$\chi^2$, $Q$, and $\mdot$} than multiplying the stellar velocity dispersion by a factor of two
\end{itemize}
The first degeneracy is illustrated in \tab{bestfit1} and \tab{bestfit2}.
{ In the case of NGC~5055, NGC~3351, NGC~3521, NGC~4736, NGC~7331, NGC~2403, and NGC~4214 although either a division of $\alpha_{0}$ by two or a multiplication of $\xi$ by two is favoured by most best-fit models, the second degeneracy mentioned above becomes evident when inspecting the ten models with the lowest $\chi^2$ for these galaxies. In addition, the best-fit models of NGC~7793 needed a decreased metallicity ($2 \times \alpha_0$) with respect to the leaky box model (Eq.~\ref{molecule}).}
The lower metallicities are consistent with observations (\citealt{2010ApJS..190..233M}, \citealt{2015ApJ...812...39S}) 
and can be understood as a consequence of external accretion of metal-poor gas.

The division of $\alpha_0$ by two is equivalent to a doubling of the metallicity, which is in conflict with existing
metallicity measurements (e.g., \citealt{2010ApJS..190..233M}, \citealt{2019ApJ...887...80K}, \citealt{2020ApJ...893...96B}). 
A higher value of $\xi$ can be justified in the following way:
\citet{1998ApJ...500...95T} have shown by modeling SN explosions
in different environments that the kinetic energy of the remnants is about ten percent of the total SN energy irrespective of the
density and metallicity of the ambient medium. The SN energy
input into the ISM is $E^{\rm kin}_{\rm SN} \sim  10^{50}$~ergs. The integrated number of SNe type II in the Galaxy is taken to 
be $\dot{N}_{\rm SN} \sim 1/60$~yr$^{-1}$ (\citealt{2021NewA...8301498R}). The Galactic star
formation rate is taken to be  $M_* = 1.6$~M$_{\odot}$yr$^{-1}$ (\citealt{2015ApJ...806...96L}). 
With a kinetic to total SN energy fraction of $16$\,\% one obtains  $\xi=  9.2 \times 10^{-8}$~(pc/yr)$^2$,
a factor two higher than the value used by \citet{2003A&A...404...21V} and \citet{2011AJ....141...24V}.
The improved large-scale model presented in Appendix~\ref{app:model}
is equivalent to the model used by \citet{2011AJ....141...24V} provided that $\xi=  9.2 \times 10^{-8}$~(pc/yr)$^2$.

\section{Results \label{sec:results}}

Unlike the model presented in \citet{2017A&A...602A..51V}, the current model is able to produce radial profiles of the infrared fluxes and molecular line emission. In this way we can directly compare the observational radial profiles to our models. 
{Typical uncertainties of the derived $\mdot$ and $Q$ are $0.3$~dex and $0.2$~dex, respectively. This translates into uncertainties of
$2$-$3$~km\,s$^{-1}$ for the turbulent velocity dispersion of the ISM and $0.1$-$0.2$~dex for the CO and HCN conversion factors. 

In this section we focus on the results of NGC~6946 as an example; the results for the other galaxies are available in the Appendix~\ref{appendix}. The best-fit models presented in \tab{bestfit1} are computed using our molecular emission recipe {(Sect.\ref{sec:brightco}) to fix the values of $\delta$, $\xi$, $\alpha_0$, and $\gamma$, and to produce radial profiles of $Q$ and $\mdot$. We then recalculated the molecular line emission of our best-fit models using RADEX instead of our molecular emission recipe to produce the final radial profiles.
Since the use of RADEX is time consuming, we could not use it for the $Q$-$\mdot$ grid calculations.
As a consistency check we used RADEX for the $Q$-$\mdot$ grid calculations for NGC~6946 with the best-fit values of $\delta$, $\xi$,
$\alpha_0$, and $\gamma$ given in Table~\ref{bestfit1}. The resulting RADEX} best-fit model yields a
better fit to the CO(1-0) and CO(2-1) line emission, a less good fit to the SFR in the inner disk, a lower $Q$ parameter ($Q =1$), a 
comparable $\mdot$, and a somewhat lower gas velocity dispersion compared to the model where RADEX was used a posteriori.
Since these changes are within the uncertainties of the model, we are confident that our derived model parameters are meaningful.

\subsection{NGC~6946} \label{ngc6946s}

The best-fit models for NGC~6946 obtained with our molecular emission calculation recipe are presented in \fig{ngc6946}(c) and \fig{ngc6946}(d). The best-fit models using RADEX for the molecular line emission with and without CO(1-0) in the $\chi^2$ minimization are compared in \fig{ngc6946}(a) and \fig{ngc6946}(b), respectively. The total $\chitot$ without $\chicoa$ is $40$, much lower than the one obtained when $\chicoa$ is included in $\chitot$ ($\chitot=2102$ ; see \tab{bestfit1}). The very small errors given for the CO(1-0) EMPIRE data obtained via stacking tend to favor the CO(1-0) profile over the other quantities and lead to larger $\chitot$ values. The $\chihi$ with $\chicoa$ in $\chitot$ is 182 compared to $\chihi=7$ without $\chicoa$ in $\chitot$. The difference is mainly due to the central part of the galaxy ($R \la 2$~kpc) that is not well reproduced by the model. Likewise, $\chisfr=7$ is much lower in the model without $\chicoa$ than in the model with $\chicoa$ in $\chitot$ ($\chisfr=1355$) because of the poor fit within the central 2~kpc. Both H{\sc i} and SFR radial profiles are reproduced within the error bars from 2~kpc to the edge of the disk with or without $\chicoa$ in $\chitot$. The best-fits of the CO(2-1) radial profiles remain fairly comparable with or without $\chicoa$ in $\chitot$ ($\chicob=97$ instead of 25). 

The comparison of the model and observed infrared profiles shows that without CO(1-0) the model is much better at reproducing the radial profiles in the central part of the galaxy. Beyond 2~kpc, the 250 and 500~$\mu$m profiles are also well reproduced without CO(1-0). The profile of the 100$\mu$m emission, on the other hand, is not well reproduced by the model with or without $\chicoa$ in $\chitot$. The radial profiles of the model 250 and 500~$\mu$m profiles are somewhat steeper than the observed profiles
in the model including $\chicoa$ in the $\chi^2$ minimization. The free parameters $\beta$ and GDR, do not modify the slope of the far-IR profiles.

\begin{figure*}[!h]

  \begin{subfigure}{.5\textwidth}
    \centering
    \caption{NGC 6946 best model with $\chi^2_{\rm CO10}$ (RADEX)}
    \includegraphics[width=1.\linewidth]{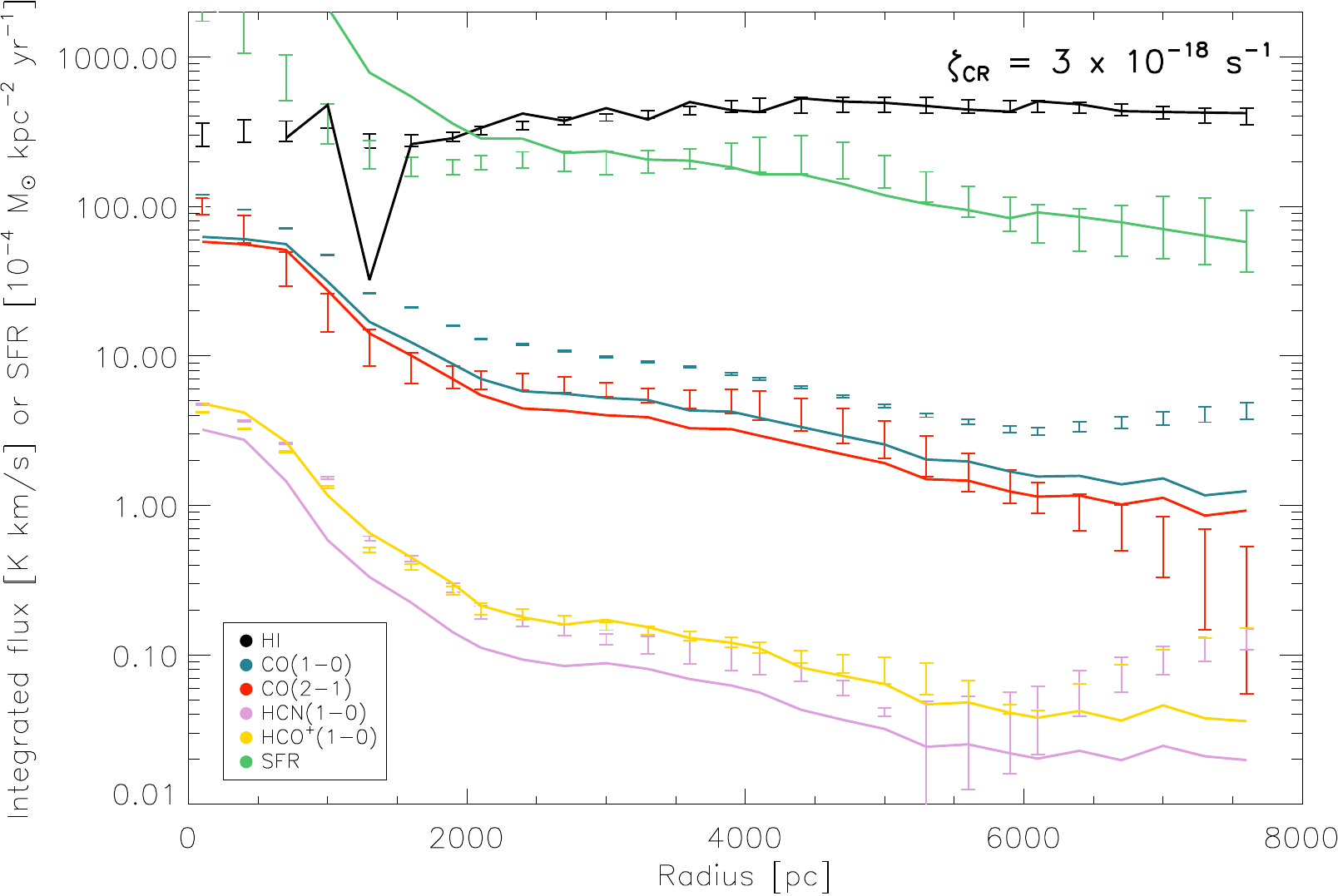}
  \end{subfigure}%
  \begin{subfigure}{.5\textwidth}
    \centering
    \caption{NGC 6946 best model without $\chi^2_{\rm CO10}$ (RADEX)}
    \includegraphics[width=1.\linewidth]{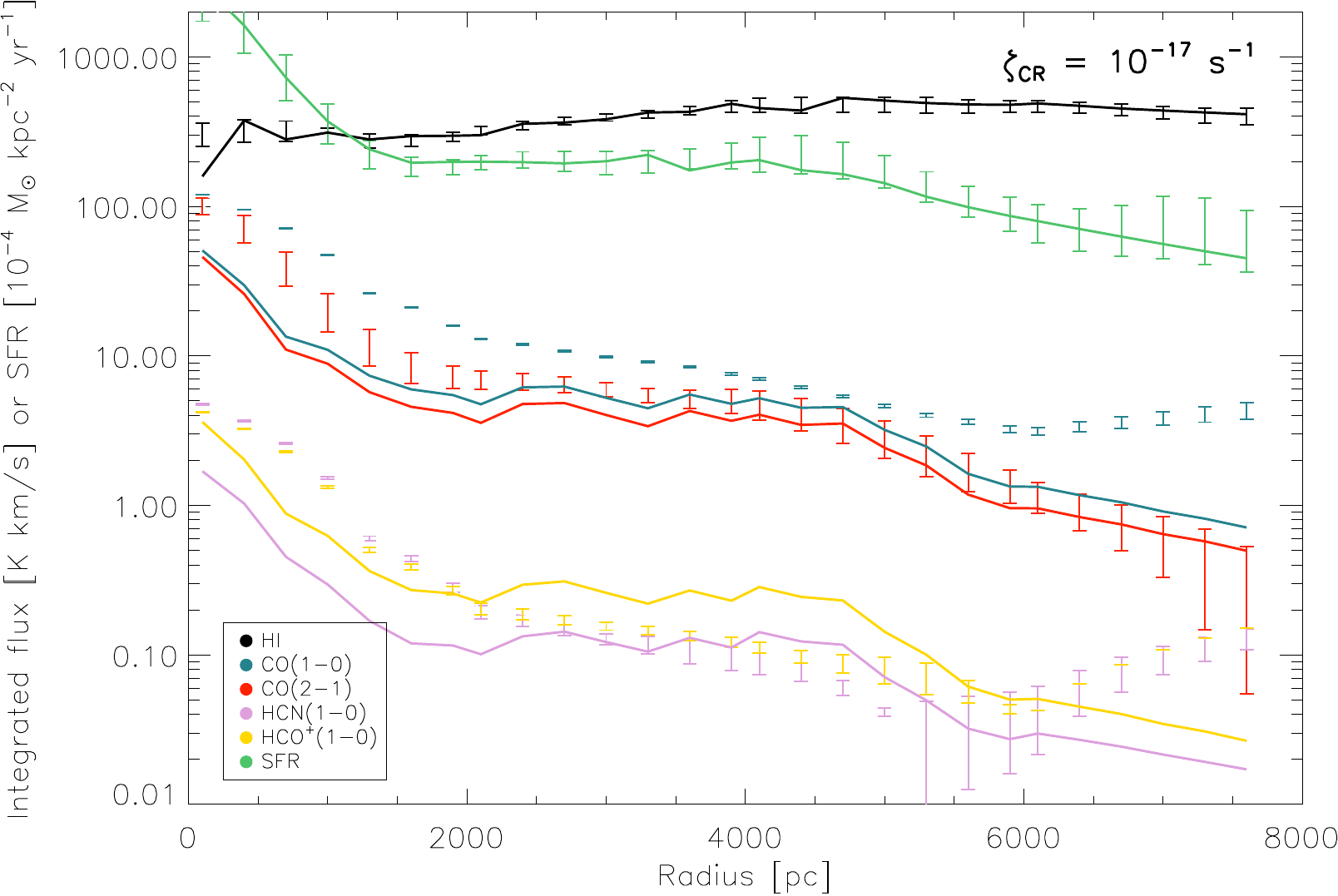}
  \end{subfigure}
  
  \begin{subfigure}{.5\textwidth}
    \centering
    \caption{NGC 6946 best-fit models with $\chi^2_{\rm CO10}$}
    \includegraphics[width=1.\linewidth]{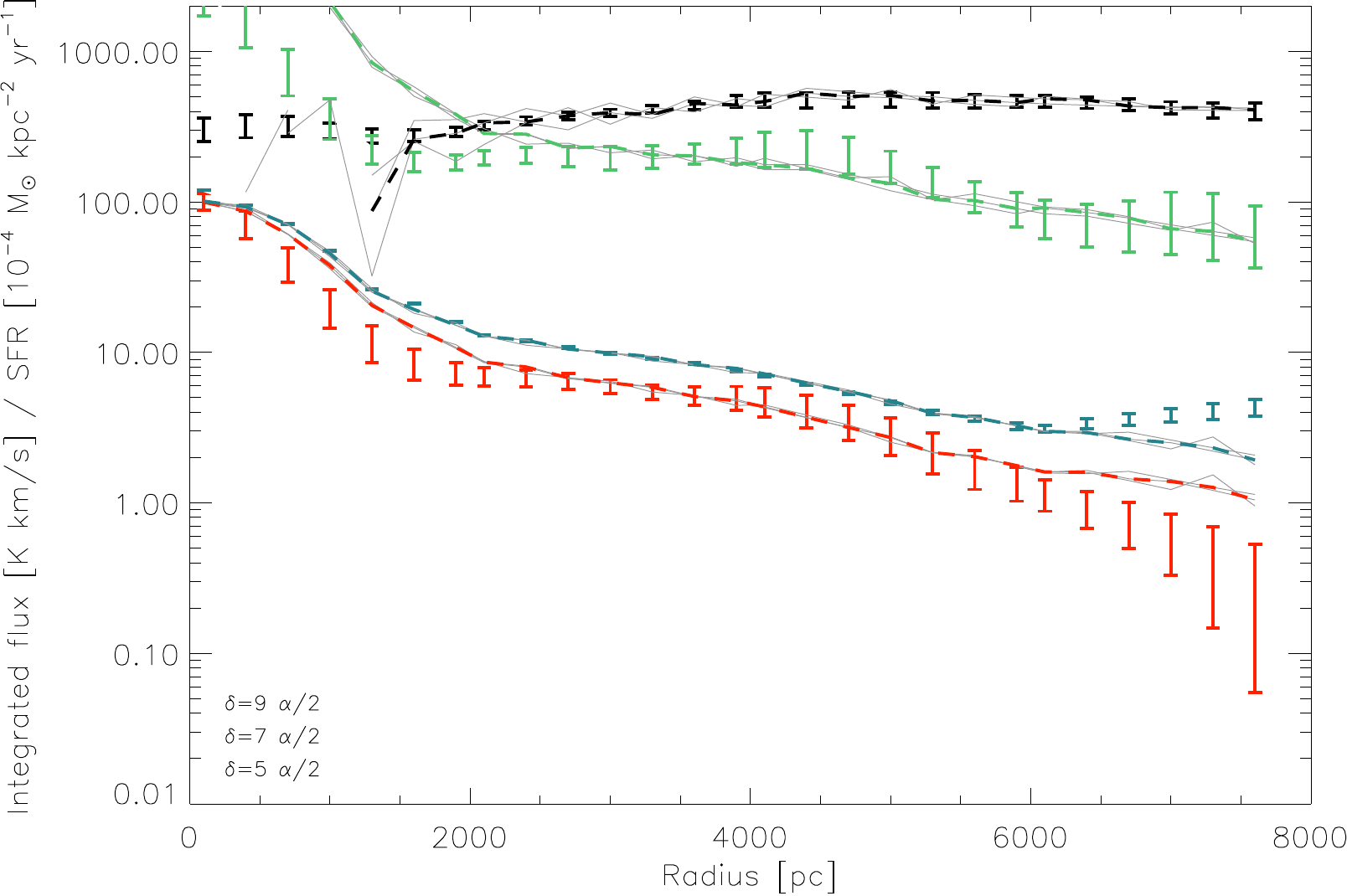}
  \end{subfigure}%
  \begin{subfigure}{.5\textwidth}
    \centering
    \caption{NGC 6946 best-fit models without $\chi^2_{\rm CO10}$}
    \includegraphics[width=1.\linewidth]{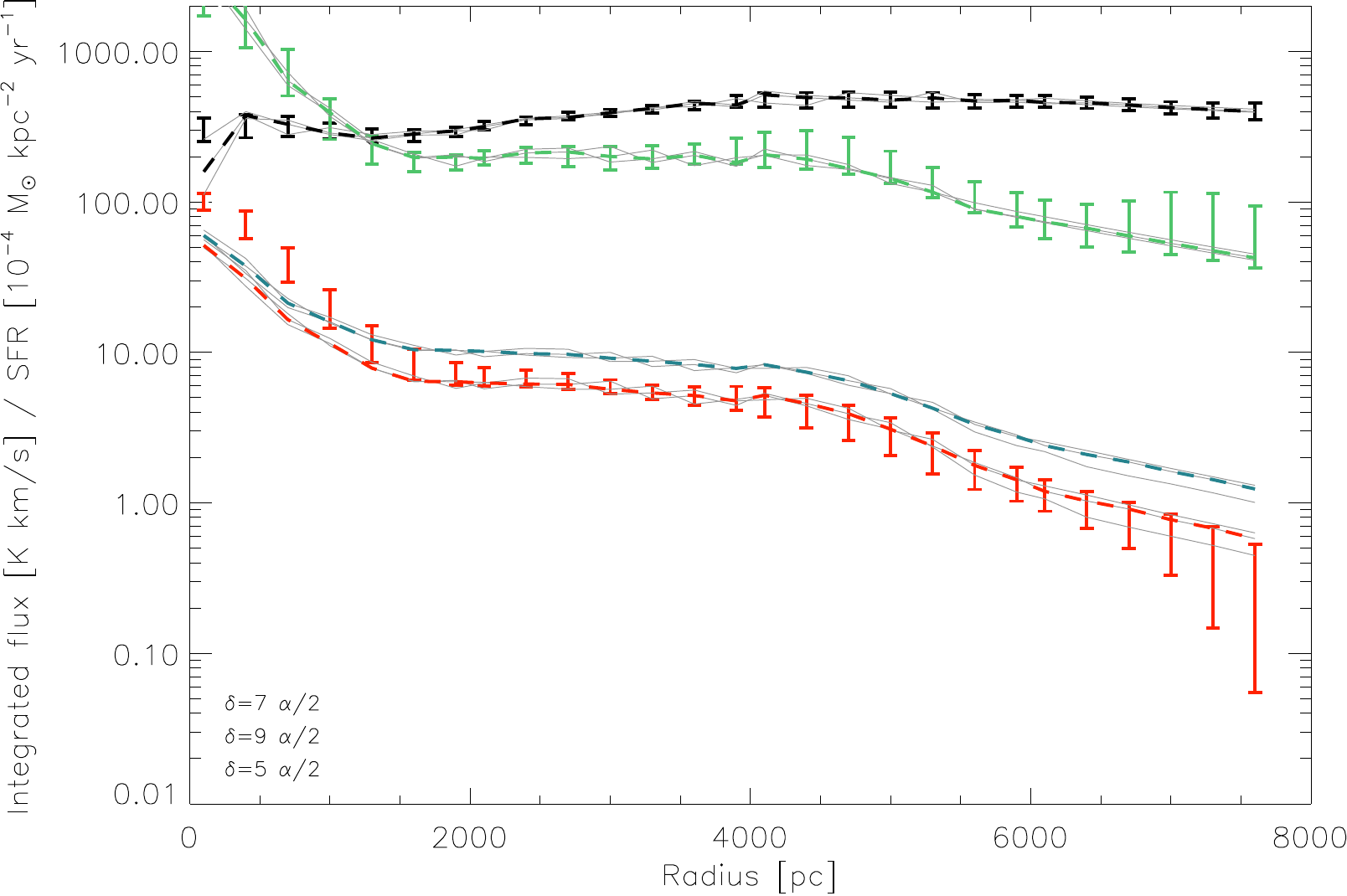}
  \end{subfigure}

  \begin{subfigure}{.5\textwidth}
    \centering
    \caption{NGC 6946 best-fit infrared profiles with $\chi^2_{\rm CO10}$}
    \includegraphics[width=1.\linewidth]{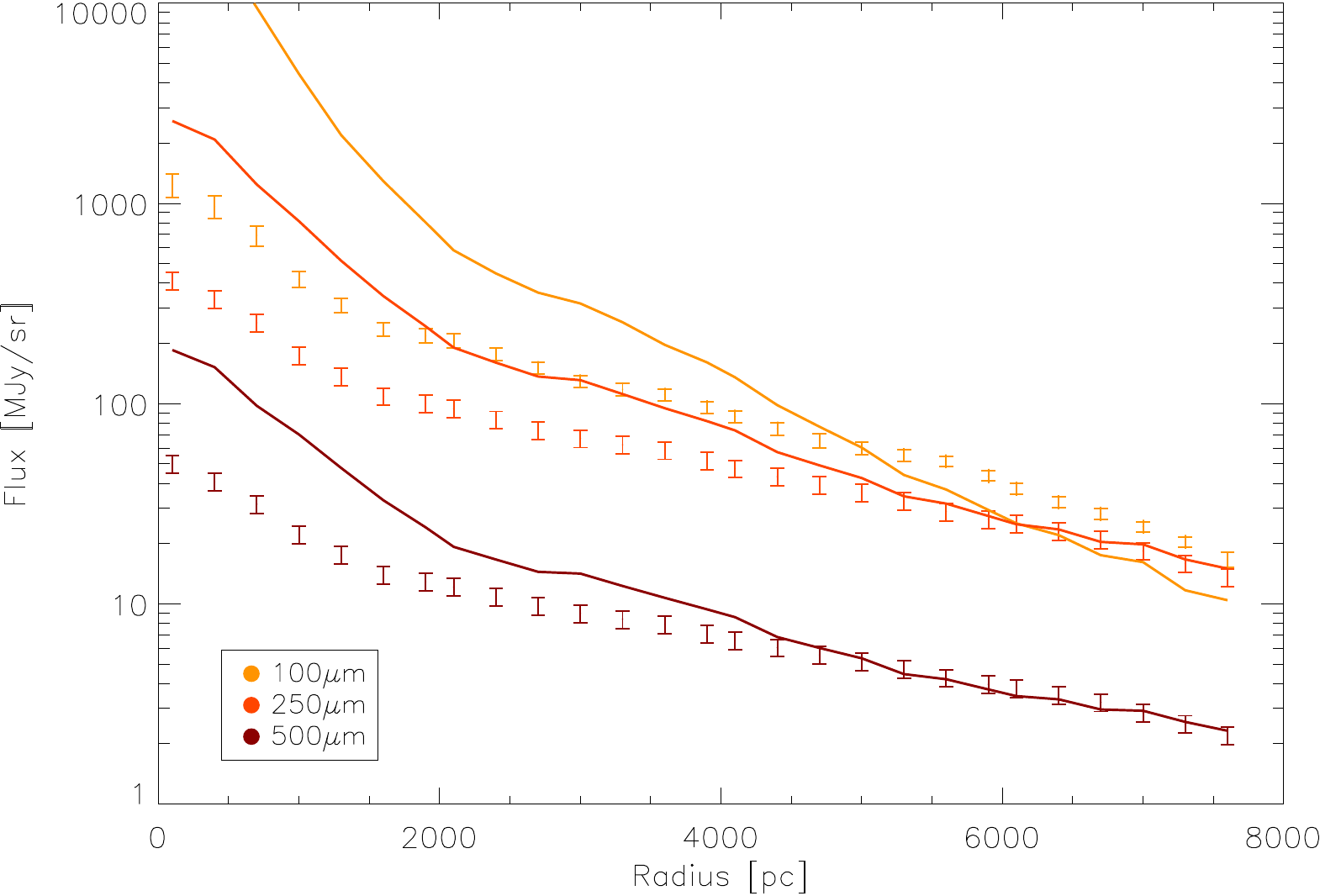}
  \end{subfigure}%
  \begin{subfigure}{.5\textwidth}
    \centering
    \caption{NGC 6946 best-fit infrared profiles without $\chi^2_{\rm CO10}$}
    \includegraphics[width=1.\linewidth]{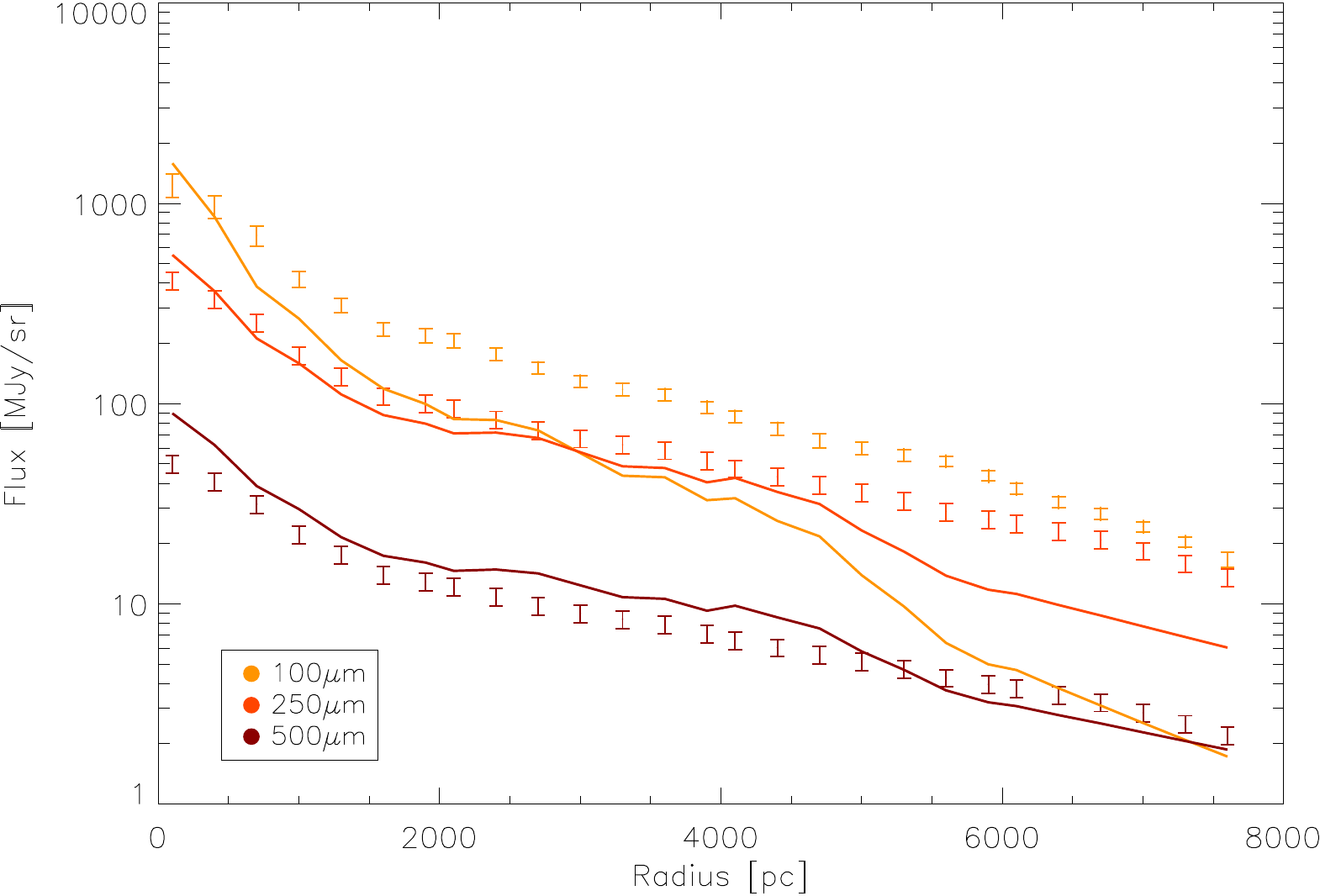}
  \end{subfigure}
  
  \caption{NGC~6946. Errors bars correspond to the observations. Solid lines correspond the median results of the best-fit models, thin grey lines to the the best-fit models. {The values of the secondary parameters for these best-fit models are given in the lower left corner of the middle panels.} (a) Best-fit model including $\chi^2_{\rm CO10}$ using RADEX; (b) best-fit model without $\chi^2_{\rm CO10}$ using RADEX; (c) best-fit models including $\chi^2_{\rm CO10}$; (d) best-fit models without $\chi^2_{\rm CO10}$; (e) infrared profiles of the best-fit model including $\chi^2_{\rm CO10}$; (f) infrared profiles of the best model without $\chi^2_{\rm CO10}$. {The cosmic ray ionization rate $\ion$
is given for each model in the upper panels.}}
  \label{ngc6946}
  \end{figure*}
  
  \begin{figure*}[!h]

    \begin{subfigure}{.48\textwidth}
      \centering
      \caption{NGC 6946 best model with $\chi^2_{\rm CO10}$}
      \includegraphics[width=1.\linewidth]{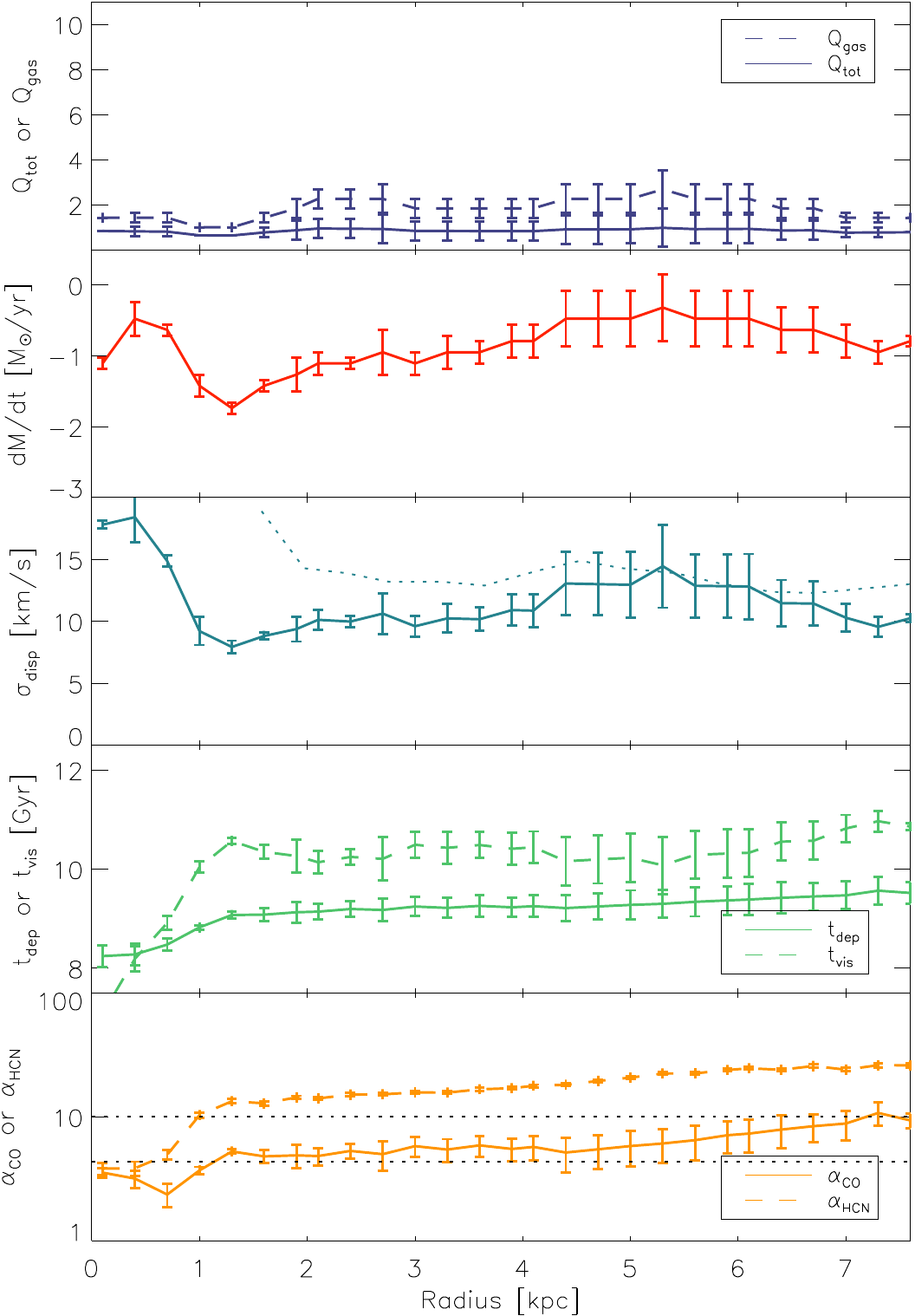}
    \end{subfigure}%
    \begin{subfigure}{.48\textwidth}
      \centering
      \caption{NGC 6946 best model without $\chi^2_{\rm CO10}$}
      \includegraphics[width=1.\linewidth]{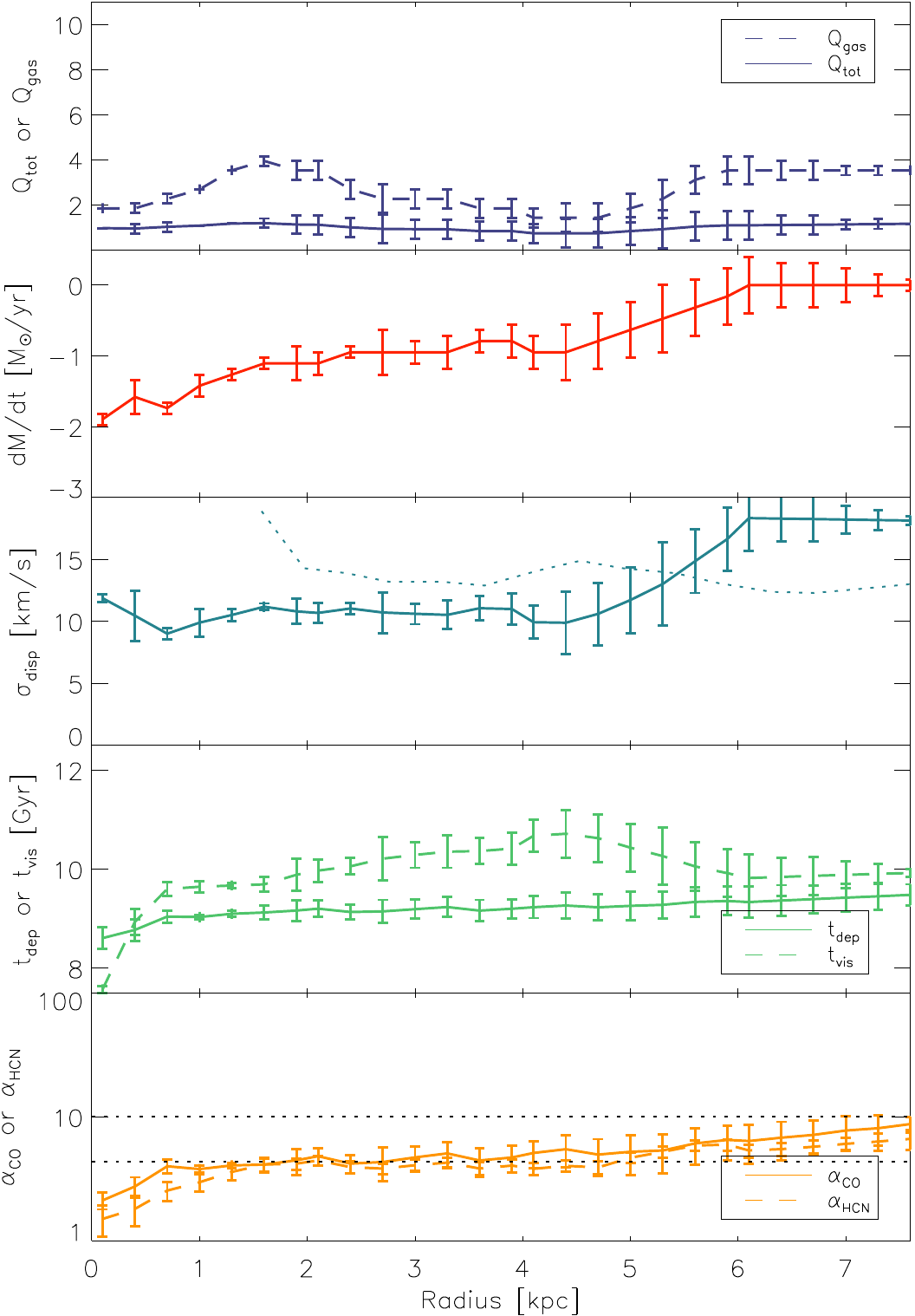} 
    \end{subfigure}

    \caption{NGC~6946. Radial profiles of the main physical quantities with and without $\chico$ in calculations. First panels (dark blue): total (solid) and gaseous (dashed) Toomre $Q$ parameter. Second panels (red): mass accretion rate $\dot{M}$. Third panels (blue): velocity dispersion profiles $\sigma_{\rm disp}$ (solid) with the observed H{\sc i} moment~2 linewidth (\citealt{2013AJ....146..150C}; dotted). Fourth panels (green): depletion ($t_{\rm dep}$; solid) and viscous ($t_{\rm vis}$; dashed) timescales. Fifth panel (yellow): CO-to-H$_2$ ($\alpha_{\rm CO}$; solid) and HCN-to-dense gas ($\alpha_{\rm HCN}$; dashed) conversion factors. Dotted lines correspond to the Galactic value of $\alpha_{\rm CO}^{\rm MW} = 4.36~\rm{M_\odot\,pc^2\,(K\,km\,s^{-1})^{-1}}$ and the Gao \& Solomon value $\alpha_{\rm HCN}^{\rm MW} = 10~\rm{M_\odot\,pc^2\,(K\,km\,s^{-1})^{-1}}$.}
    \label{ngc6946b}
    \end{figure*}

The main improvement provided by the use of the CO(1-0) is the fit of the HCN(1-0) and HCO$^+$(1-0) profiles with a significantly lower $\chi^2_{\rm dense} = \chihcn + \chihco$, $5524$ instead of $20430$ (see \tab{bestcr}). However, $\chihi$ and $\chisfr$ increase by a factor of about 25 and 200, respectively, when $\chicoa$ is included in $\chitot$. The model is therefore not able to reproduce the infrared (and thus the SFR) and the dense molecular line emission simultaneously in the central part of the galaxy.

The physical parameters $Q$ and $\mdot$ associated with the best-fit models are presented in \fig{ngc6946b}(a and b). The Toomre parameter $Q_{\rm gas}$ increases from around unity near the galaxy center to $Q_{\rm gas} = 2-3$ towards the outer disk for both $\chi^2$ calculations. The accretion rate $\mdot$ increases monotonically from 1 to 6~kpc by a factor of $\sim 10$ from the center to the edge of the disk. The combination of these two parameters leads to a {total velocity dispersion $\sigma_{\rm disp}=\sqrt{v_{\rm turb}^2+c_{\rm s}^2}$, which} increases slightly from the canonical value of 10~km\,s$^{-1}$ at 1~kpc to more 13~km\,s$^{-1}$ at 6~kpc, \fig{ngc6946b}(c and d). The model without $\chicoa$ in $\chitot$ shows velocity dispersions higher than 15~km\,s$^{-1}$ at radii larger than 6~kpc.

We compared the model velocity dispersion profiles to the observed profiles based on THINGS and HERACLES data presented by \citet{2013AJ....146..150C}. 
\citet{2015AJ....150...47I} found a constant velocity dispersion around the canonical value of 10 km\,s$^{-1}$ up to R$_{25}$ in NGC~6946.
The model velocity dispersion profile including $\chicoa$ in the $\chi^2$ minimization is consistent with the observed profiles \fig{ngc6946b}(c) and \fig{ngc6946b}(d).
The model without $\chicoa$ in $\chi^2$ significantly overestimates the velocity dispersion for $R > 6$~kpc.
We conclude that this increase of the model velocity dispersion of the order of 8~km\,s$^{-1}$ is not physical. 
This issue is further discussed in \sect{QMV}.

 The profiles of the gas depletion and viscous timescales of the best-fit models are presented in \fig{ngc6946b}(e and f). 
{The viscous timescale is defined as:
\begin{equation}
    t_{\rm vis}(R) = \frac{R^2}{\sqrt{3} \vturb \ldriv}\ ,
\end{equation}
where $\ldriv$ is the turbulent driving length.}
The gas depletion time corresponds to the time that a galaxy takes to locally convert its gas into stars. The timescale of radial gas transport is given by the viscous timescale. By comparing these two times, it is possible to estimate whether or not a galaxy can maintain its star formation via radial gas transport without external accretion. For the best-fit models with and without $\chicoa$ in the $\chitot$ calculations the gas depletion time is almost constant beyond 2~kpc with $\taudep$~$\sim~1$-$2$~Gyr, which is consistent with the average value for spiral galaxies (\citealt{2008AJ....136.2846B}, \citealt{2008AJ....136.2782L}, \citealt{2021MNRAS.501.4777E}). The viscous timescale is much longer than the gas depletion time for all radii. Therefore there must be external accretion onto the galactic disk to maintain the observed star-formation rate for more than a few Gyr. 

We computed the CO and HCN conversion factor profiles of NGC~6946, \fig{ngc6946b}(g and h). For both $\chi^2$ calculations, the CO-to-H$_2$ conversion factors increase from the galaxy center to the edge of the CO emission distribution, with a mean $\aco$ close to the Galactic value of $\acomw$ = 4.36~$\aunit$ (\citealt{2013ARA&A..51..207B}). 

{The HCN-to-dense gas conversion factor, defined as the ratio between the flux I$_{\rm HCN}$ and the molecular gas formed in clouds with density over $n=3 \times 10^{4}$~cm$^{-3}$,  of the model with $\chicoa$ in $\chitot$ is about a factor of two higher than the value of $\alpha_{\rm HCN} = 10~ \aunit$ (\citealt{2004ApJ...606..271G}) and consistent with the value found by \citet{2010ApJS..188..313W}.
Since the model underpredicts the observed HCN emission by a factor of two, the model HCN conversion factor is
probably overestimated by the same factor.
On the other hand, the HCN-to-dense gas conversion factor of the model without $\chicoa$ in $\chitot$ is about a factor of two lower than 
the value of \citet{2004ApJ...606..271G}.
The difference is mainly caused by higher HCN abundances at a cosmic ray ionization rate of $\ion=10^{-17}$~s$^{-1}$
compared to the HCN abundances at a three times lower $\ion$. We estimate the HCN conversion factor to be
$\alpha_{\rm HCN} = 5$-$10~ \aunit$.
In both models the HCN-to-dense gas conversion factor decreases within the inner kpc.}

\subsection{The galaxy sample} 

In this section we review the results obtained for the other galaxies of the sample. The $\chicoa$ and $\chicob$
are those calculated with excitation temperatures obtained from  the method outlined in Sect.~\ref{sec:brightco}.
The corresponding figures, which show the integrated line emission based on the results of RADEX, are available in Appendix~\ref{appendix}, 
ordered in the same way as in \tab{bestfit1} and \tab{bestfit2}.

{We first discuss the galaxies for which EMPIRE data are available (\tab{bestfit1}). { Whereas most of these galaxies have
a $\chitot$ including $\chicoa$ between $200$ and $400$, the $\chitot$ of NGC~3627 and NGC~6946 are $1053$ and $2102$, respectively.}
As discussed in Sect.~\ref{ngc6946s}, the high $\chitot$ of NGC~6946 is caused by an overestimation of the central star formation
rate. In NGC~3627, the star formation and H{\sc i} radial profiles could not be fitted in a satisfactory way. The model without 
$\chicoa$ reproduces the available observations significantly better than the model with 
$\chicoa$ in the $\chitot$ calculations. We thus adopted the model with $\chicoa$ in $\chitot$ for all EMPIRE galaxies except NGC~3627.

For all EMPIRE galaxies the model IR profiles are broadly consistent with observations. In general, the model
IR profiles are steeper than the observed IR profiles. This effect becomes stronger at smaller wavelengths.
The preferred CR ionization rate is $\ion = 10^{-18}$~s$^{-1}$.

For all EMPIRE galaxies except NGC~6946 the Toomre parameter $Q_{\rm gas}$ decreases with radius. 
The highest central $Q_{\rm gas}$ are found in NGC~628 and NGC~5055 ($Q \sim 8$). 
In NGC~3627 the Toomre parameter decreases from $Q_{\rm gas}=4$-$6$ in the galaxy center to $Q_{\rm gas}=1.5$-$2$ at $R=5$~kpc.
The NGC~5194 model yields a Toomre parameter, which decreases from $Q_{\rm gas}=2$ in the center to about unity at $R=1.5$~kpc

The radial profiles of the mass accretion rate generally increase with radius. In NGC~628 and NGC~3184
the mass accretion rate increases in the inner $2$~kpc and stays approximately constant at larger radii.
The mass accretion rate is about constant $\mdot \sim 0.1$~M$_\odot$~yr$^{-1}$ in NGC 3627 and NGC~5055.
In NGC~5194 the mass accretion rate is about $\mdot=10^{-2}$~M$_\odot$~yr$^{-1}$ in the inner disk and increases to almost 
$10^{-1}$~M$_\odot$~yr$^{-1}$ at $R=5.5$~kpc 

The resulting velocity dispersions are about 10~km\,s$^{-1}$ in NGC~628, NGC~3184, NGC~3627, and NGC~5055.
They are consistent with observations when available.
In NGC~5194, $\sigma_{\rm disp} \sim 8$~km\,s$^{-1}$, significantly lower than the observed velocity dispersion of $\sim 15$~km\,s$^{-1}$. { The high observed velocity dispersion could be due to non-circular motions induced by the gravitational interaction with NGC~5195.}
The strong increase of the model velocity dispersion at $R=7$~kpc is not observed.

The CO conversion factors are fairly constant or increase slightly with radius. Whereas NGC~628 and NGC~3184 show a CO 
conversion factor about twice as high as the Galactic value, those of NGC~3627 and NGC~5055 are about Galactic.
In NGC~5194 the CO conversion factor increases slightly from the Galactic value at $R=1$~kpc to $1.5$ times this value at $R=4$~kpc.
Like CO conversion factors, the HCN conversion factors are fairly constant or increase slightly with radius.
The HCN conversion factor is close to the Gao \& Solomon value in NGC~5055 and somewhat lower than the Gao \& Solomon value in 
NGC~628, NGC~3184, and NGC~5194. The lowest HCN conversion factor is found in NGC~3627.
All conversion factors, except the CO conversion factor in the outer disk of NGC~3184, are located within the observed ranges 
($3 \leq \alpha_{\rm HCN} \leq 30~\rm{M_{\odot}\,pc^2\,(K\,km\,s^{-1})^{-1}}$, \citealt{2018MNRAS.479.1702O};
$2 \leq \alpha_{\rm HCN} \leq 10~\rm{M_{\odot}\,pc^2\,(K\,km\,s^{-1})^{-1}}$, \citealt{2013ARA&A..51..207B}). 
In four out of six EMPIRE galaxies the CO conversion factor exceeds the HCN conversion factor.

Whereas the radial profiles of the gas depletion timescale are relatively flat in all EMPIRE galaxies, that of the
viscous timescales increase with radius. In all EMPIRE galaxies the viscous timescale exceeds the gas depletion timescale
for $R > 1$~kpc.}

For the rest of the massive galaxies ($M_* > 10^{10}$~M$_{\odot}$), where no EMPIRE data are available, the model is able to fit the observational profiles in a satisfactory way. The $\chitot$ of NGC~2841, NGC~3198 and NGC~4736 are the lowest of all galaxies, with values below 30. With a value of $270$, NGC~3521 shows the highest $\chitot$ (including only $\chicob$) of the whole galaxy sample. This is mainly due to an underestimation of the model CO(2-1) integrated line emission. Nevertheless, we consider that this is still an acceptable fit.
The resulting radial profiles of $Q_{\rm gas}$ and $\mdot$ are similar to those described above.

The model velocity dispersion profiles are approximately flat except for those of NGC~3521 and NGC~7331 where the velocity dispersion 
suddenly increases by $\sim 8$~km\,s$^{-1}$ in the outer disk. We think that this is a limitation of the model which needs to significantly increase the value of $\mdot$ in the regions where the SFR changes from a molecular gas dominated regime to an atomic gas dominated one (i.e. where the CO flux becomes negligible compared to the H{\sc i} flux) to fit the observational profiles satisfactorily (see Sect.\ref{QMV}).
The viscous timescales are significantly higher than the gas depletion times in all galaxies.
The CO conversion factors are approximately constant around one half and twice the Galactic value. Only NGC~3351 shows a 
CO conversion factor that exceeds twice the Galactic value.

{For most of the} low-mass galaxies ($M_* < 10^{10}$~M$_{\odot}$) where no EMPIRE data are available, the $\chitot$ are considerably higher than those found for the high-mass galaxies, with values ranging from about 3000 for NGC~2976 to more than 7000 for NGC~925. For each of the  low-mass galaxies, the fit of the CO(2-1) and SFR profiles becomes worse in the outer galactic disk. {Whereas the molecular depletion times of the inner disks are similar to that commonly observed in star-forming spiral galaxies 
(\citealt{2008AJ....136.2846B}, \citealt{2008AJ....136.2782L}), they are significantly higher in the outer disks of the low-mass galaxies, {except for NGC~7793}. 
The model cannot accommodate these high molecular depletion times, which leads to the high $\chitot$.
We can only speculate that gas compression due to external accretion might be the cause of such a large amount of molecular gas in the outer disks of low-mass galaxies, which forms stars with a significantly lower efficiency than in the inner disk.} 
A high Toomre parameter $Q_{\rm gas} \sim 8$ is
found in the inner disks of NGC~2403, NGC~2967, NGC~4214, {and NGC~7793} decreasing to $Q_{\rm gas} \sim 1.5$-$2$ in the outer disks.
The observed IR profiles of all low-mass galaxies are well reproduced by the model. In contrast to the high-mass galaxies,
the radial profiles of the model gas velocity dispersion slightly decrease with radius. Where measured, the model profiles are
consistent with the observed profiles. The viscous timescales exceed the gas depletion time in  NGC~2403, NGC~2967, NGC~4214, {and NGC~7793}
beyond $\rm{R}~\sim~2~\rm{R}_{\rm d}$, {the radius within which half of the galaxy’s luminosity is contained.} 
The CO conversion factors all increase with radius. In the outer disks they are significantly higher than the
Galactic value.

\subsection{General results}

In this section we summarize and generalize the results obtained for all galaxies. 

\subsubsection{Toomre parameters $Q$, accretion rate $\mdot$ and turbulent velocity dispersion $\vturb$ \label{QMV}}

{The Toomre $Q$ parameter can be interpreted as a stability criterion, where disks with $Q < 1$
are subject to fragmentation (see Eq.\ref{toomre}). It can also be interpreted as a measure of the
gas mass with respect to the maximum stable gas mass ($Q_{\rm gas}=1$) for a given angular velocity and gas velocity dispersion.}
The radial profiles of the Toomre parameters $Q_{\rm gas}$ generally decrease with radius {(\fig{QM})}.
We found $Q_{\rm gas} > 5$ in the central parts of {six} galactic disks. In the outer parts of the disks $Q_{\rm gas}$
rarely exceeds $Q_{\rm gas}=3$. An exception is NGC~2841, where $Q_{\rm gas} > 4$ in the entire gas disk.
The total (gas and stars) Toomre parameter $Q_{\rm tot}$ exceeds unity in the inner disk of a significant number of galaxies. 
$Q_{\rm tot}$ is close to unity in the outer disks, except for NGC~628 and NGC~2841. 

In all massive galaxies except NGC~3351, the mass accretion rate increases with radius from $\mdot \sim 10^{-2}$~M$_{\odot}$yr$^{-1}$ in the center to $\mdot \sim 10^{-1}$~M$_{\odot}$yr$^{-1}$ at the edge of the CO distribution {(\fig{QM})}. Three low-mass galaxies show the opposite trend. In addition, NGC~925 {and NGC~7793 have roughly} constant mass accretion rates. Our mass accretion rates ($\sim 0.1$~M$_{\odot}$yr$^{-1}$) are significantly lower than those derived by
\citet{2016MNRAS.457.2642S}, which can reach up to $\sim 1$~M$_{\odot}$yr$^{-1}$. {They are consistent with the net
mass accretion rates derived by \citet{2021}: in their sample of 54 local spiral galaxies all galactic disks show some degree of radial 
gas flows, with radial velocities typically of a few km\,s$^{-1}$, but these flows do not seem to have a preferential direction. As a consequence, the
average radial velocity and mass flow rate across the sample at a given radius are nearly constant and close to zero.}
 \begin{figure}[!ht]
  \center
  \begin{subfigure}{0.48\textwidth}
    \centering
    \caption{Massive galaxies with EMPIRE data.}
    \includegraphics[width=1.\linewidth]{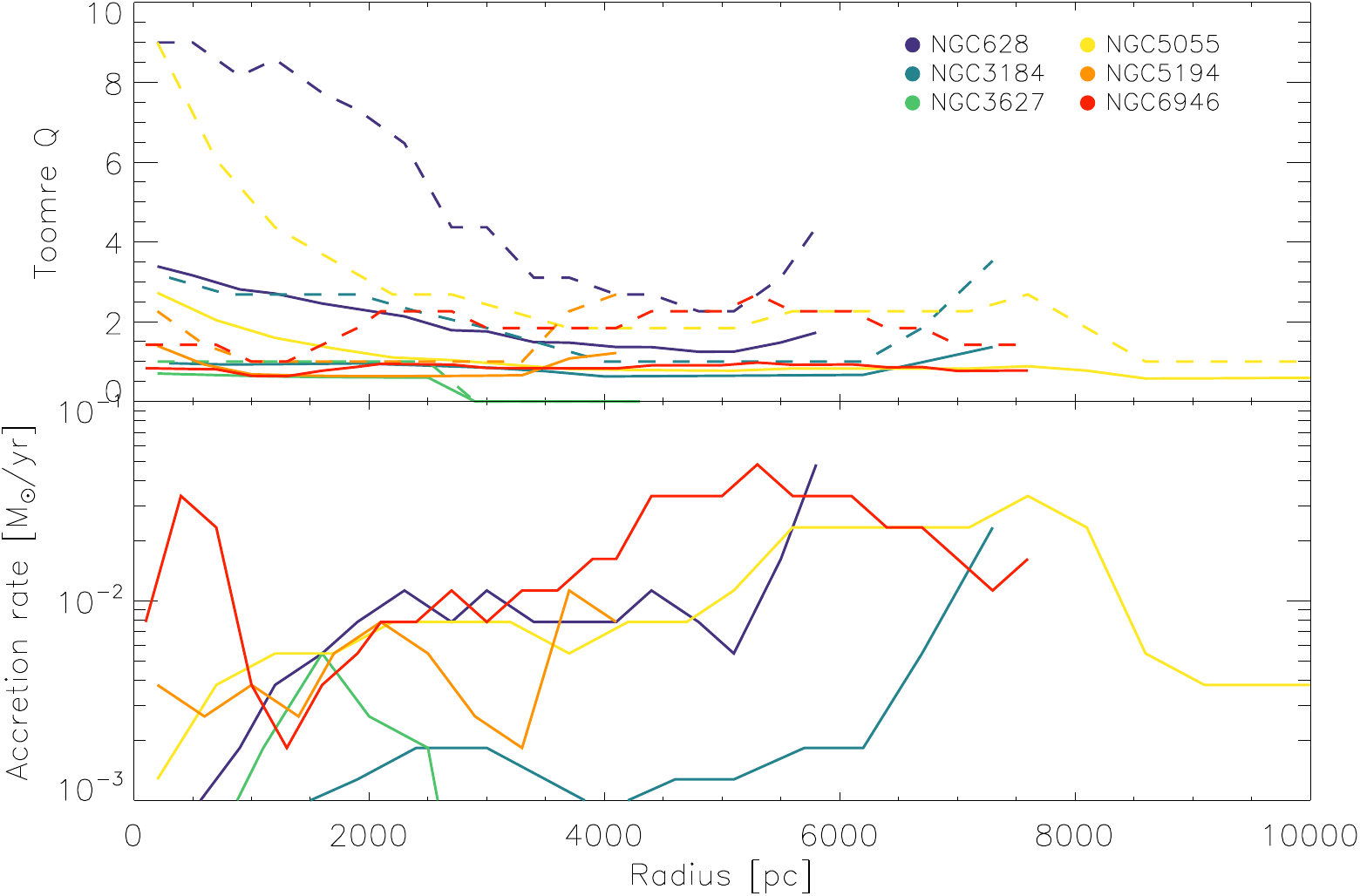}
    
  \end{subfigure}%
  
  \begin{subfigure}{0.48\textwidth}
    \centering
    \caption{Massive galaxies without EMPIRE data.}
    \includegraphics[width=1.\linewidth]{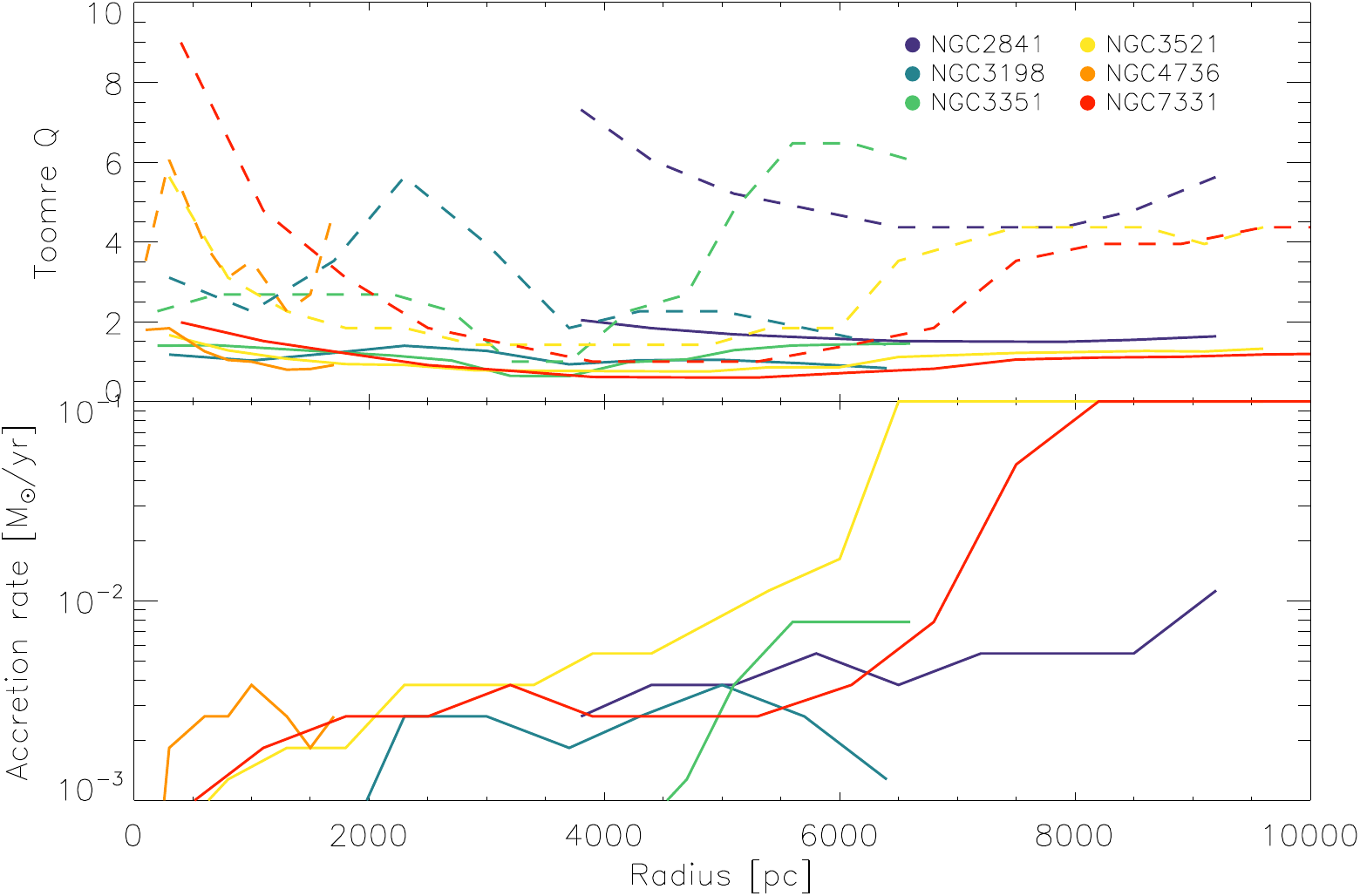}
    
  \end{subfigure}%
  
  \begin{subfigure}{0.48\textwidth}
    \centering
    \caption{Low-mass galaxies without EMPIRE data.}
    \includegraphics[width=1.\linewidth]{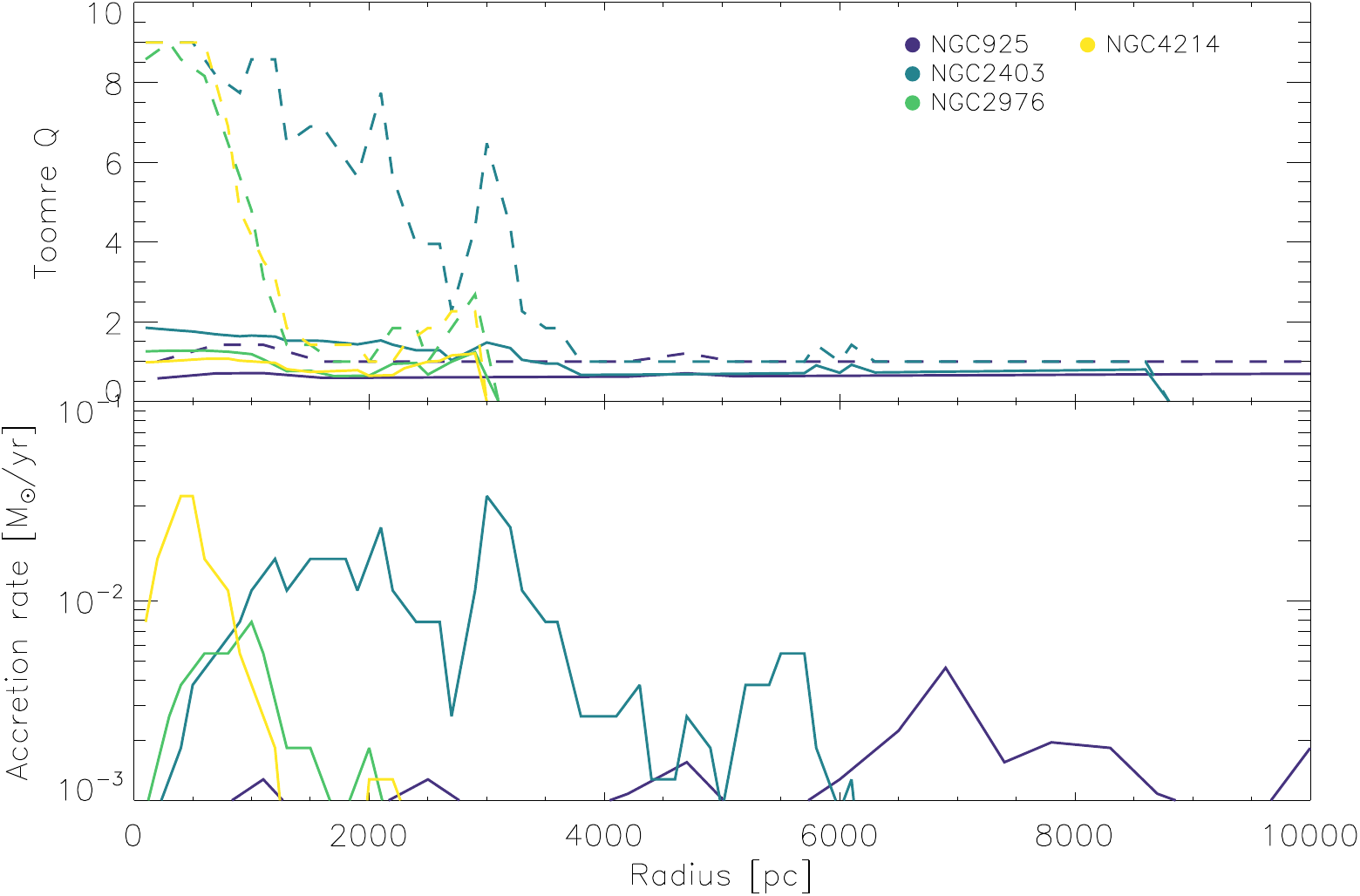}
   
  \end{subfigure}
  
  \caption{Toomre parameter $Q$ (solid: $Q_{\rm tot}$; dashed: $Q_{\rm gas}$) and mass accretion rate $\mdot$ of the high-mass galaxies with EMPIRE data (upper panel), high-mass galaxies without EMPIRE data (middle panel),
and low-mass galaxies (lower panel; $M_* < 10^{10}$~M$_{\odot}$).}
  \label{QM}
  \end{figure}

The Toomre parameter $Q$ and the accretion rate $\mdot$ are partially degenerate: an increase in either parameter leads to an increase in the turbulent velocity dispersion of the gas. In the massive galaxies, an increase of the accretion rate is accompanied by a decrease of the Toomre parameter, leading to a relatively constant velocity dispersion around the canonical value of $\vturb \sim 10$~km\,s$^{-1}$  (\fig{vdisp}). 
{The velocity dispersions of the less massive galaxies decrease with radius.
We need to take into account the finite lifetime of molecular clouds ($f_{\rm mol}^{\rm life}$) for the 
molecular gas fraction (Eq.~\ref{eq:lifetime}). 
Models without the finite lifetime systematically lead to radially increasing profiles of the gas velocity dispersion, which is contrary to observations.}
The slight increase in the velocity dispersion profiles close to the edge of the CO distribution can be explained by the difficulty of the model to fit regions with very low molecular gas surface densities {in the outer galactic disks} where the gas is dominated by the atomic phase. To fit the H{\sc i} and SFR observational profiles, the accretion rate has to be significantly increased within the model, resulting in an increase of the velocity dispersion. However, this result is contrary to observations (\citealt{2013AJ....146..150C}, \citealt{2015AJ....150...47I}). We observe the strongest discrepancy between our velocity dispersion profiles and the observed profiles for NGC~3521 and NGC~7331. In these galaxies, the velocity dispersion increases sharply at $R \sim 6$-$7$~kpc.

This behavior is caused by a change in the regime of energy injection. Whereas in the inner disk the energy injection by SNe dominates, the energy injection through accretion (\eq{sfr3}) is comparable to that of SNe in the outer disks of NGC~3521, and NGC~7331. Indeed, model calculations without the additional term of energy injection due to accretion show no sudden increase of the gas velocity dispersion in the outer disk (dashed lines in the middle panel of Fig.~\ref{vdisp}).
The sudden increase of the velocity dispersion can also be avoided by a decrease of the gravitational potential of the galactic disk. This can be achieved by a local increase of the stellar velocity dispersion by a factor of two compared to our best fit models. Since NGC~7331 already needs an increase of the stellar velocity dispersion to fit the available data, a further increase might not be realistic. Alternatively, it is expected that an increase of the metallicity in the outer disk also leads to a decrease of the gas velocity dispersion. However, this would flatten the metallicity gradient of these galaxies, which is not consistent with the observed metallicity gradient (\citealt{2010ApJS..190..233M}). {As a third possibility the star-formation timescale might be increased in the outer disk \citep{2011AJ....141...24V}. 
These authors replaced the local free-fall time by the molecule formation time as the relevant timescale for star formation in the outer galactic disk.}
All these modifications increase the molecular gas depletion time.

\begin{figure}[!ht]
  \center
  \begin{subfigure}{0.48\textwidth}
    \centering
    \caption{Massive galaxies with EMPIRE data.}
    \includegraphics[width=1.\linewidth]{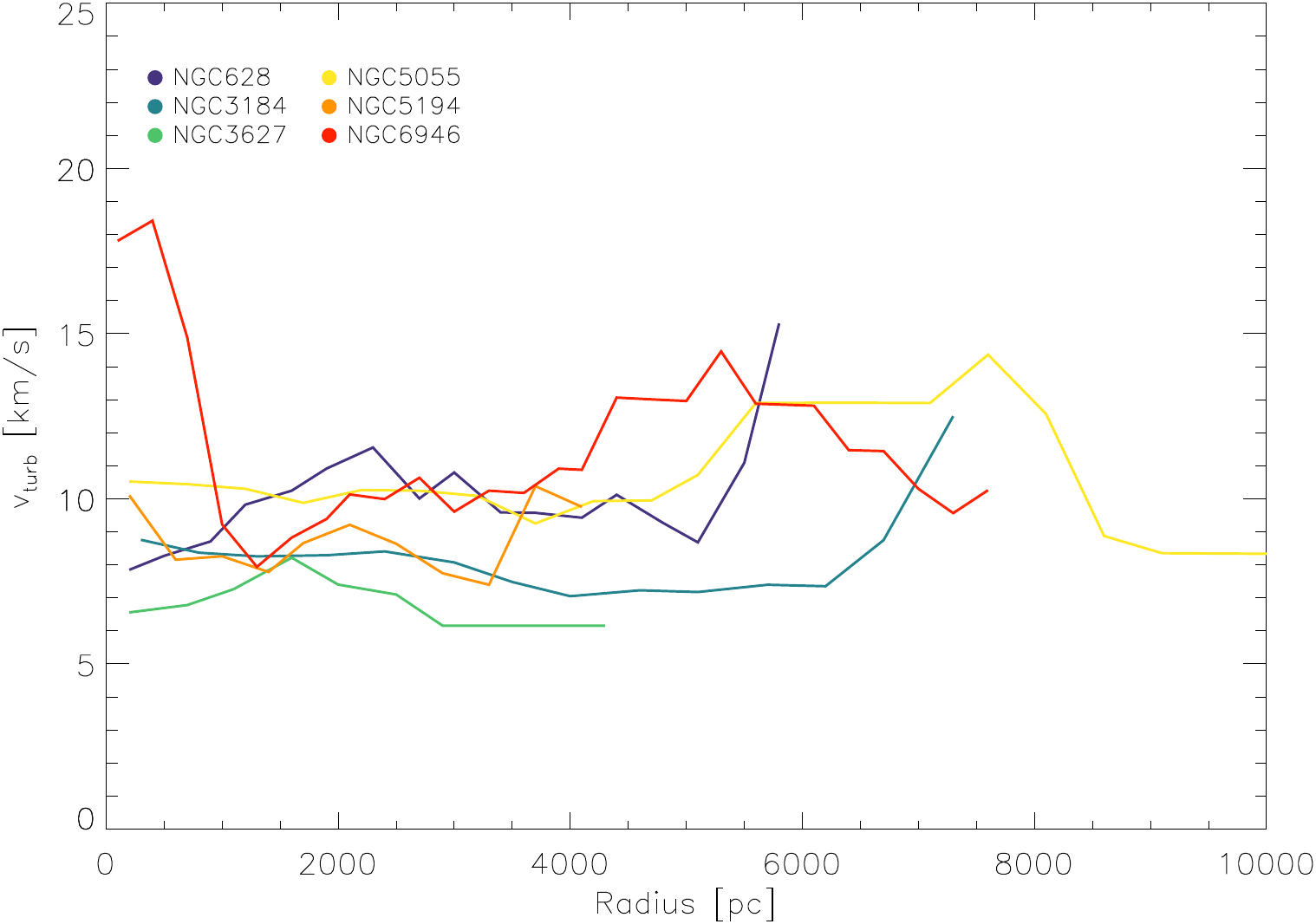}
  \end{subfigure}%
  
  \begin{subfigure}{0.48\textwidth}
    \centering
    \caption{Massive galaxies without EMPIRE data.}
    \includegraphics[width=1.\linewidth]{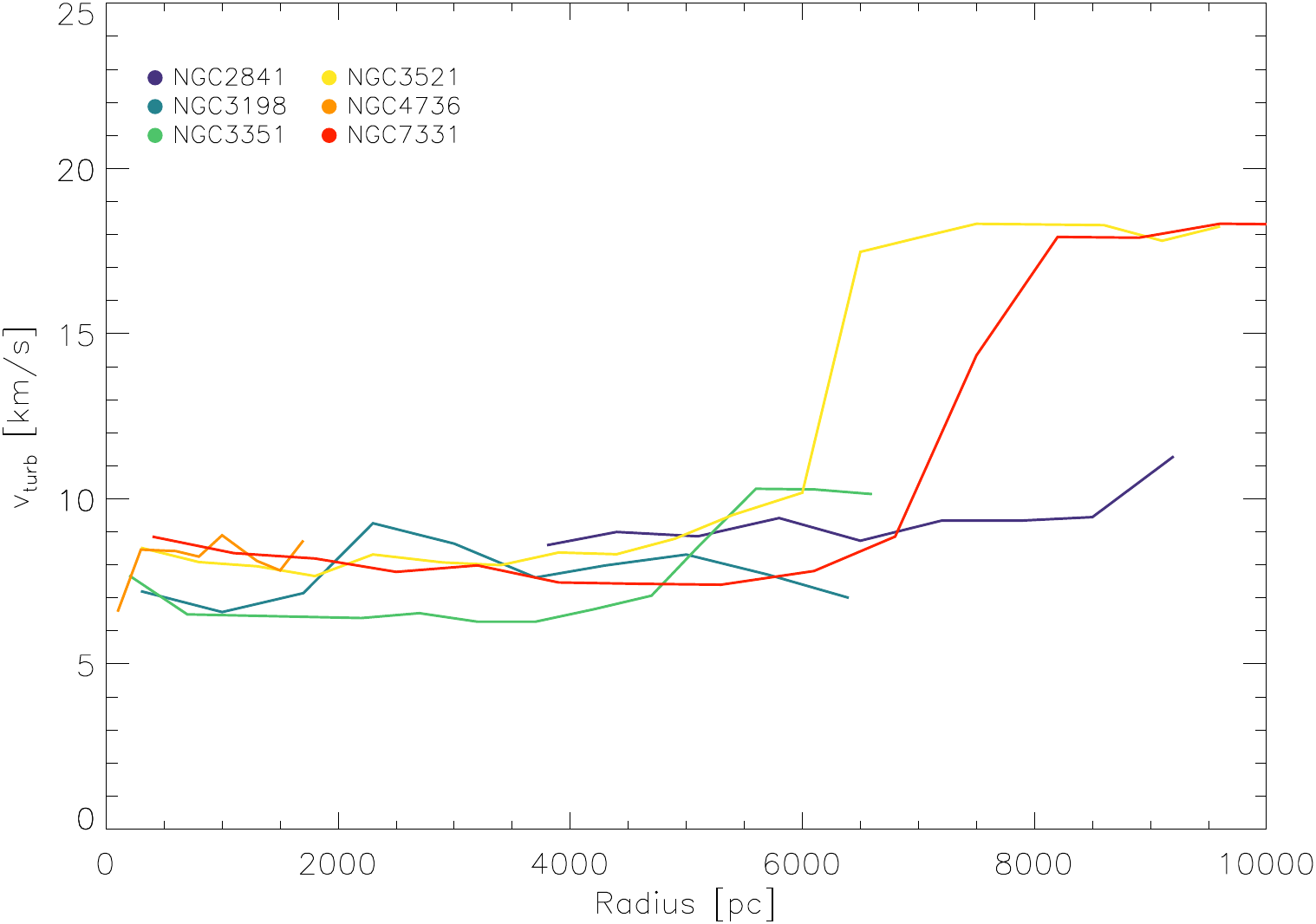}
  \end{subfigure}%
  
  \begin{subfigure}{0.48\textwidth}
    \centering
    \caption{Low-mass galaxies without EMPIRE data.}
    \includegraphics[width=1.\linewidth]{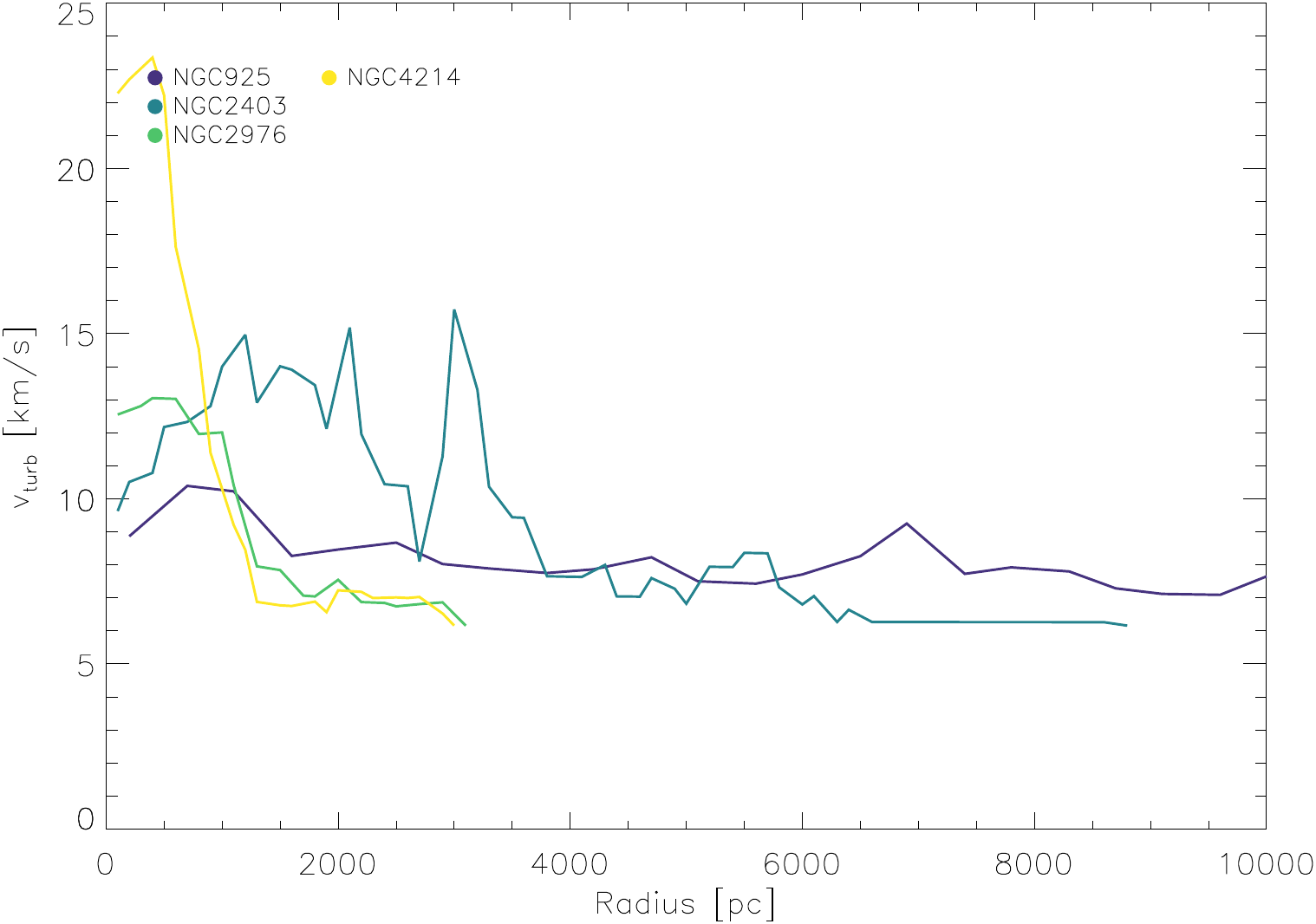}
  \end{subfigure}
  
  \caption{Gas velocity dispersion profiles $\sigma_{\rm disp}=\sqrt{v_{\rm turb}^2+c_{\rm s}^2}$ of the high-mass galaxies with EMPIRE data (upper panel), high-mass galaxies without EMPIRE data (middle panel),
and low-mass galaxies (lower panel; $M_* < 10^{10}$~M$_{\odot}$). The dashed lines for NGC~3521 and NGC~7331 models without the term for turbulent energy injection due to the gain of potential energy (\eq{sfr3}).}
  \label{vdisp}
  \end{figure}

We compared the profiles of the Toomre parameter $Q$ and the mass accretion rate $\mdot$ of the model with the values presented by \citet{2011AJ....141...24V} where the gas velocity dispersion was included in the $\chi^2$ calculations and 
the radial profile of $Q$ was given as input and $\mdot$ was constant for the entire disk. In addition, \citet{2011AJ....141...24V} assumed
a Galactic CO conversion factor. The mass accretion rate was determined by the total star-formation rate of the galactic disks. 
In particular, our model of NGC~5194 led to a significantly better fit to the available observations than the \citet{2011AJ....141...24V} model. 

The models of 6 out of {17} galaxies have more than three times lower mass accretion rates than those of the \citet{2011AJ....141...24V}
model for the following reasons: in NGC~5055, NGC~4736, and NGC~7331 \citet{2011AJ....141...24V} used the observed H{\sc i} velocity dispersions of 
(\citealt{2009AJ....137.4424T}), which are more  than $5$~km\,s$^{-1}$ higher than those obtained by our model. Given that the observed velocity dispersions are upper
limits because of the beam smearing of non-circular motions, we are confident that our model yields more realistic mass accretion
rates for these galaxies. NGC~3351 shows a particularly low mass accretion rate. Whereas our model preferred $Q_{\rm gas} \sim 2.5$,
\citet{2011AJ....141...24V} set $Q=8$. Again, our gas velocity dispersion is significantly smaller than that of \citet{2011AJ....141...24V}. The same observation is true for NGC~925 and NGC~2976. We conclude that $Q_{\rm gas}$, $\mdot$, and thus $\sigma_{\rm disp}$ might have been overestimated by
\citet{2011AJ....141...24V}.

\subsubsection{Ionization rate $\ion$ \label{ioniz}}

Without any source of ionization, the network of reactions responsible for the great richness of the interstellar medium cannot be initiated. The dominant mechanism responsible for the ionization of molecules in dense clouds is cosmic ray ionization. The cosmic ray ionization rate $\ion$ was estimated by many studies, resulting in a range from more than $\rm 10^{-16}~s^{-1}$ for diffuse regions to a few $10^{-18}$~s$^{-1}$ for the densest gas clouds (e.g. \citealt{1968ApJ...152..971S}, \citealt{1986ApJS...62..109V}, \citealt{2007ApJ...671.1736I}). \citet{2012ApJ...745...91I} investigated the value of $\ion$ in diffuse regions and found that the rate varies around an average value of $\ion\,=\,3 \times 10^{-16}\,\rm s^{-1}$. Within dense molecular clouds, the cosmic ray ionization rates cover a range of about two orders of magnitude ($10^{-18} - 10^{-16}$~s$^{-1}$) and are subject to considerable uncertainties (\citealt{2009A&A...501..619P}). \citet{2006PNAS..10312269D} suggested that $\ion = 10^{-17}$~s$^{-1}$ corresponds to the lower limit of the ionization rate in dense regions. In our model $\ion$ is mainly determined by the HCO$^+$ abundance. In addition, a decrease of $\ion$ leads to a slight decrease of the CO emission and a slight variation of the HCN line emission. The model HCO$^+$-to-CO emission ratio is approximately proportional to $\ion$. NGC~628 shows the lowest HCO$^+$-to-CO ratio. The comparison of the best-fit models for NGC~628 with the different values of $\ion$ is presented in \fig{rates}. We found that only the model using an ionization rate of $\ion = 10^{-18}~\rm{s}^{-1}$ is able to reproduce the observed HCO$^+$-to-CO and HCN-to-HCO$^+$ ratio. This value is at the lower end of the range given by \citet{2009A&A...501..619P}.
\begin{figure}[!ht]
    \centering
    \includegraphics[width=1.\linewidth]{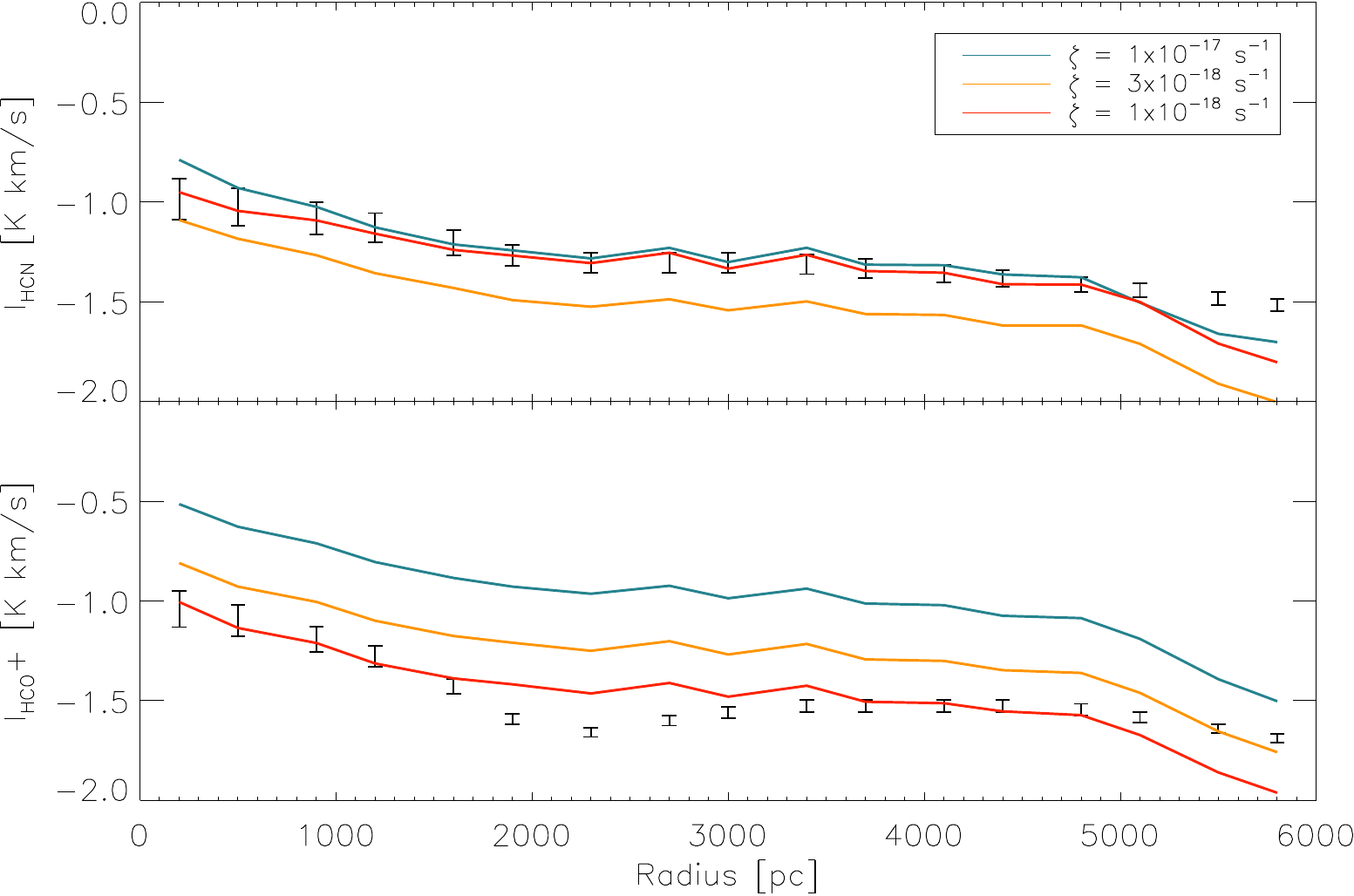}
  \caption{NGC~628. Best-fit models of HCN and HCO$^+$ integrated line emission with different CR ionization rates $\zeta_{\rm CR}$.}
  \label{rates}
  \end{figure}
\paragraph{}
For all galaxies included in the EMPIRE survey, we varied the value of $\ion$ to determine the CR ionization rate that provides the best fit to the observed emission line profiles. The $\chi^2_{\rm dense} = \chi^2_{\rm HCN} + \chi^2_{\rm HCO^+}$ for the different ionization rates $\ion=10^{-17}$, $3\,\times\,10^{-18}$ and, $10^{-18}$~s$^{-1}$ are presented in \tab{bestcr}. For most galaxies, a cosmic ray ionization rate smaller than $\ion=10^{-17}~\rm{s}^{-1}$ is needed. 

 \subsubsection{Molecular line emission as a function of density \label{whichdensity}}

 Star formation occurs in the densest regions of the ISM, within giant molecular clouds (GMCs). In general the SFR correlates with the available amount of molecular gas, whose main tracer is CO emission (\citealt{1959ApJ...129..243S}). CO emission traces gas with densities n$\,\gtrsim 100\,\rm cm^{-3}$. Gas of higher densities can be detected in the HCN and HCO$^+$ lines (e.g. \citealt{2004ApJ...606..271G}). \citet{2004ApJ...606..271G} defined the dense gas fraction as the ratio between the gas mass with densities $\rm n(H_2) > 3 \times\,10^4 cm^{-3}$ and the total gas mass. They found a linear relationship between the infrared and HCN luminosities, which they interpreted as a linear relation between the star-formation rate and the dense gas mass. \citet{2015PASP..127..266M} and \citet{2017A&A...605L...5K} questioned the hypothesis that HCN is mainly emitted by dense gas. 
Based on observations of the Orion A cloud, they found that HCN is already emitted from regions with a density of approximately $\rm n(H_2) \sim 10^3\,cm^{-3}$. For the calculations of the HCN-to-dense gas model conversion factor, we use the definition of the dense gas fraction of \citet{2004ApJ...606..271G}. To investigate at which densities most of the HCN is emitted, we show the brightness temperatures as a function of gas density for the CO(2-1), HCN(1-0) and HCO$^+$(1-0) lines at three galactic radii for the EMPIRE galaxies in \fig{density}. 
The dependence of the brightness temperatures on galactic radius is weak.
{ The model CO brightness temperatures of the different galaxies are quite uniform. The CO brightness temperature reaches the typical sensitivity of ALMA observation of $\sim 0.1$~K at densities between 20 and 30 $\rm{cm}^{-3}$. Likewise, the HCN and HCO$^+$ brightness temperatures are homogeneous except for NGC~3627, which shows higher HCN brightness tempatures for densities $500 \lesssim n \lesssim 1000\ \rm{cm}^{-3}$ at $\rm R = 2 R_d$ and $100 \lesssim n \lesssim 1000\ \rm{cm}^{-3}$ at $\rm R = 3 R_d$.} 
\begin{figure*}[!ht]
  \centering
  \includegraphics[width=0.8\hsize]{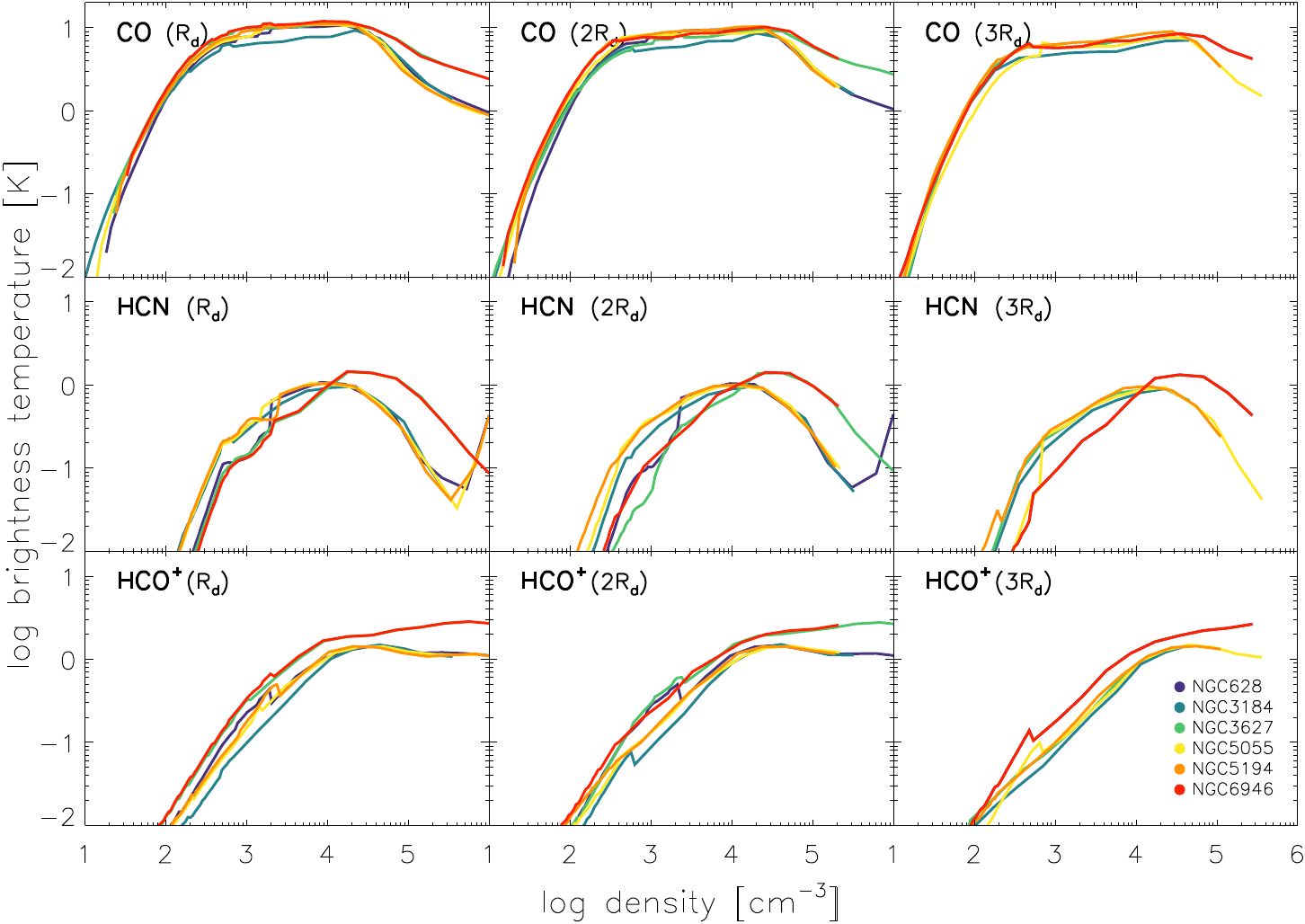}
  \caption{CO(2-1), HCN(1-0), and HCO$^+$(1-0) brightness temperature as a function of gas density. From left to right, the panels correspond to 1 to 3 times the characteristic length scale of the stellar disk R$_{\rm d}$. }
  \label{density}
\end{figure*}

In the model, the CO(2-1) brightness temperature reaches its maximum at a few $100$~cm$^{-3}$. The CO brightness temperature exceed 1~K for densities higher than $\sim 10^2$~cm$^{-3}$, those of HCN and HCO$^+$ at densities higher than $\sim 10^3$~cm$^{-3}$. The HCN brightness temperature increases for density from 10$^2$ to 10$^3$~cm$^{-3}$ and then stays constant at about 1~K. The HCO$^+$ gradually increases from $10^2$ to $10^4$~cm$^{-3}$. 

{We used the brightness temperatures to calculate} the integrated line emission using \eq{ccoo}. The resulting integrated line emission as a function of density is presented in \fig{density2}. The peak of the integrated line emission around $n \sim 10^3$~cm$^{-3}$ occurs close to the transition between the diffuse and self-gravitating gas in our analytical model. The additional factor of $0.6$ (Sect.~\ref{sec:massfrac}) decreases the integrated line emission of self-gravitating clouds. In the inner part of the galaxy ($R \sim R_{\rm d}$, {the disk scale length}) the CO line emission mostly originates in diffuse, i.e. non self-gravitating, gas of density $n < 10^3~\rm{cm}^{-3}$. A non-negligible fraction of the HCN and HCO$^+$ emission originates from gas at these densities as well. 

By integrating the line intensities of \fig{density2}, we determined the density at which half of the line flux is emitted {(Table~\ref{half})}. At $R=R_{\rm d}$, we found average values of n$_{(1/2)} \sim 200$-$500$~cm$^{-3}$ for CO(2-1), n$_{(1/2)} \sim 1000$-$3000$~cm$^{-3}$ for HCN(1-0), and about n$_{(1/2)}$~$\sim 1000$-$3000$~cm$^{-3}$ for HCO$^+$. {None but one of the mid-flux densities significantly changes with radius by more than a factor of three.} The HCN and HCO$^+$ mid-flux density thresholds are low compared to the thresholds usually assumed in the literature in dense gas clouds {$3 \times 10^4$~cm$^{-3}$; Gao \& Solomon 2004)} These high densities are based on assumed HCN and HCO$^+$ brightness temperature of about 30~K (\citealt{2004ApJ...606..271G}). {\citet{2017ApJ...835..217L} found that lognormal gas distributions with low mean densities and small widths generate most of their HCN emission from low densities $\la 10^3$-$10^4$~cm$^{-3}$. If a power law tail is added to the distribution most HCN is emitted by gas with densities exceeding
$10^4$~cm$^{-3}$. \citet{2018MNRAS.479.1702O} calculated the HCN emission from simulated dense, star-forming cores and found that HCN emission traces gas with a luminosity-weighted mean number density of $\sim 10^4$~cm$^{-3}$. 
Based on 100~pc resolution observations of NGC~3627, \citet{2021MNRAS.506..963B} found that HCN and HCO$^+$ to CO(2–1) line ratios show greater 
scatter than $^{13}$CO  and C$^{18}$O to CO(2-1) line ratios, which might suggest that they trace densities above the mean molecular gas density.
On the other hand}, \citet{2016ApJ...823..124L} showed that the HCN and HCO$^+$ lines can be excited and detected in diffuse gas because the emission brightness at the limit of detectability is independent of the critical density and varies only as the $n(H)$-$N({\rm HCN})$ product.
\begin{figure*}[!ht]
  \centering
  \includegraphics[width=0.8\hsize]{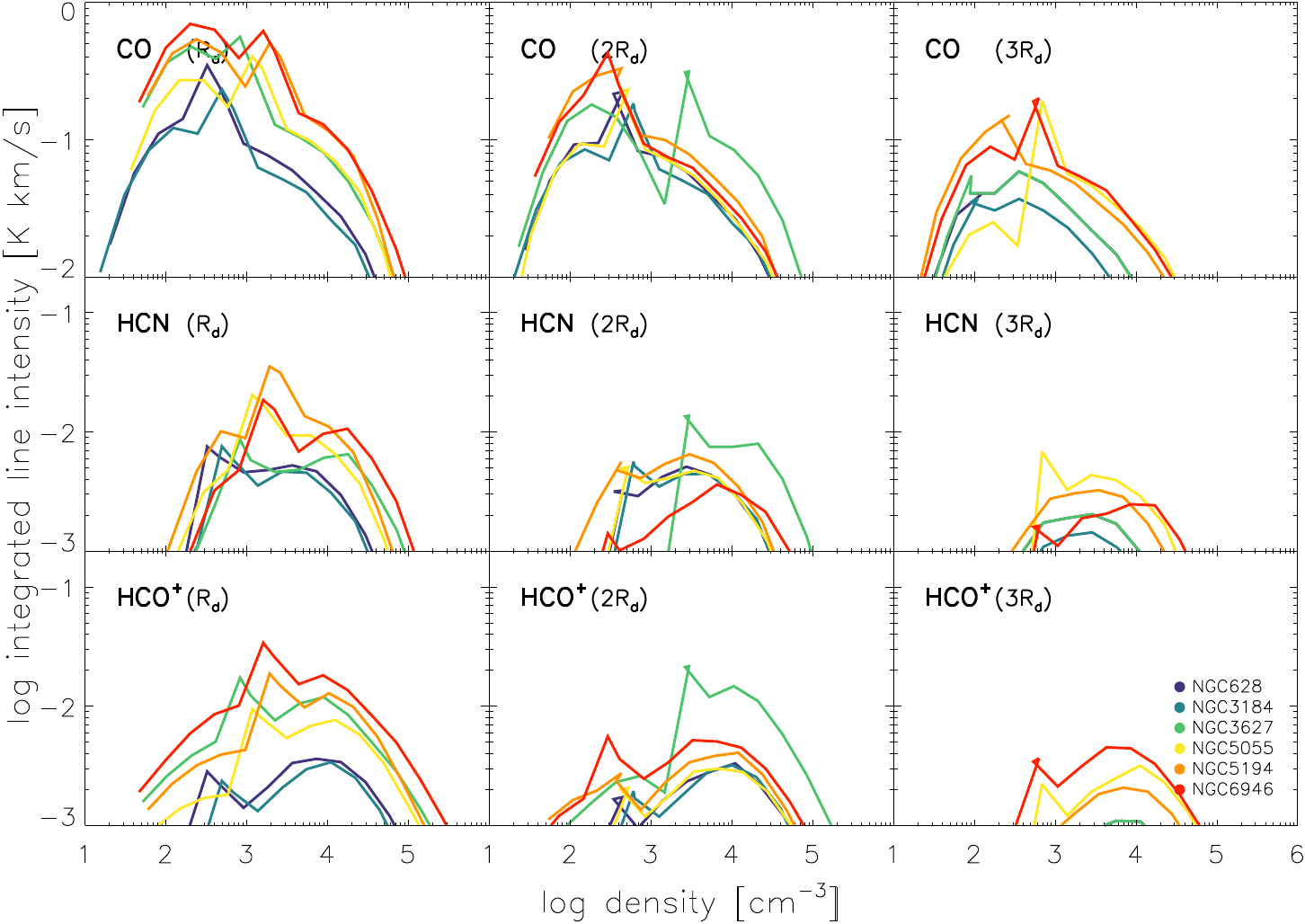}
  \caption{CO(2-1), HCN(1-0) and HCO$^+$(1-0) integrated line intensity as a function of gas density. From left to right, panels correspond to 1 to 3 times the characteristic length scale of the disk R$_{\rm d}$. }
  \label{density2}
\end{figure*}
We also computed the fractional integrated line intensity for $n > 10^3$, $10^{3.5}$, $10^4$, and $10^{4.5}$~cm$^{-3}$ (Table~\ref{tab:fracint11}).
A significant fraction of the HCN ($25$-$30$\,\%) and HCO$^+$ ($15$-$20$\,\%) emission stems from gas with densities exceeding $n=10^4$~cm$^{-3}$.
{\citet{2017A&A...599A..98P} studied the ratio of the molecular line fluxes originating from the dense gas within the Orion~B giant molecular cloud. 
Based on $0.04$~pc resolution observations within a field of view of $5.6 \times 7.5$~pc,
they found that the fraction of the total flux emitted from the densest regions with $n > 7.3 \times 10^{3}\rm cm^{-3}$ corresponds to 8\% for the CO, 18\% for the HCN, and 16\% for the HCO$^+$. 
Since the mean density $n_{\rm H_2} \sim 400$~cm$^{-3}$ \citep{2017A&A...599A..98P} of the Orion B cloud is comparable to the density of selfgravitating clouds in our model, the comparison of their flux fractions to the corresponding model flux fractions is relevant for the HCN and HCO$^+$ emission.
We calculated the flux ratios based on the model profiles with the same density thresholds. The results are presented in \tab{gasfrac}. 
There is little variation of the flux ratios at the three different galactic radii. 
{ Our CO and HCO$^+$ flux ratios at $R=3 \times R_{\rm d}$ (roughly solar radius) are comparable to the values given by \citet{2017A&A...599A..98P} whereas our HCN flux ratio is about $50$\,\% higher than the ratio found by these authors.
Within our model the flux fraction depends on the probability density distribution of the gas density and the brightness temperature of the molecular lines},
which in turn depends on the gas dispersion velocity, density, molecular gas surface density, and temperature.
A sizable fraction of the model CO emission is produced by dense, non-selfgraviting gas at larger scales than those of giant molecular clouds.
It is somewhat surprising that the CO flux fraction within a giant molecular cloud, as Orion B, is comparable to our model CO flux fraction.}

\begin{table*}[!ht]
  \caption{Density thresholds where 50\% of the total flux is emitted, in cm$^{-3}$.}
  \label{half}
  \begin{center}
   \renewcommand{\arraystretch}{1.2}
   \addtolength{\tabcolsep}{-4.pt}
  \begin{tabular}{c|ccc|ccc|ccc}
     \hline
     \hline
      Galaxy & \begin{tabular}{c}  \\ CO(2-1) \end{tabular} & \begin{tabular}{c} R$_{\rm d}$ \\ HCN(1-0) \end{tabular} & \begin{tabular}{c} \\ HCO$^+$(1-0) \end{tabular}
      & \begin{tabular}{c}  \\ CO(2-1) \end{tabular} & \begin{tabular}{c} 2 R$_{\rm d}$ \\ HCN(1-0) \end{tabular} & \begin{tabular}{c} \\ HCO$^+$(1-0) \end{tabular}
      & \begin{tabular}{c}  \\ CO(2-1) \end{tabular} & \begin{tabular}{c} 3 R$_{\rm d}$ \\ HCN(1-0) \end{tabular} & \begin{tabular}{c} \\ HCO$^+$(1-0) \end{tabular} \\

      \hline
     NGC628&         377&        1241&        4961&         275&        2200&        2200&         461&        1843&        1843\\
     NGC3184&         153&         688&        2751&         247&        1263&        1263&         180&         359&        1435\\
     NGC3627&         193&        1310&        2620&         243&        1197&        1197&        1367&        1367&        2734\\
     NGC5055&         315&        1557&        6227&         426&        3402&        1701&         704&        5630&        2815\\
     NGC5194&         309&        1630&        2717&         251&        1378&         689&         267&        1065&        1065\\
     NGC6946&         247&        1839&        2202&         238&        1633&         817&         537&        1074&        1074\\
      \hline
  \end{tabular}

  \end{center}     
\end{table*}

\begin{table*}[!ht]
  \caption{Percentage of total flux originating from dense clouds ($\rm{n} > 7.3 \times 10^3$ cm$^{-3}$,  \citealt{2017A&A...599A..98P}).}
  \label{gasfrac}
  \begin{center}
   \renewcommand{\arraystretch}{1.2} 
   \addtolength{\tabcolsep}{-4.pt}
  \begin{tabular}{c|ccc|ccc|ccc} 
     \hline
     \hline
      Galaxy & \begin{tabular}{c}  \\ CO(2-1) \end{tabular} & \begin{tabular}{c} R$_{\rm d}$ \\ HCN(1-0) \end{tabular} & \begin{tabular}{c} \\ HCO$^+$(1-0) \end{tabular}
      & \begin{tabular}{c}  \\ CO(2-1) \end{tabular} & \begin{tabular}{c} 2 R$_{\rm d}$ \\ HCN(1-0) \end{tabular} & \begin{tabular}{c} \\ HCO$^+$(1-0) \end{tabular}
      & \begin{tabular}{c}  \\ CO(2-1) \end{tabular} & \begin{tabular}{c} 3 R$_{\rm d}$ \\ HCN(1-0) \end{tabular} & \begin{tabular}{c} \\ HCO$^+$(1-0) \end{tabular} \\

      \hline
NGC628&           7\%&          43\%&          26\%&           7\%&          38\%&          26\%&           8\%&          34\%&          27\%\\
NGC3184&           4\%&          31\%&          18\%&           6\%&          30\%&          21\%&           4\%&          20\%&          16\%\\
NGC3627&           4\%&          35\%&          19\%&           7\%&          32\%&          24\%&           9\%&          25\%&          22\%\\
NGC5055&           8\%&          42\%&          26\%&           5\%&          34\%&          21\%&          10\%&          40\%&          28\%\\
NGC5194&           7\%&          37\%&          22\%&           4\%&          29\%&          19\%&           3\%&          21\%&          14\%\\
NGC6946&           6\%&          40\%&          24\%&           3\%&          27\%&          16\%&           7\%&          31\%&          23\%\\
      \hline
      Orion~B & & & & & & & 8\% & 18\% & 16\%  \\
      \hline
  \end{tabular}

  \end{center}     
\end{table*}


\subsubsection{Conversion factors}

In most galaxies the radial profiles of the CO conversion factor increase monotonically with radius {(\fig{cofa})}. 
The conversion factors range between half and twice the Galactic value in most of the galaxies.  
Notable exceptions are NGC~3351 and the outer disks of NGC~2976 and NGC~4214 where the CO conversion factors are
higher than twice the Galactic value.
Our CO conversion factors are about $50$\,\% higher than those found by \citet{2013ApJ...777....5S} (Table~\ref{tab:coweigh}).
There is a big discrepancy for NGC~4736 where our CO weighted conversion factor is five times higher than that of
\citet{2013ApJ...777....5S}. {Likewise, our model CO conversion factor in the inner kpc of NGC~3351 is about a factor of 
five higher than that of \citet{2022ApJ...925...72T} based on 100~pc resolution $^{12}$CO, $^{13}$CO, and C$^{18}$O observations 
and assuming a CO abundance of $3 \times 10^{-4}$. A lower CO abundance caused by a low metallicity as suggested by
\citet{2007MNRAS.382..251D} decreases the difference between the \citet{2022ApJ...925...72T} and our conversion factors.}
\begin{table}[!ht]
  \caption{CO-weighted conversion factors}
  \label{tab:coweigh}
  \begin{center}
   \renewcommand{\arraystretch}{1.2}
  \begin{tabular}{ccc}
     \hline
     \hline
      Galaxy & Mean $\aco^{\rm Model}$ & Mean $\aco^{\rm Sandstrom}$\\
      \hline
      NGC~628 & 6.4 & 5.1 \\
      NGC~3184 & 5.9 & 6.3 \\
      NGC~3627 & 5.5 & 1.8 \\
      NGC~5055 & 5.9 & 3.7 \\
      NGC~5194 & 5.7 & - \\ 
      NGC~6946 & 3.1 & 1.8 \\
      NGC~2841 & 9.0 & 5.7 \\
      NGC~3198 & 15.8 & 11.9 \\
      NGC~3351 & 6.7 & 2.9 \\
      NGC~3521 & 7.9 & 7.3 \\
      NGC~4736 & 5.4 & 1.1 \\
      NGC~7331 & 9.0 & 10.7 \\ 
      NGC~925 & 11.1 & 10.0 \\ 
      NGC~2403 & 8.9 & - \\ 
      NGC~2976 & 6.7 & 4.7 \\ 
      NGC~4214 & 5.5 & - \\ 
      NGC~7793 & 9.3 & - \\
      \hline
  \end{tabular}
   \end{center}
     
\end{table}    
The models suggest that the conversion factors are slightly sub-Galactic in the center and exceed the Galactic value by up to a factor of 2 at the edge of the CO-emitting disk. For the HCN-to-dense gas conversion factor, the model yields values between half and twice the Gao \& Solomon value. As for $\aco$, the radial profiles tend to increase slightly from the galaxy center to the edge of the molecular disk. 

\begin{figure}[!ht]
  \center
  \begin{subfigure}{0.48\textwidth}
    \centering
    \caption{Massive galaxies with EMPIRE data.}
    \includegraphics[width=1.\linewidth]{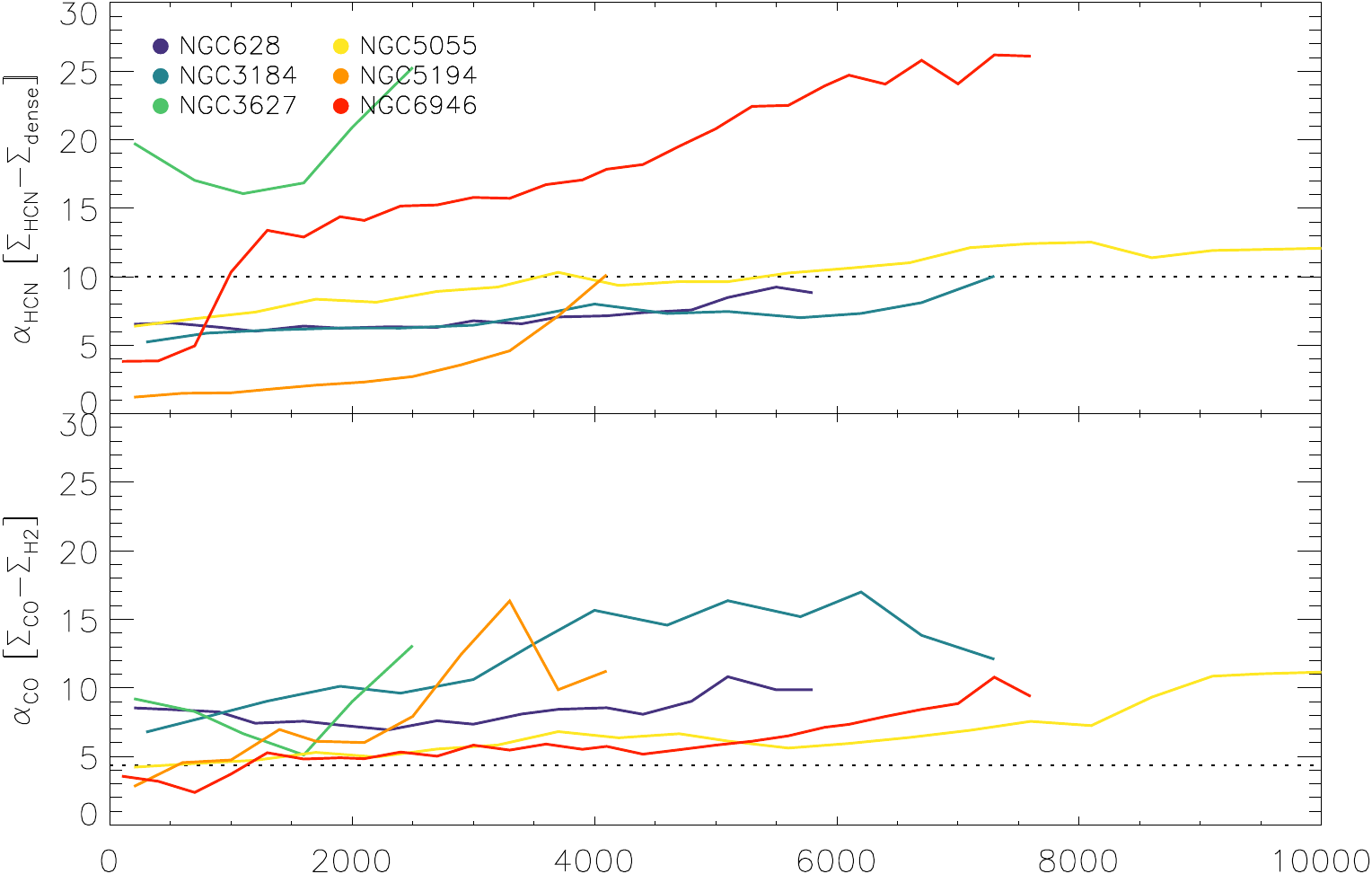}
  \end{subfigure}%
  
  \begin{subfigure}{0.48\textwidth}
    \centering
    \caption{Massive galaxies without EMPIRE data.}
    \includegraphics[width=1.\linewidth]{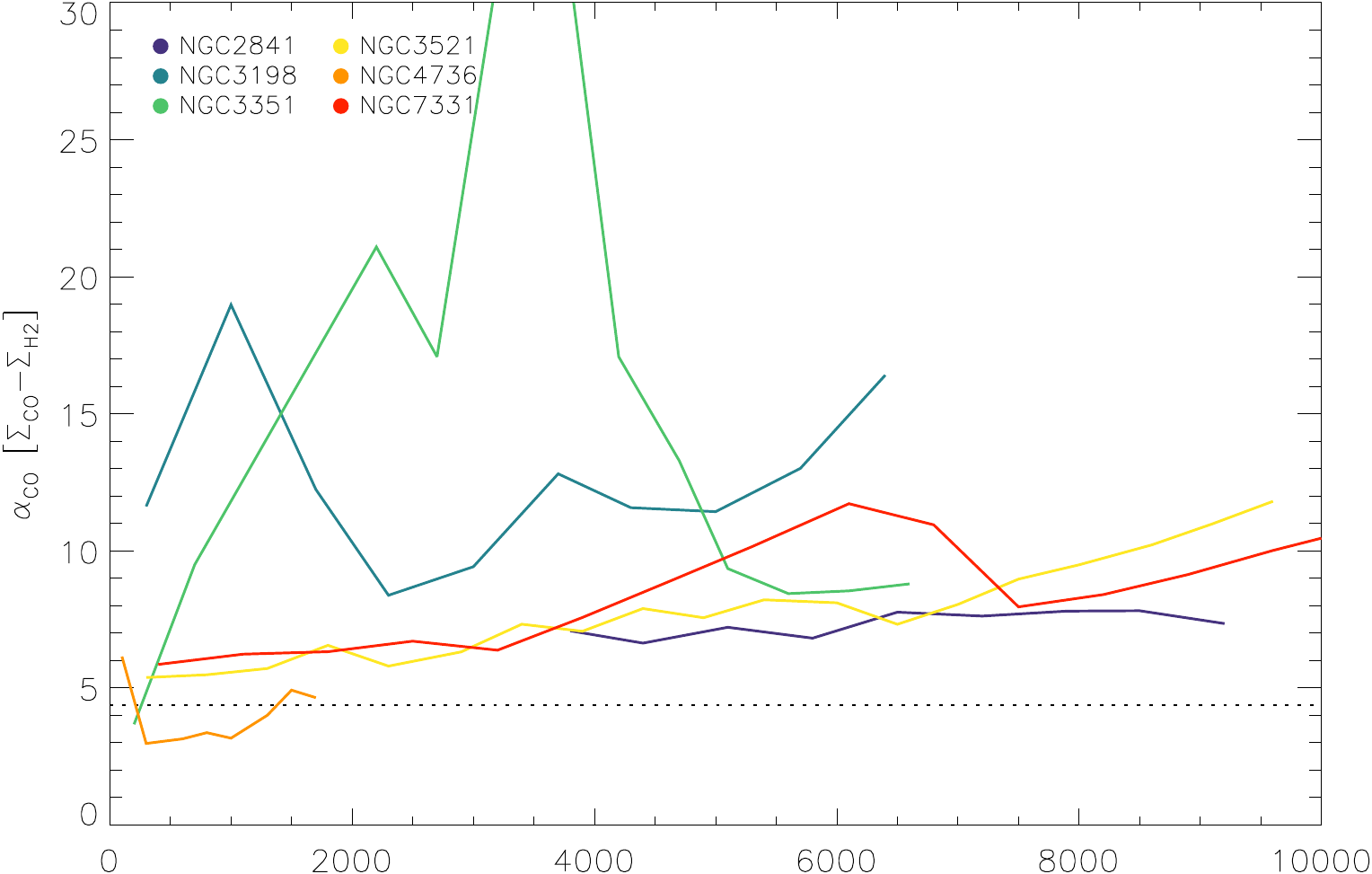}
  \end{subfigure}%
  
  \begin{subfigure}{0.48\textwidth}
    \centering
    \caption{Low-mass galaxies without EMPIRE data.}
    \includegraphics[width=1.\linewidth]{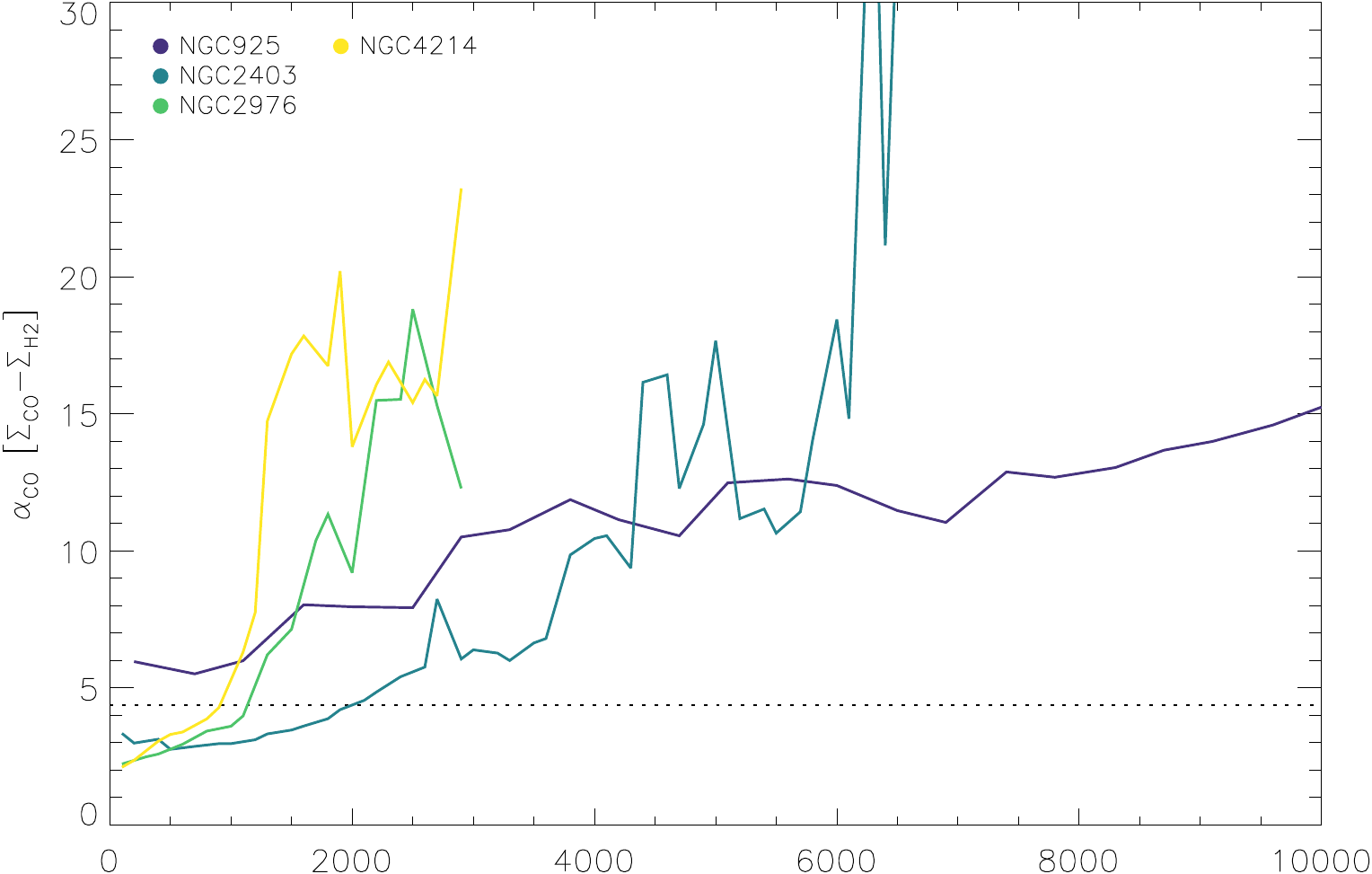}
  \end{subfigure}
  
  \caption{CO-to-H$_2$ ($\alpha_{\rm CO}$) and HCN-to-dense gas ($\alpha_{\rm HCN}$) conversion factors of the high-mass galaxies with EMPIRE data (upper panel), high-mass galaxies without EMPIRE data (middle panel),
and low-mass galaxies (lower panel; $M_* < 10^{10}$~M$_{\odot}$). {Dense gas corresponds to $\rm{n} > 3 \times 10^{4} \rm{cm}^{-3}$}}.
  \label{cofa}
  \end{figure}

\subsubsection{Gas depletion and viscous timescales \label{sec:depvis}}

The radial profiles of the gas depletion and viscous timescale are presented in \fig{time}. 
For all massive galaxies, except NGC~628, the viscous timescales exceed the depletion timescales for $R>R_{\rm d}$ by a factor of about 10. For NGC~628, $\tauvis$ is particularly small in the inner $2.5$~kpc, and close to the gas depletion timescale of $2$-$3$~Gyr. 

For the low-mass galaxies, with $M_* < 10^{10}$~M$_{\odot}$, the model yields $\tauvis \leq \taudep$ up to $R \sim 2 R_{\rm d}$ for NGC~4214, NGC~2403, {and NGC~7793} and up to $R \sim R_{\rm d}$ for NGC~2976. This suggests that for low-mass galaxies, star formation can be maintained by radial gas accretion within the disk from the center to $\rm{R}~\sim~2~\rm{R}_{\rm d}$ (see also \citealt{2011AJ....141...24V}). On the other hand, for massive galaxies ($M_* > 10^{10}$~M$_{\odot}$) external gas accretion is needed to supply star-formation with fresh gas, especially in the inner disk region. In the absence of external gas accretion, the gas surface density of the inner disk will significantly decrease within a few Gyr leading to
a high $Q_{\rm gas}$ and to a $Q_{\rm tot}$ exceeding unity (see Sect.~\ref{QMV}). Condensation of halo gas on
galactic fountains (e.g. \citealt{2017ASSL..430..323F}) might be an important source of gas accretion onto the inner galactic disk.

\begin{figure}[!ht]
  \center
  \begin{subfigure}{0.48\textwidth}
    \centering
    \caption{Massive galaxies with EMPIRE data.}
    \includegraphics[width=1.\linewidth]{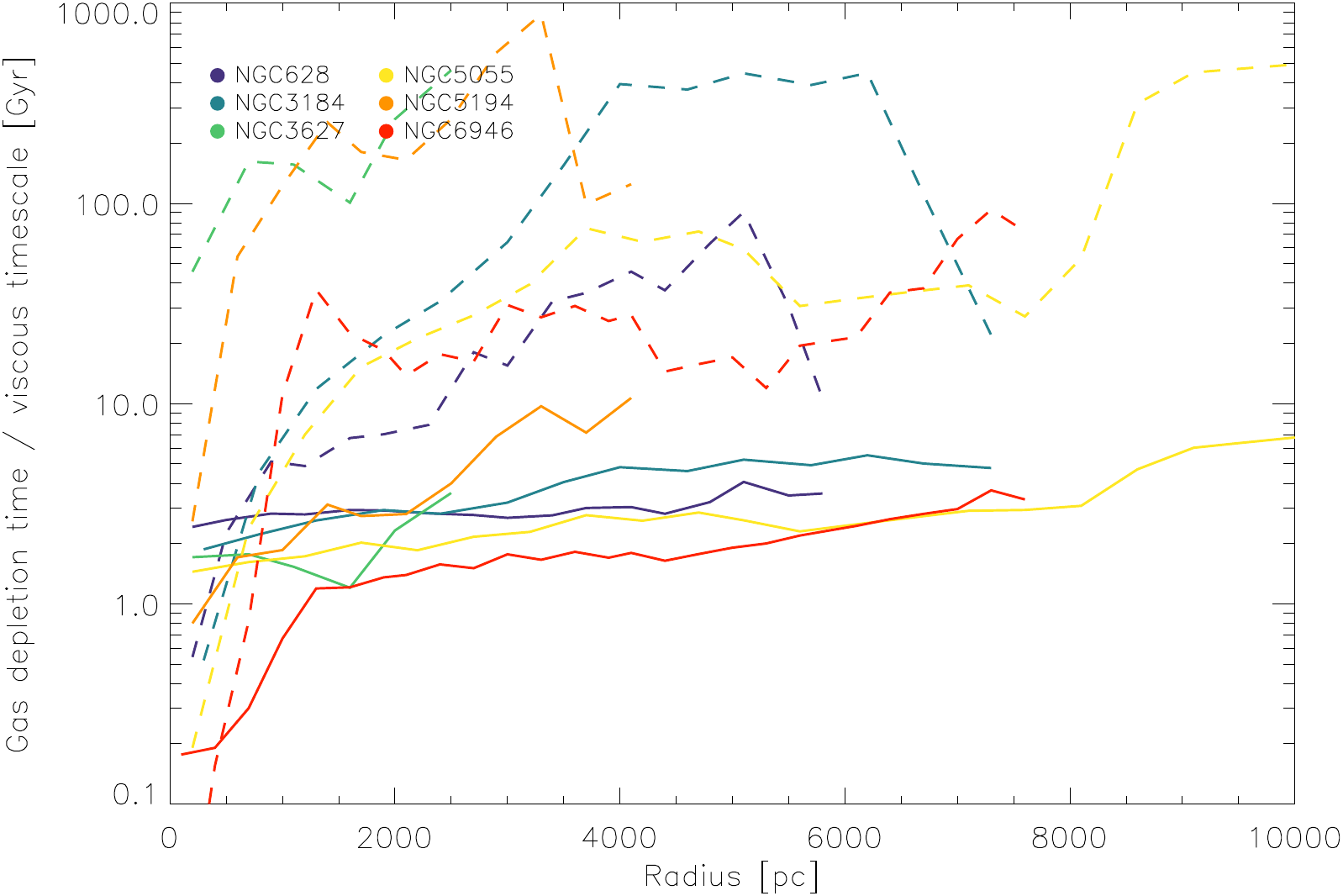}
  \end{subfigure}%
  
  \begin{subfigure}{0.48\textwidth}
    \centering
    \caption{Massive galaxies without EMPIRE data.}
    \includegraphics[width=1.\linewidth]{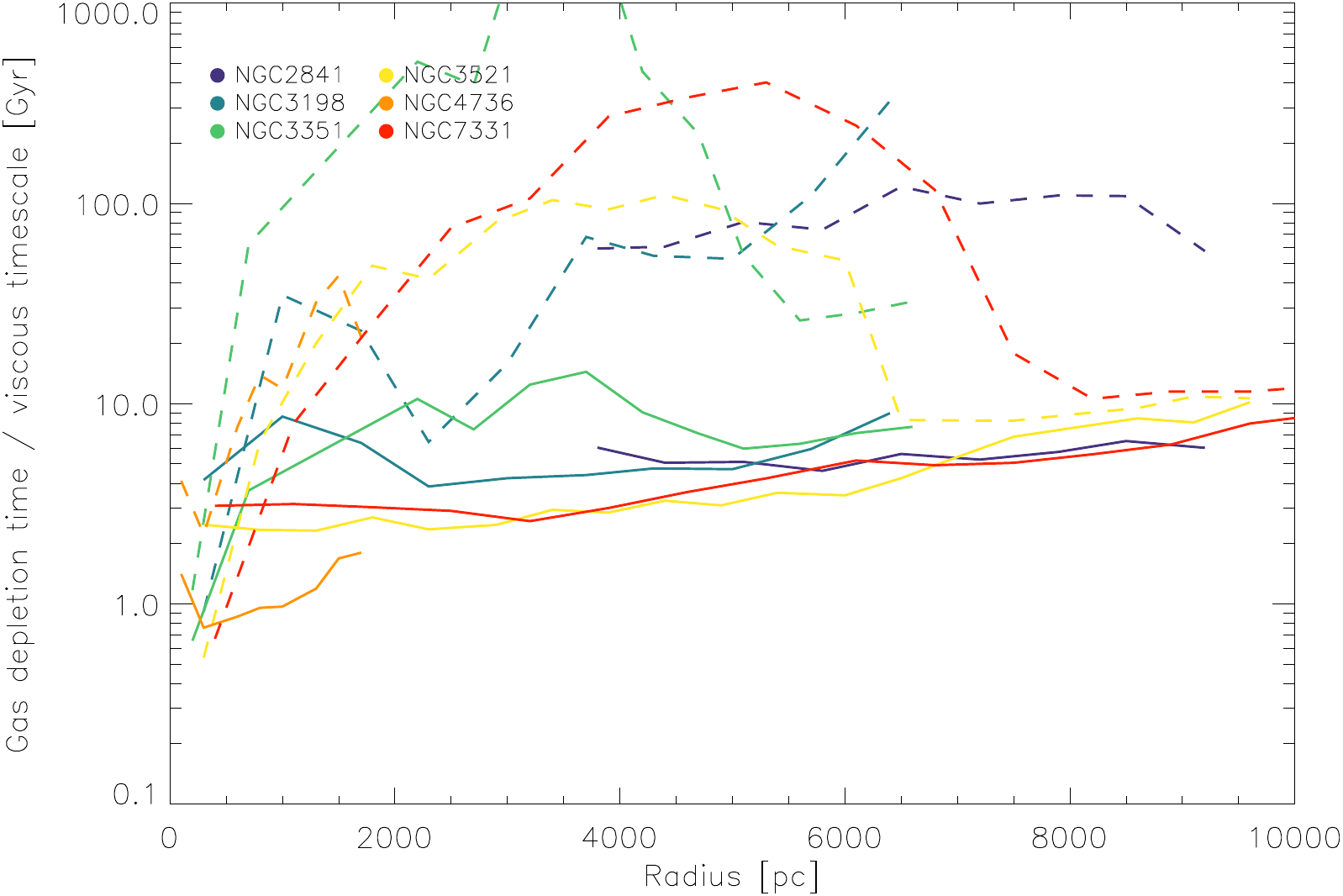}
  \end{subfigure}%
  
  \begin{subfigure}{0.48\textwidth} 
    \centering
    \caption{Low-mass galaxies without EMPIRE data.}
    \includegraphics[width=1.\linewidth]{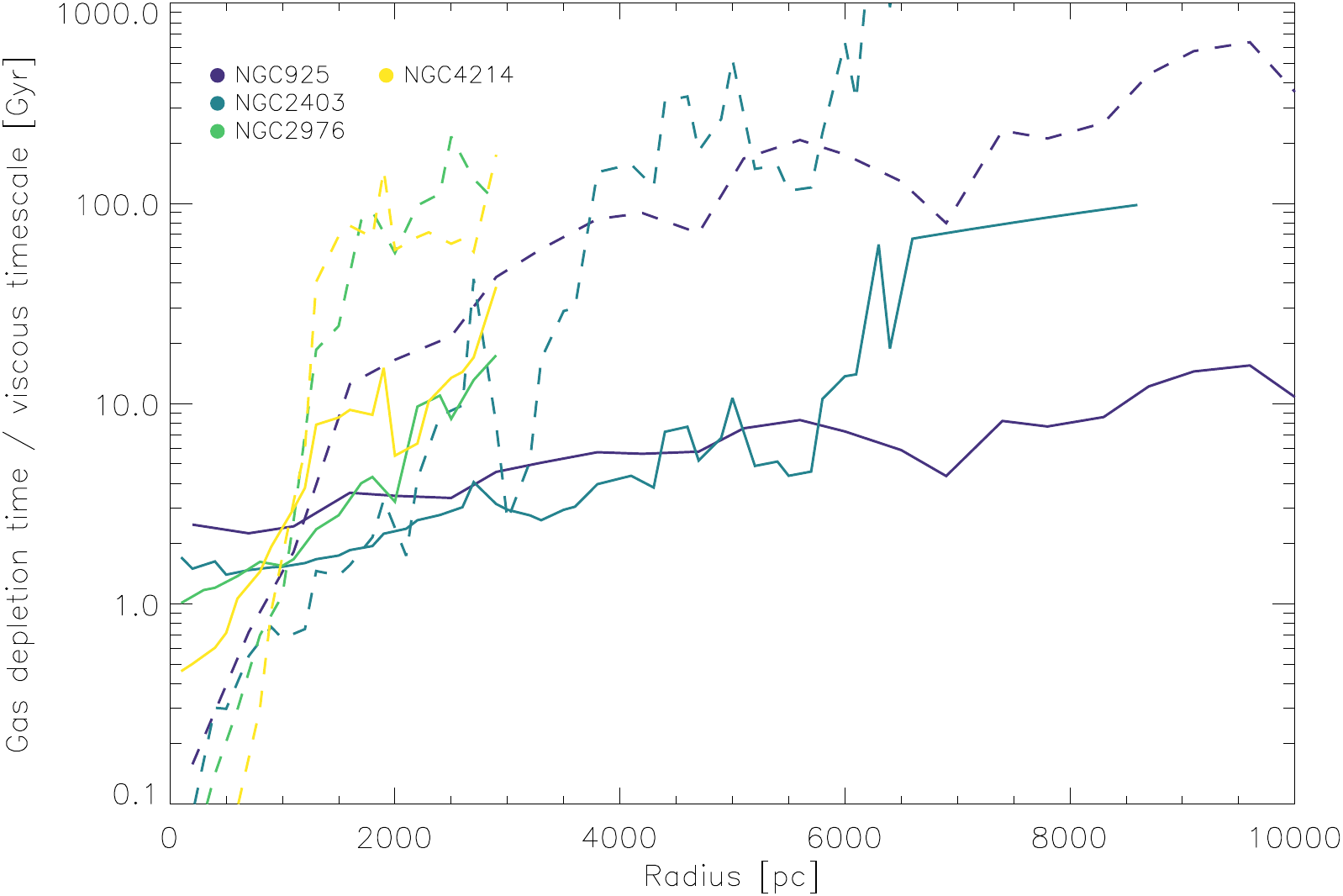}
  \end{subfigure}
  
  \caption{Gas depletion (solid lines) and viscous timescale (dashed lines) of the high-mass galaxies with EMPIRE data (upper panel), high-mass galaxies without EMPIRE data (middle panel), and low-mass galaxies (lower panel; $M_* < 10^{10}$~M$_{\odot}$).}
  \label{time}
  \end{figure}

\section{Discussion \label{sec:discussion}}

\subsection{Low cosmic ray ionization rates $\ion$}

As discussed in \sect{ioniz}, our model requires particularly low ionization rates to reproduce the observed HCO$^+$ radial profiles. Our initial assumption of $\ion=10^{-17}$~s$^{-1}$ corresponds already to the lower limit suggested by \citet{2006PNAS..10312269D}. In general, lower CR ionization rates are derived in gas of higher densities (\citealt{2009A&A...501..619P}, \citealt{2017ApJ...845..163N}). For four out of six {EMPIRE} galaxies (NGC~628, NGC~3184, NGC~5055, and NGC~5194) the lowest CR ionization rate of $\ion = 10^{-18}$s$^{-1}$ is required. The first two galaxies have particularly low gas surface densities below 20~M$_\odot$/pc$^2$. {We can only speculate that these galaxies have a low magnetic field strength and a significant fraction of the ionizing cosmic ray particles escape from the galactic gas disk in the vertical direction. Energy equipartition between the cosmic ray particles
and the magnetic field is commonly assumed in spiral galaxies (e.g., \citealt{2005AN....326..414B}). Cosmic ray particles and magnetic fields are supposed
to be strongly coupled and to exchange energy until equilibrium is reached. A lower magnetic field can thus confine less cosmic ray particles.}

\subsection{The CO conversion factor of NGC~6946}

\citet{2020ApJ...903...30B} estimated the $\aco$ conversion factor based on [C{\sc II}] and CO radial profiles of NGC~6946 using the \citet{2017MNRAS.470.4750A} model. They obtained an increasing trend for $\aco$ with galactocentric radius, in broad agreement with the metallicity gradient. This behavior is consistent with prior observational works based on a combination of H{\sc i}, CO, and IR data involving the gas-to-dust ratio (e.g. \citealt{1995ApJ...452..262S}, \citealt{1997A&A...326..963B}, \citealt{2013ApJ...777....5S}). Our radial profile of $\aco$ has approximately the same shape as that of \citet{2020ApJ...903...30B} but is about a factor of two higher. \citet{2013ApJ...777....5S} determined the CO conversion factor via
\begin{equation}
  \aco = \frac{1}{I_{\rm CO}} \left( {\rm GDR}\,\Sigma_{\rm D} - \Sigma_{HI}\right)\ ,
\end{equation}
simultaneously solving for $\aco$ and the GDR by assuming that the GDR is approximately constant on kpc scales. With our inferred gas-to-dust ratio of GDR~$\sim 200$ for NGC~6946 (see Table \ref{betagdr}), the CO conversion factor of \citet{2013ApJ...777....5S} would increase to a value close to the Galactic conversion factor. {We conclude that our radial profile of the CO conversion factor is consistent within a factor of two with
that of \citet{2013ApJ...777....5S} and independent of the gas-to-dust ratio.}

\subsection{The effects of CO photo-dissociation in low-mass galaxies}

In an externally irradiated gas cloud, a significant H$_2$ mass may lie outside
the CO region, that is {dark in CO} in the outer regions of the cloud where the gas phase carbon resides in C or C$^+$.
In this region, H$_2$ self-shields or is shielded by dust from UV photo-dissociation, whereas CO is photo-dissociated.
Following \citet{2010ApJ...716.1191W}, the dark gas mass fraction for a cloud of constant density is
\begin{equation}
\label{eq:fdg}
f_{\rm DG}=\frac{M_{\rm H_2}-M_{\rm CO}}{M_{\rm H_2}}=1-\big(1 - \frac{2 \Delta A_{\rm V, DG}}{A_{\rm V}}\big)^3
,\end{equation}
with
\begin{equation}
\begin{split}
\Delta A_{\rm V,DG}=&0.53-0.045\,{\rm ln}\big(\frac{\dot{\Sigma_*}/(10^{-8}~{\rm M}_{\odot}{\rm pc}^{-2}{\rm yr}^{-1})}{n_{\rm cl}}\big)-\\
&0.097\,{\rm ln}\big(\frac{Z}{Z_{\odot}}\big)
\end{split}
,\end{equation}
and $A_{\rm V}=2\,(Z/Z_{\odot})N_{\rm cl}/(1.9 \times 10^{21}~{\rm cm}^{-2})$ where $N_{\rm cl}$ is the H$_2$ column density.
{The mass fraction of CO-emitting gas} is then:
\begin{equation} \label{fco}
  f_{\rm CO}=f_{\rm H_2} \, \left(1 - \frac{2 \Delta A_{\rm V, DG}}{A_{\rm V}}\right)^3 \ .
\end{equation}

We tested the influence of the CO photo-dissociation on the observed CO brightness temperature for the low-mass galaxies where the effect is expected to be important (\fig{kfmol}). In these galaxies we detected no significant effect. This is consistent with the fact that {most of} these galaxies are relatively CO-bright, meaning that CO emission is detected beyond $\rm{R}~\sim~2~\rm{R}_{\rm d}$. 
{Even in the CO-dim galaxy NGC~7793 CO photo-dissociation only leads to a decrease of the CO emission by $\sim 50$\,\% in the inner disk and
by a factor of about two in the outer disk. This translates into an increase of the CO flux by a factor of $1.7$ in the absence of CO photodissociation.}

\begin{figure*}[!ht]
  \centering
  \includegraphics[width=0.9\hsize]{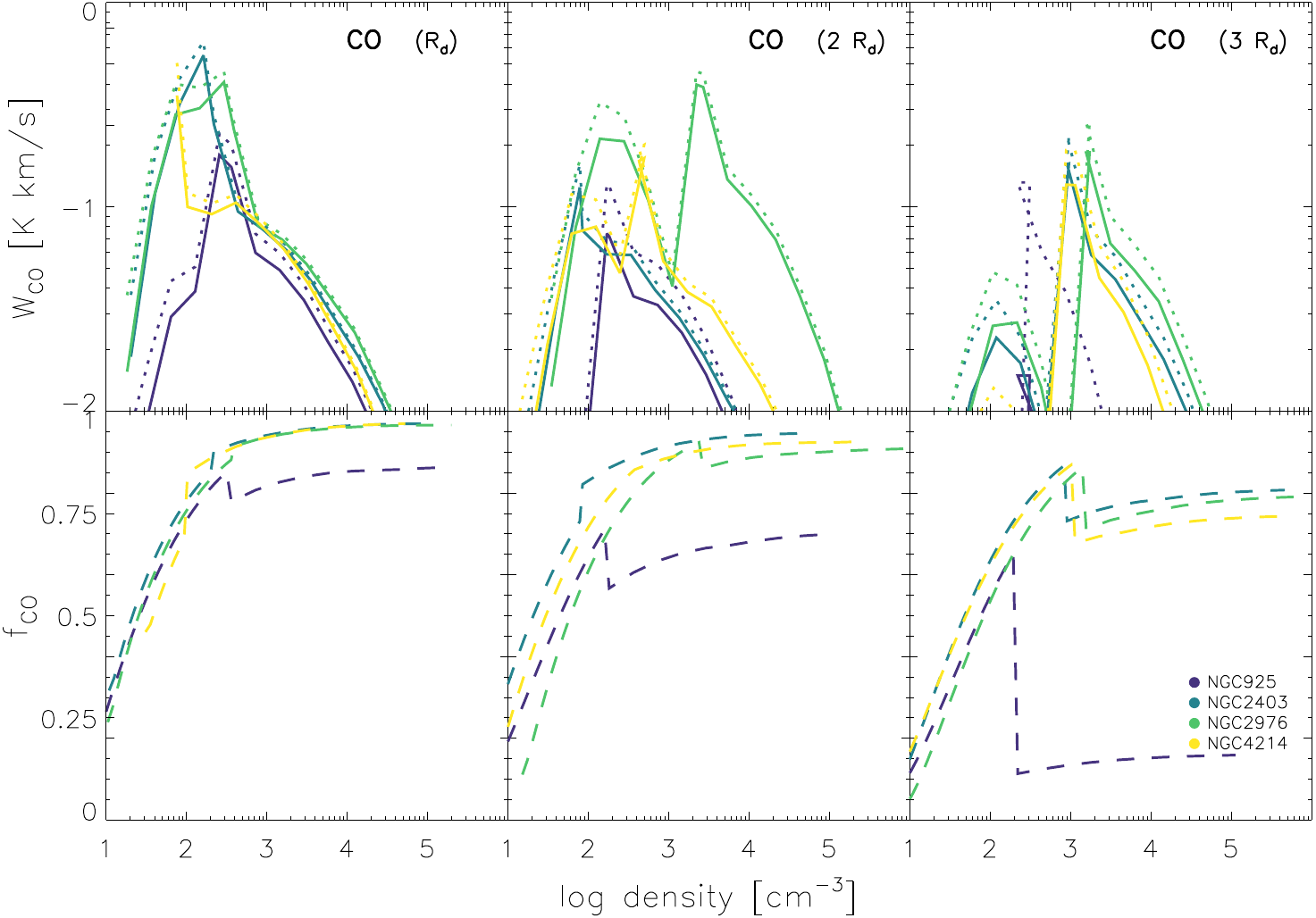}
  \caption{Upper panels: integrated CO line intensity as a function of density for the four low-mass galaxies. Dotted lines: with, solid lines: without CO photo-dissociation. Lower panels: {mass fraction of CO-emitting gas as a function of density in the presence of CO photo-dissociation} (\eq{fco}).}
  \label{kfmol}
\end{figure*}

\subsection{Anomalous gas in NGC~2403 \label{sec:n2403}}

\citet{2011AJ....141...24V} suggested that only low-mass galaxies ($M_* < 10^{10}$~M$_{\odot}$) can sustain their star-formation rate through radial gas accretion within the galactic disk. Within the framework of our model, this is only possible within $\rm{R}~\sim~2~\rm{R}_{\rm d}$ {(Sect.~\ref{sec:depvis} and Fig.~\ref{time})}. In order to sustain star-formation through radial gas accretion within the thin gas disk, the gas mass has to be transported in a different way from large radii to $\rm{R}~\sim~2~\rm{R}_{\rm d}$. In NGC~2403, \citet{2002AJ....123.3124F} found a faint extended and kinematically anomalous gas component. The cold H{\sc i} disk is surrounded by a thick and clumpy H{\sc i} layer characterised by slow rotation and infall motion (10-20~km\,s$^{-1}$) towards the center. The gas surface density of the anomalous component is about 1~M$_\odot$ pc$^{-2}$. The mass accretion rate is given by:
\begin{equation}
  \mdot = 2 \pi \nu \Sigma = 2 \pi R v_{\rm R} \Sigma \sim 0.3 ~\rm{M_\odot/yr},
\end{equation}
{where $\nu$ is the gas viscosity (Eq.~\ref{eq:visc3}) and $v_{\rm R}$ the radial velocity}, see also \citet{2002ApJ...578..109F}. This is exactly the radial mass accretion rate found by our model at $\rm{R}~\sim~2~\rm{R}_{\rm d}$ (\fig{fig:2403}). Therefore, we suggest that the fuel for star-formation reaches this radius from outside via a thick disk component. From $\rm{R}~\sim~2~\rm{R}_{\rm d}$ inward the gas is transported radially through turbulent gas viscosity, fully sustaining star-formation. We can only speculate that a comparable mechanism acts in the other low-mass galaxies in our sample.

\section{Conclusions \label{sec:conclusions}}

The model of a turbulent clumpy gas accretion disk of \citet{2017A&A...602A..51V} was used to calculate the SFR, H{\sc i}, CO(1-0), CO(2-1), HCN(1-0), and HCN$^+$ radial profiles 
of a sample of 17 nearby spiral galaxies observed by the THINGS and HERACLES surveys. {A subsample of six galaxies was observed by the EMPIRE survey.
The large-scale properties of the model are} the gas surface density,  density,  disk  height,  turbulent  driving  length  scale, velocity  dispersion,  
gas  viscosity,  volume  filling  factor,  and molecular  fraction.  Small-scale  properties  are  the  mass,  size and density of clouds, and the associated timescales crossing, free-fall, and H$_2$ formation 
timescales of the most massive self-gravitating gas clouds. These quantities depend on the stellar surface density, the angular velocity $\Omega$, the disk radius $R$, and three 
free parameters, which are the Toomre parameter $Q$ of the gas, the mass accretion rate $\dot{M}$, and the ratio $\delta$ between the driving length scale of turbulence  
and the cloud size. In addition, Galactic scaling laws for the dependence of the gas density and velocity dispersion
on the cloud size were included in the model. The gas temperature was calculated through the equilibrium between turbulent mechanical heating and molecular line cooling.
The molecular abundances were determined via NAUTILUS \citep{2009A&A...493L..49H} and the molecular line emission via RADEX \citep{2007A&A...468..627V}. 

For different values of $\delta$ we determined the free parameters $Q$ and $\dot{M}$ at each galactic radius using three independent measurements: the 
neutral gas (H{\sc i}), molecular gas (CO), and SFR (FUV + 24$\mu$m). 
{We did not consider gas disks that are unstable to fragmentation ($Q<1$).}
The model also yields the FIR radial profiles, which can be directly compared to Herschel data as a validation of our SFR recipe based on FUV and 24$\mu$m data.  
Furthermore, we determined the cosmic ray ionization rate by comparing the model HCN and HCO$^+$(1-0) emission to EMPIRE observations
\citep{2019ApJ...880..127J}.

Based on the model fitting  of  the  radial  profiles  of  the  total  SFR, H{\sc i}, CO, HCN, and HCO$^+$ we conclude that
\begin{enumerate} 
\item 
the Toomre parameter $Q_{\rm tot}$ exceeds unity in the inner disk of a significant number of galaxies (Fig.~\ref{QM}).
In two galaxies, $Q_{\rm tot}$ also exceeds unity in the outer disk. Thus, in spirals galaxies $Q_{\rm tot}=1$ is not {ubiquitous}.
\item
The finite lifetime of molecular clouds ($f_{\rm mol}^{\rm life}$) has to be taken into account for the 
molecular gas fraction (Sect.~\ref{QMV}). 
\item
The model gas velocity dispersions are consistent with the observed H{\sc i} velocity dispersions (Fig.~\ref{vdisp}). 
\item
{In all but one} of the six galaxies observed by the EMPIRE survey the model cosmic ray ionization rate is found to be significantly smaller than 
$\zeta_{\rm CR}=10^{-17}$~s$^{-1}$ (Table~\ref{bestcr}). For four out of six galaxies we found  $\zeta_{\rm CR}=10^{-18}$~s$^{-1}$.
\item
Within our model HCN and HCO$^+$ is already detectable in relatively low-density gas ($\sim 1000$~cm$^{-3}$; Fig.~\ref{density2}). 
The model HCN(1–0) and HCO$^+$(1–0) line emission traces densities of several $10^3$~cm$^{-3}$ (Table~\ref{half}).
\item
The derived CO conversion factor of most of the massive galaxies ($M_* > 10^{10}$~M$_{\odot}$) is close to the Galactic value and increases slightly with galactic radius (Fig.~\ref{cofa}).
\item
In low-mass galaxies ($M_* < 10^{10}$~M$_{\odot}$) the CO conversion factor can increase steeply with radius (Fig.~\ref{cofa}).
\item
CO-dark gas mass is not relevant in the CO-bright low-mass galaxies (Fig.~\ref{kfmol}). {In the CO-dim galaxy NGC~7793 the inclusion
of CO photodissociation leads to a $\sim 50$\,\% lower CO emission in the inner disk and an about two times lower CO emission in the outer disk.}
\item
The derived HCN conversion factors are consistent with $\ahcn=10~\aunit$ of \citet{2004ApJ...606..271G} within a factor of two and increase with galactic radius (Fig.~\ref{cofa}).
\item
In all galaxies the molecular gas depletion timescale ranges between 1 and 5~Gyr (Fig.~\ref{time}).
\item
In almost all massive galaxies the viscous timescale greatly exceeds the star formation timescale (Fig.~\ref{time}). In the absence of gas accretion from
the galactic halo, the galaxies will undergo starvation (\citealt{1980ApJ...237..692L}), where the cold gas is exhausted by the star formation activity of the galaxy. This naturally leads to $Q_{\rm tot} > 1$.
\item
In the low-mass galaxies the viscous timescale is smaller than the star formation timescale for $R \la 2 R_{\rm d}$. Thus, the star formation rate can be sustained by radial mass inflow within the galactic disk within $\rm{R}~\sim~2~\rm{R}_{\rm d}$ (Fig.~\ref{time}). 
\item
We suggest that the fuel for star-formation reaches $\rm{R}~\sim~2~\rm{R}_{\rm d}$ from {the outskirts of the galaxy} via a thick gas disk component as in NGC~2403 (Sect.~\ref{sec:n2403}).
\end{enumerate}

The combination of the large-scale model of a turbulent clumpy star-forming galactic gas disk together with local scaling relation based on Galactic 
observations is thus able to simultaneously reproduce SFR, IR, H{\sc i}, CO, HCN, and HCO$^+$ radial profiles of local spiral galaxies.
The resulting gas velocity dispersions and CO and HCN conversion factors are in agreement with those found in the literature.
As a potential next step, the model can be applied to high-resolution observations of ultra-luminous infrared galaxies (ULIRGs) and
high-z star-forming galaxies.

\begin{acknowledgements}
We would like to thank Adam Leroy for his comments, which helped to improve the article and J.~den~Brok for making the
PHANGS CO(2-1) radial profiles of NGC~3627 and NGC~5194 available to us. {We also thank the anonymous referee for their useful comments, which helped to significantly improve the article.}
\end{acknowledgements}

\bibliographystyle{aa}
\bibliography{aa}

\appendix

\section{Detailed description of the model \label{app:model}}

The analytical model of a turbulent clumpy star-forming galactic disk has a large-scale and a small-scale part.

\subsection{Large-scale part}

In this subsection we describe in detail the physics that govern the large-scale part of the model. 
\subsubsection{Hydrostatic equilibrium}
Within the model the ISM is considered as a single turbulent gas in vertical hydrostatic equilibrium. The turbulent pressure is defined as $P_{\rm turb} = \rho \sigmaturb^2$, where $\sigmaturb$ is the total velocity dispersion that takes into account both, the velocity dispersion caused by turbulence $\vturb$ and a constant thermal velocity $v_{\rm therm} = \vtherm = 6~\rm{km~s^{-1}}$ such as $\sigmaturb = \sqrt{\vturb^2 + \vtherm^2}$. Following \citet{1989ApJ...338..178E}, we can establish the first main equation of the model: 
\begin{equation} \label{pressure}
    P_{\rm turb} = \rho \sigmaturb^2 = \frac{\pi}{2} G \Sigma \left( \Sigma + \Sigma_\star \frac{\sigmaturb}{\gamma \vdisp } \right)\ , 
\end{equation}
where $\rho$ is the midplane density, $\Sigma$ is the total gas surface density, $\Sigma_\star$ is the stellar surface density, $\vdisp$ is the vertical stellar velocity dispersion, {and $\gamma$ is a fudge factor}. Given the stellar surface density and the stellar length scale of the disk $l_\star$, the vertical stellar velocity dispersion can be computed following \citet{2002ASPC..275...47K}:
\begin{equation} \label{eq:vdisp}
    \vdisp = \sqrt{ 2 \pi G \Sigma_\star \frac{l_\star}{7.3}}\ .
\end{equation}

\subsubsection{Energy transfer by turbulence}

In galaxies ISM turbulence is mainly maintained by stellar feedback such as supernova explosions, ionizing radiation or stellar winds (\citealt{2004RvMP...76..125M}, \citealt{2004ARA&A..42..211E}). Within the framework of our model we consider supernova explosions as the dominant source of energy. Turbulence is expected to form eddies with a typical size $\ldriv$ (the turbulent driving length scale) at the origin of the formation of the densest gas clouds. The SNe energy is cascaded from the largest to the smallest scales. The energy per unit time, which is carried by turbulence, is:
\begin{equation}
    \dot{E} \simeq -\dot{E}_{\rm SN} = - \frac{\rho \nu }{2} \int \frac{\vturbb^2}{\ldriv^2}dV \ ,
\end{equation}
where $\nu$ is the viscosity of the gas defined as $\nu = \vturbb \ldriv$ with the 3D turbulent velocity dispersion $\vturbb = \sqrt{3}\,\vturb$. If we define the surface density of the gas as $\Sigma = \rho H$ and assume the integration over the volume $\int \rm{d}V = V = AH$, we can connect the energy input into the ISM by SNe directly to the SFR with the assumption of a constant initial mass function as:
\begin{equation} \label{sfr2}
    \frac{\dot{E}_{SN}}{\Delta A} = \frac{\Sigma \nu }{2} \frac{\vturbb^2}{\ldriv^2} = \xi \dot{\Sigma}_\star \ ,
\end{equation}
where $\dot{E}_{SN}$ is the energy injected by the supernova explosions, $\Delta A$ is the unit area, and $\sigmastar$ is the star-formation rate. The factor $\xi$ relates the energy injection by supernova explosions to the star formation rate. It is considered as constant and its canonical value was estimated from observations in the Milky Way, given $\xi = 4.6 \times 10^{-8}\ \rm pc^2/yr^2$ (\citealt{2003A&A...404...21V}). In the presence of high disk mass accretion rates, the energy injection through the gain of potential energy can be important. In this case, \eq{sfr2} becomes
\begin{equation} \label{sfr3}
  \frac{\Sigma \nu }{2} \frac{\vturbb^2}{\ldriv^2} = \xi \dot{\Sigma}_\star + \frac{1}{2 \pi} \dot{M} \Omega^2 \ ,
\end{equation}
where $\mdot$ is the mass accretion rate and $\Omega$ is the angular velocity.

\subsubsection{Viscosity and accretion}

The model is based on the assumption that turbulence redistributes the angular momentum within the disk via gas viscosity. The galactic gas disk is treated as a turbulent clumpy accretion disk, where angular momentum is transported outwards permitting the gas to move inwards. Assuming a continuous and non-zero external gas mass accretion $\dot{\Sigma}_{\rm ext}$, the simplified time evolution of the disk surface density is given by
\begin{equation}
    \frac{\partial \Sigma}{\partial t} \sim \frac{\nu \Sigma}{R^2} - \dot{\Sigma}_\star + \dot{\Sigma}_{\rm ext}\,.
\end{equation} 
The mass accretion rate within the disk is
\begin{equation} \label{mdot}
    \dot{M} = - 2\pi R \Sigma v_{\rm rot} = \frac{1}{v_{\rm rot}} \frac{\partial}{\partial R}\left( 2\pi \Sigma R^3 \frac{d\Omega}{dR}\right) \,.
\end{equation}
With the approximation $\partial/\partial R \sim R$ and $v_{\rm rot} = \Omega R$, one obtains:
\begin{equation}
    {\nu \Sigma = - \frac{\dot{M}}{2 \pi R} \Omega \left( \frac{d\Omega}{dR} \right)^{-1}}\, ,
\end{equation}
where the viscosity of the gas $\nu$ is defined as
\begin{equation}
 \label{eq:visc3}
    \nu = \sqrt{3} \vturb \ldriv\ .
\end{equation}
Contrary to \citet{2011AJ....141...24V}, where it was assumed that $\dot{\Sigma}_{\rm ext}\,=\,\dot{\Sigma}_\star$ and thus $\partial \Omega/\partial t\,=\,0$, and $\mdot = {cte}$, the mass accretion derived from the observational radial profiles varies with radius. We are mostly interested in the viscous timescale $\tauvis\,\sim\,R^2/\nu$ (see \sect{ngc6946s}).

\subsubsection{Self-gravitating clouds and star-formation \label{sec:clumpiness}}

The clumpiness of the model gas disk implies that the density of a single gas cloud $\rho_{\rm cl}$ depends directly on the average density of the disk $\rho$. In the model, these two quantities are linked by the volume filling factor $\phiv$, such that $\rho_{\rm cl} = \phiv^{-1} \rho$. 
Following \citet{2011AJ....141...24V} and \citet{2021A&A...645A.121V}, the star formation rate per unit volume is given by
\begin{equation} \label{starformation}
\rhosfr = \phiv \rho \tauffcl^{-1} 
\end{equation}
For self-gravitating clouds with a Virial parameter of unity, the turbulent crossing time $\tauturbcl$ equals twice the free-fall time $\tauturbcl$ (\citealt{2021A&A...645A.121V}):
\begin{equation} \label{tautau}
    \tauturbcl =\frac{\sqrt{3}}{2}\frac{l_{\rm cl}}{ v_{\rm turb,cl}} = 2 \tauffcl = \sqrt{\frac{3\pi\phiv}{32G\rho}}\ ,
\end{equation}
where $l_{\rm cl}$ and $v_{\rm turb,cl}$ are the size and the turbulent 3D velocity dispersion of a single gas cloud, respectively. Following Larson's law (\citealt{1981MNRAS.194..809L}), we can simplify the expression of the turbulent crossing time: 
\begin{equation} \label{larson}
    \frac{\sqrt{3}}{2}\frac{l_{\rm cl}}{ v_{\rm turb,cl}} =  \frac{\sqrt{3}}{2} \frac{\ldriv}{\vturb \sqrt{\delta}}\ ,
\end{equation}
where $\delta$ is the scaling between the driving length scale and the size of the
largest self-gravitating structures, such as $\delta = \ldriv/l_{\rm cl}$. All these considerations lead us to the second expression of the SFR in the model: 
\begin{equation} \label{starformation2}
    \rhosfr = \frac{4\sqrt{\delta}}{\sqrt{3}} \phiv \rho \frac{\vturb}{\ldriv}
\end{equation}
and $\dot{\Sigma}_*=\rhosfr \, l_{\rm driv}$.
This SFR recipe is close to the prescription suggested by \citet{2012ApJ...745...69K}
\begin{equation}
\label{eq:krumholz}
\dot{\Sigma}_*=f_{\rm H_2}\,\epsilon_{\rm ff} \frac{\Sigma}{t_{\rm ff}}\ ,
\end{equation}
where $\epsilon_{\rm ff}$ is the star formation efficiency per free-fall time. 
The relevant size scale for the density entering $t_{\rm ff}$ is that corresponding to the outer scale of the turbulence 
that regulates the SFR, which corresponds to $l_{\rm driv}$ in our model.
For a consistency check we directly compared the two SFR prescriptions and found that both have comparable slopes and normalizations (Fig.~\ref{fig:compasfr}).
\begin{figure*}[!ht]
  \centering
  \includegraphics[width=\hsize]{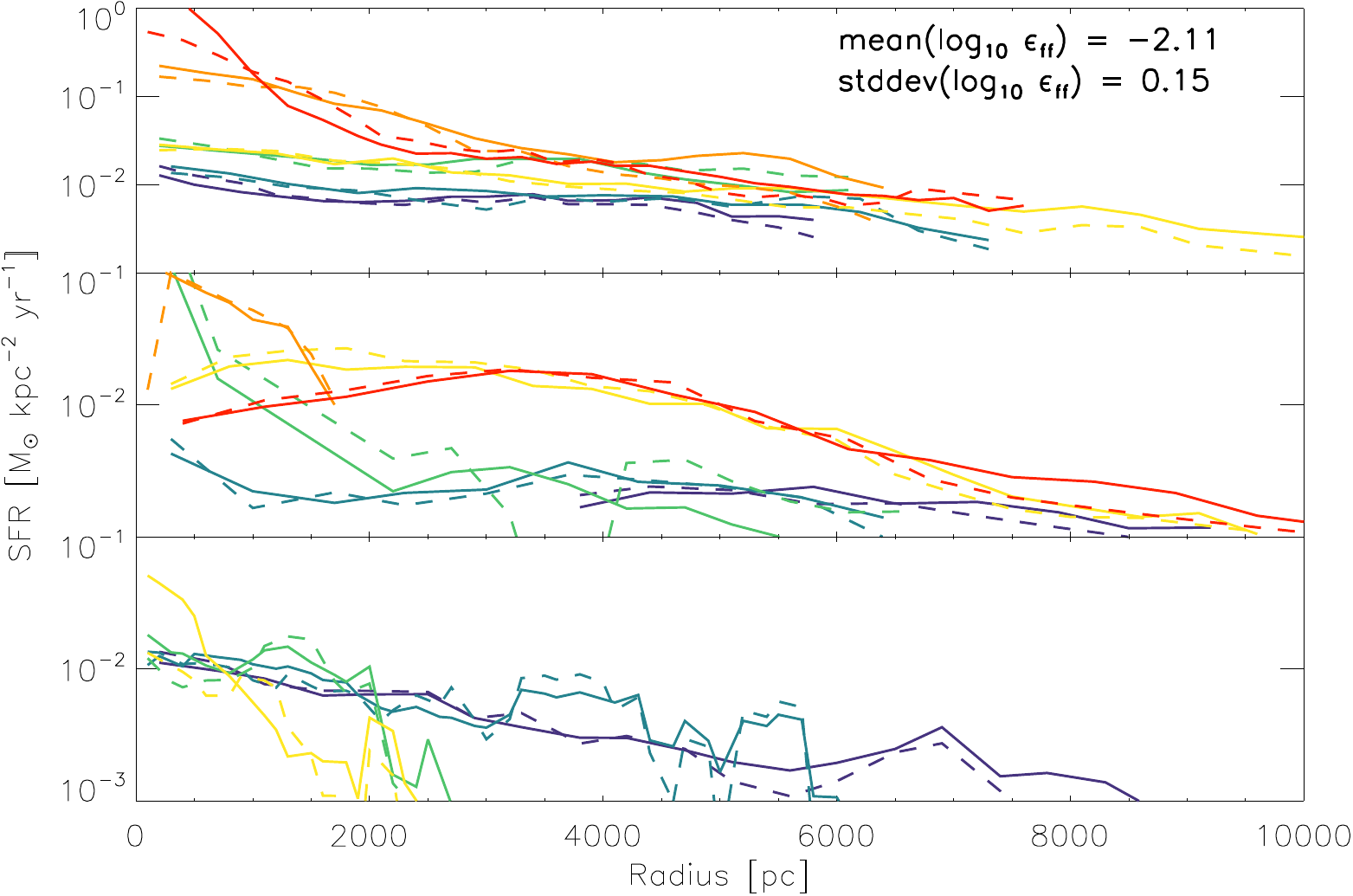}
  \caption{Star formation rate per unit area as a function of galactic radius. Upper panel: massive galaxies with EMPIRE data.
  Middle panel: massive galaxies without EMPIRE data. Lower panel: low-mass galaxies. Solid lines: our model SFR prescription
  (Eq.~\ref{starformation}). Dashed lines: \citealt{2012ApJ...745...69K} prescription; Eq.~\ref{eq:krumholz}.}
  \label{fig:compasfr}
\end{figure*}
We calculated the star formation efficiency per free-fall time by injecting our model SFR into Eq.~\ref{eq:krumholz} 
for each galaxy and found $\langle \epsilon_{\rm ff} \rangle = 0.8$\,\% for our sample, which is consistent with the values 
measured by \citet{2018ApJ...861L..18U} and \citet{2019ApJ...880..127J}.

\subsubsection{Metallicity}

One of the main assumptions of the model is that molecular clouds are relatively short-lived, appearing and disappearing in a cloud crossing time. They might not reach chemical equilibrium within their lifetime, which is about the free-fall time. We defined the characteristic time of H$_2$ formation as
\begin{equation} \label{taumol}
    \tau_{\rm mol} = \frac{\alpha}{\rho_{\rm cl}}\ ,
\end{equation}
where $\alpha$ is the constant of molecule formation that depends on the gas metallicity and temperature (\citealt{1996ApJ...468..269D}). The metallicity of the model is estimated using a leaky box model with an effective yield based on the gas fraction
\begin{equation} \label{molecule}
    \alpha = \alpha_0 \times \left( \ln{ \left( \frac{\Sigma_\star + \Sigma}{\Sigma}\right)}  \right)^{-1}
\end{equation}
where $\alpha_0 = 3.6 \times 10^7~\rm{yr~M_\odot~pc^{-3}}$. The relation between the metallicity and the constant of molecule formation is the following: 
\begin{equation} \label{metallicity}
    \frac{Z}{Z_\odot} = \frac{\alpha_\odot}{\alpha}\ ,
\end{equation}
where $\alpha_\odot = 2.2 \times 10^7~\rm{yr~M_\odot~pc^{-3}}$ (\citealt{1997ARA&A..35..179H}). The metallicity can be linked to the effective yield $y_{\rm eff}$ defined as $y_{\rm eff}$ = Z/ln(1/$f_{\rm gas})$, where $f_{\rm gas}$ is the gas fraction. \citet{2011AJ....141...24V} showed that these
metallicities are consistent with those measured by \citet{2010ApJS..190..233M}.

\subsubsection{Gas fragmentation and Toomre parameter Q}

One of the main parameters of the model is the Toomre parameter $Q$ (\citealt{1964ApJ...139.1217T}). It describes the stability of the gas disk regarding radial gas fragmentation: 
\begin{equation} \label{toomre}
    Q = \frac{\sigmaturb \kappa}{\pi G \Sigma}\ ,
\end{equation}
where $\kappa$ is the epicyclic frequency
\begin{equation}
    \kappa = \sqrt{4 \Omega^2 + R \frac{d\Omega^2}{dR}}\ .
\end{equation}
The Toomre $Q$ parameter is also used as a measure of the gas content of the disk, with $Q = 1$ for the maximum disk gas mass.

\subsection{Small-scale part}

In this subsection the physics that govern the small-scale part of the model are described. This involves scaling relations for the density and velocity dispersion of the gas clouds. 

\subsubsection{Mass fraction \label{sec:massfrac}}

The small-scale is divided into two distinct sub-scales, non-self-gravitating and self-gravitating gas clouds of density $\rho_{\rm cl}$. For each density, the mass fraction of the gas is determined by a lognormal probability distribution function (\citealt{1997MNRAS.288..145P}):
\begin{equation}
    p(x)dx = \frac{1}{x\sqrt{2 \pi \sigma^2}}\exp{\left(-\frac{(\ln{x + \sigma^2/2})^2}{2\sigma^2}\right)}\rm{d}x\ ,
\end{equation}
where $x = \rho_{\rm cl} / \rho$ is the overdensity and the standard deviation $\sigma$ is determined by the Mach number $\mathcal{M}$
\begin{equation}
    \sigma^2 \simeq \ln \left( {1 + (\mathcal{M}/2)^2} \right)\ .
\end{equation}
The mass fraction of gas with overdensities exceeding $x$ is thus defined as:
\begin{equation} \label{massfrac}
    \frac{\Delta M}{M} = \frac{1}{2} \left( 1 + \rm{erf}\left( \frac{\sigma^2 - 2\ln{x}}{2^\frac{3}{2}\sigma} \right) \right)\ .
\end{equation}
Based on the results of \citet{2014ApJ...780..173B}, who found that the fraction of the GMC mass residing within regions with densities higher than $\rm n \sim 10^3 cm^{-3}$ is about 10\%, and following \citet{2017A&A...599A..98P}, the mass fraction of self-gravitating cloud was decreased by a constant factor of 0.6. The inclusion of this factor was necessary to fit the observed integrated CO, HCN and HCO$^+$ line fluxes (\citealt{2017A&A...602A..51V}).

\subsubsection{ISM scaling relations \label{sec:scaling}}

We assumed different scaling relations for the two density regimes: (i) for non-selfgravitating clouds, we adopted the 
scaling relations found for galactic H{\sc i} by \citet{1983Ap&SS..93...37Q}: $\rho_{\rm cl} \propto l^{-2}$, $v_{\rm turb,cl} \propto l^{1/3}$, and 
thus $v_{\rm turb,cl} = v_{\rm turb}(\rho_{\rm cl}/\rho)^{-1/6}$, where $v_{\rm turb}$ and $\rho$ are the turbulent velocity and the density 
of the disk, respectively. Since the minimum density considered in this work is $\sim 100$~cm$^{-2}$, the maximum turbulent velocity of diffuse clouds 
is $\sim v_{\rm turb}/2 \sim 5$~km\,s$^{-1}$.
(ii) For self-gravitating clouds, we adopt the scaling relations of \citet{2010A&A...519L...7L}: $\rho_{\rm cl} \propto l^{-1.4}$,
$v_{\rm turb,cl} \propto l^{1/2}$.
As described in Sect.~\ref{sec:clumpiness}, the scale of the largest self-gravitating clouds $l_{\rm cl}$ is smaller
than the turbulent driving length scale $l_{\rm driv}$ by a factor $\delta=l_{\rm driv}/l_{\rm cl}$.
We assume that the turbulent velocity dispersion of the largest self-gravitating clouds of density $\rho_{\rm sg}$
is $v_{\rm turb,cl} = v_{\rm turb}/\sqrt{\delta}$, where $v_{\rm turb}$ is the velocity dispersion of the disk.
Furthermore, we assume $\rho_{\rm cl} \propto l^{-1}$ and 
$v_{\rm turb,cl} = v_{\rm turb}/\sqrt{\delta}(\rho_{\rm cl}/\rho_{\rm sg})^{-1/2} \propto l^{\frac{1}{2}}$ (\citealt{1987ApJ...319..730S}).

\subsubsection{Molecule abundances from chemical network \label{chem}}

For the determination of the H$_2$ column density of a gas cloud, we take into account
(i) the photo-dissociation of H$_2$ molecules and (ii) the influence of the finite cloud lifetime on the H$_2$ formation.
For the photo-dissociation of H$_2$ molecules, we follow the approach of \citet{2008ApJ...689..865K} and \citet{2009ApJ...693..216K}. In a second step, 
we take into account the molecular fraction due to the finite lifetime of the gas cloud
\begin{equation} 
\label{eq:lifetime}
f_{\rm mol}^{\rm life}=t_{\rm ff}^{\rm cl}/t_{\rm mol}^{\rm cl}/(1+t_{\rm ff}^{\rm cl}/t_{\rm mol}^{\rm cl})\ .
\end{equation}
{If the H$_2$ formation timescale $t_{\rm mol}$ is longer than the free-fall timescale, the gas cloud will be mainly atomic during its lifetime.
The inclusion of the cloud lifetime decreases the molecular gas surface density and thus the CO emission in the outer disks.
The best fit to the CO data then leads to a lower $Q$ and/or $\mdot$ resulting in a lower gas velocity dispersion compared
to the model without the inclusion of the cloud lifetime.}
The total molecular fraction of a cloud is $f_{\rm mol}=f_{\rm mol}^{\rm life} \times f_{\rm mol}^{\rm diss}$. The molecular fraction
due to the finite lifetime $f_{\rm mol}^{\rm life}$ has the highest influence on $f_{\rm mol}$ at large galactic radii.

The abundances of the different molecules are determined using the gas-grain code NAUTILUS presented in \citet{2009A&A...493L..49H}. This code computes the abundances of chemical species as a function of time by solving the rate equations for a network of reactions. The input parameters are the crossing time, density, gas temperature, UV flux, cosmic ray ionization rate, and the initial elemental abundances of the model gas clouds. We assume a
UV flux that is proportional to the large-scale star formation rate and a constant cosmic ray ionization rate. The gas temperature of the clouds
is calculated by the equilibrium between gas heating and cooling (Sect.~\ref{sec:heatcool}).

\subsubsection{Heating and cooling mechanisms \label{sec:heatcool}}

Gas heating is provided via two distinct mechanisms: (i) turbulent mechanical heating and (ii) radiation heating via cosmic rays ionization. The main mechanisms for gas cooling are CO and H$_2$ line emission. 

The thermal balance of gas and dust is determined by gas-dust collisions in the model. To determine this balance, we simultaneously solved the following equations:
\begin{equation} \label{gascooling}
    \Gamma_{\rm{g}} - \Lambda{\rm{g}} - \Lambda_{\rm{gd}} = 0
\end{equation}
and
\begin{equation} \label{dustcooling}
    \Gamma_{\rm{d}} - \Lambda_{\rm{d}} + \Lambda_{\rm{gd}} = 0 \ ,
\end{equation}
where $\Lambda_{\rm{d}}$ is the dust cooling rate, $\Lambda_{\rm{g}}$ corresponds to the molecular line cooling, $\Gamma_{\rm{d}}$ is the radiative heating of dust grains, $\Gamma_{\rm{g}}$ the heating via turbulence and cosmic rays, and $\Lambda_{\rm{gd}}$ the dust cooling energy transfer between dust and gas due to collisions.

\subsubsection{Brightness temperatures and molecular line emission \label{sec:brightco}}

The molecular line emission is computed using the brightness temperature formalism. The difference in brightness temperature between on- and off- positions is given by
\begin{equation} \label{ta}
    \Delta T^*_{\rm A} = (1 - e^{-\tau}) \frac{h\nu}{k} \left(  \frac{1}{e^{h\nu/kT_{\rm ex}}-1} - \frac{1}{e^{h\nu/kT_{\rm bg}}-1} \right)
\end{equation}
where $\tau$ is the optical depth of the line, $\nu$ the frequency of the observations, $h$ and $h$ the Planck and Boltzmann constants, and $T_{\rm ex}$ and $T_{\rm bg}$ the excitation and background brightness temperatures. 

For simplicity only a single collider (H$_2$) is considered. The excitation temperature is
\begin{equation}
\frac{1}{T_{\rm ex}}=\big(\frac{1}{T_{\rm g}}+(\frac{A_{ul}}{n q_{ul}}\frac{T_{\rm bg}}{T_*})\frac{1}{T_{\rm bg}}\big)/(1+\frac{A_{ul}}{nq_{ul}}\frac{T_{\rm bg}}{T_*})\ ,
\end{equation}
where $T_*=h \nu_{ul}/k$, $n$ is the gas density, $nq_{ul}$ the collisional de-excitation rate, and $A_{ul}$ the Einstein coefficients of 
the transition $ul$. The background brightness temperature $T_{\rm bg}$ is the sum
of the effective emission temperatures of the galaxy's dust $T_{\rm eff\ dust}$ and the cosmic background at the 
galaxy redshift $T_{\rm CMB}$ (see Eq.~17 of \citealt{2013ApJ...766...13D}). 
For optically thin transitions the ratio of the radiative and collisional rates is the ratio of 
the density to the critical density for a given transition
\begin{equation}
n_{\rm crit}=\frac{A_{ul}}{q_{ul}}\ .
\end{equation}

We consider two-level molecular systems, in which the level populations are determined by
a balance of collisions with H$_2$, spontaneous decay and line photon absorption, and stimulated emission with
$\tau > 1$.  The molecular abundances were calculated by NAUTILUS (see Sect.~\ref{chem}).
For simplicity, we neglected the hyperfine structure of HCN. 

The rotation constants, Einstein coefficients, and collision rates were taken from the Leiden Atomic and 
Molecular Database (LAMDA; \citealt{2005yCat..34320369S}). The CO collision rates were provided by \citet{2010ApJ...718.1062Y}.
The HCN collision rates were taken from the He--HCN rate coefficients calculated by \citet{2010MNRAS.406.2488D}, 
scaled by a factor of $1.36$ to go to HCN--H$_2$ (see \citealt{1976ApJ...205..766G}). The HCO$^+$ collision rates were taken from \citet{1999MNRAS.305..651F}.

We compared the brightness temperatures obtained by our approximate recipe to those calculated with the statistical equilibrium radiative transfer code RADEX from \citet{2007A&A...468..627V}. RADEX is a one-dimensional non-LTE radiative transfer code that uses the escape probability formulation assuming an isothermal and homogeneous medium without large-scale velocity fields. We systematically specify when the results presented were obtained using RADEX.

The integrated line emission, in K km\,s$^{-1}$, is computed using the following expression:
\begin{equation} \label{ccoo}
  W = 2.35 \sum_{i=1}^N (\Delta T_{\rm A}^{\star})_i  (\phi_{\rm A})_i (v_{\rm turb,cl})_i (\frac{\Delta M}{M})_i\ ,
\end{equation}
where $T_{\rm A}^{\star}$ is the brightness temperature, $\phi_{\rm A}$ is the surface filling factor, $\frac{\Delta M}{M}$ is the mass fraction as defined in \eq{massfrac}, and v$_{\rm turb,cl}$ is the turbulent velocity of the cloud at a given density. The factor $2.35$ links the turbulent velocity to the linewidth. CO photo-dissociation is taken into account following \citet{2010ApJ...716.1191W}.

\subsubsection{Interstellar radiation field and cosmic ray ionization}

The stellar radiation field is set by the SFR (Sect.~\ref{sec:sfr1}) and stellar mass radial profiles $\Sigma_*$:
\begin{equation}
\frac{F}{F_0}=k \times \big( \frac{\dot{\Sigma_*}}{10^{-8}~{\rm M}_{\odot}{\rm pc}^{-2}{\rm yr}^{-1}} + \frac{\Sigma_*}{40~{\rm M_{\odot}pc^{-2}}} \big)\ , 
\end{equation}
where $F_0=5.3 \times 10^{-3}$~ergs\,cm$^{-2}$s$^{-1}$ (\citealt{2001ApJ...557..736G}).
The constant cosmic ray ionization rate is constrained by the observed HCO$^+$(1–0) emission.
In practice, we calculated models with different ionization rates and chose the value that yielded the best fit to the HCO$^+$(1-0) radial profile.

\subsubsection{Thermal dust emission \label{sec:dustemission}}

The dust temperature $T_{\rm d}$ of a gas cloud of a given density and size illuminated by a local mean radiation field is calculated by
solving Eq.~\ref{dustcooling}. With the dust mass absorption coefficient of  $\kappa(\lambda)=\kappa_0(\lambda_0/\lambda)^{\beta}$, the
dust optical depth is
\begin{equation}
\tau(\lambda)= \kappa(\lambda)\,\Sigma_{\rm cl} (GDR)^{-1}\ ,
\end{equation}
where $\Sigma_{\rm cl}$ is the cloud surface density in g/cm$^2$. We used $\kappa_0(250~\mu{\rm m})=4.8$~cm$^2$g$^{-1}$ (\citealt{2001ApJ...554..778L}, 
\citealt{2012ApJ...745...95D}) and $\beta=1.5$. The absorption coefficient at $160$~$\mu$m is $\kappa_0(160~\mu{\rm m})=9.4$~cm$^2$g$^{-1}$.
This value is consistent with but at the lower end of the range found in the literature  $\kappa_0(160~\mu{\rm m})=10$-$15$~cm$^2$g$^{-1}$
(\citealt{2001ApJ...548..296W}, \citealt{2004ApJS..152..211Z}, \citealt{2007ApJ...657..810D}, \citealt{2014ApJ...797...85G}).

The infrared emission at a given wavelength at a given galactic radius
$R$ is calculated in the following way:
\begin{equation}
\label{eq:idust}
I_{\rm dust}(\lambda)=\sum_{\rm i=1}^{\rm N}  \big(\Phi_{\rm A} \big)_i\, (\frac{\Delta M}{M})_i\, \big(1-\exp(-\tau(\lambda))\big)_i B(\lambda,T_{\rm d})_i\ ,
\end{equation}
where $B(\lambda,T_{\rm d})$ is the Planck function and $\Phi_{\rm A}=1.5\,(\Delta M/M)\,(\Sigma/\Sigma_{\rm cl})$ the area filling factor.
The factor $1.5$ takes into account that the mean cloud surface density is $1.5$ times lower than the surface density in the cloud center
$\Sigma_{\rm cl}=\rho_{\rm cl}l_{\rm cl}$.

\section{Best-fit models \label{appendix}}

The model and observed H{\sc i}, SFR, and molecular line and IR emission radial profiles are presented in the upper part 
of each figure for the different galaxies of our sample. The model parameters and derived physical quantities
are shown in the lower parts of the figures. 
\begin{figure*}
\center
\begin{subfigure}{.4\textwidth}
  \centering
  \caption{NGC~628 best model with $\chi^2_{\rm CO10}$}
  \includegraphics[width=1.\linewidth]{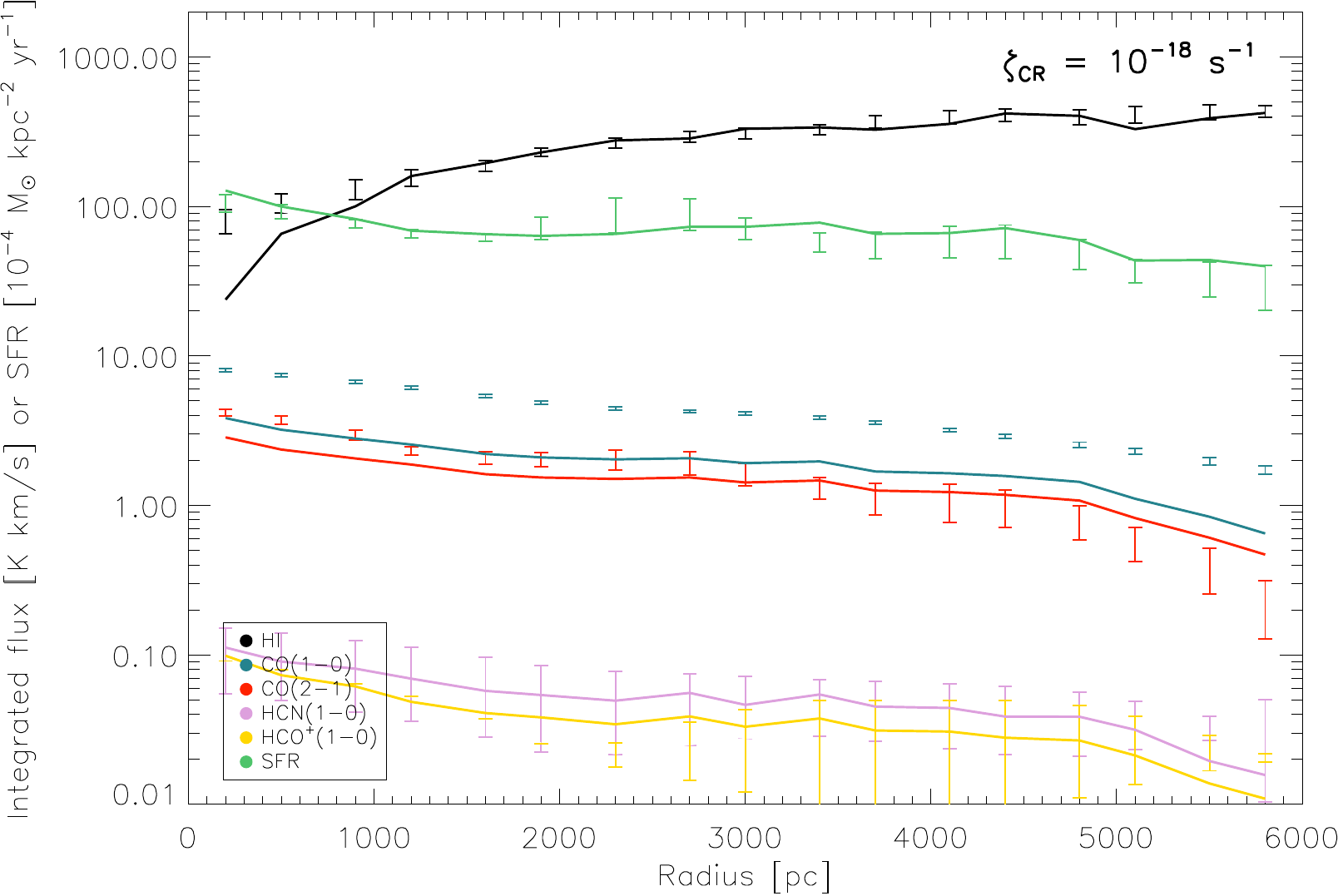}
\end{subfigure}%
\begin{subfigure}{.4\textwidth}
  \centering
  \caption{NGC~628 best model without $\chi^2_{\rm CO10}$}
  \includegraphics[width=1.\linewidth]{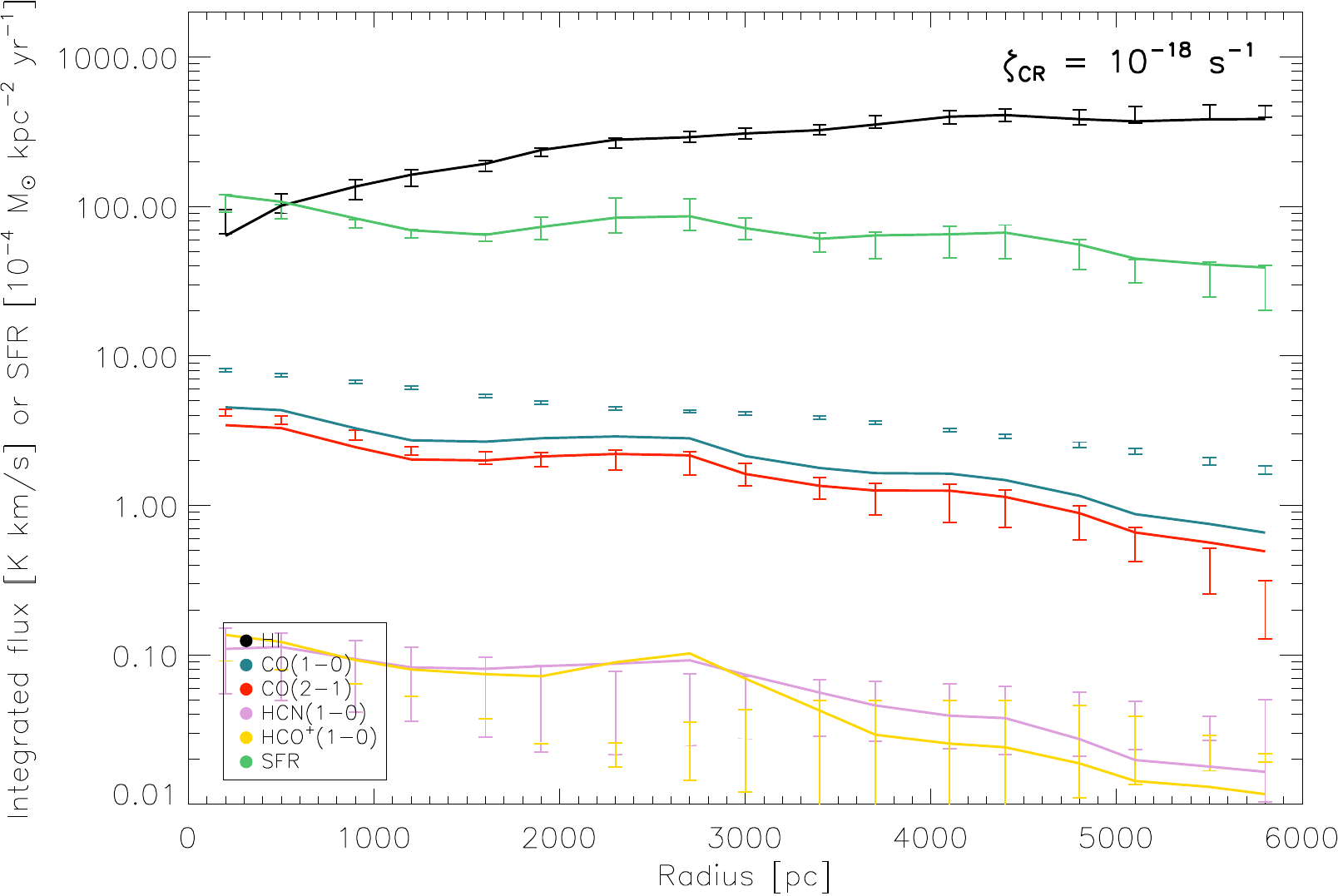}
\end{subfigure}

\begin{subfigure}{.4\textwidth}
  \centering
  \caption{NGC~628 infrared profiles with $\chi^2_{\rm CO10}$}
  \includegraphics[width=1.\linewidth]{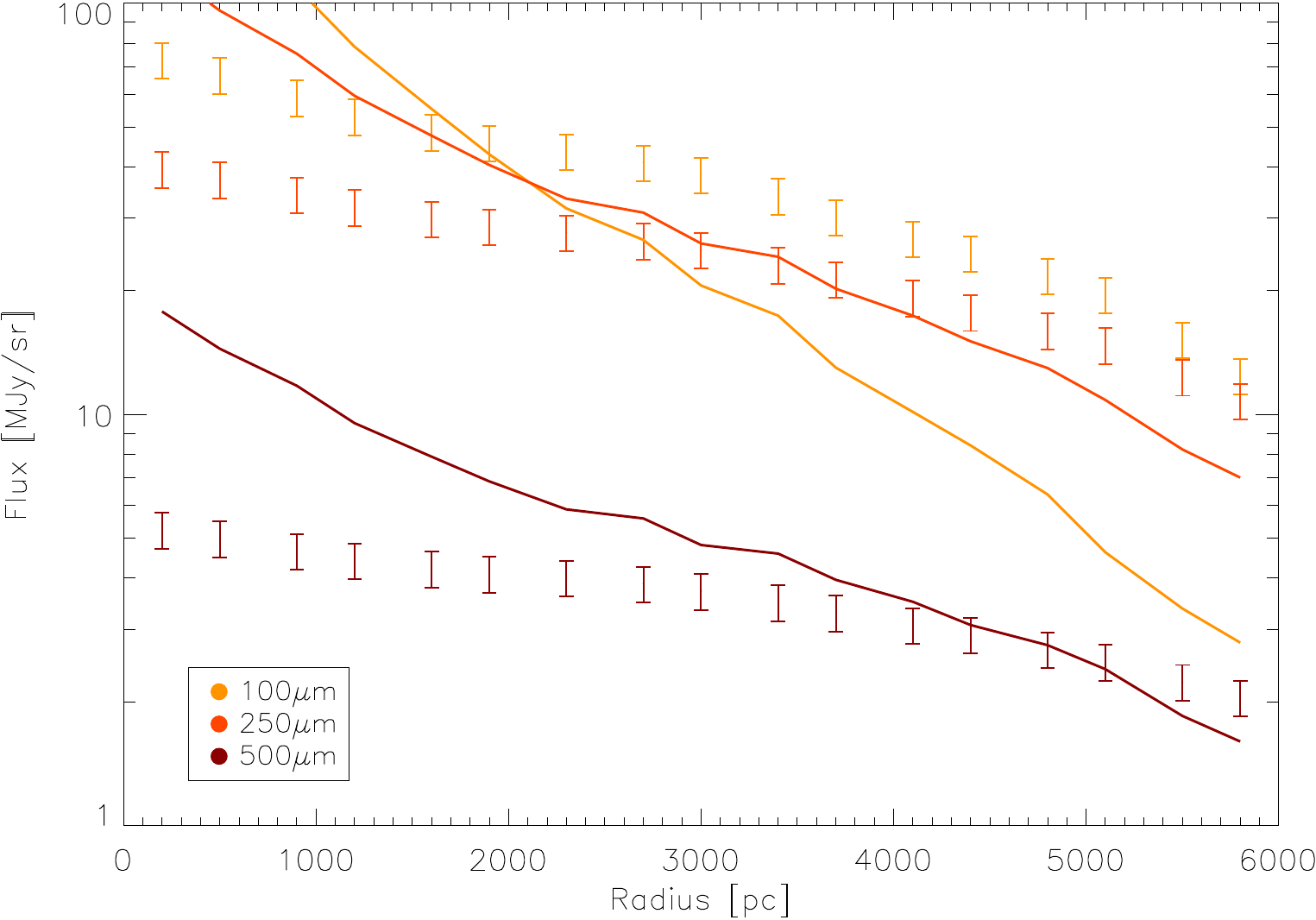}
\end{subfigure}%
\begin{subfigure}{.4\textwidth}
  \centering
  \caption{NGC~628 infrared profiles without $\chi^2_{\rm CO10}$}
  \includegraphics[width=1.\linewidth]{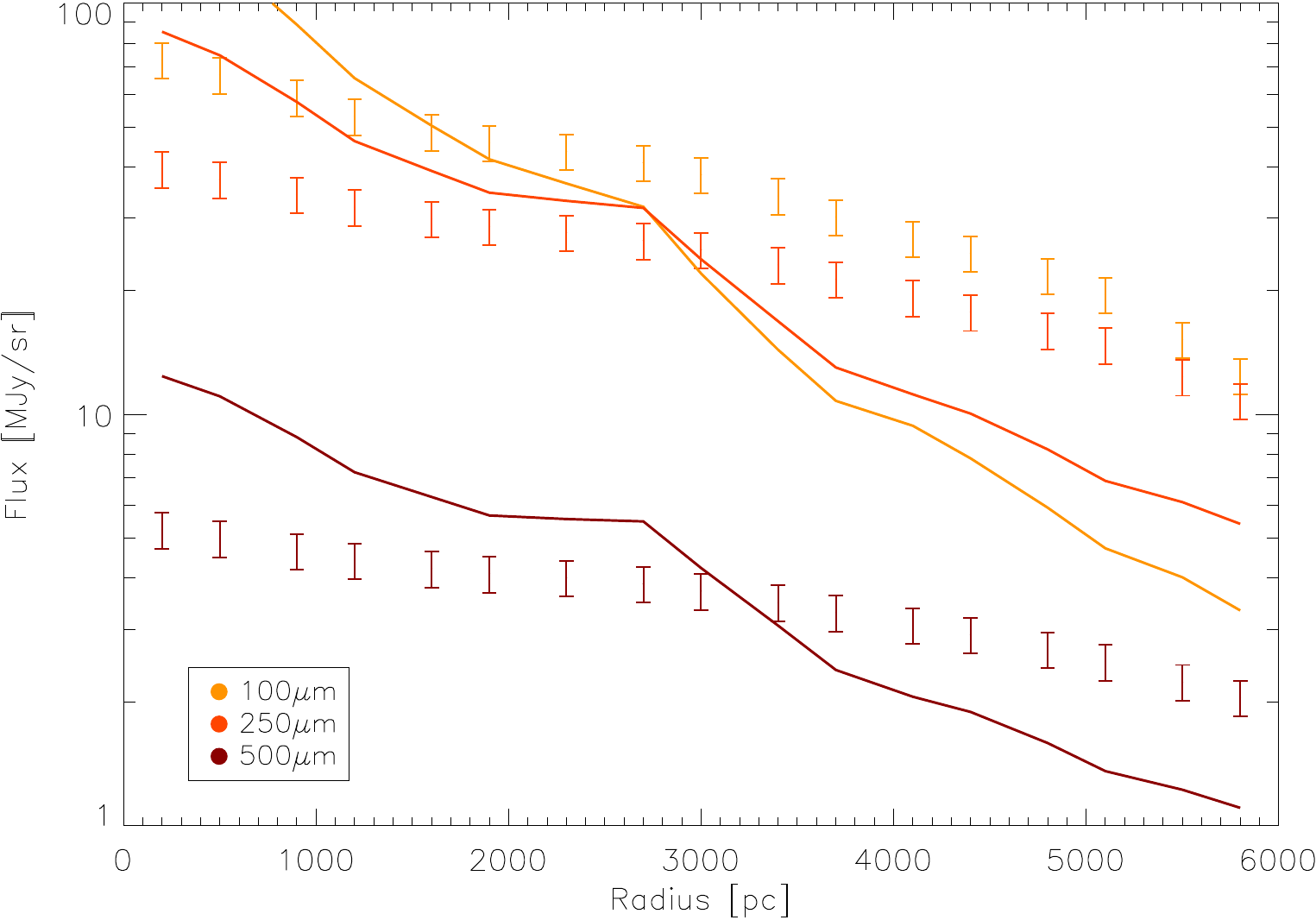}
\end{subfigure}

\begin{subfigure}{.4\textwidth}
  \centering
  \caption{Main properties with $\chi^2_{\rm CO10}$.}
  \includegraphics[width=1.\linewidth]{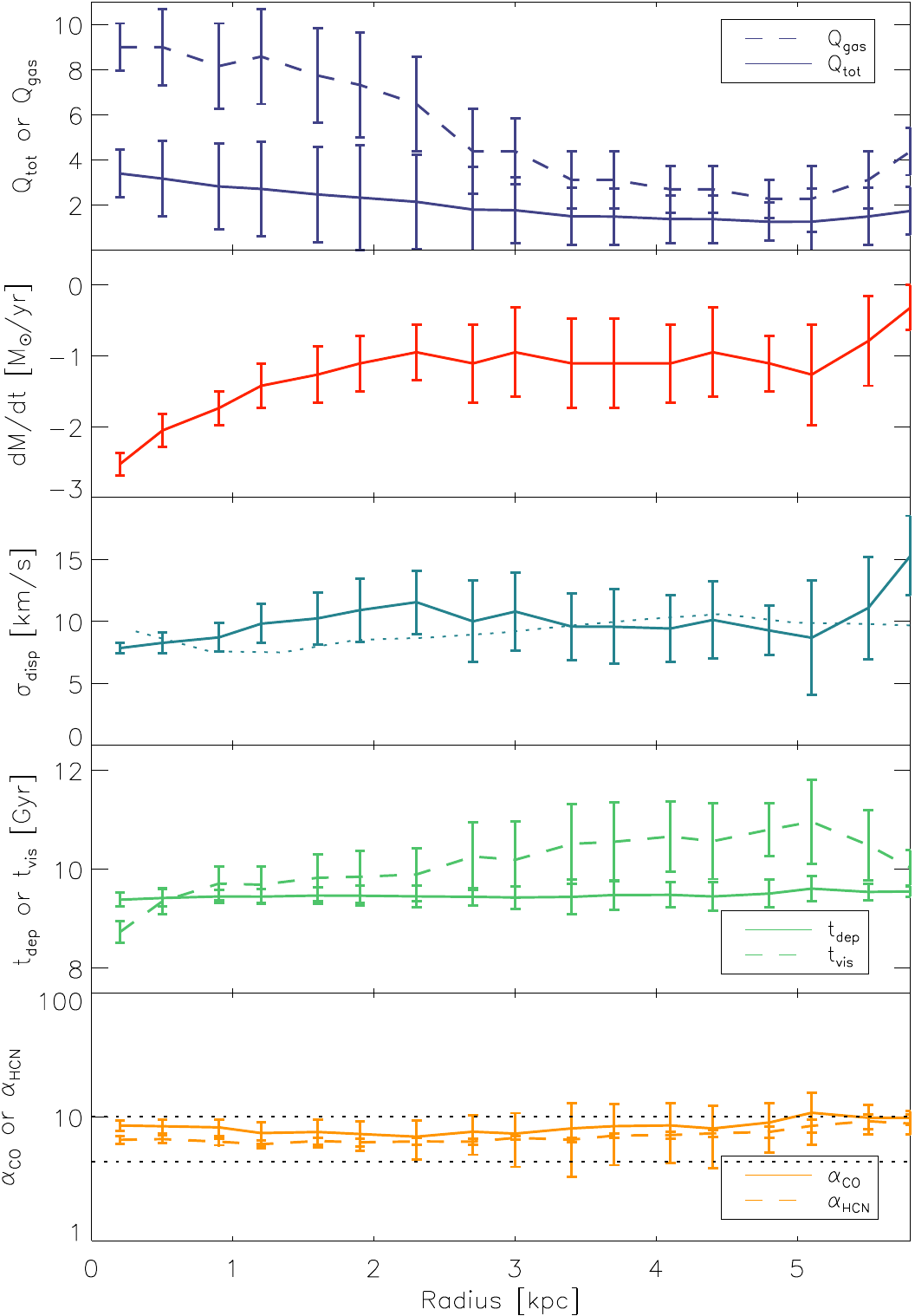}
\end{subfigure}%
\begin{subfigure}{.4\textwidth}
  \centering
  \caption{Main properties without $\chi^2_{\rm CO10}$.}
  \includegraphics[width=1.\linewidth]{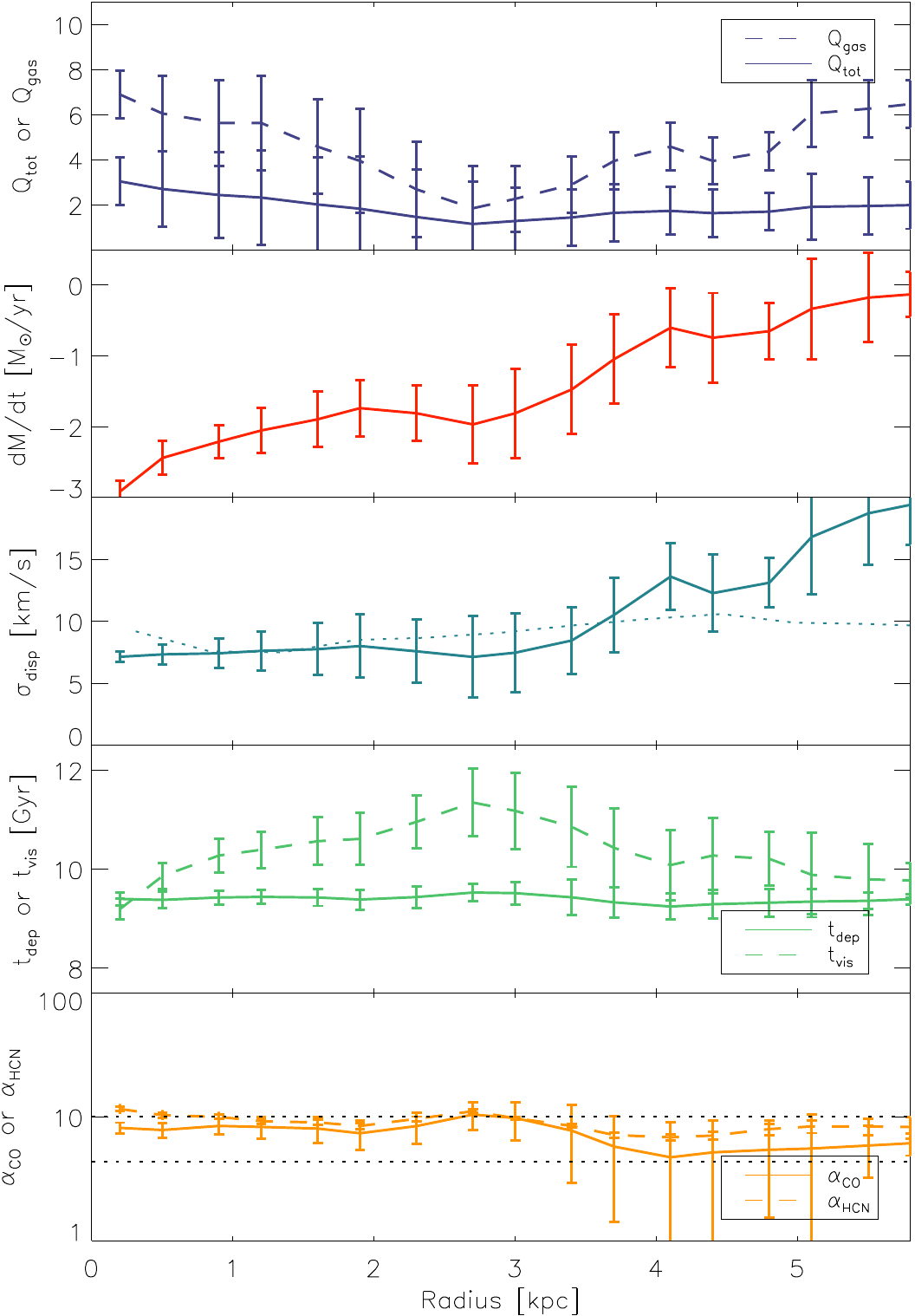}
\end{subfigure}

\caption{NGC~628 best-fit models, infrared profiles and radial profiles of main physical quantities.
See Figs.~\ref{ngc6946} and \ref{ngc6946b} for explanations.}
\label{fig:n628}
\end{figure*}

\begin{figure*}
  \center
  \begin{subfigure}{.45\textwidth}
    \centering
    \caption{NGC~3184 best model with $\chi^2_{\rm CO10}$}
    \includegraphics[width=1.\linewidth]{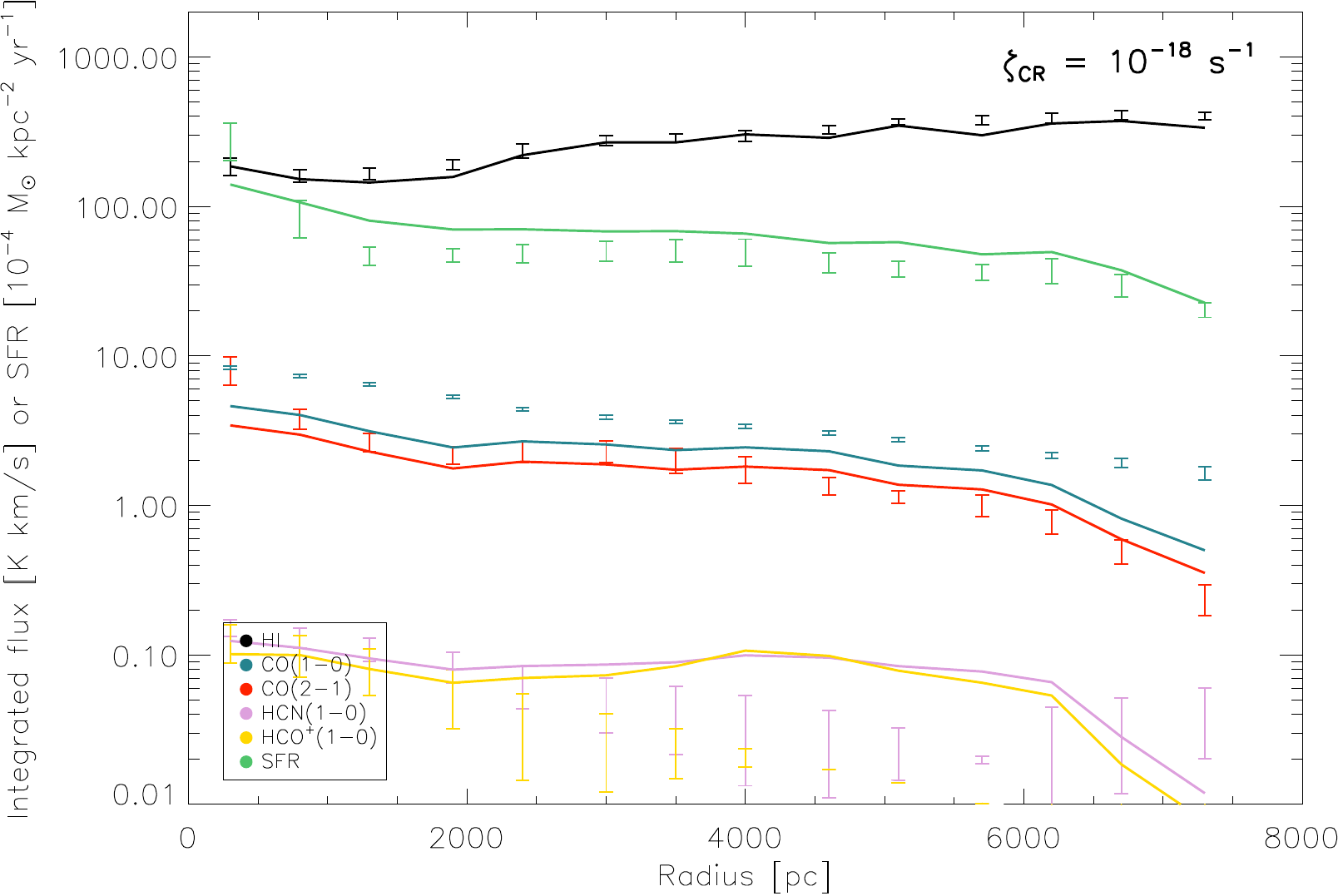}
  \end{subfigure}%
  \begin{subfigure}{.45\textwidth}
    \centering
    \caption{NGC~3184 best model without $\chi^2_{\rm CO10}$}
    \includegraphics[width=1.\linewidth]{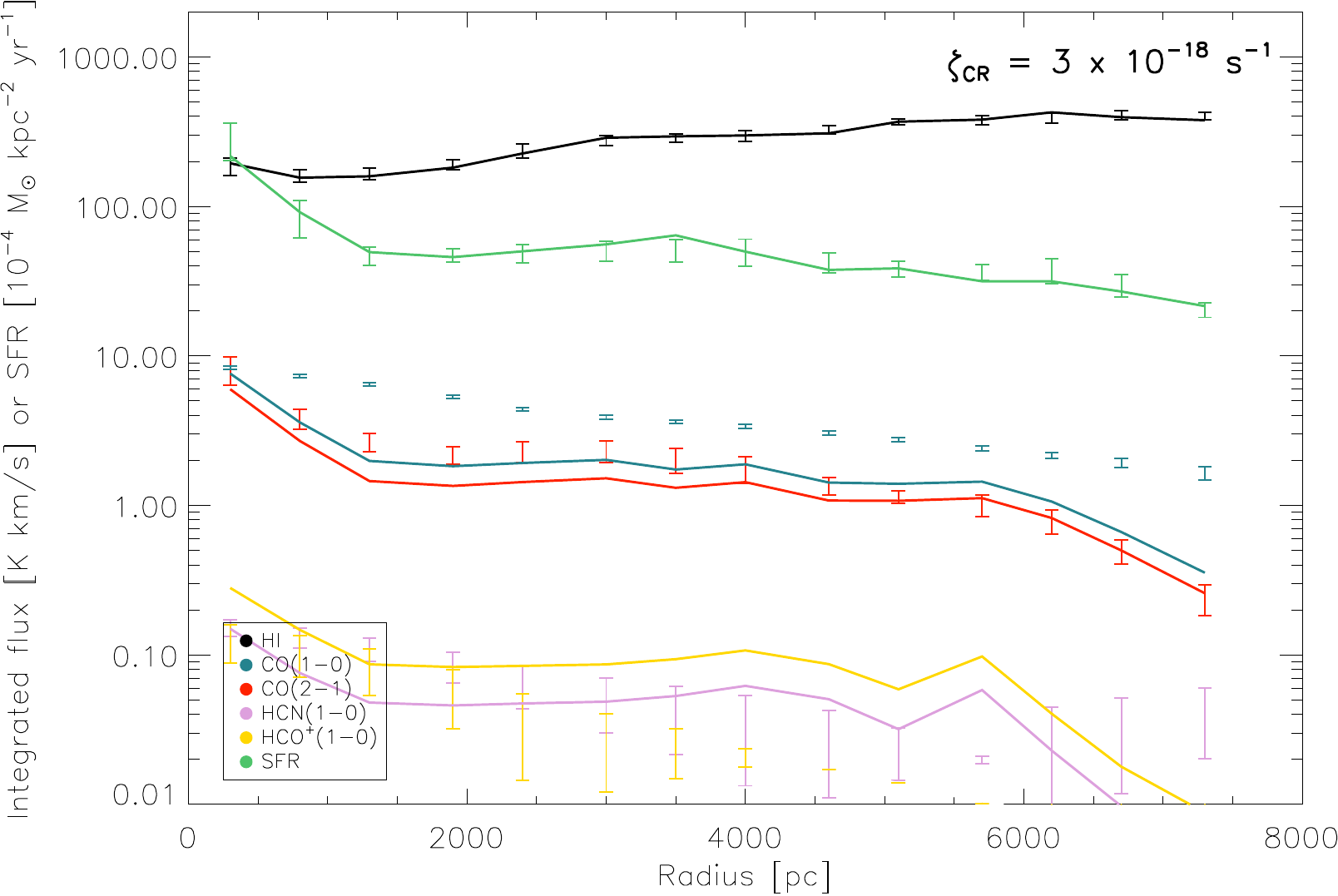}
  \end{subfigure}
  
  \begin{subfigure}{.45\textwidth}
    \centering
    \caption{NGC~3184 infrared profiles with $\chi^2_{\rm CO10}$}
    \includegraphics[width=1.\linewidth]{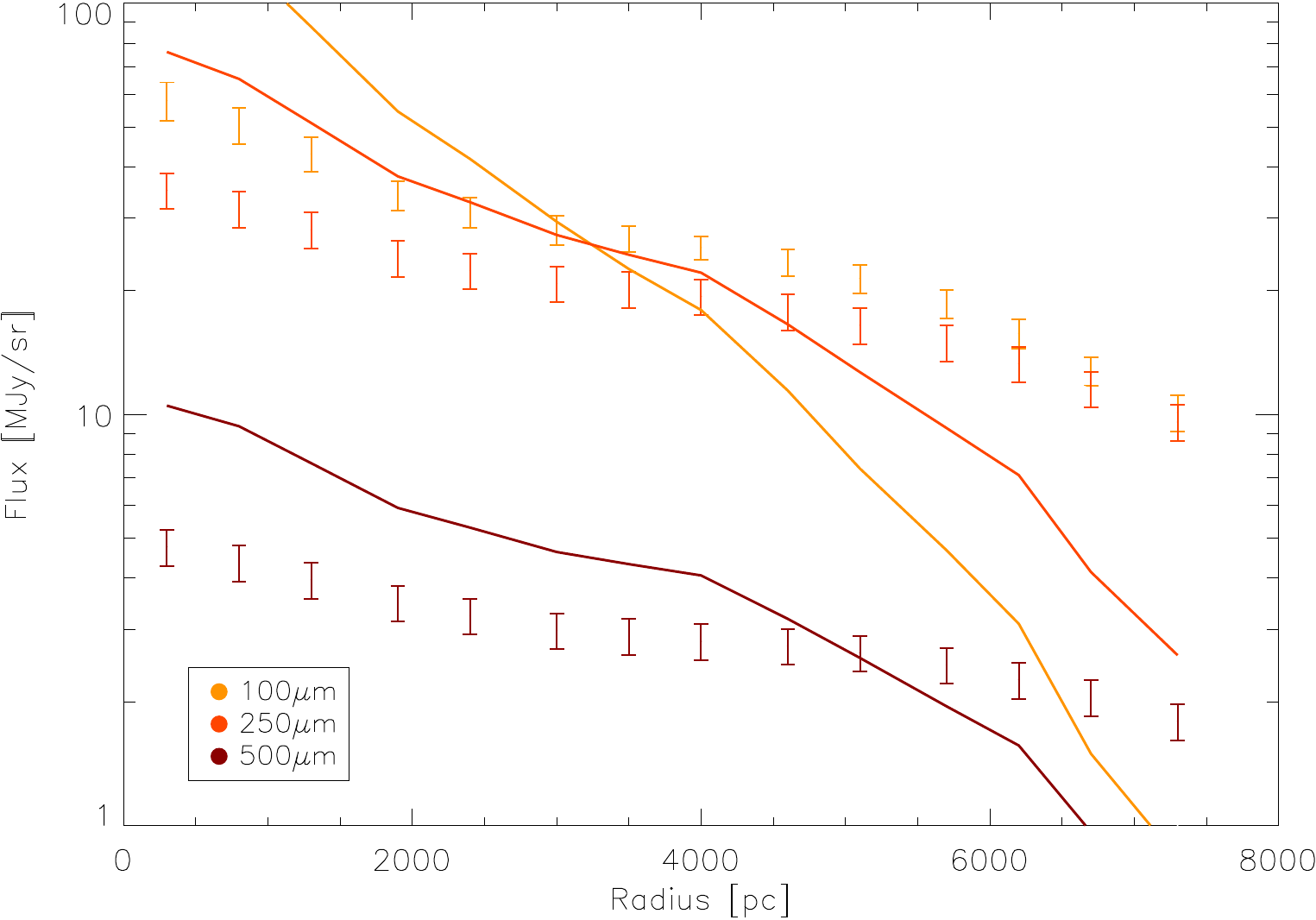}
  \end{subfigure}%
  \begin{subfigure}{.45\textwidth}
    \centering
    \caption{NGC~3184 infrared profiles without $\chi^2_{\rm CO10}$}
    \includegraphics[width=1.\linewidth]{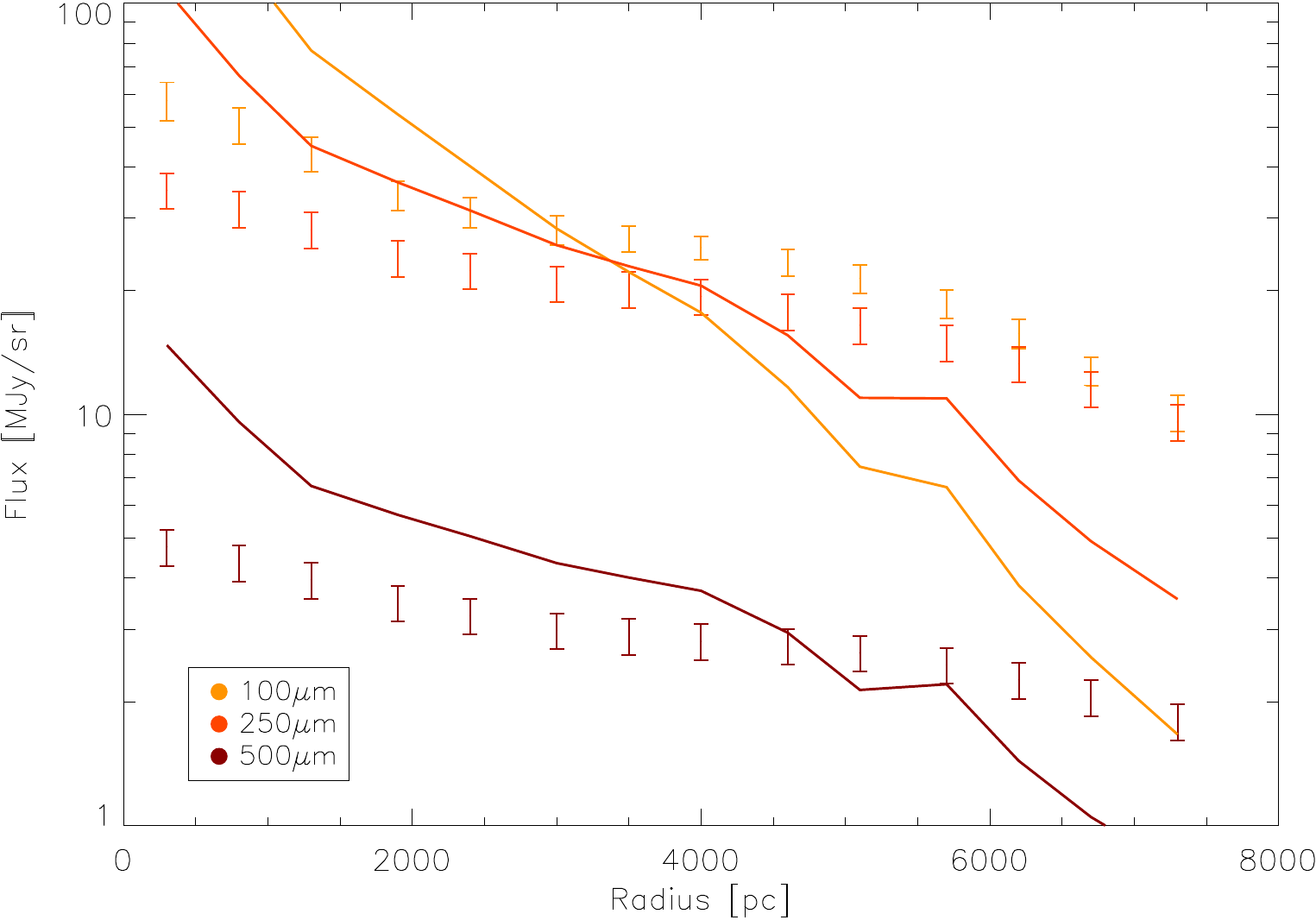}
  \end{subfigure}
  
  \begin{subfigure}{.45\textwidth}
    \centering
    \caption{Main properties with $\chi^2_{\rm CO10}$.}
    \includegraphics[width=1.\linewidth]{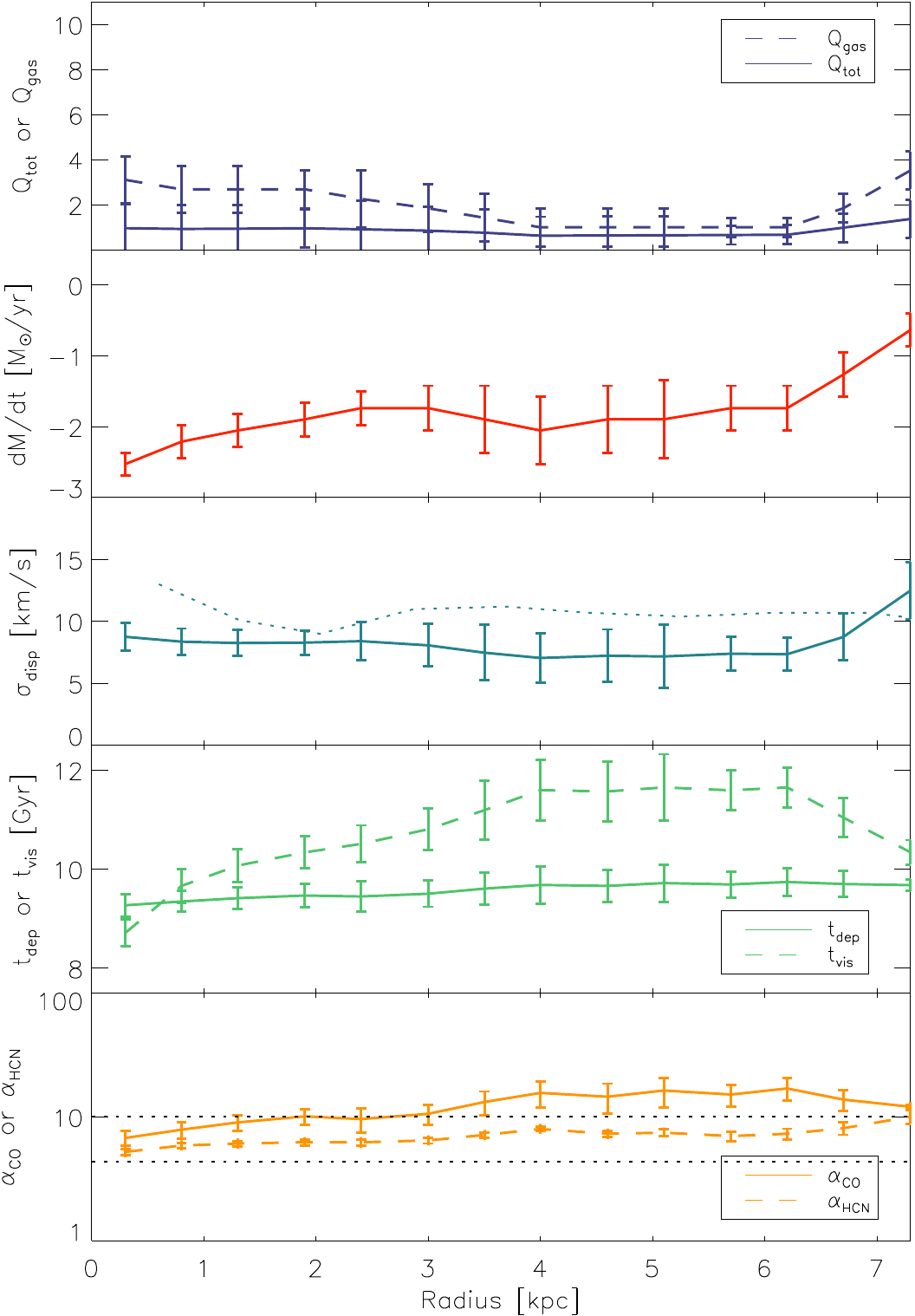}
  \end{subfigure}%
  \begin{subfigure}{.45\textwidth}
    \centering
    \caption{Main properties without $\chi^2_{\rm CO10}$.}
    \includegraphics[width=1.\linewidth]{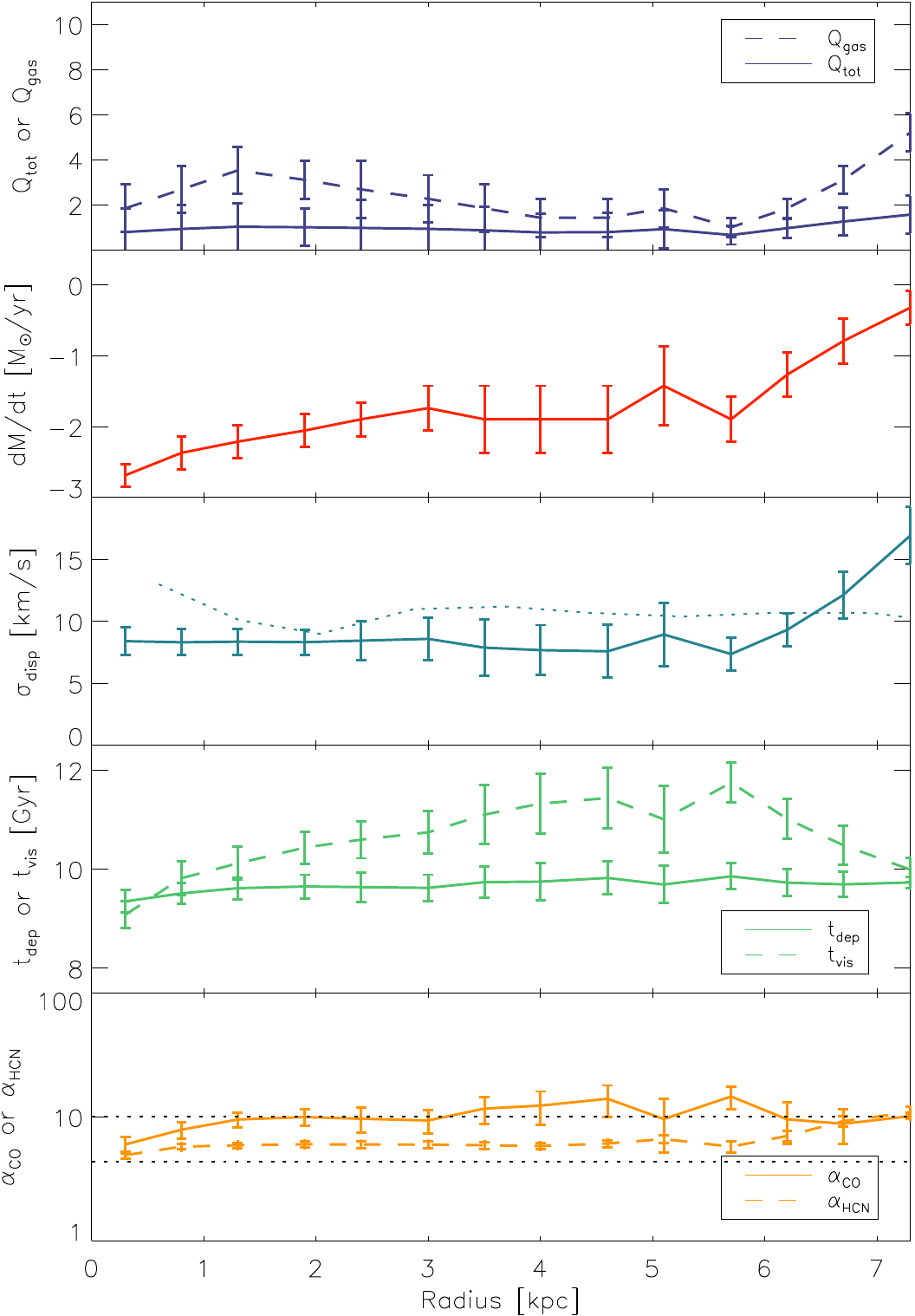}
  \end{subfigure}
  
  \caption{Same as Fig.\ref{fig:n628} for NGC~3184.}
  \label{fig:n3184}
  \end{figure*}

  \begin{figure*}
    \center
    \begin{subfigure}{.45\textwidth}
      \centering
      \caption{NGC~3627 best model with $\chi^2_{\rm CO10}$}
      \includegraphics[width=1.\linewidth]{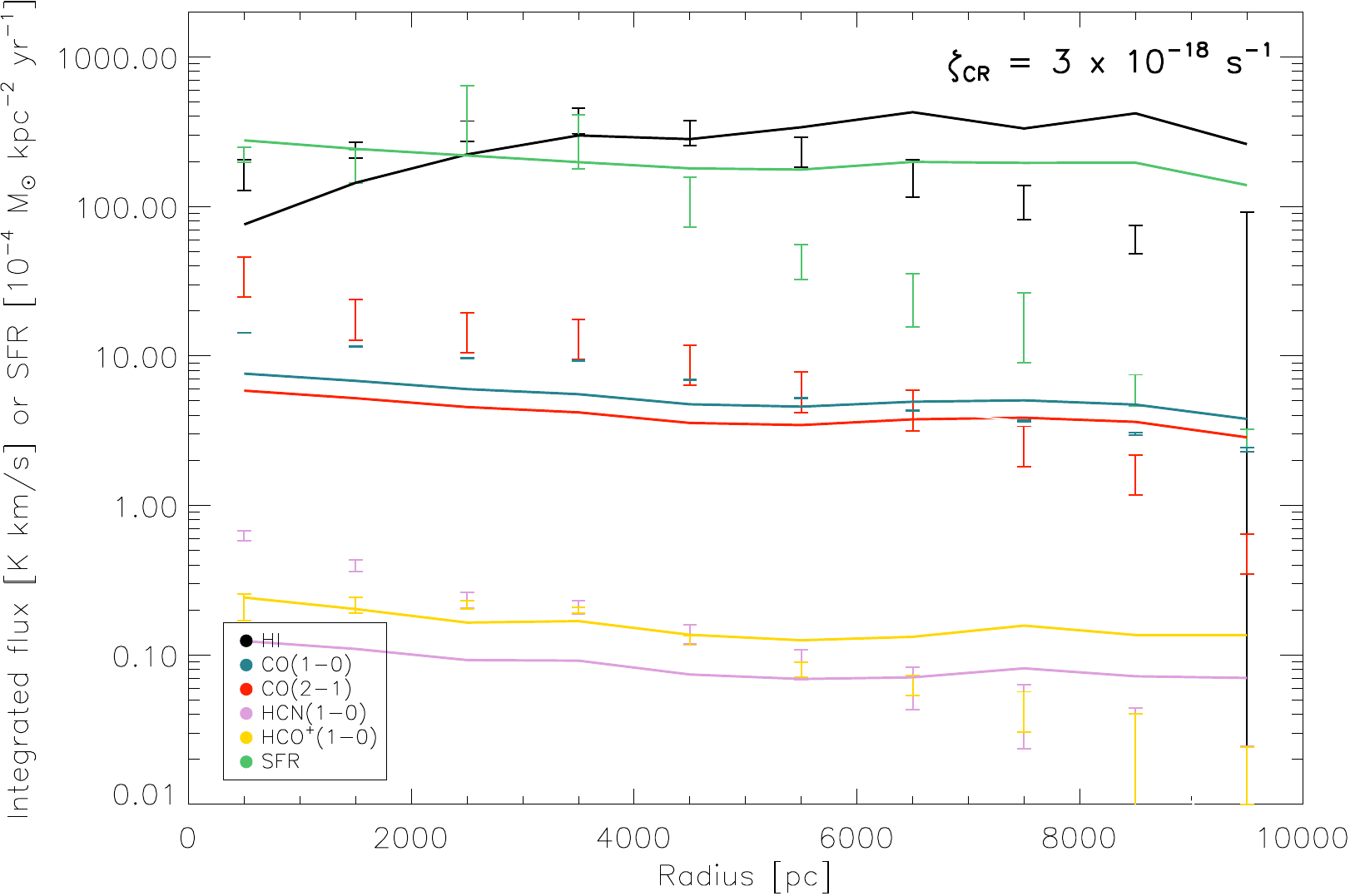}
    \end{subfigure}%
    \begin{subfigure}{.45\textwidth}
      \centering
      \caption{NGC~3627 best model without $\chi^2_{\rm CO10}$}
      \includegraphics[width=1.\linewidth]{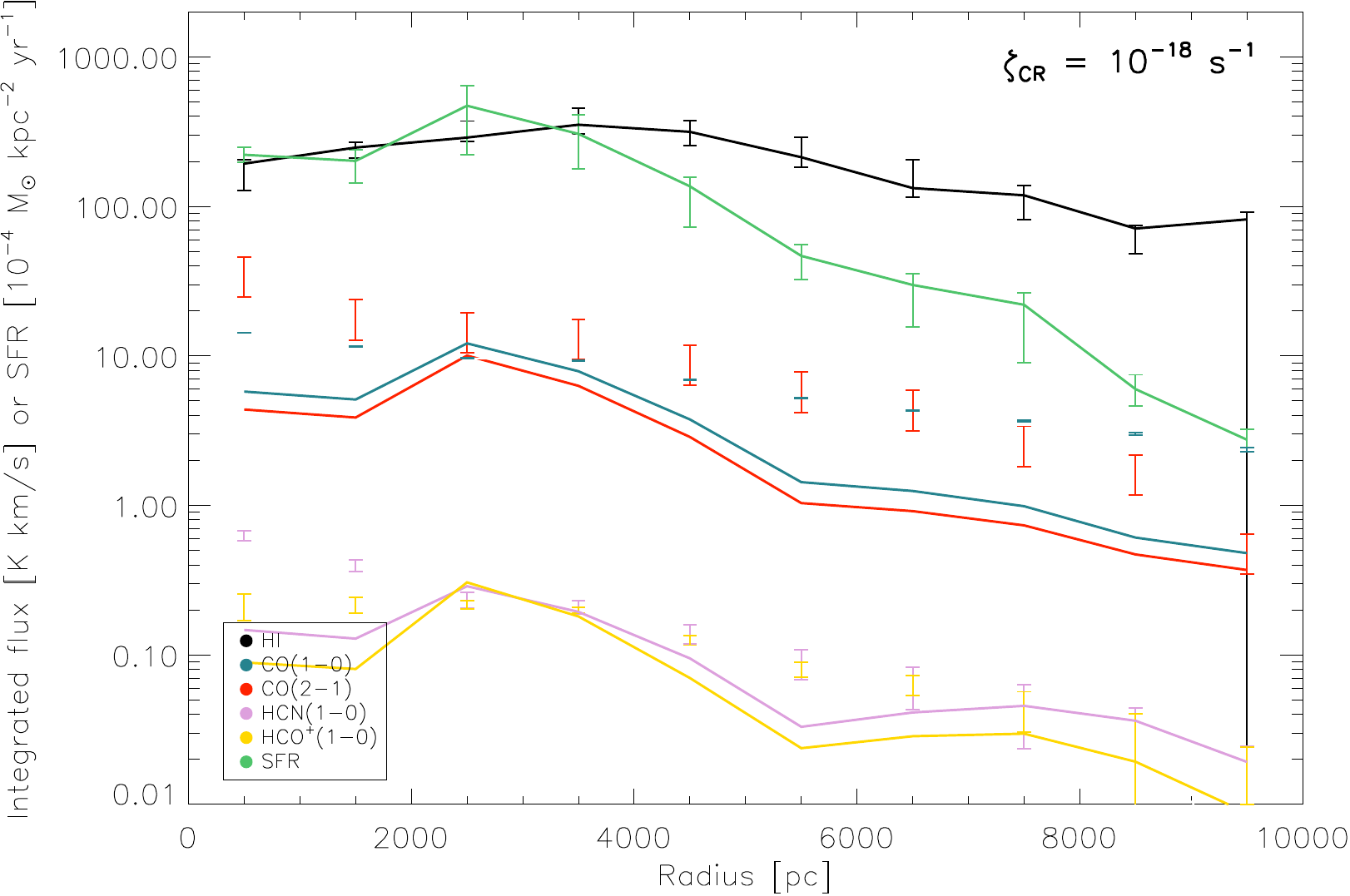}
    \end{subfigure}
    
    \begin{subfigure}{.45\textwidth}
      \centering
      \caption{NGC~3627 infrared profiles with $\chi^2_{\rm CO10}$}
      \includegraphics[width=1.\linewidth]{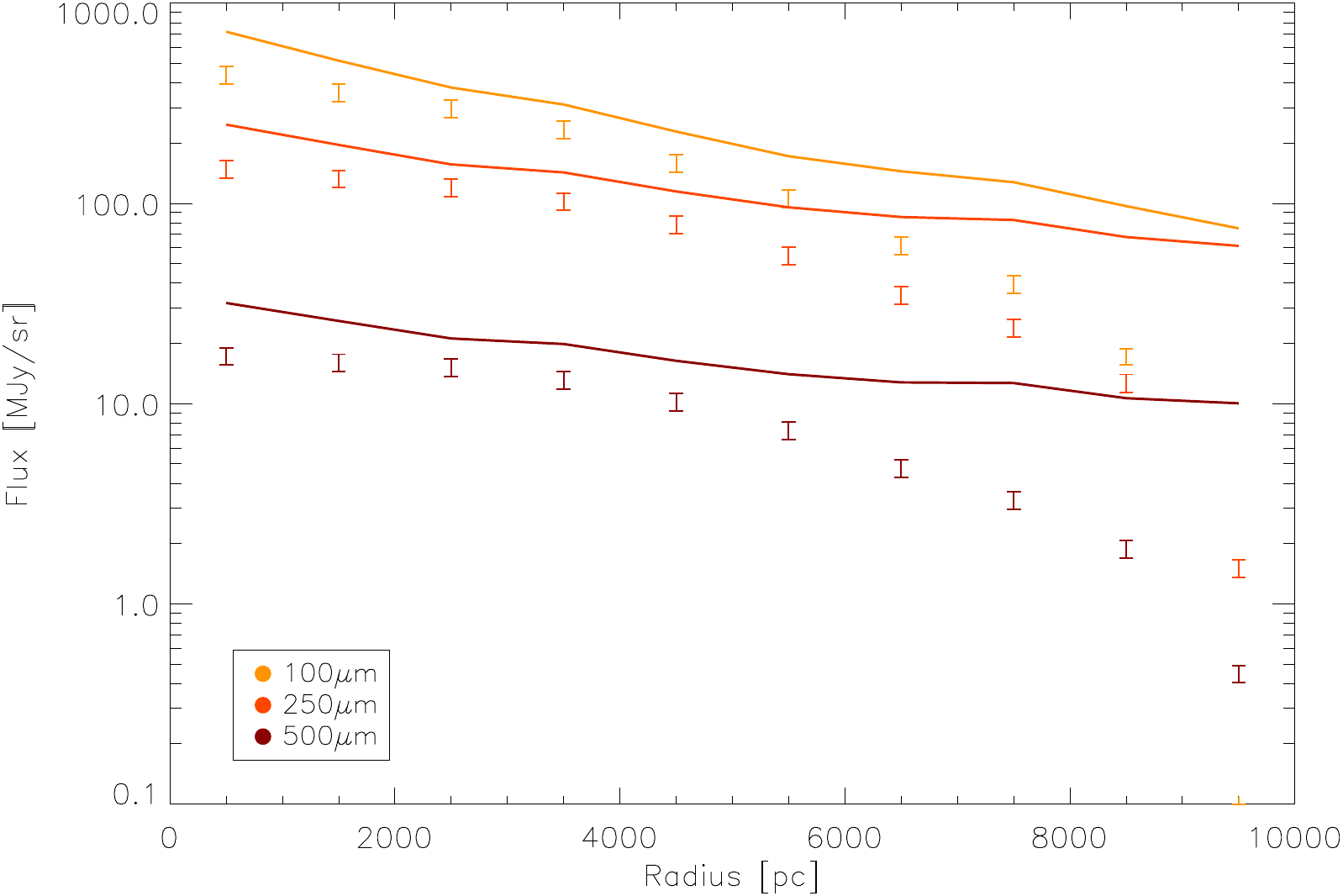}
    \end{subfigure}%
    \begin{subfigure}{.45\textwidth}
      \centering
      \caption{NGC~3627 infrared profiles without $\chi^2_{\rm CO10}$}
      \includegraphics[width=1.\linewidth]{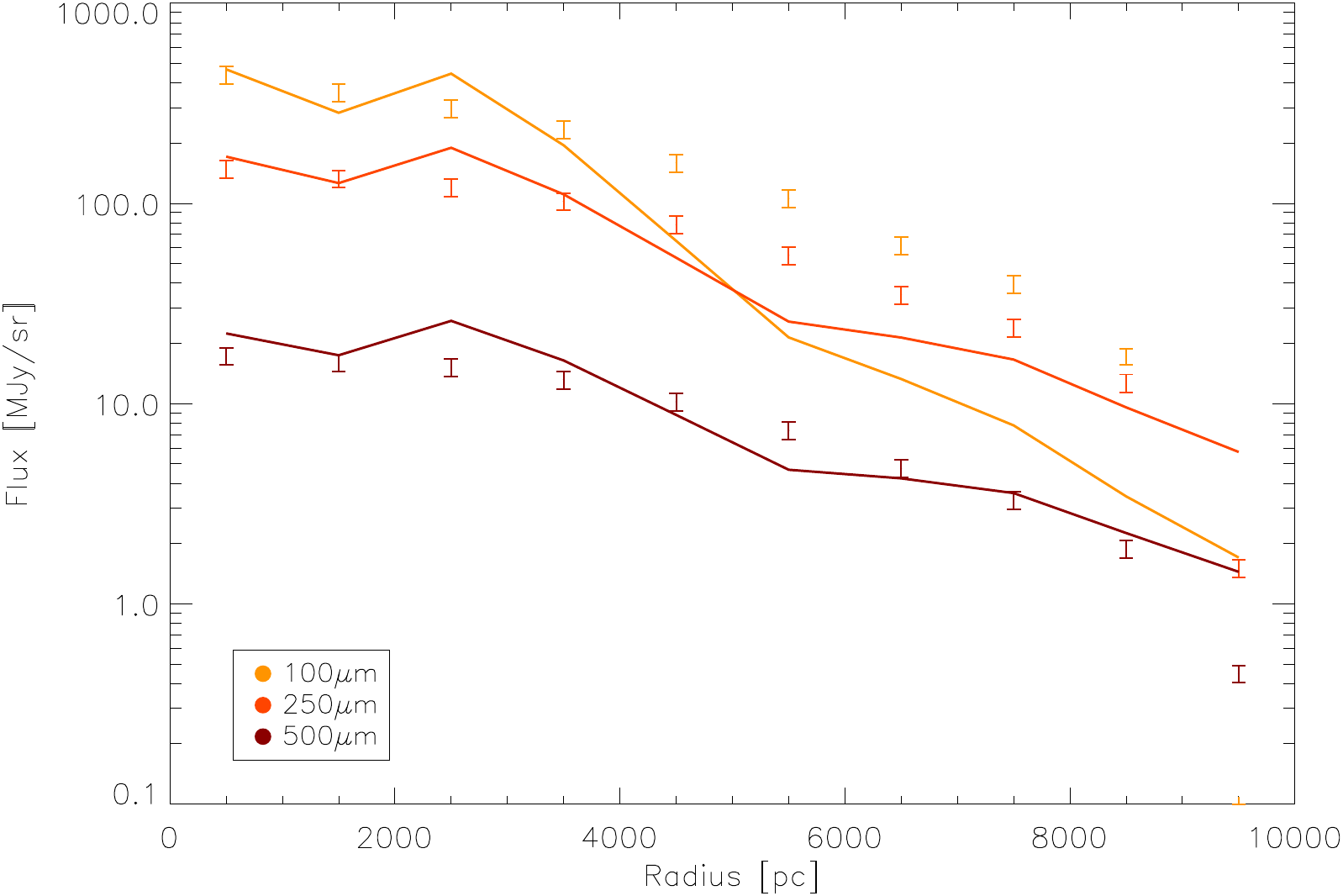}
    \end{subfigure}
    
    \begin{subfigure}{.45\textwidth}
      \centering
      \caption{Main properties with $\chi^2_{\rm CO10}$.}
      \includegraphics[width=1.\linewidth]{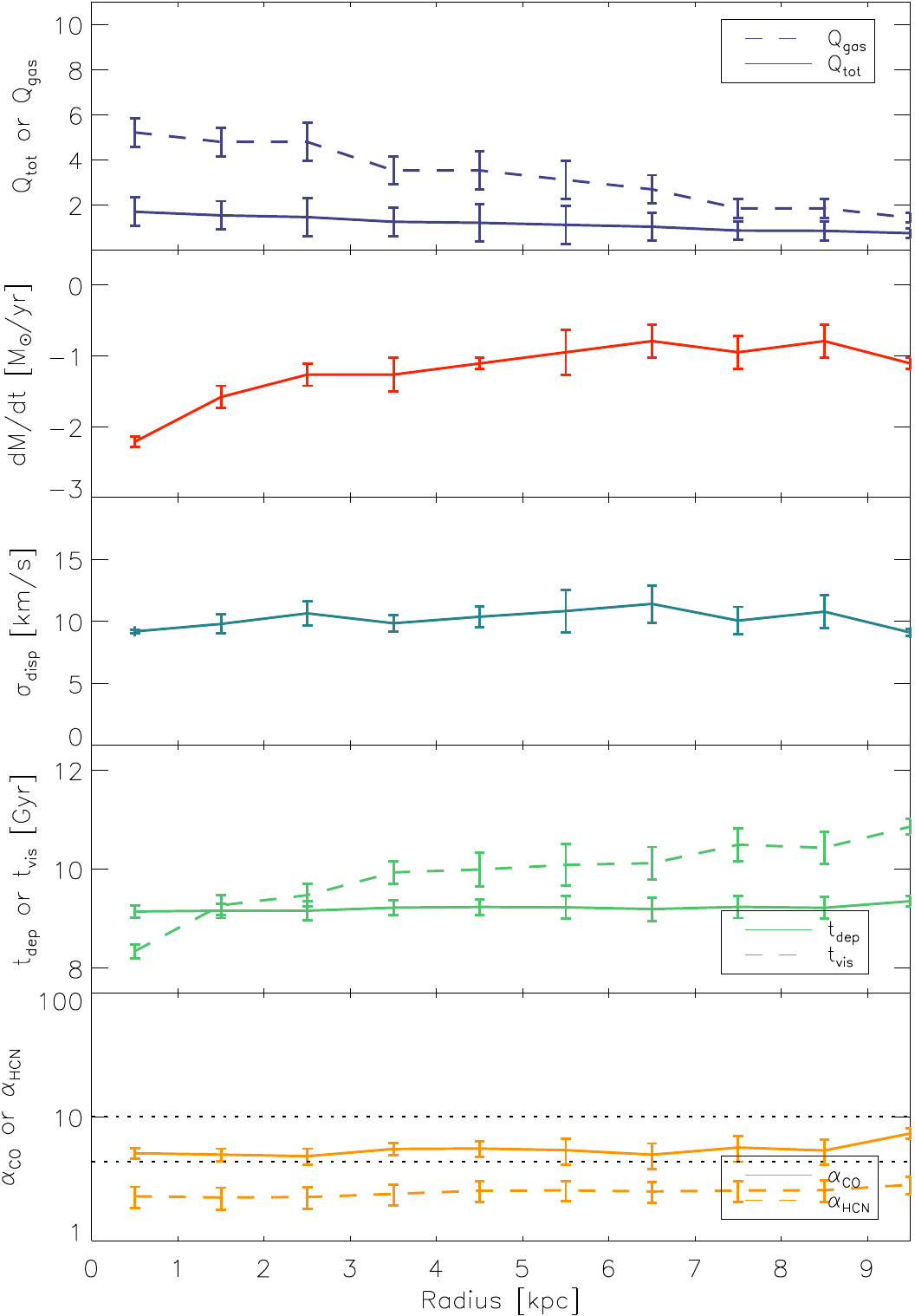}
    \end{subfigure}%
    \begin{subfigure}{.45\textwidth}
      \centering
      \caption{Main properties without $\chi^2_{\rm CO10}$.}
      \includegraphics[width=1.\linewidth]{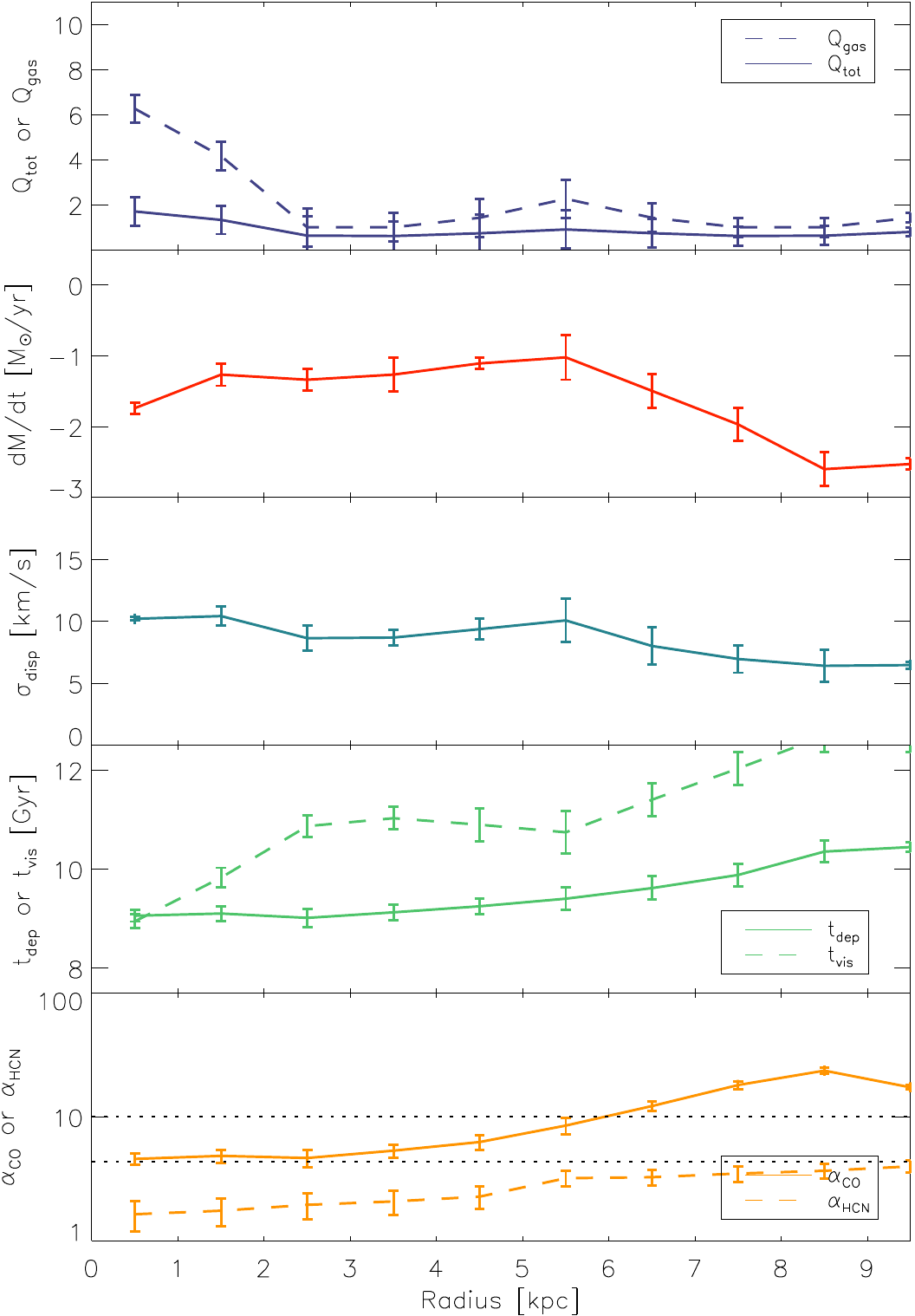}
    \end{subfigure}
    
    \caption{Same as Fig.\ref{fig:n628} for NGC~3627.}
    \label{fig:n3627}
    \end{figure*}

    \begin{figure*}
      \center
      \begin{subfigure}{.45\textwidth}
        \centering
        \caption{NGC~5055 best model with $\chi^2_{\rm CO10}$}
        \includegraphics[width=1.\linewidth]{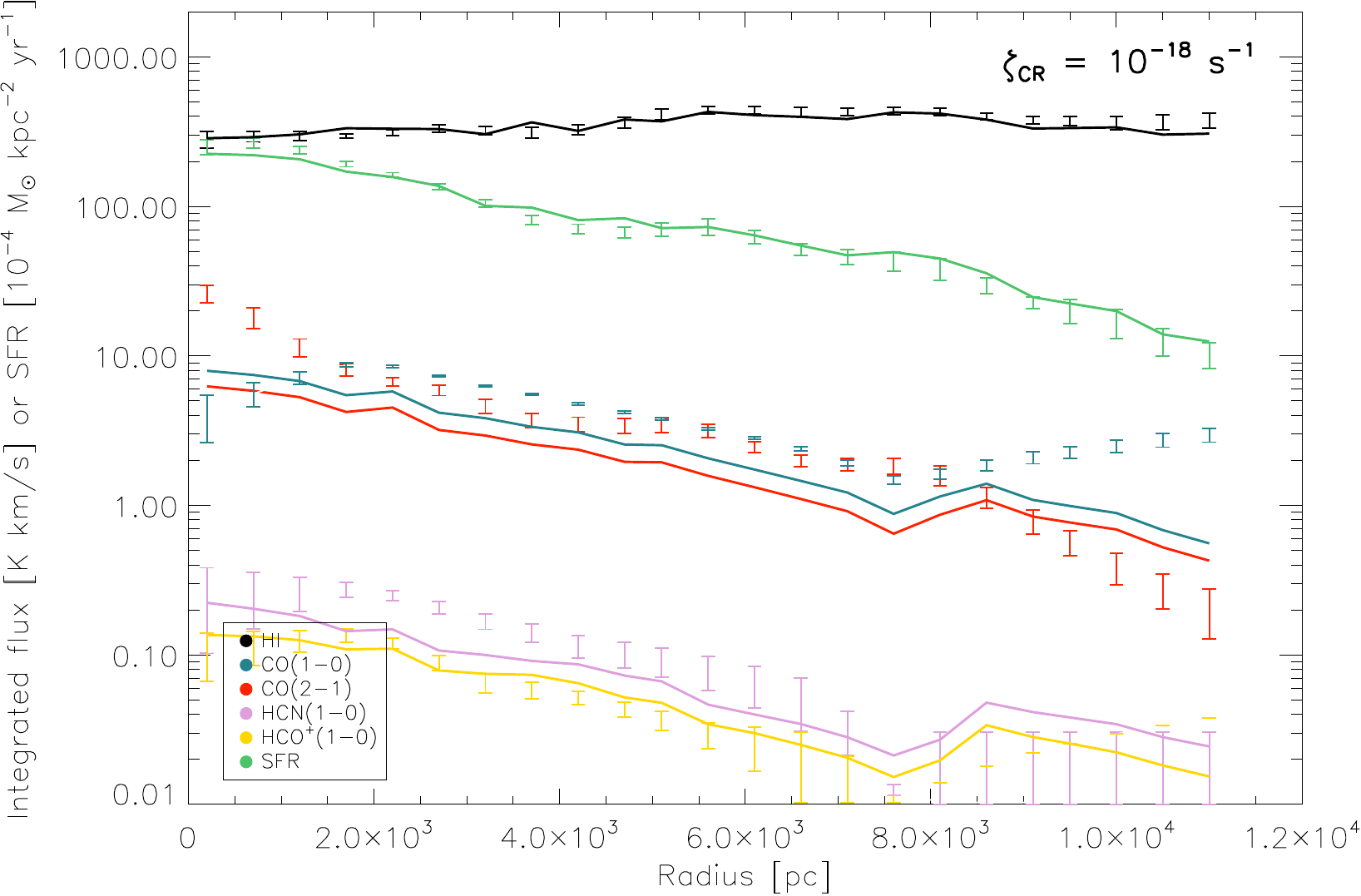}
      \end{subfigure}%
      \begin{subfigure}{.45\textwidth}
        \centering
        \caption{NGC~5055 best model without $\chi^2_{\rm CO10}$}
        \includegraphics[width=1.\linewidth]{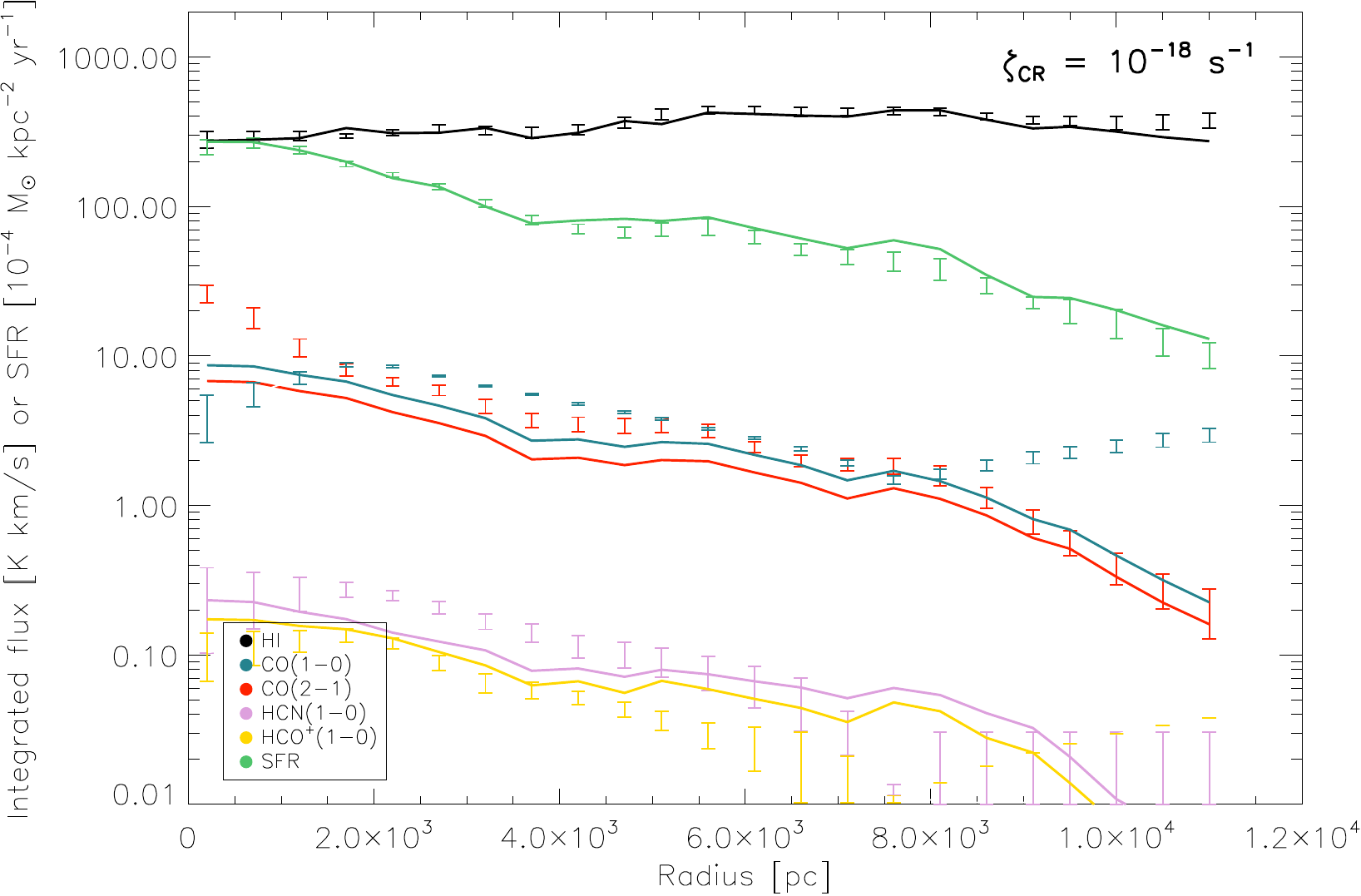}
      \end{subfigure}
      
      \begin{subfigure}{.45\textwidth}
        \centering
        \caption{NGC~5055 infrared profiles with $\chi^2_{\rm CO10}$}
        \includegraphics[width=1.\linewidth]{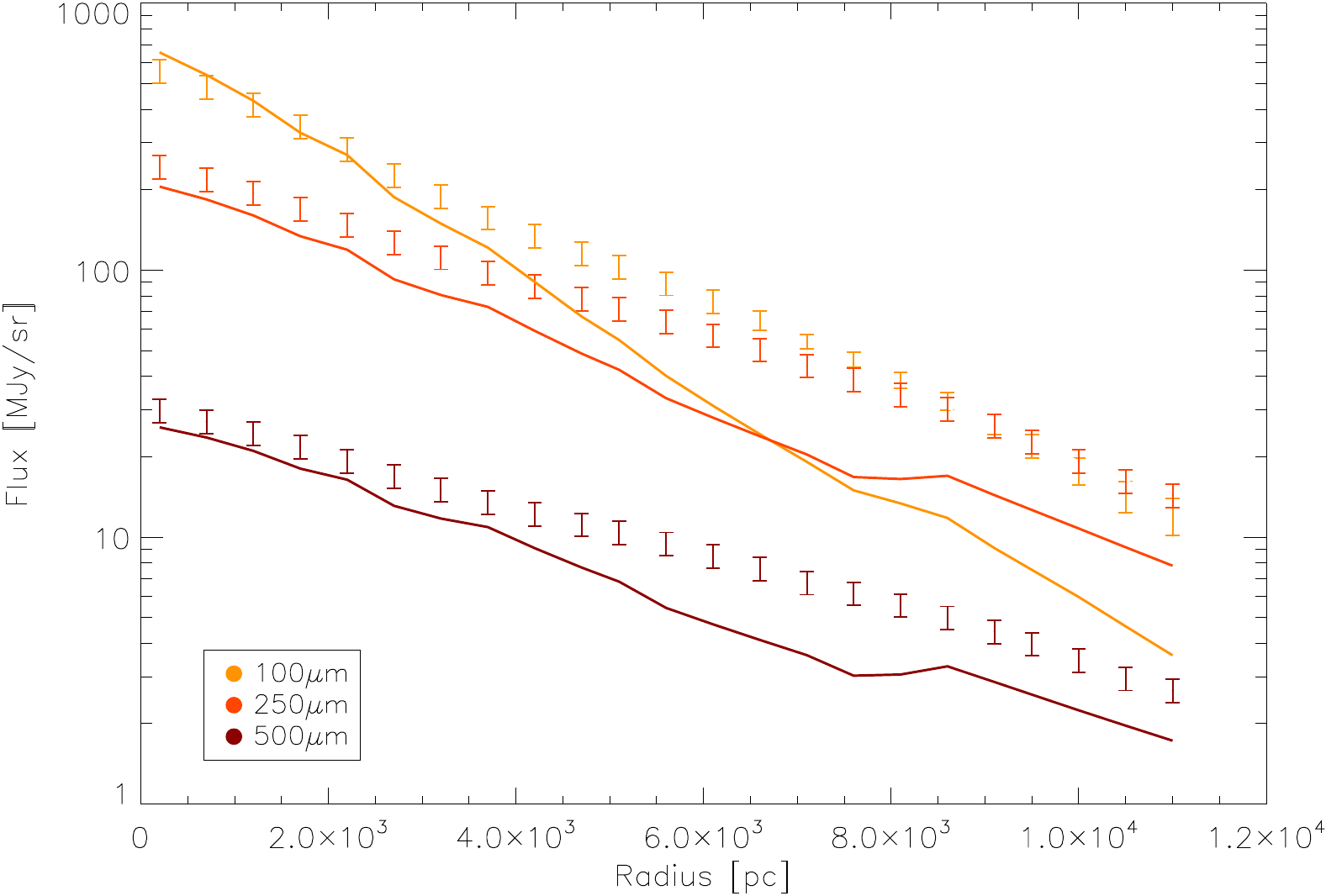}
      \end{subfigure}%
      \begin{subfigure}{.45\textwidth}
        \centering
        \caption{NGC~5055 infrared profiles without $\chi^2_{\rm CO10}$}
        \includegraphics[width=1.\linewidth]{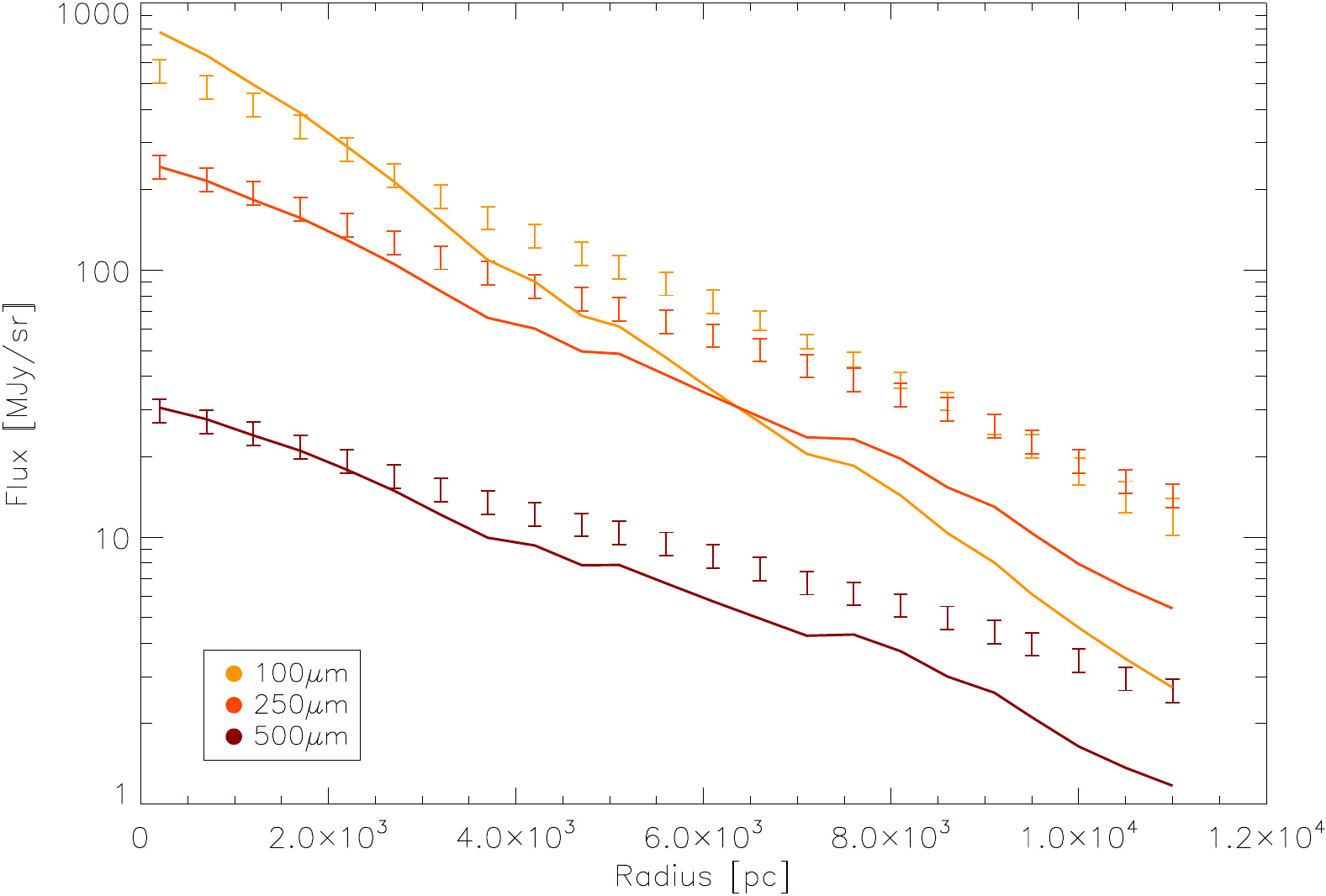}
      \end{subfigure}
      
      \begin{subfigure}{.45\textwidth}
        \centering
        \caption{Main properties with $\chi^2_{\rm CO10}$.}
        \includegraphics[width=1.\linewidth]{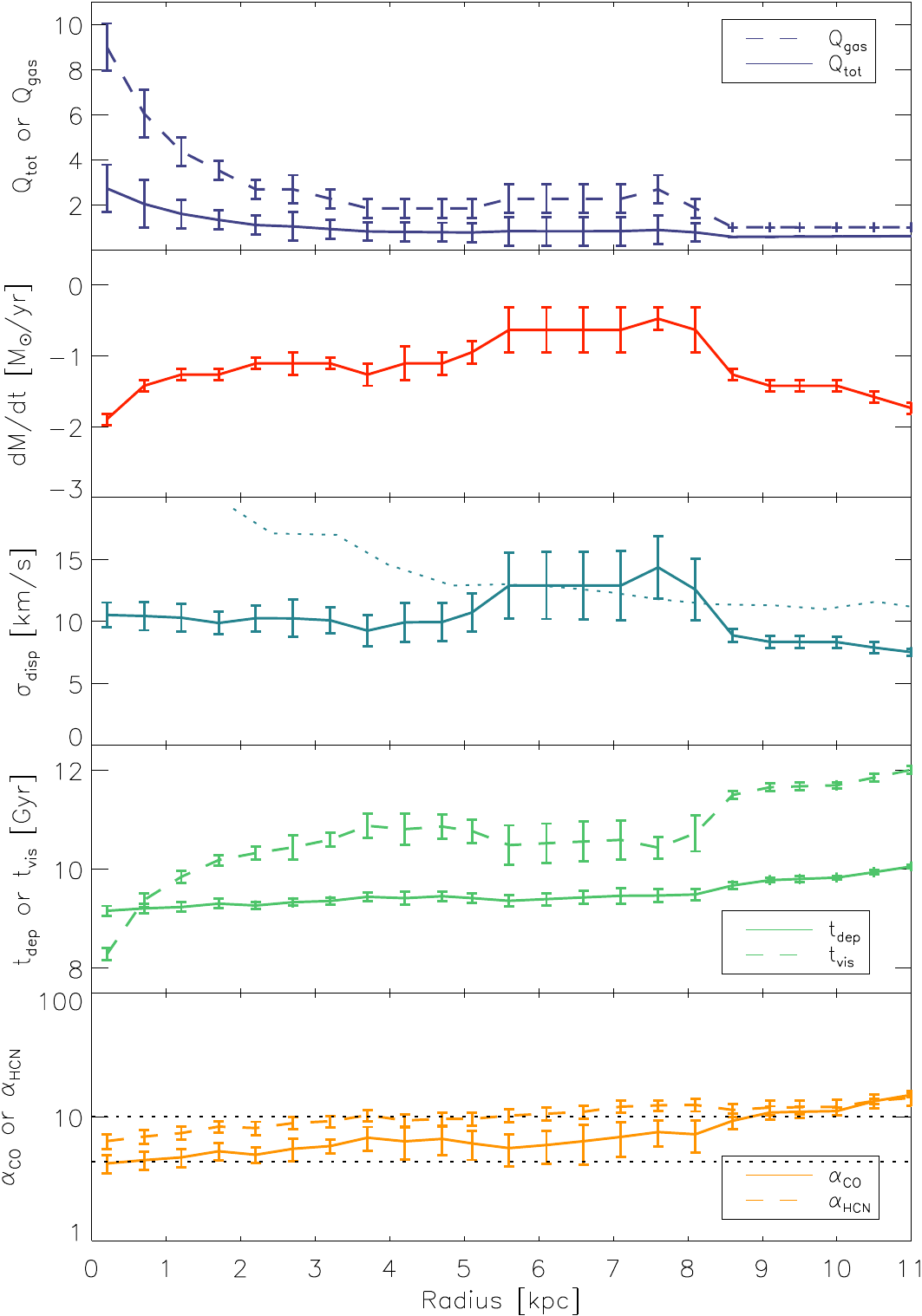}
      \end{subfigure}%
      \begin{subfigure}{.45\textwidth}
        \centering
        \caption{Main properties without $\chi^2_{\rm CO10}$.}
        \includegraphics[width=1.\linewidth]{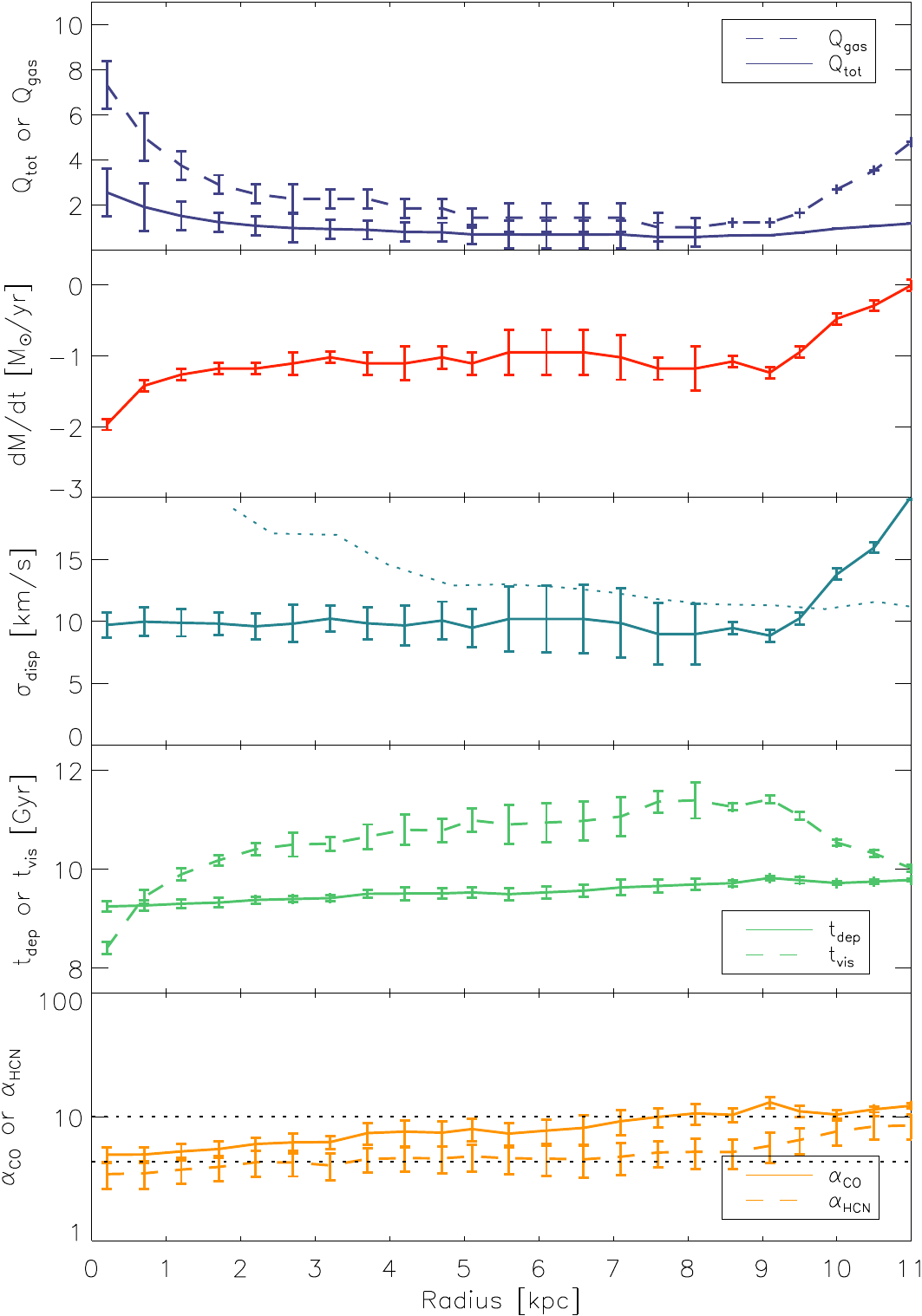}
      \end{subfigure}
      
      \caption{Same as Fig.\ref{fig:n628} for NGC~5055.}
      \label{fig:n5055}
      \end{figure*}

      \begin{figure*}
        \center
        \begin{subfigure}{.45\textwidth}
          \centering
          \caption{NGC~5194 best model with $\chi^2_{\rm CO10}$}
          \includegraphics[width=1.\linewidth]{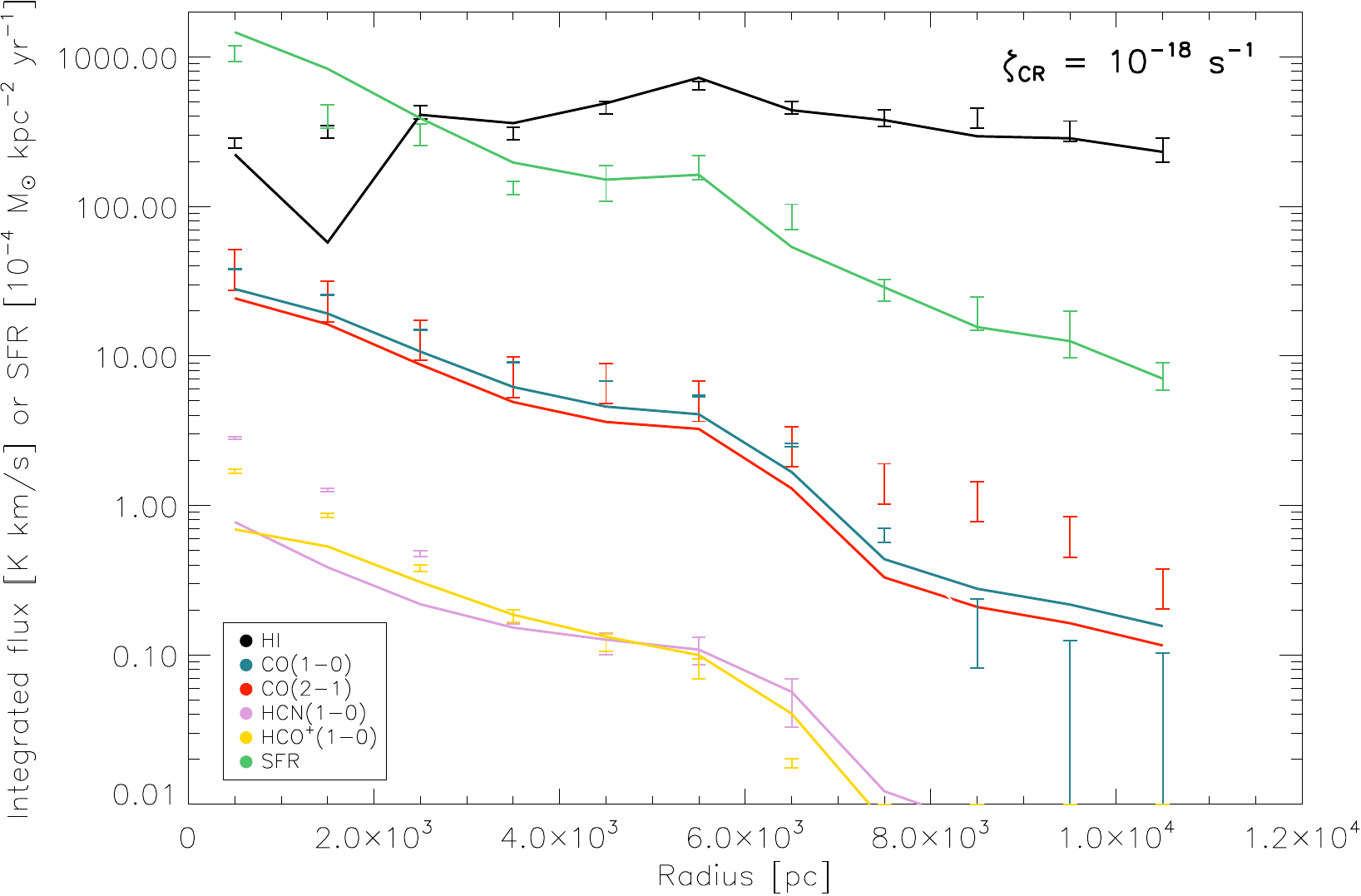}
        \end{subfigure}%
        \begin{subfigure}{.45\textwidth}
          \centering
          \caption{NGC~5194 best model without $\chi^2_{\rm CO10}$}
          \includegraphics[width=1.\linewidth]{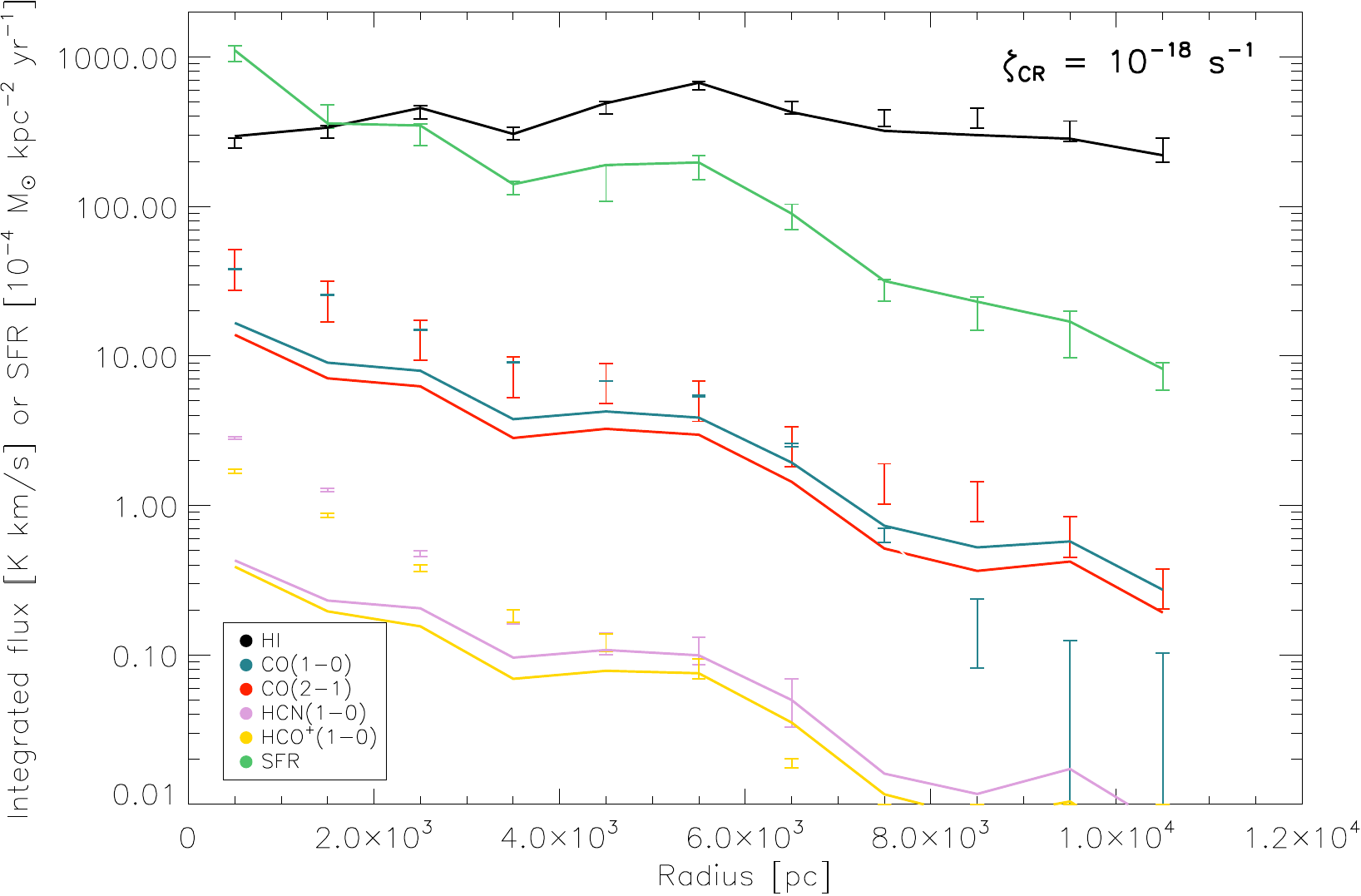}
        \end{subfigure}
        
        \begin{subfigure}{.45\textwidth}
          \centering
          \caption{NGC~5194 infrared profiles with $\chi^2_{\rm CO10}$}
          \includegraphics[width=1.\linewidth]{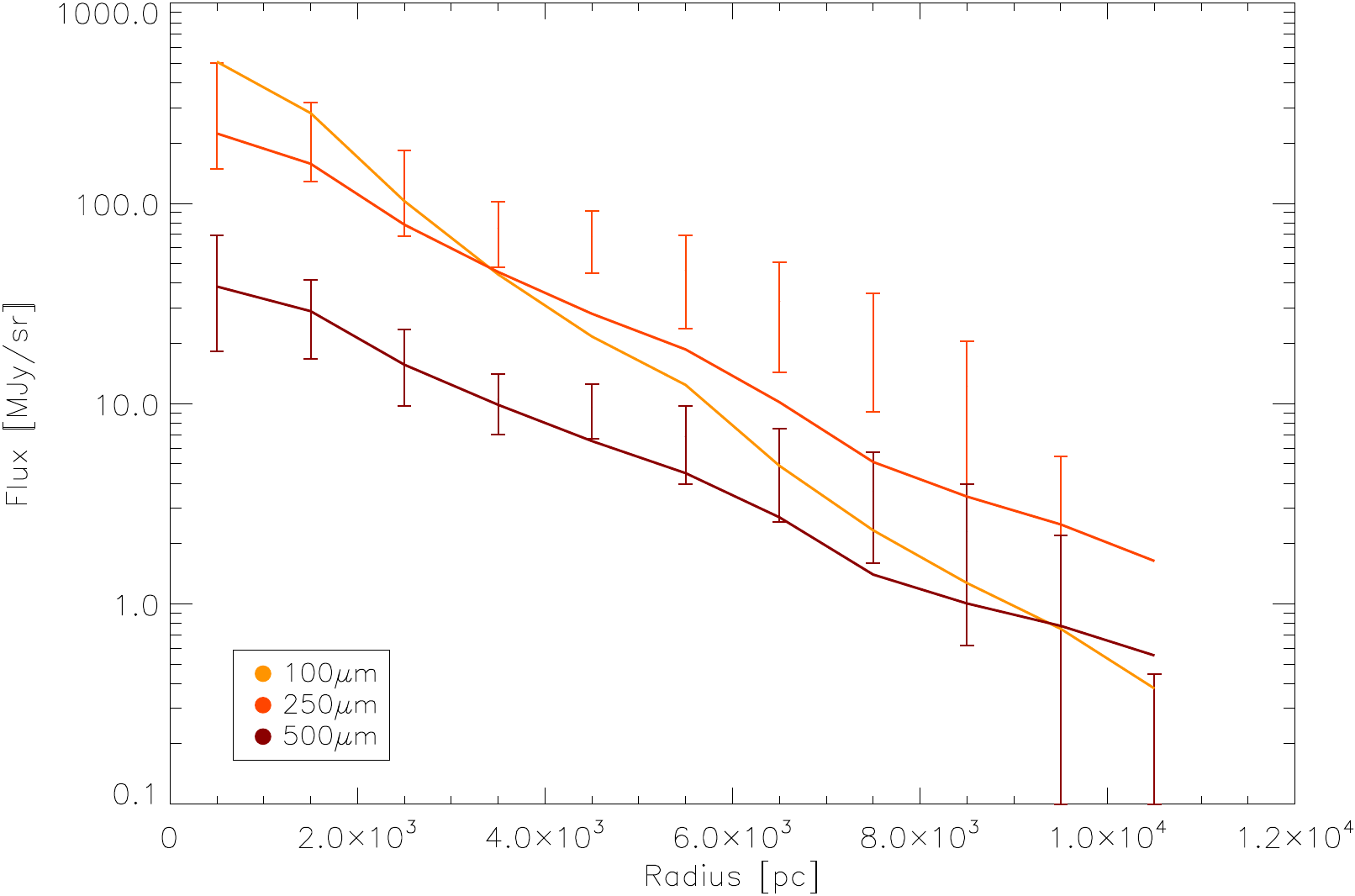}
        \end{subfigure}%
        \begin{subfigure}{.45\textwidth}
          \centering
          \caption{NGC~5194 infrared profiles without $\chi^2_{\rm CO10}$}
          \includegraphics[width=1.\linewidth]{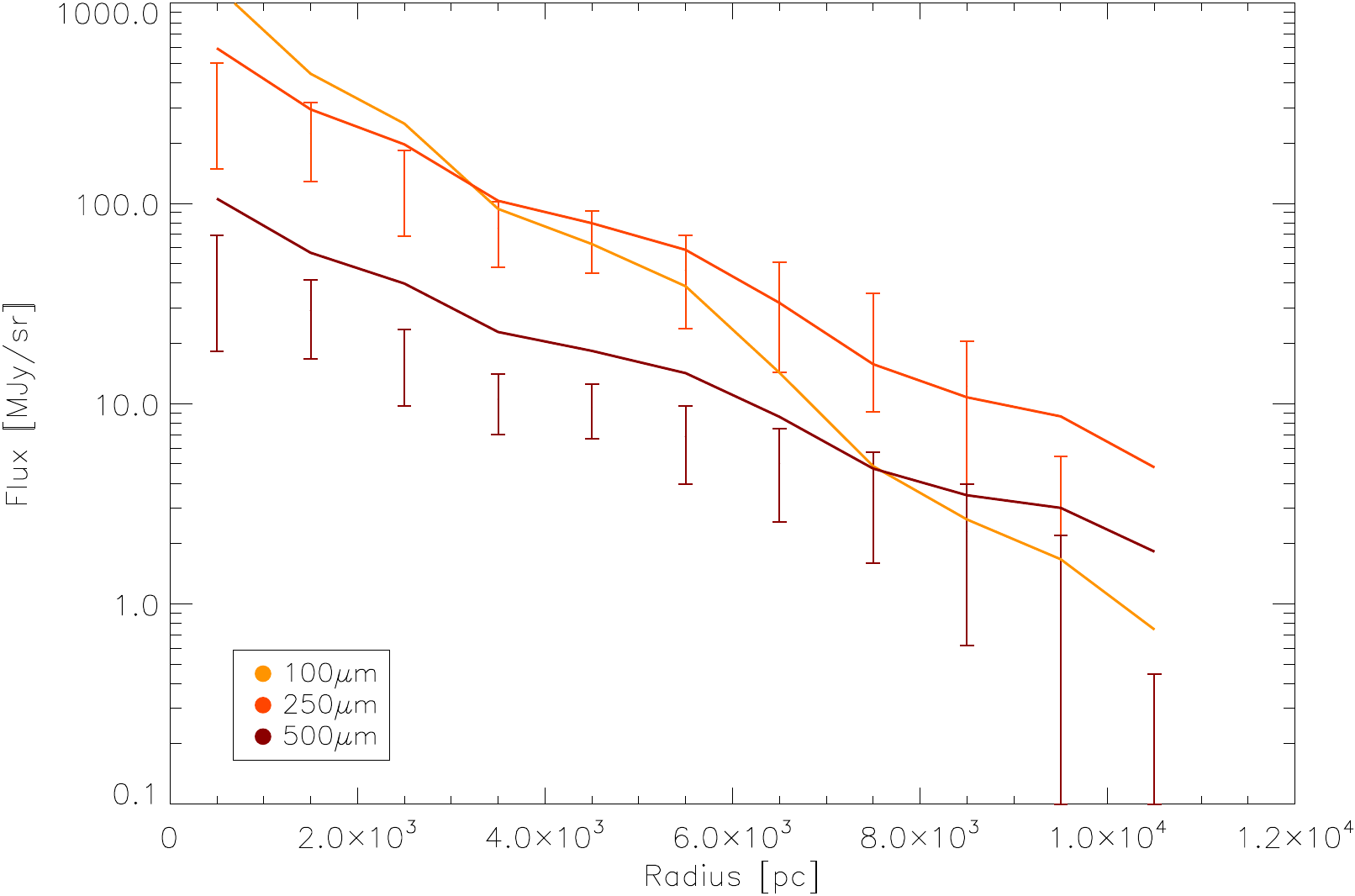}
        \end{subfigure}
        
        \begin{subfigure}{.45\textwidth}
          \centering
          \caption{Main properties with $\chi^2_{\rm CO10}$.}
          \includegraphics[width=1.\linewidth]{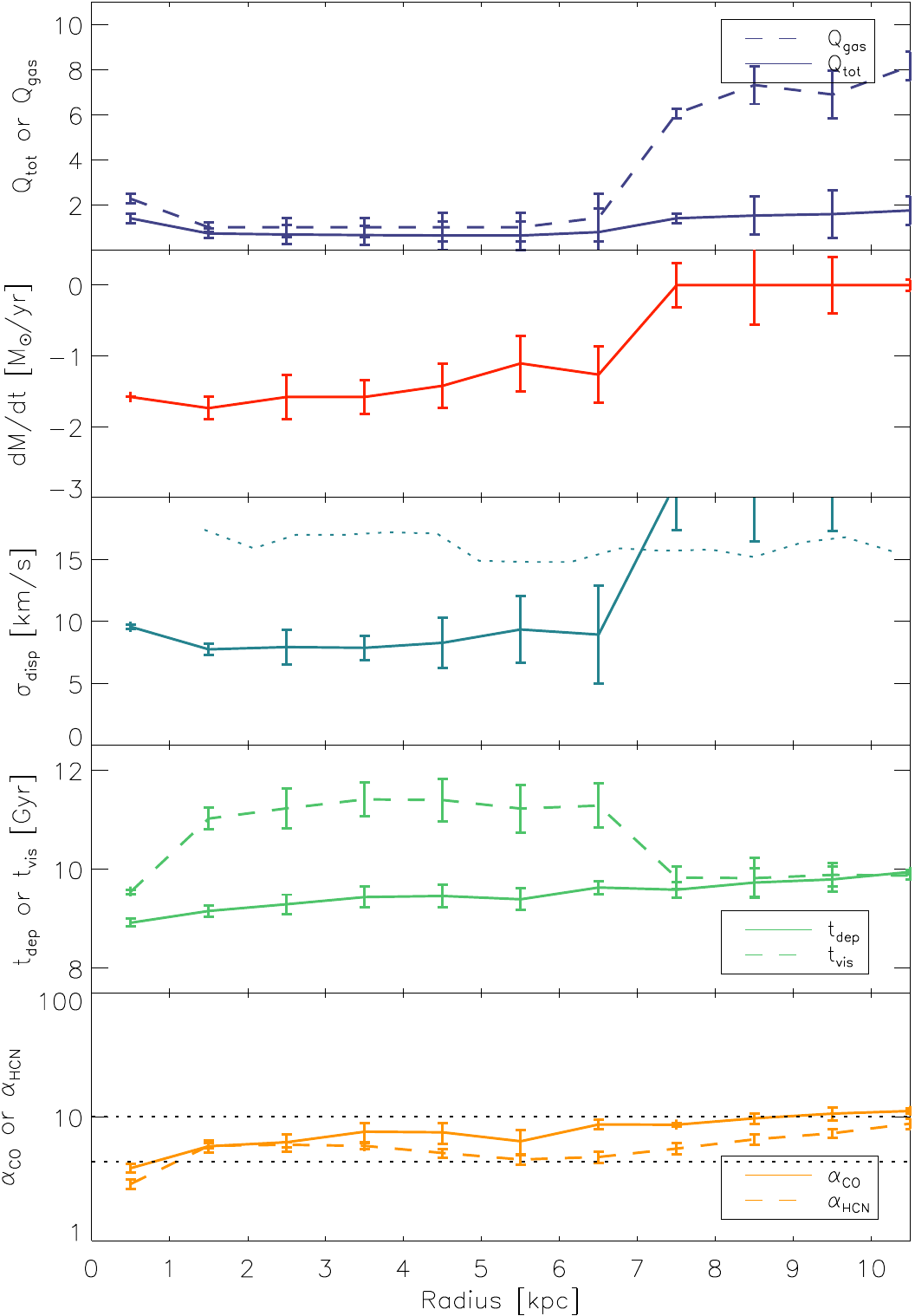}
        \end{subfigure}%
        \begin{subfigure}{.45\textwidth}
          \centering
          \caption{Main properties without $\chi^2_{\rm CO10}$.}
          \includegraphics[width=1.\linewidth]{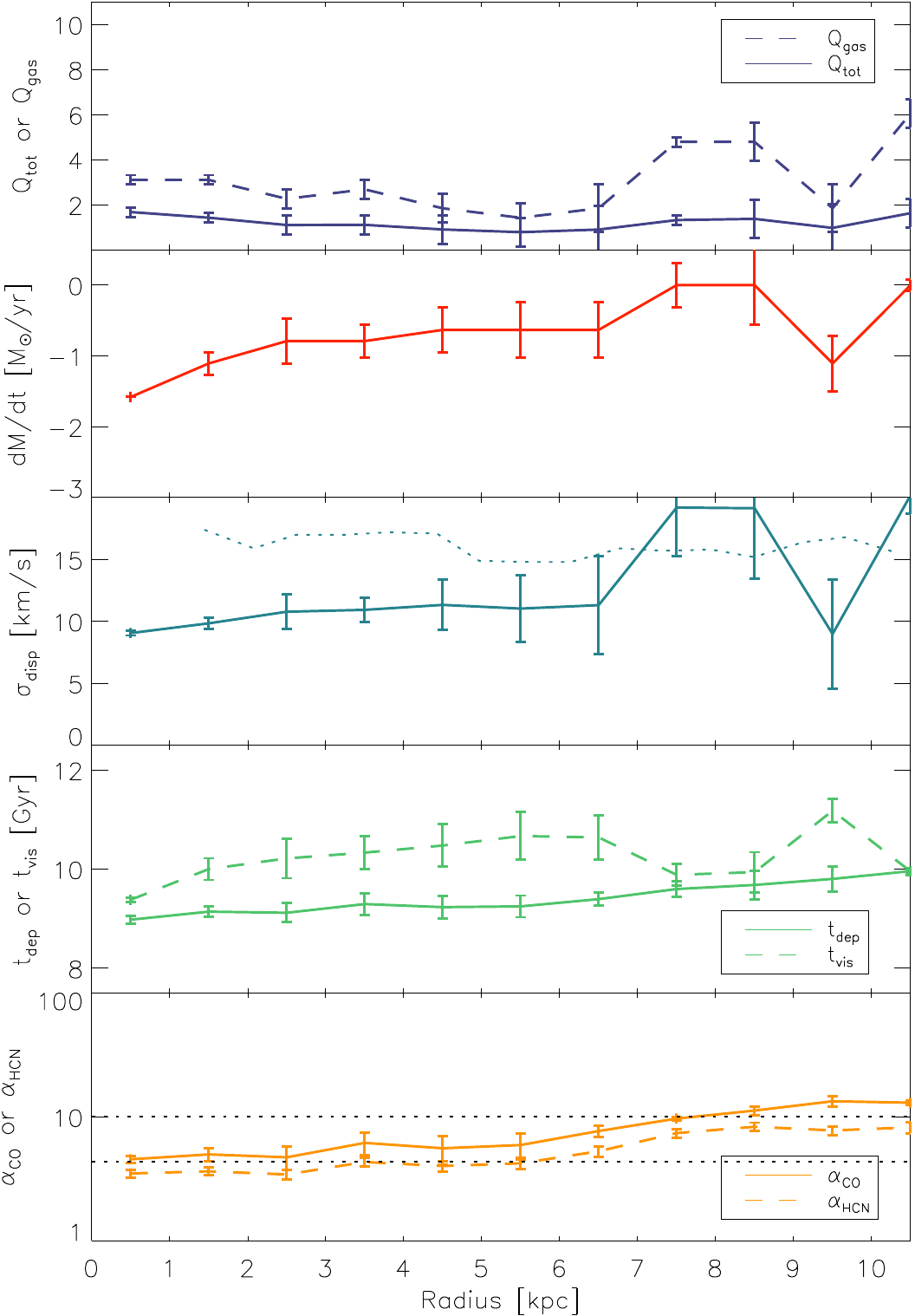}
        \end{subfigure}
        
        \caption{Same as Fig.\ref{fig:n628} for NGC~5194.}
        \label{fig:n5194}
        \end{figure*}

        \begin{figure*}
          \center
          \begin{subfigure}{.45\textwidth}
            \centering
            \caption{NGC~2841 best model}
            \includegraphics[width=1.\linewidth]{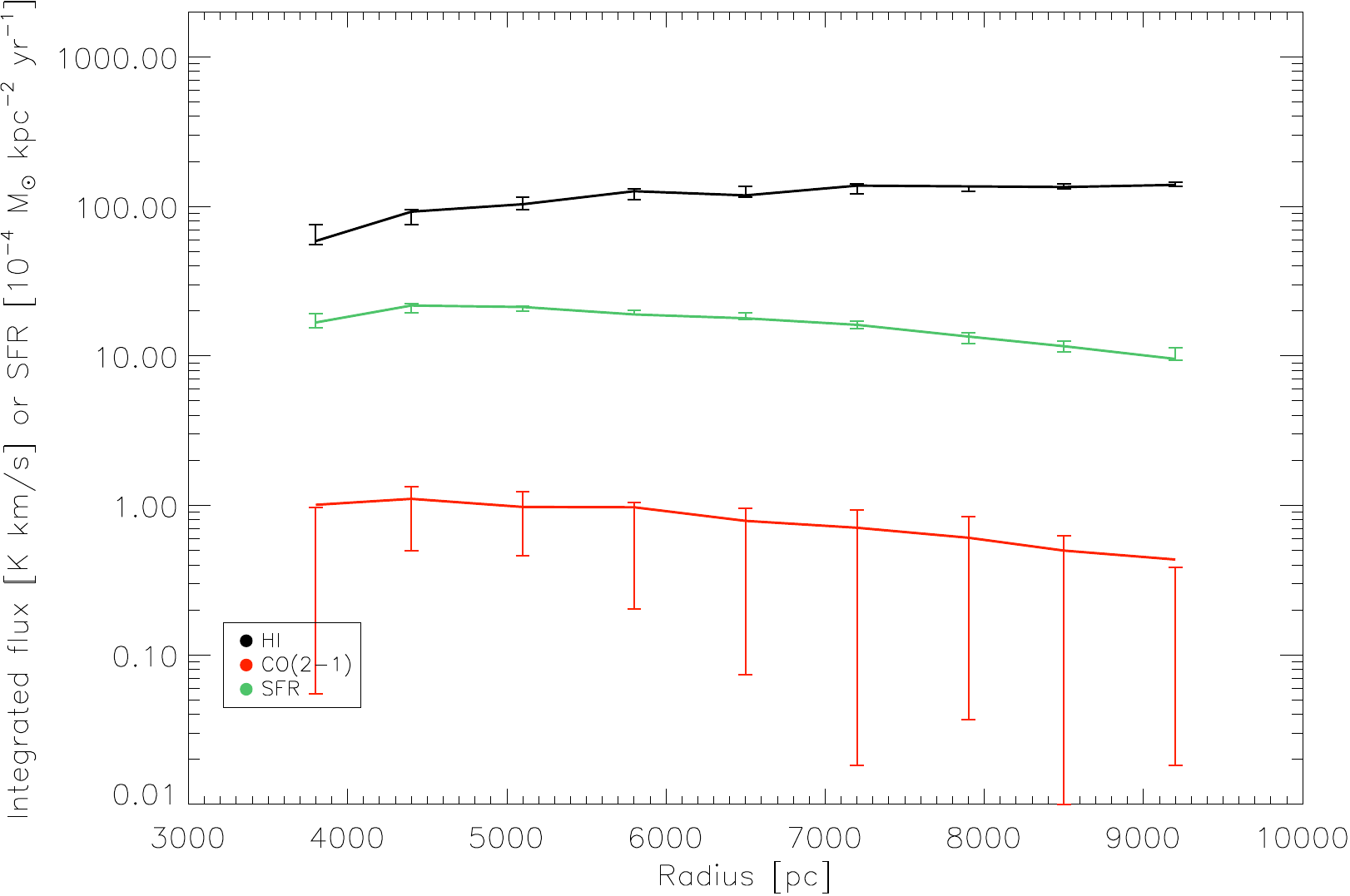}
           
          \end{subfigure}
          \begin{subfigure}{.45\textwidth}
            \centering
            \caption{NGC~2841 infrared profiles}
            \includegraphics[width=1.\linewidth]{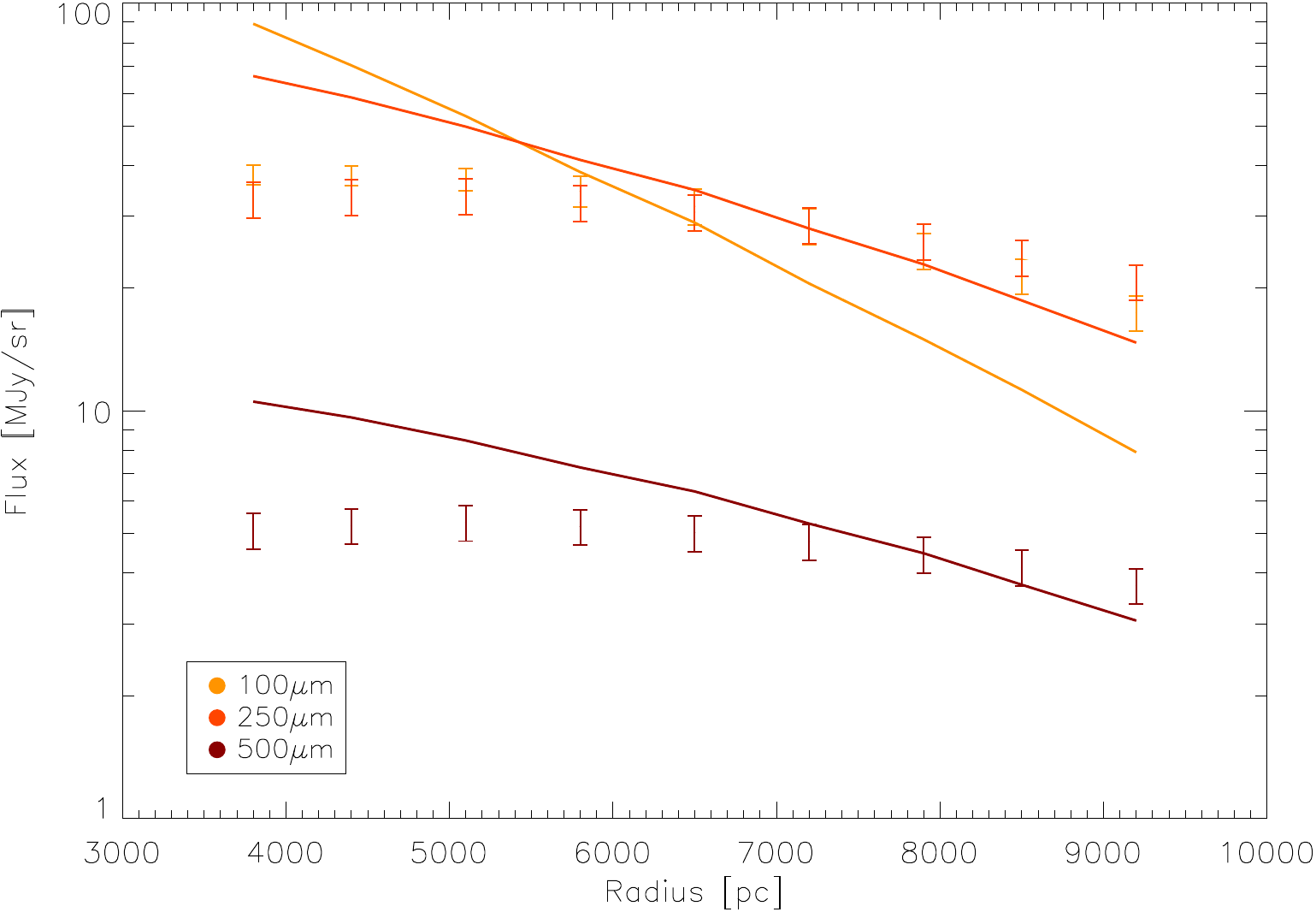}
          \end{subfigure}
          
          \begin{subfigure}{0.6\textwidth}
            \centering
            \caption{NGC~2841 main properties profiles}
            \includegraphics[width=0.8\linewidth]{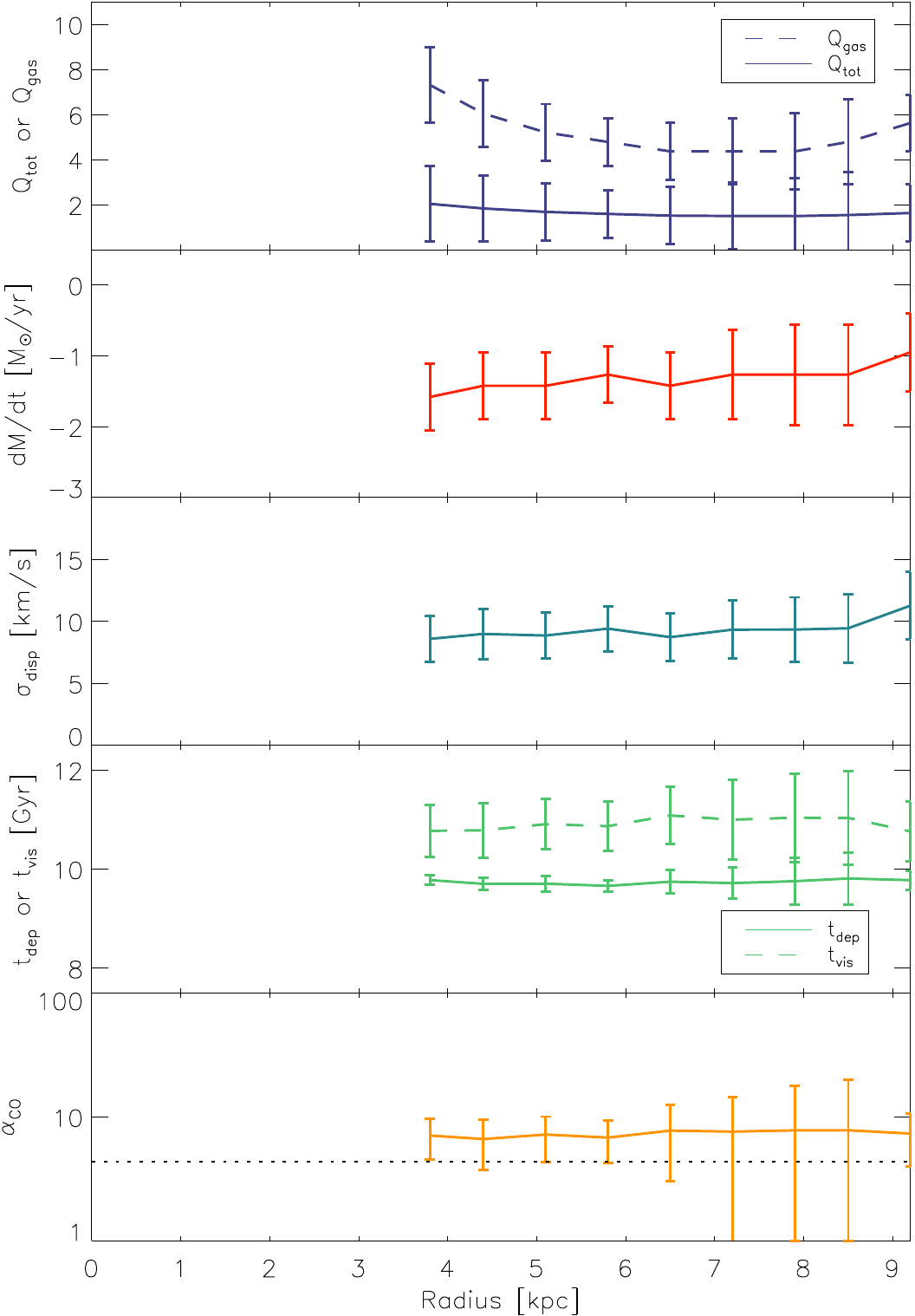}
          \end{subfigure}
           
          \caption{As Fig.~\ref{fig:n628} for NGC~2841.}
          \label{fig:fig}
          \end{figure*}

\begin{figure*}
  \center
  \begin{subfigure}{.45\textwidth}
    \centering
    \caption{NGC~3198 best model}
    \includegraphics[width=1.\linewidth]{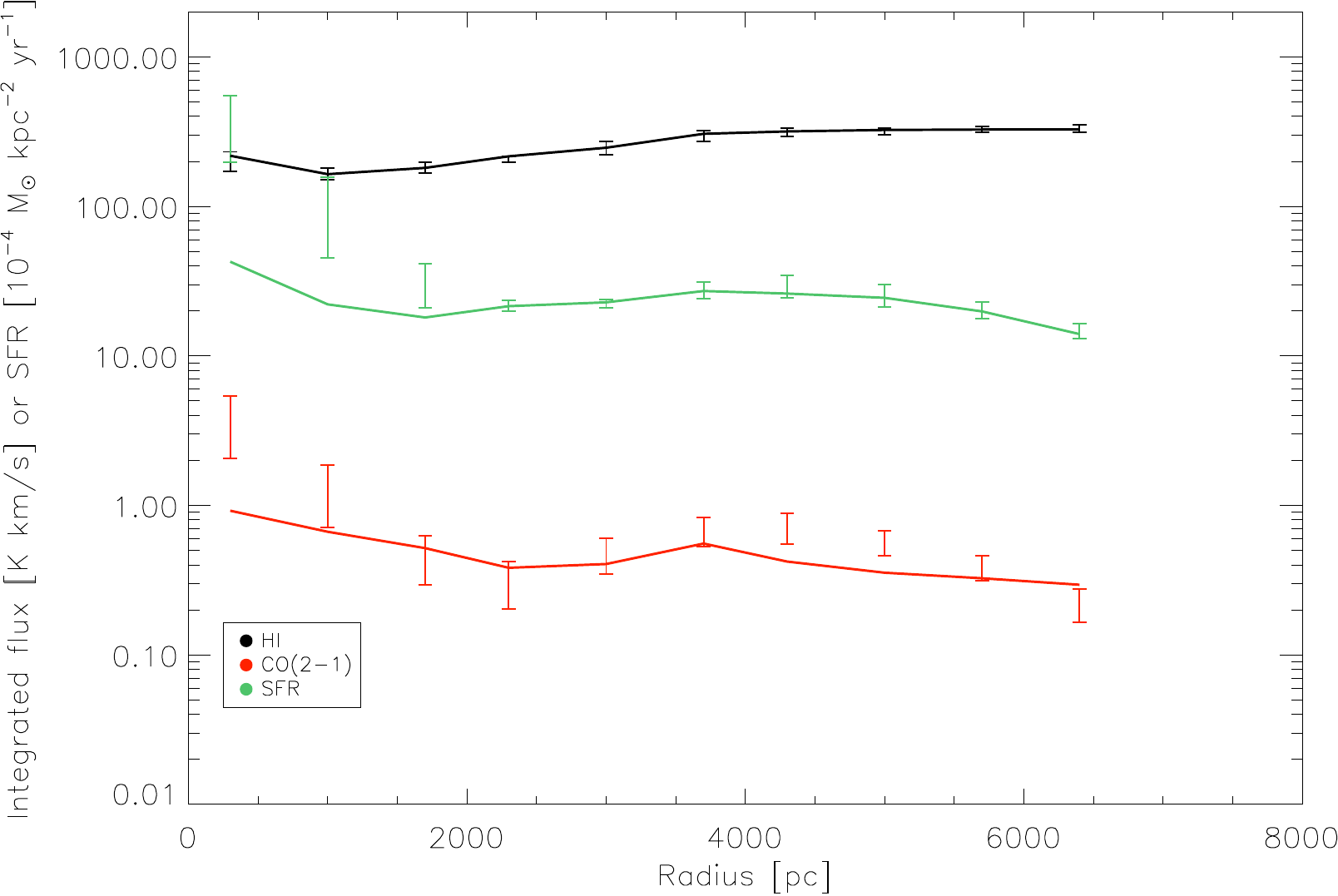}
   
  \end{subfigure}
  \begin{subfigure}{.45\textwidth}
    \centering
    \caption{NGC~3198 infrared profiles}
    \includegraphics[width=1.\linewidth]{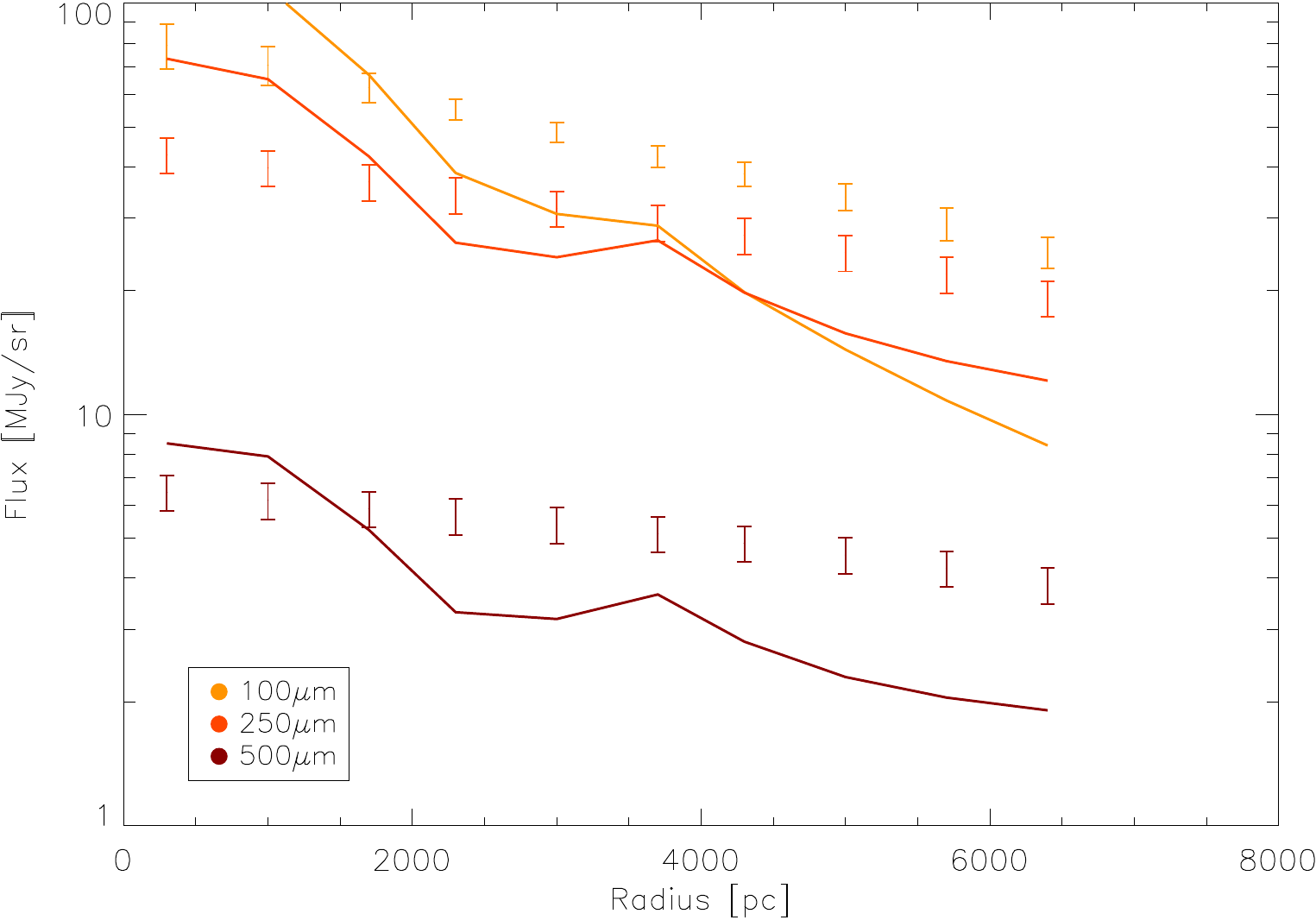}
  \end{subfigure}
  
  \begin{subfigure}{0.6\textwidth}
    \centering
    \caption{NGC~3198 main properties profiles}
    \includegraphics[width=0.8\linewidth]{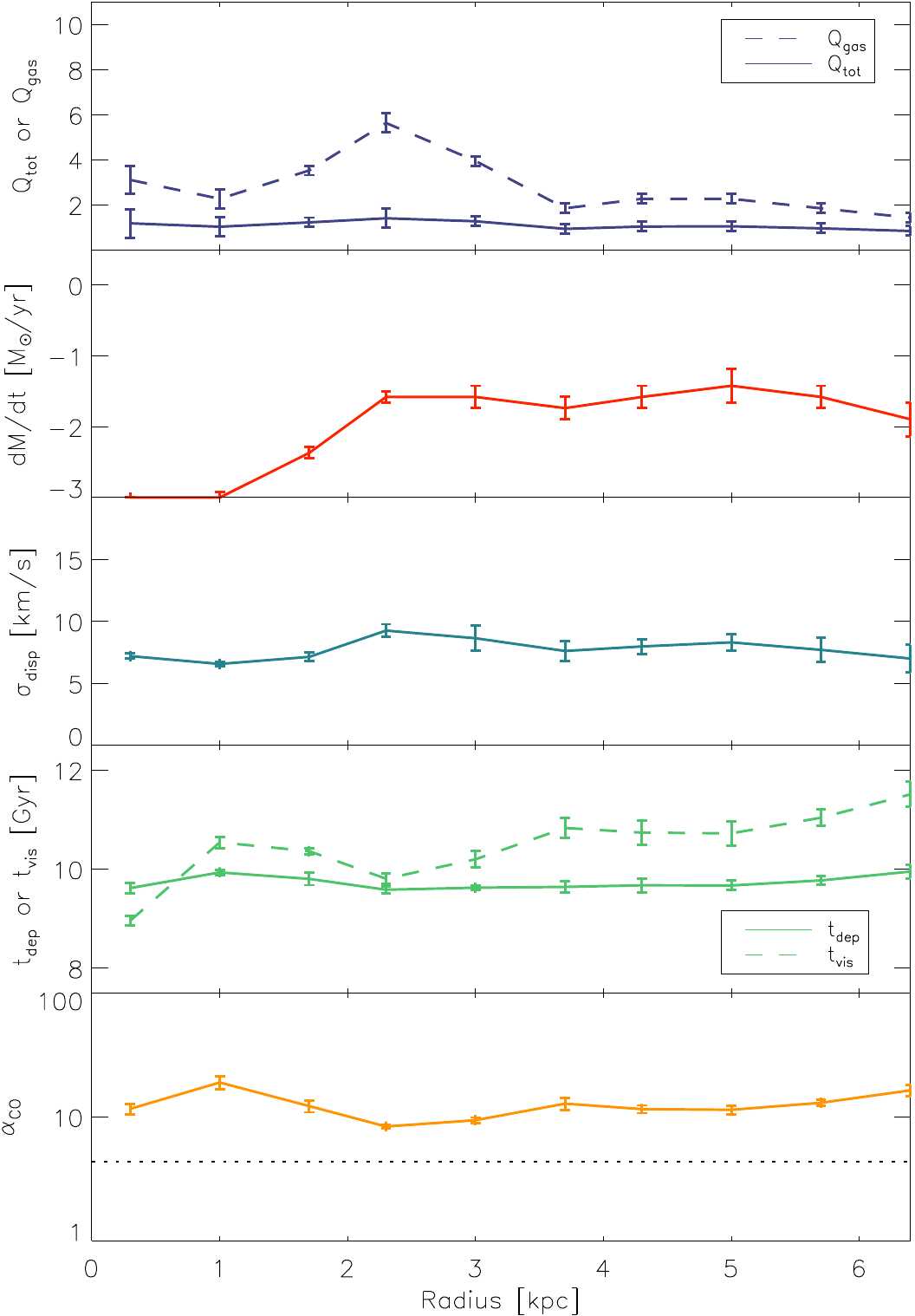}
  \end{subfigure}
   
  \caption{As Fig.~\ref{fig:n628} for NGC~3198.}
  \label{fig:fig}
  \end{figure*}
  
  \begin{figure*}
    \center
    \begin{subfigure}{.45\textwidth}
      \centering
      \caption{NGC~3351 best model}
      \includegraphics[width=1.\linewidth]{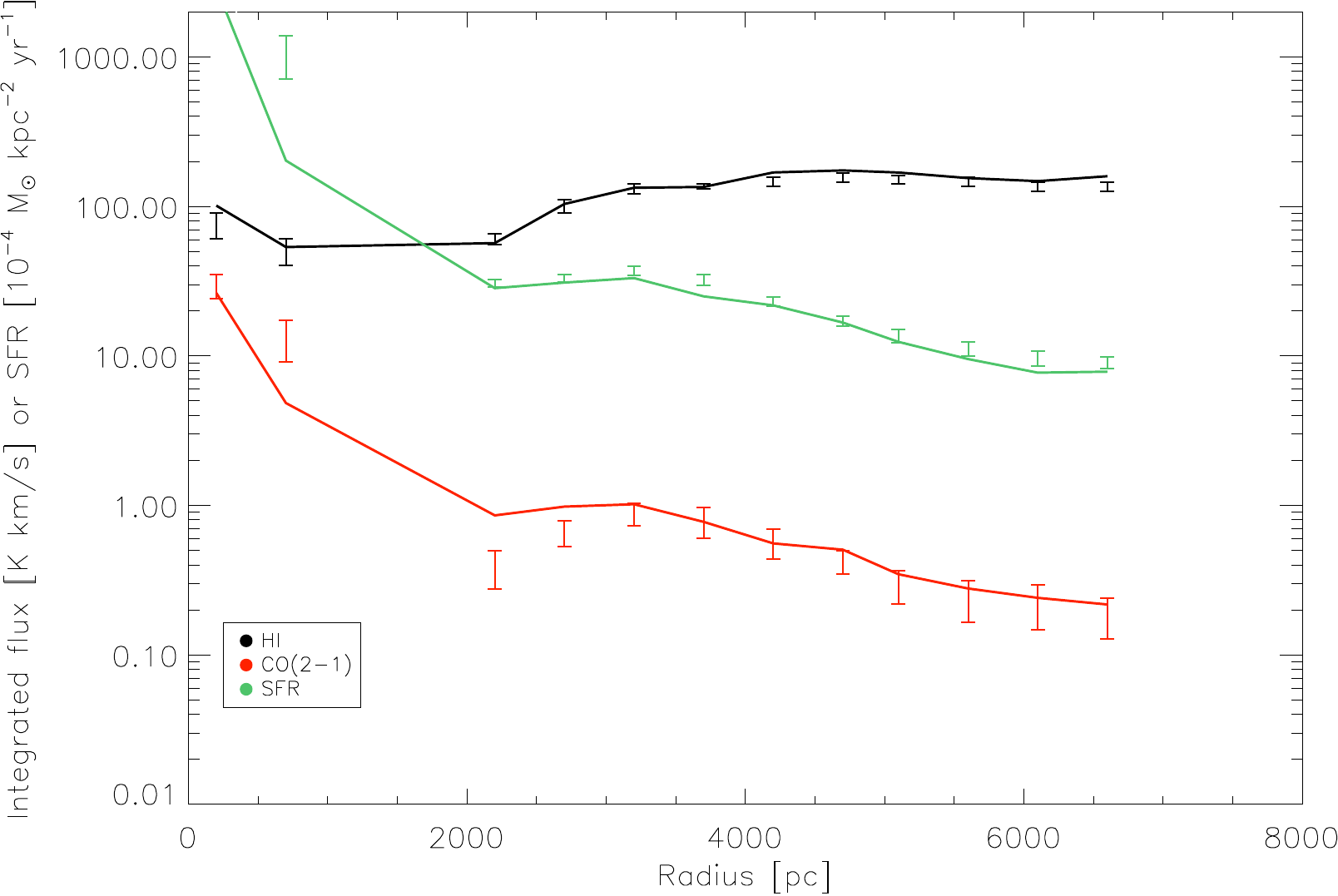}
     
    \end{subfigure}
    \begin{subfigure}{.45\textwidth}
      \centering
      \caption{NGC~3351 infrared profiles}
      \includegraphics[width=1.\linewidth]{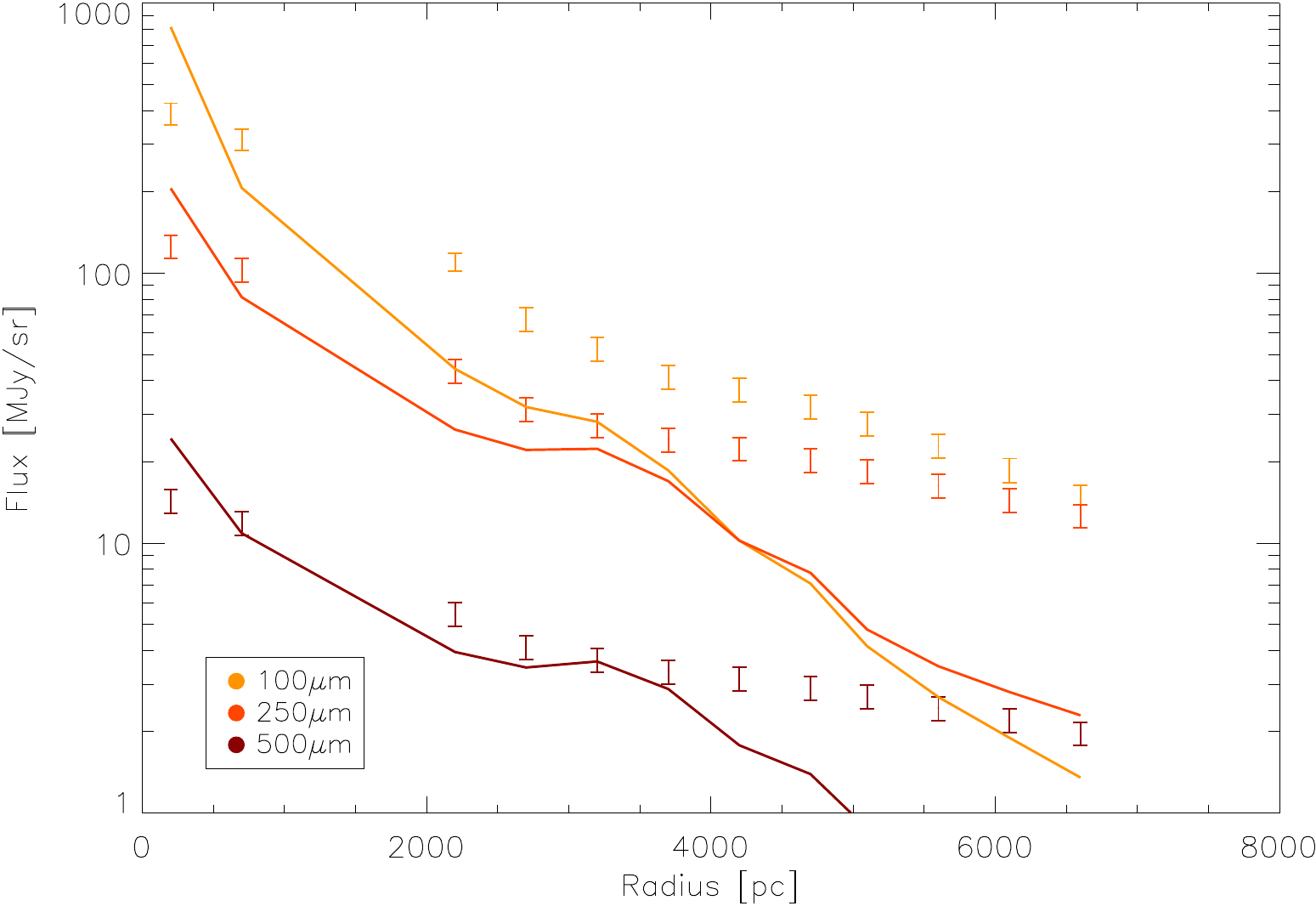}
    \end{subfigure}
    
    \begin{subfigure}{0.6\textwidth}
      \centering
      \caption{NGC~3351 main properties profiles}
      \includegraphics[width=0.8\linewidth]{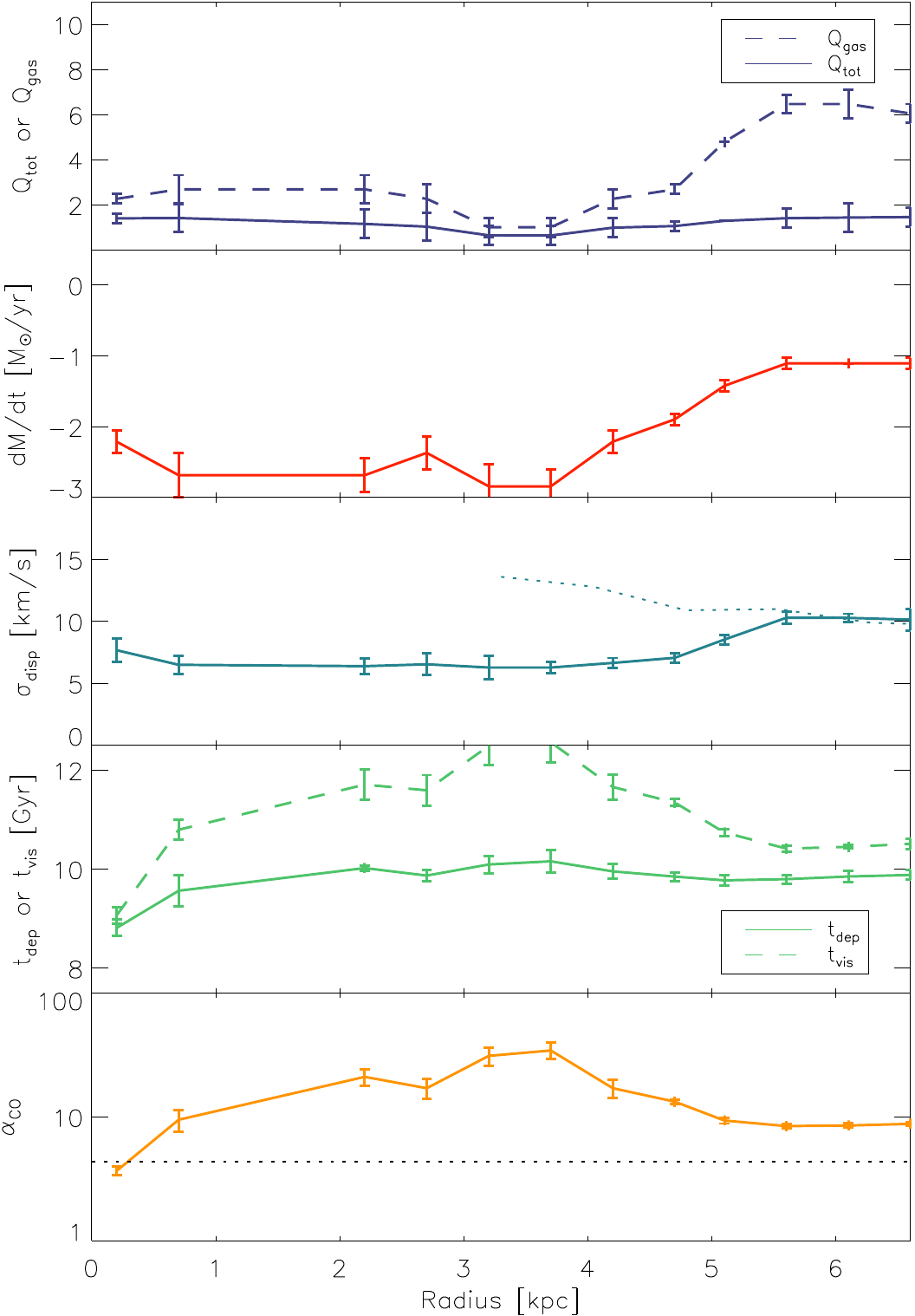}
    \end{subfigure}
     
    \caption{As Fig.~\ref{fig:n628} for NGC~3351.}
    \label{fig:fig}
    \end{figure*}
    
    \begin{figure*}
      \center
      \begin{subfigure}{.45\textwidth}
        \centering
        \caption{NGC~3521 best model}
        \includegraphics[width=1.\linewidth]{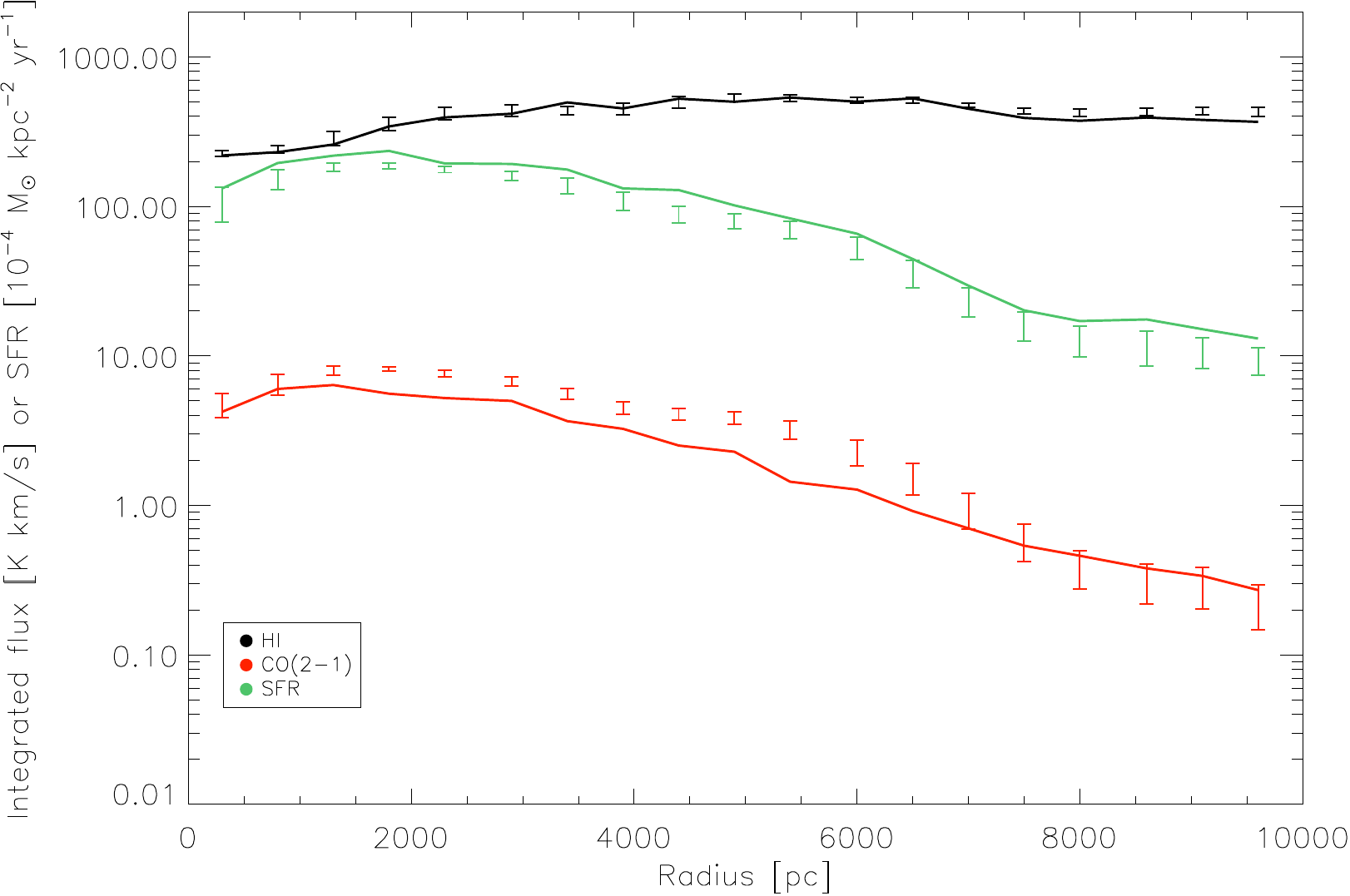}
       
      \end{subfigure}
      \begin{subfigure}{.45\textwidth}
        \centering
        \caption{NGC~3521 infrared profiles}
        \includegraphics[width=1.\linewidth]{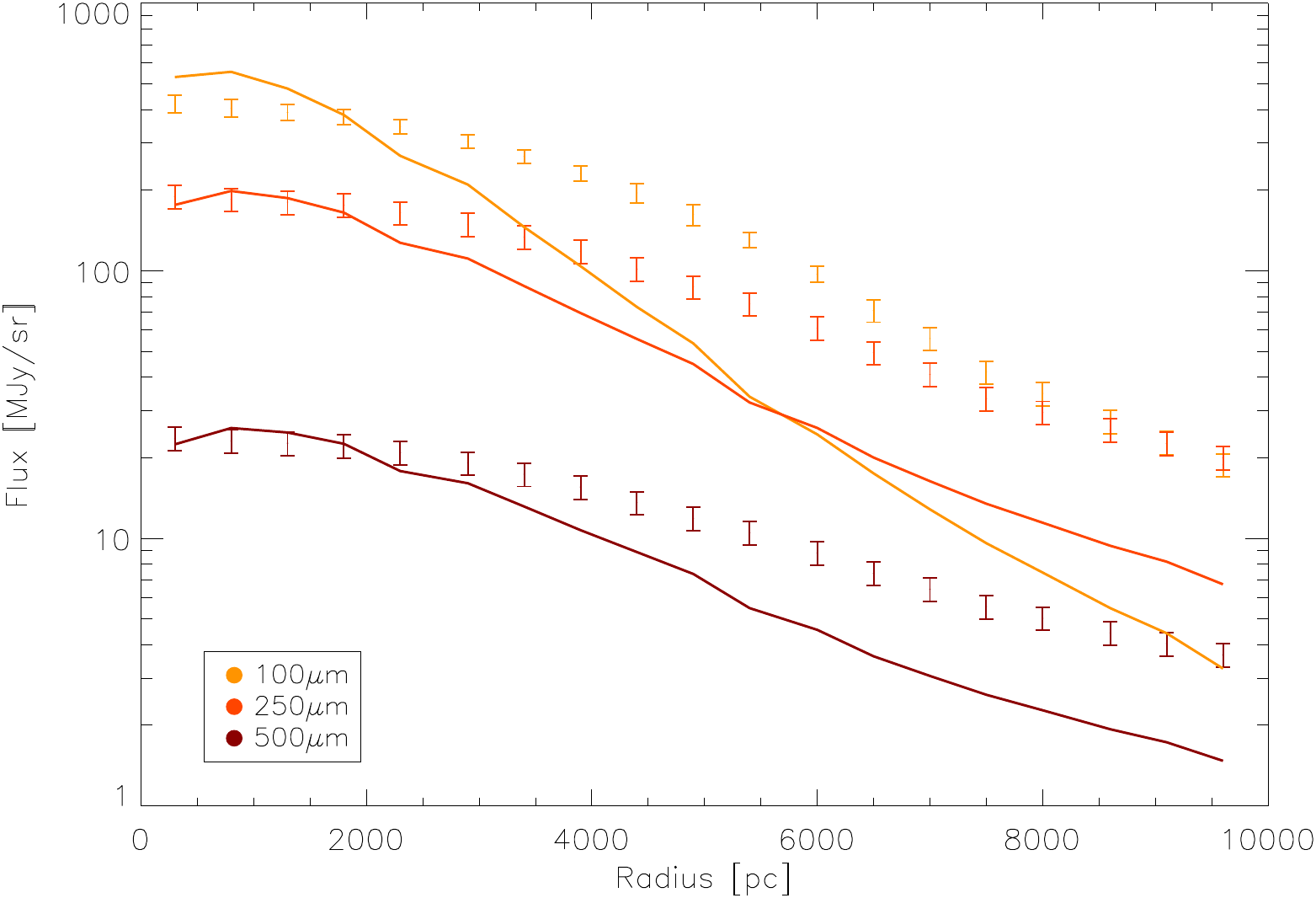}
      \end{subfigure}
      
      \begin{subfigure}{0.6\textwidth}
        \centering
        \caption{NGC~3521 main properties profiles}
        \includegraphics[width=0.8\linewidth]{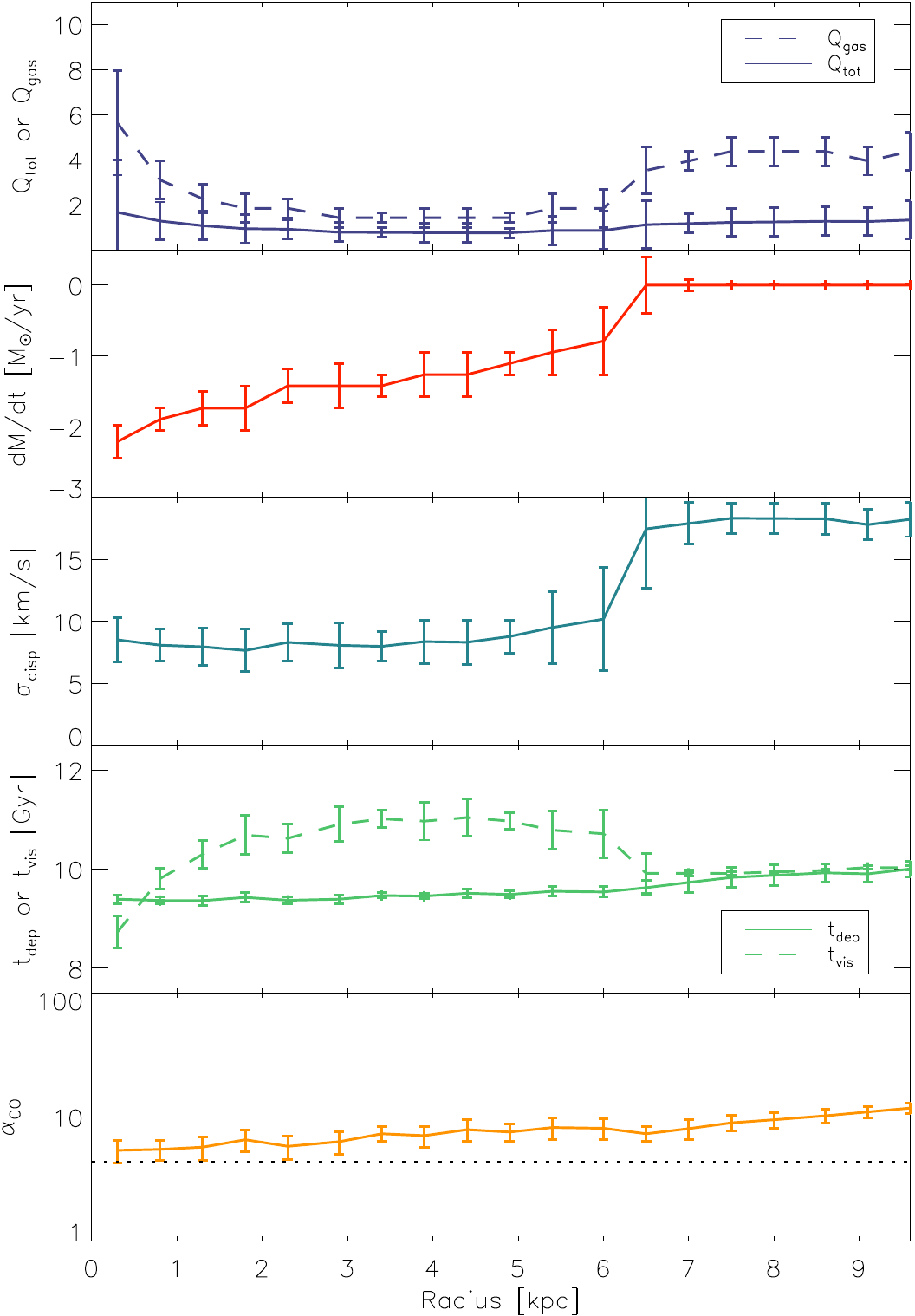}
      \end{subfigure}
       
      \caption{As Fig.~\ref{fig:n628} for NGC~3521.}
      \label{fig:fig}
      \end{figure*}

      \begin{figure*}
        \center
        \begin{subfigure}{.45\textwidth}
          \centering
          \caption{NGC~4736 best model}
          \includegraphics[width=1.\linewidth]{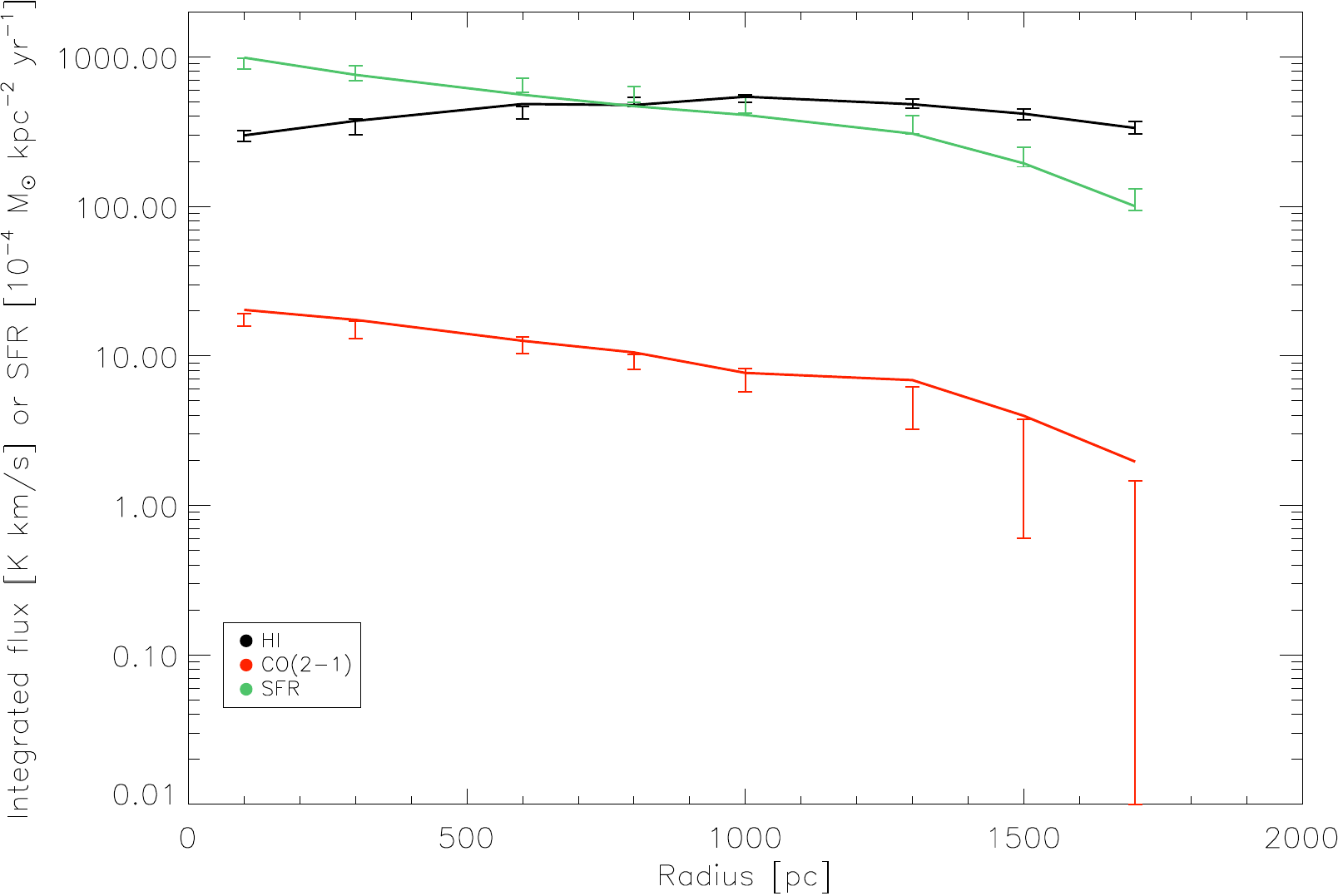}
         
        \end{subfigure}
        \begin{subfigure}{.45\textwidth}
          \centering
          \caption{NGC~4736 infrared profiles}
          \includegraphics[width=1.\linewidth]{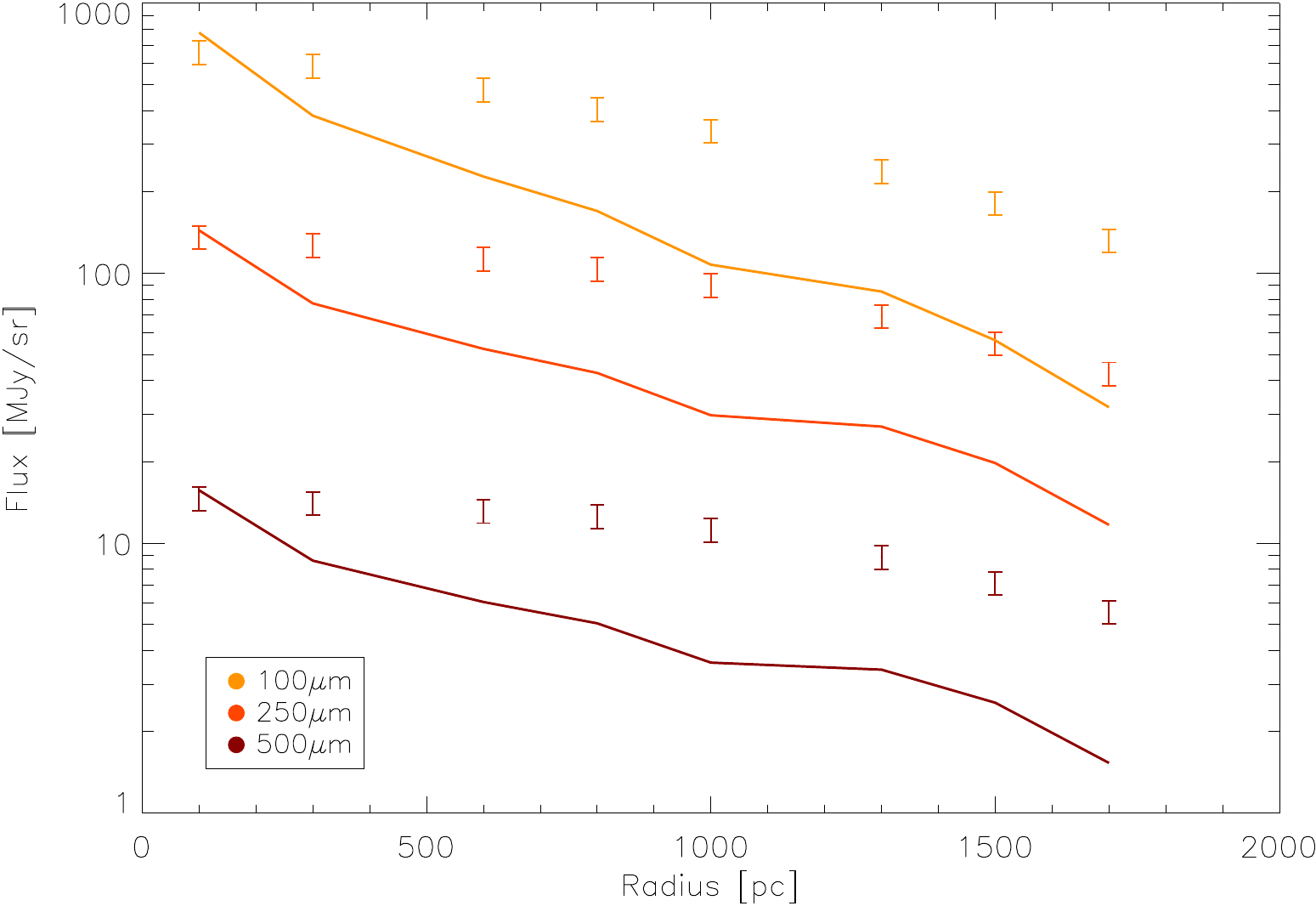}
        \end{subfigure}
        
        \begin{subfigure}{0.6\textwidth}
          \centering
          \caption{NGC~4736 main properties profiles}
          \includegraphics[width=0.8\linewidth]{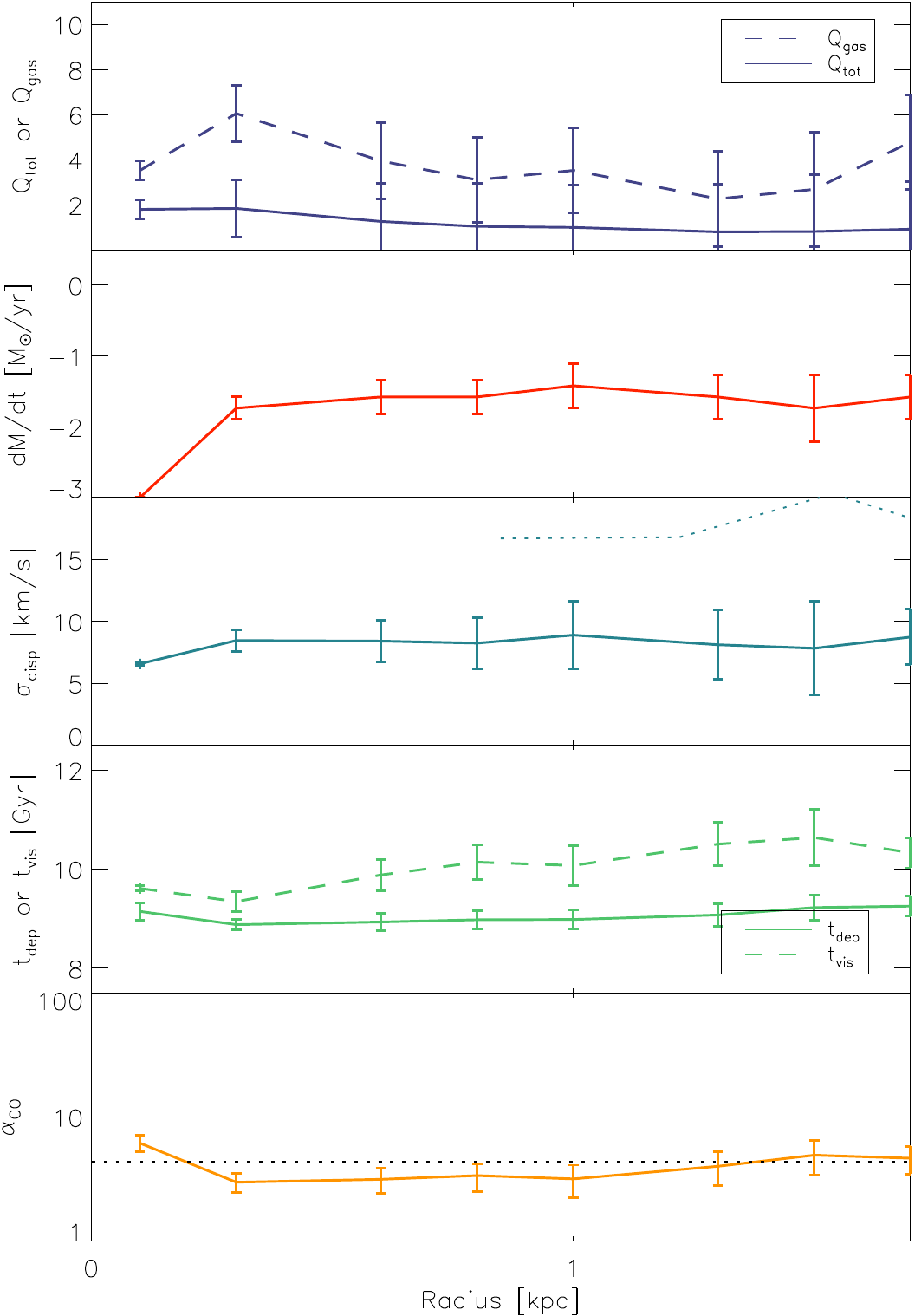}
        \end{subfigure}
         
        \caption{As Fig.~\ref{fig:n628} for NGC~4736.}
        \label{fig:fig}
        \end{figure*}

        \begin{figure*}
          \center
          \begin{subfigure}{.45\textwidth}
            \centering
            \caption{NGC~7331 best model}
            \includegraphics[width=1.\linewidth]{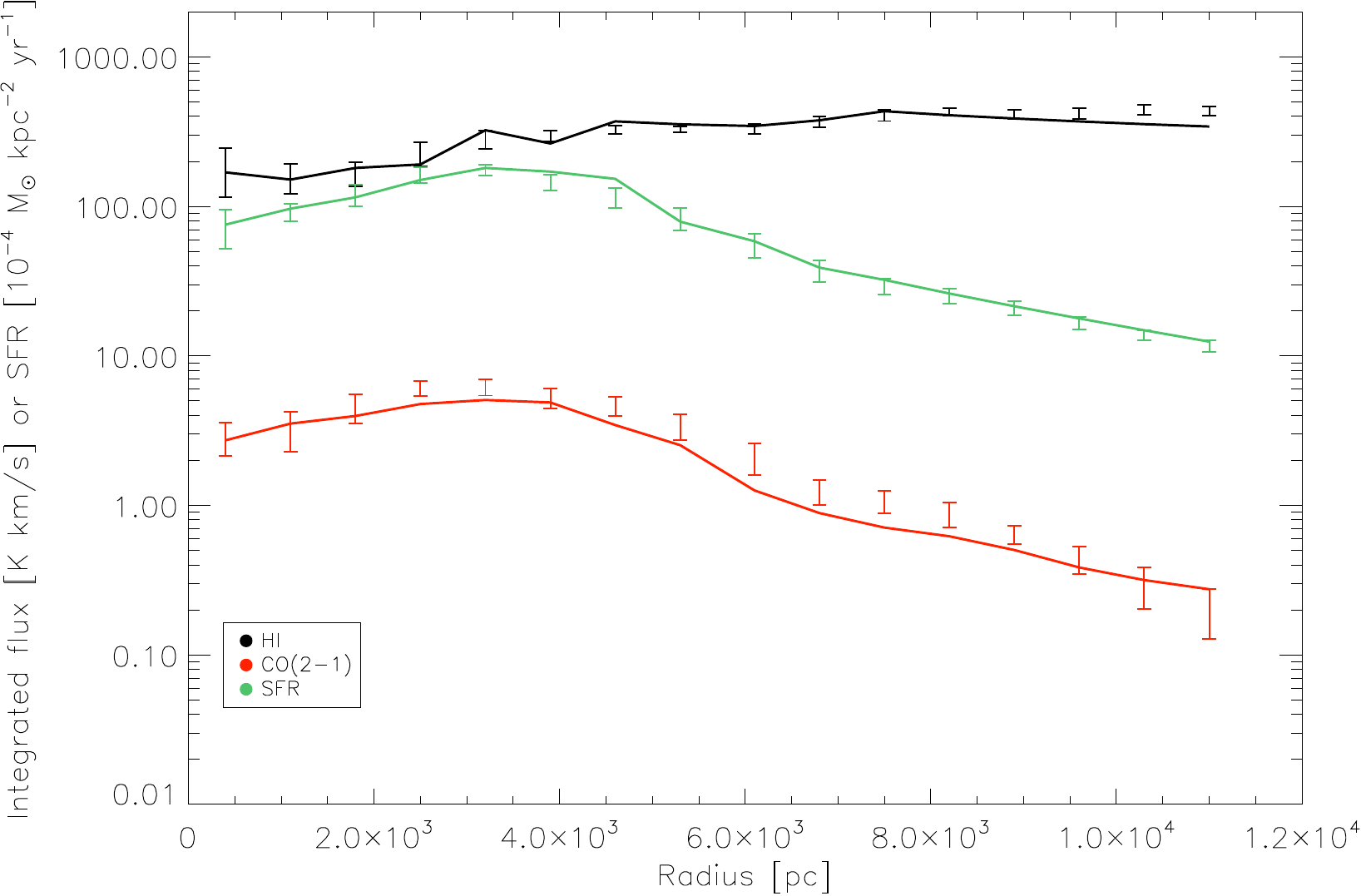}
           
          \end{subfigure}
          \begin{subfigure}{.45\textwidth}
            \centering
            \caption{NGC~7331 infrared profiles}
            \includegraphics[width=1.\linewidth]{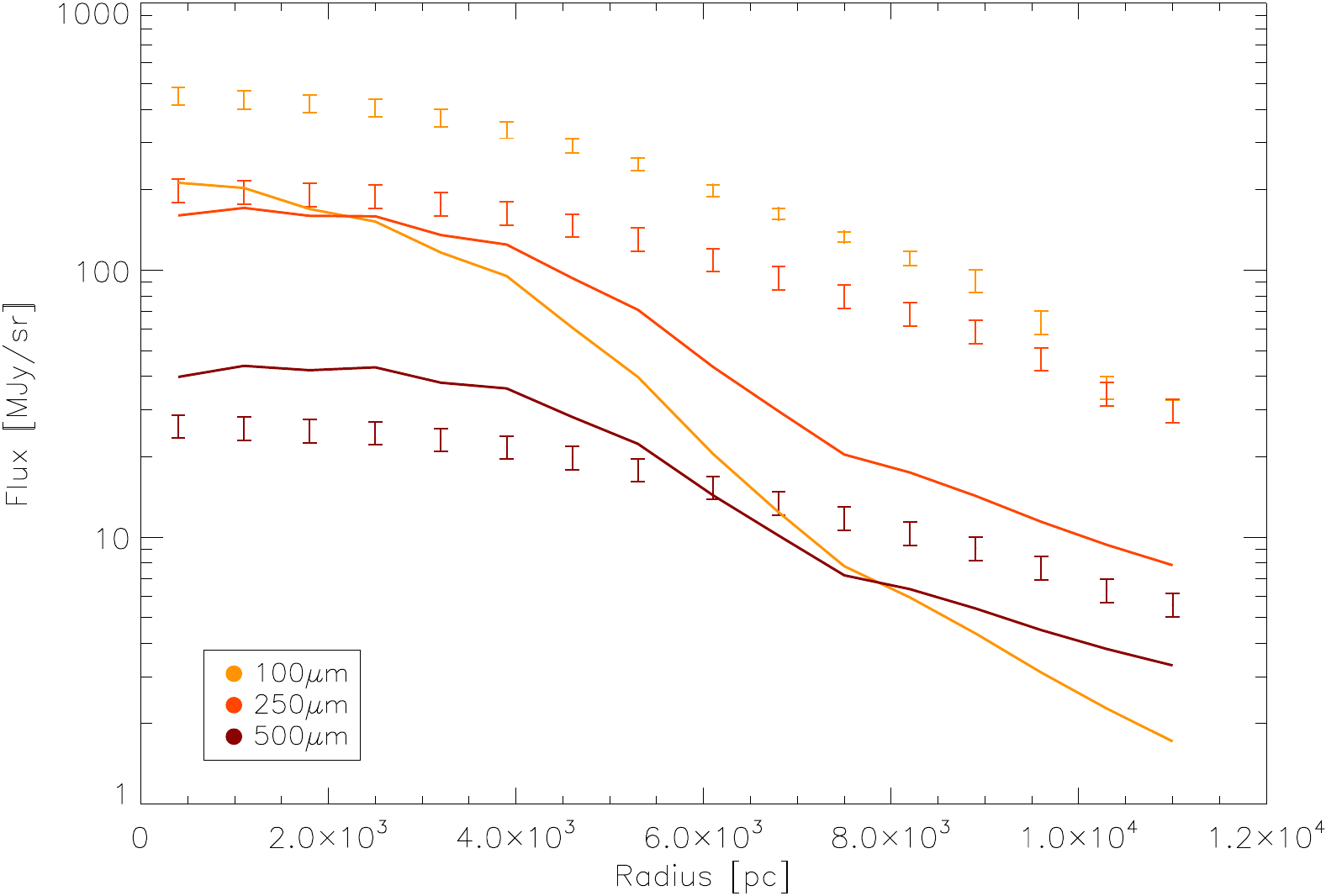}
          \end{subfigure}
          
          \begin{subfigure}{0.6\textwidth}
            \centering
            \caption{NGC~7331 main properties profiles}
            \includegraphics[width=0.8\linewidth]{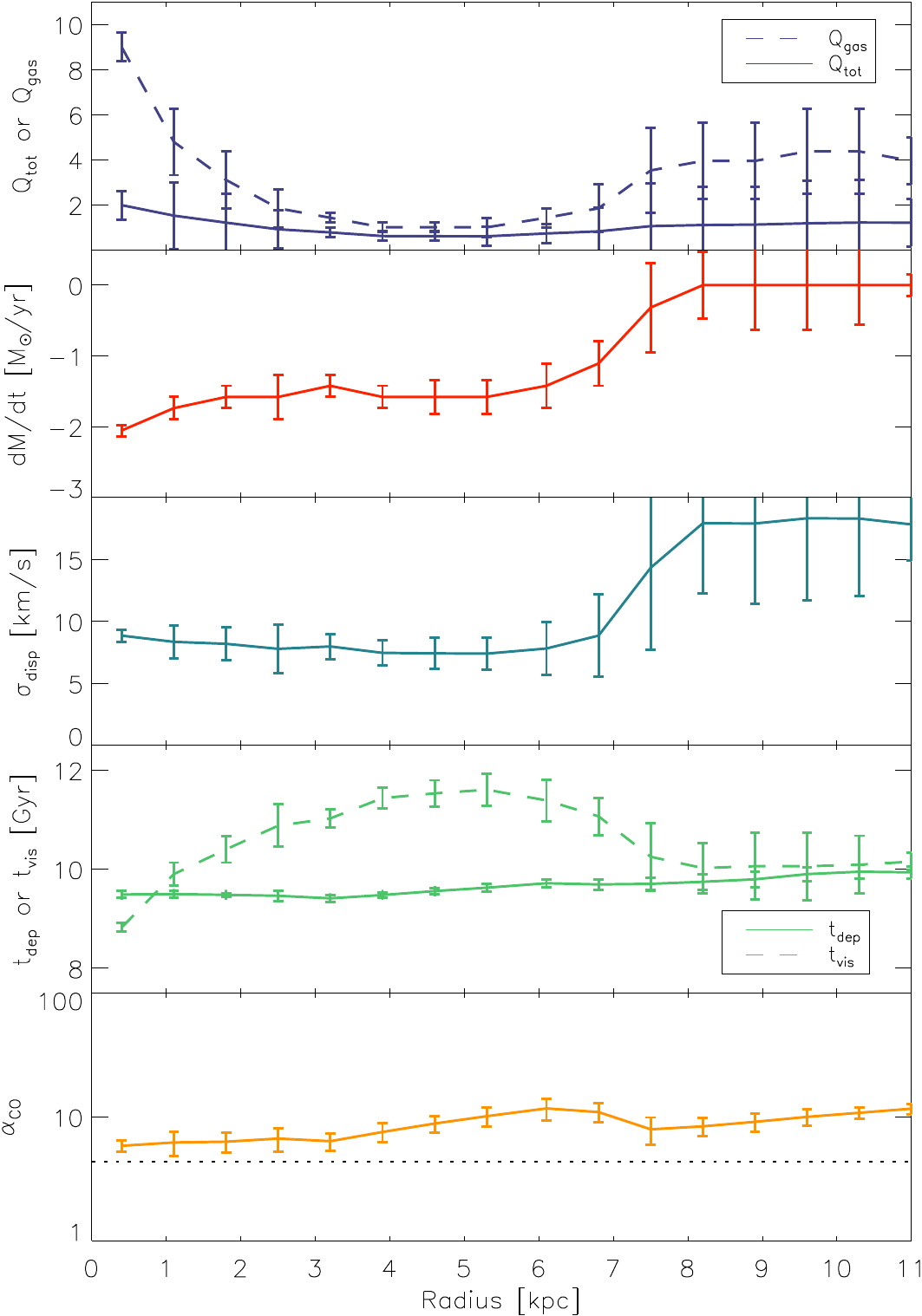}
          \end{subfigure}
           
          \caption{As Fig.~\ref{fig:n628} for NGC~7331.}
          \label{fig:fig}
          \end{figure*}

         \begin{figure*}
                  \center
                  \begin{subfigure}{.45\textwidth}
                    \centering
                    \caption{NGC~925 best model}
                    \includegraphics[width=1.\linewidth]{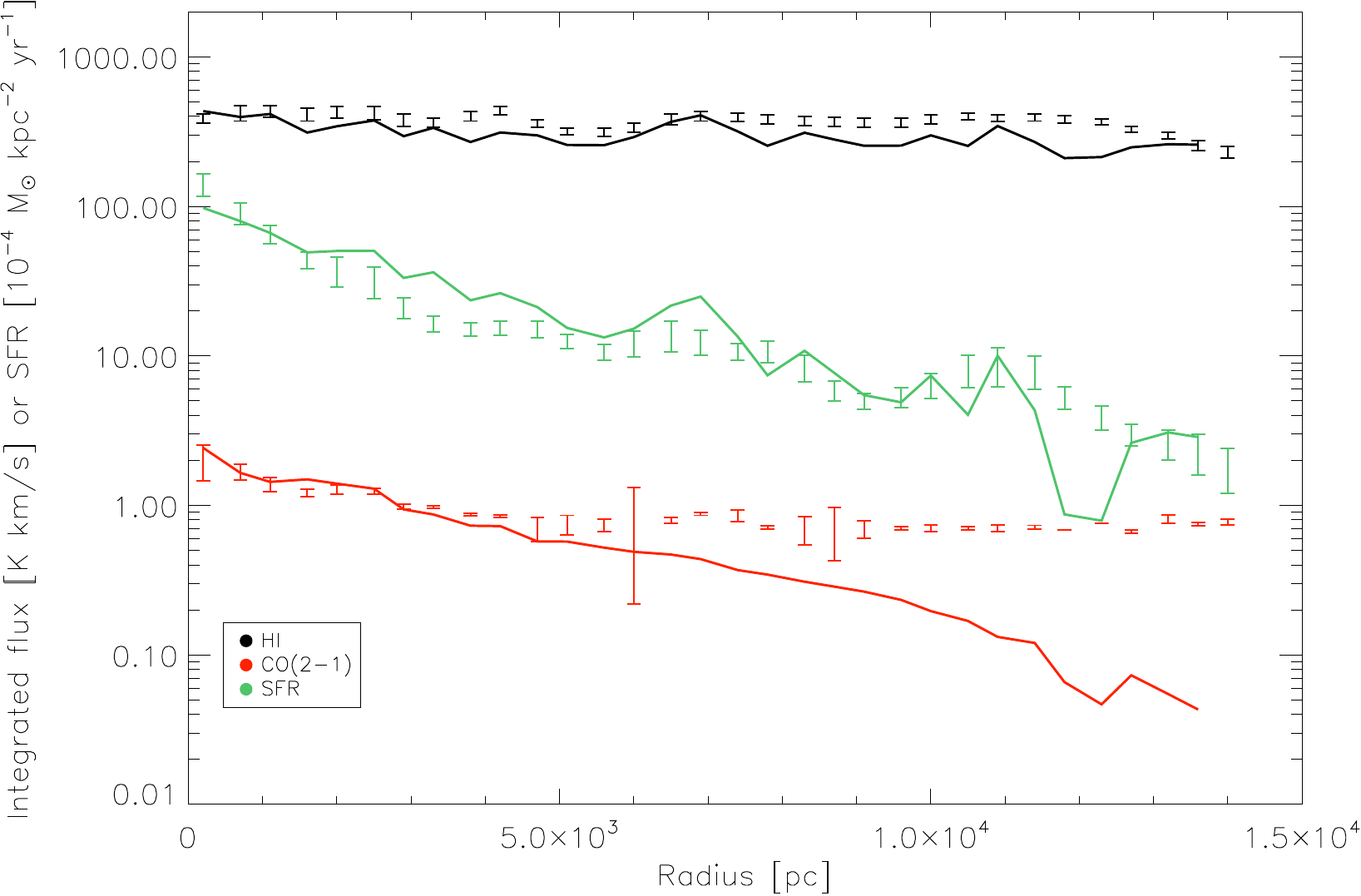}
                   
                  \end{subfigure}
                  \begin{subfigure}{.45\textwidth}
                    \centering
                    \caption{NGC~925 infrared profiles}
                    \includegraphics[width=1.\linewidth]{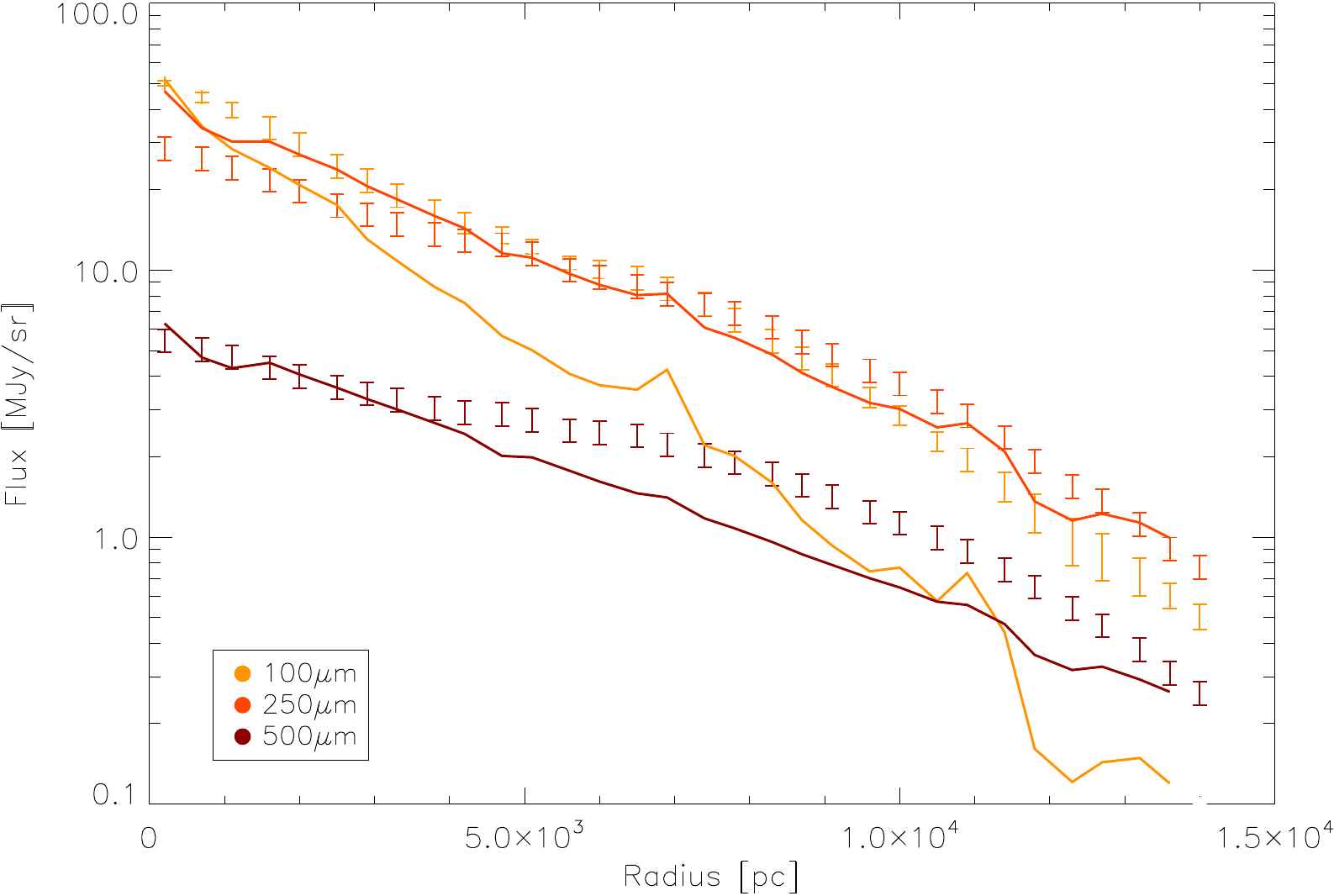}
                  \end{subfigure}
                  
                  \begin{subfigure}{0.6\textwidth}
                    \centering
                    \caption{NGC~925 main properties profiles}
                    \includegraphics[width=0.8\linewidth]{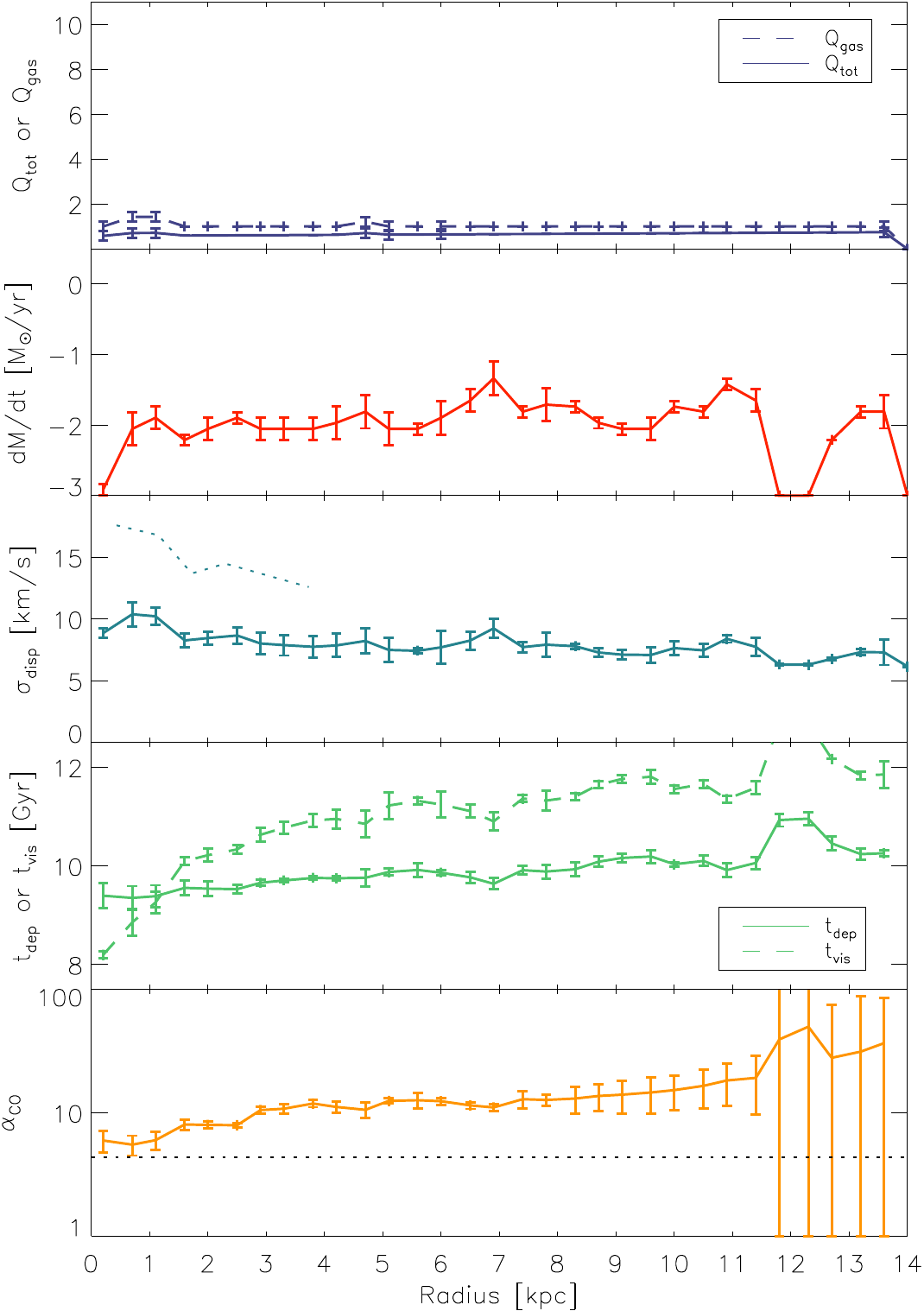}
                  \end{subfigure}
                   
                  \caption{As Fig.~\ref{fig:n628} for NGC~925.}
                  \label{fig:fig}
         \end{figure*}

          \begin{figure*}
              \center
              \begin{subfigure}{.45\textwidth}
                \centering
                \caption{NGC~2403 best model}
                \includegraphics[width=1.\linewidth]{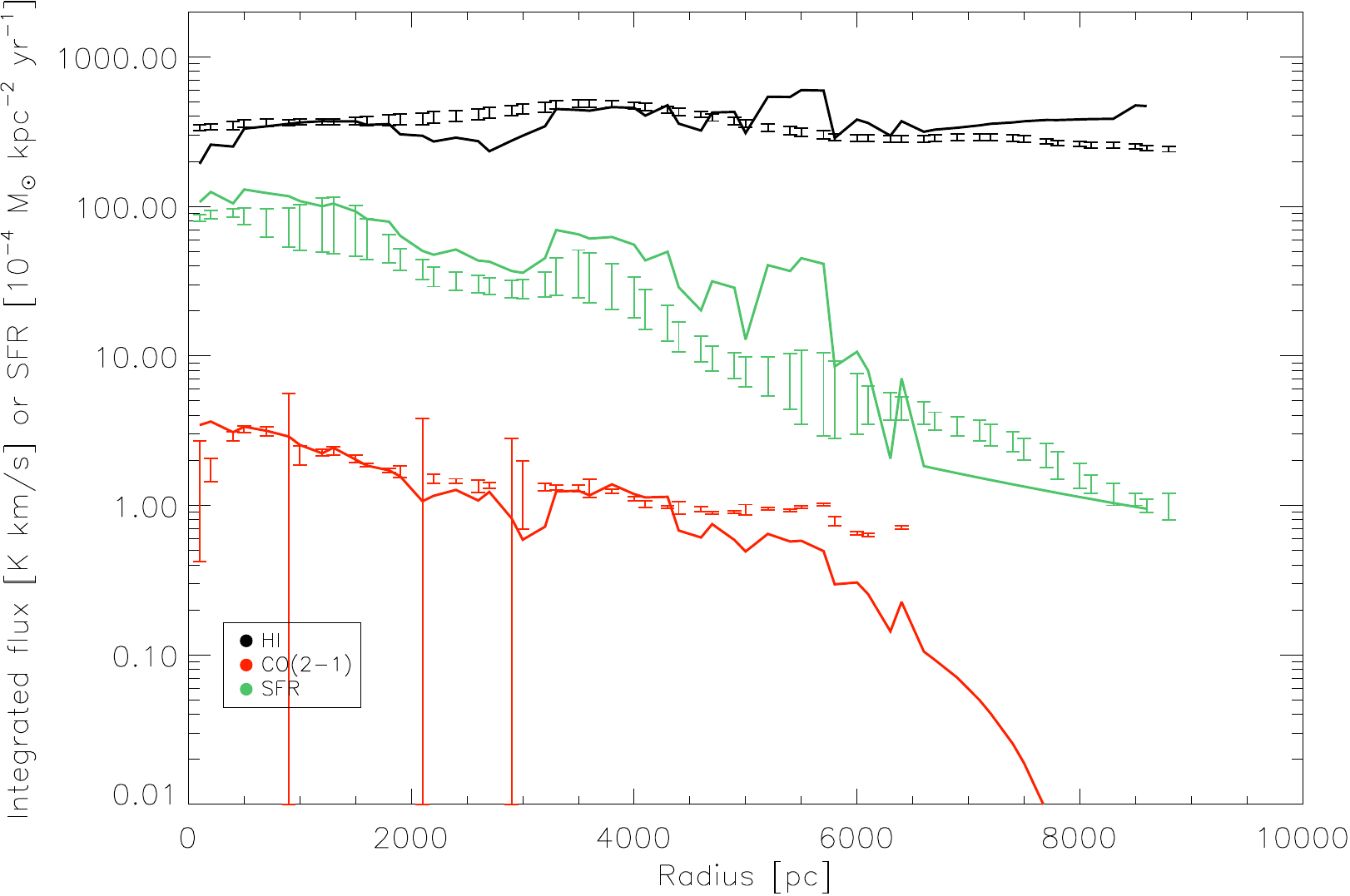}
               
              \end{subfigure}
              \begin{subfigure}{.45\textwidth}
                \centering
                \caption{NGC~2403 infrared profiles}
                \includegraphics[width=1.\linewidth]{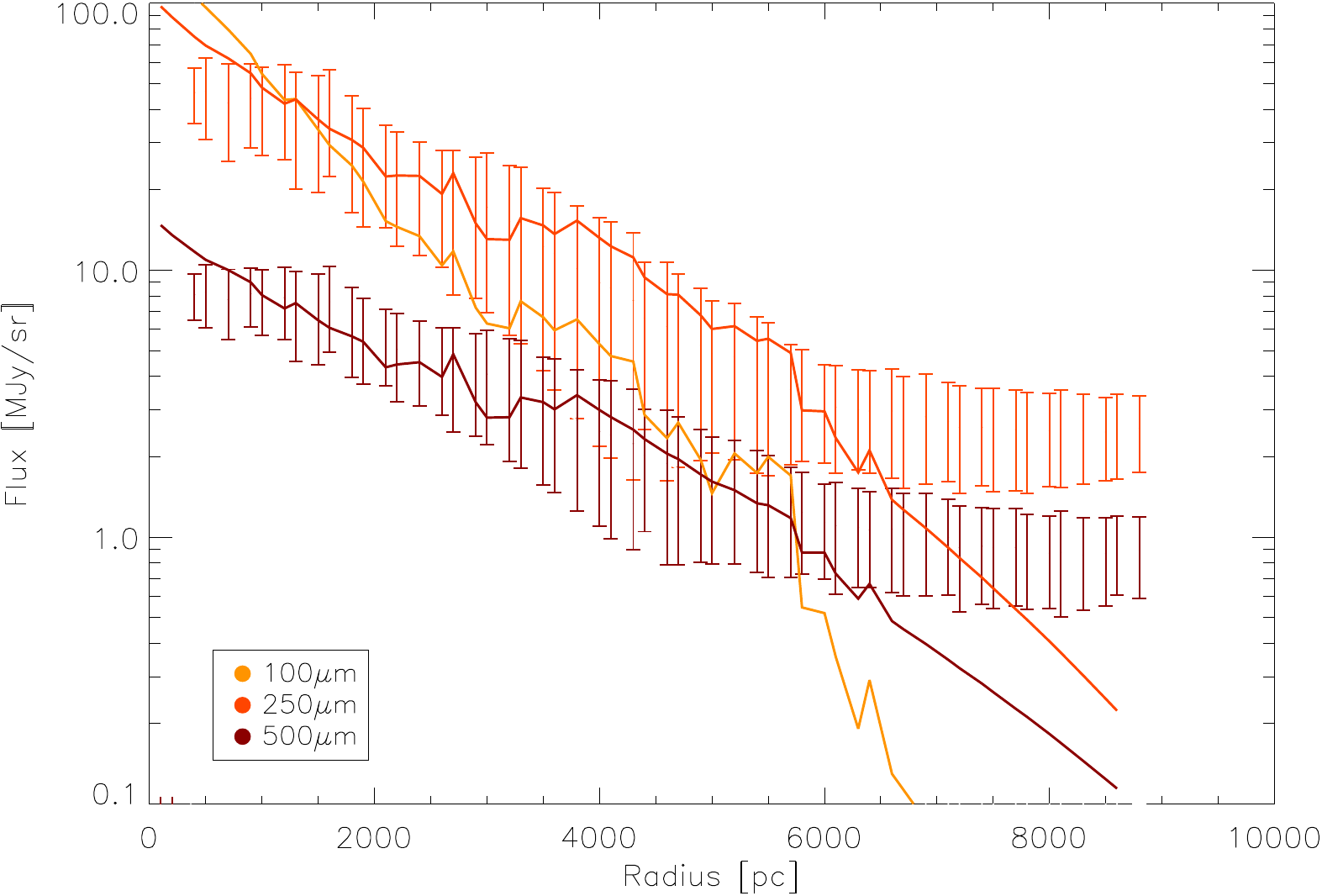}
              \end{subfigure}
              
              \begin{subfigure}{0.6\textwidth}
                \centering
                \caption{NGC~2403 main properties profiles}
                \includegraphics[width=0.8\linewidth]{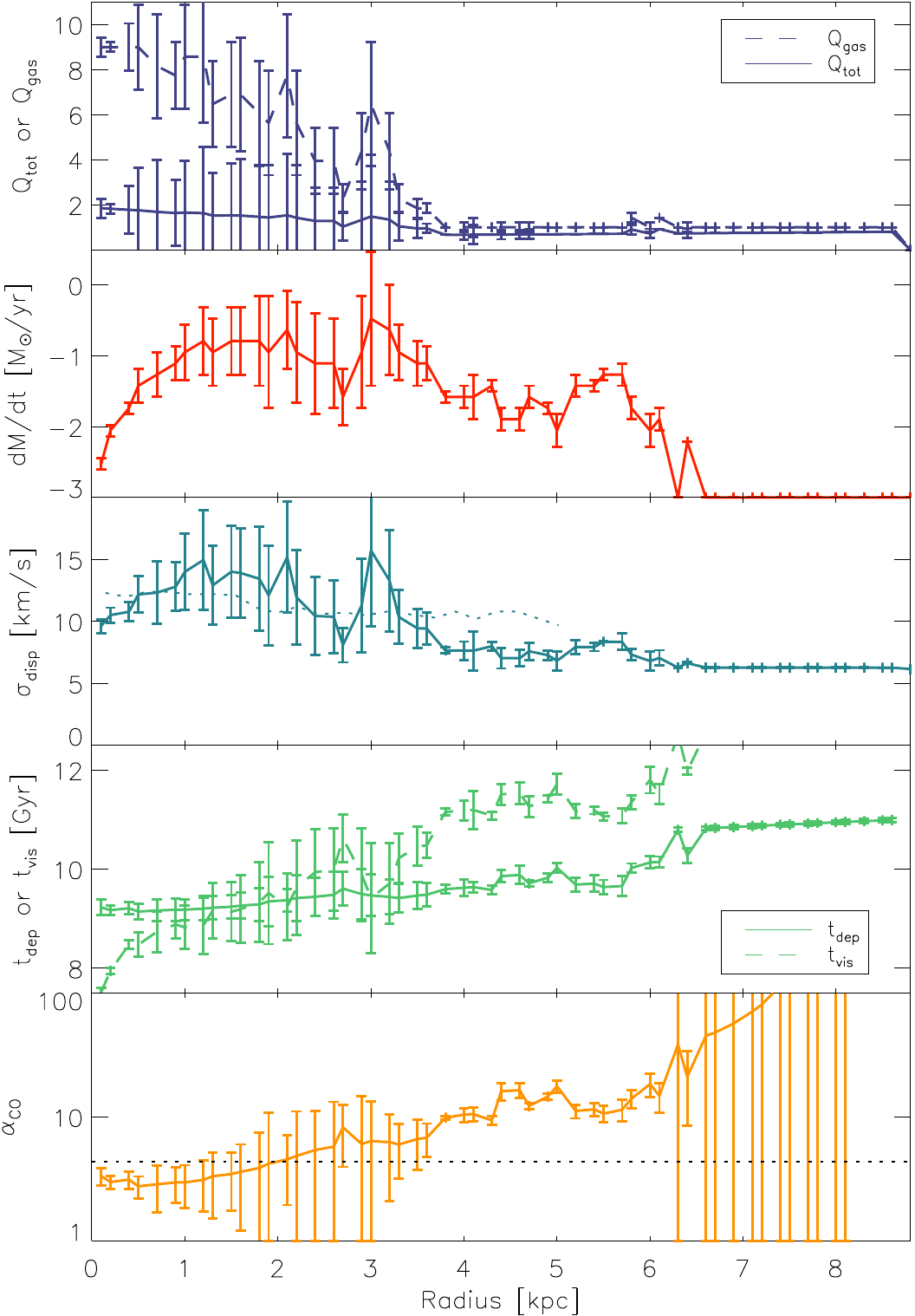}
              \end{subfigure}
               
              \caption{As Fig.~\ref{fig:n628} for NGC~2403.}
              \label{fig:2403}
              \end{figure*}

          \begin{figure*}
            \center
            \begin{subfigure}{.45\textwidth}
              \centering
              \caption{NGC~2976 best model}
              \includegraphics[width=1.\linewidth]{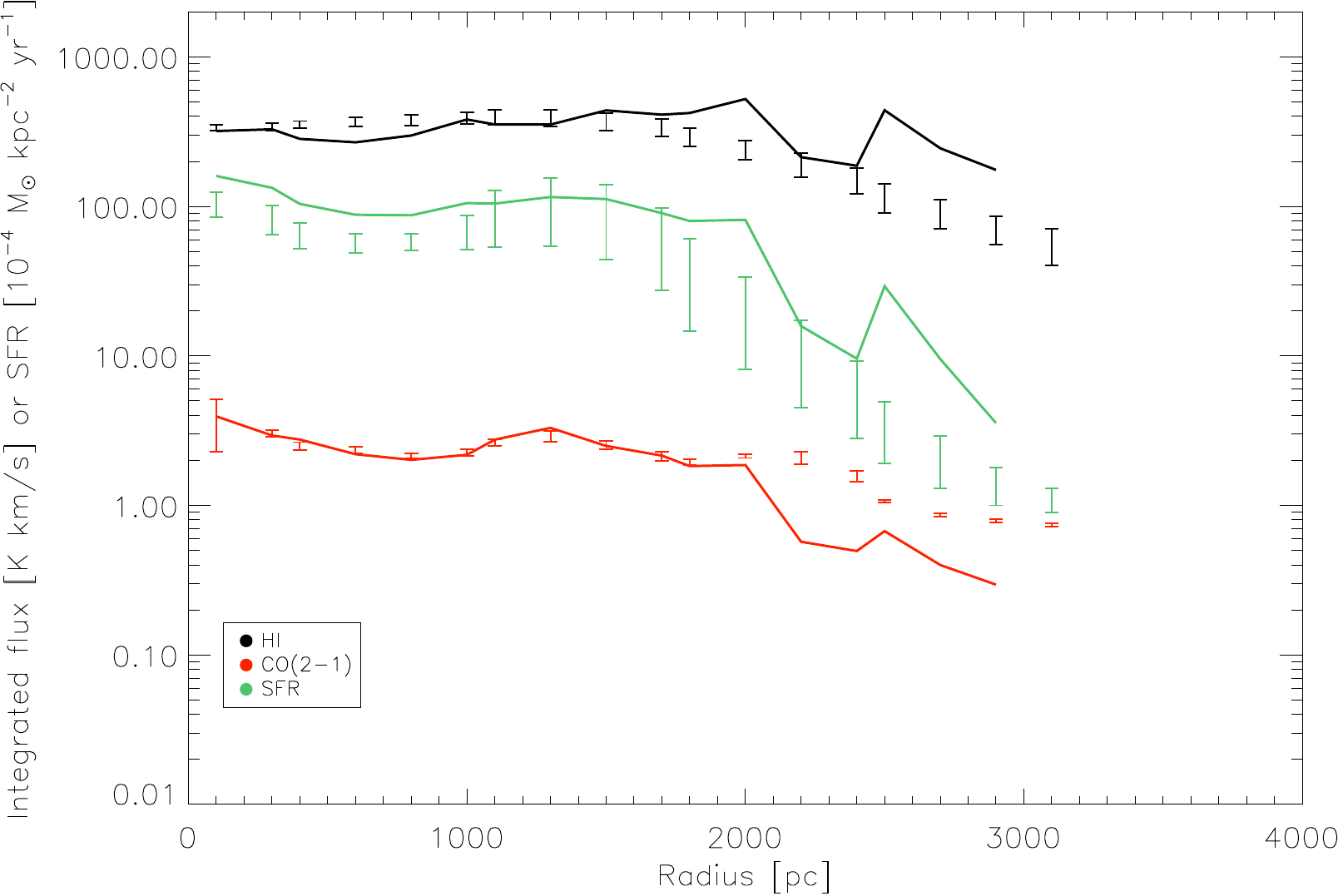}
             
            \end{subfigure}
            \begin{subfigure}{.45\textwidth}
              \centering
              \caption{NGC~2976 infrared profiles}
              \includegraphics[width=1.\linewidth]{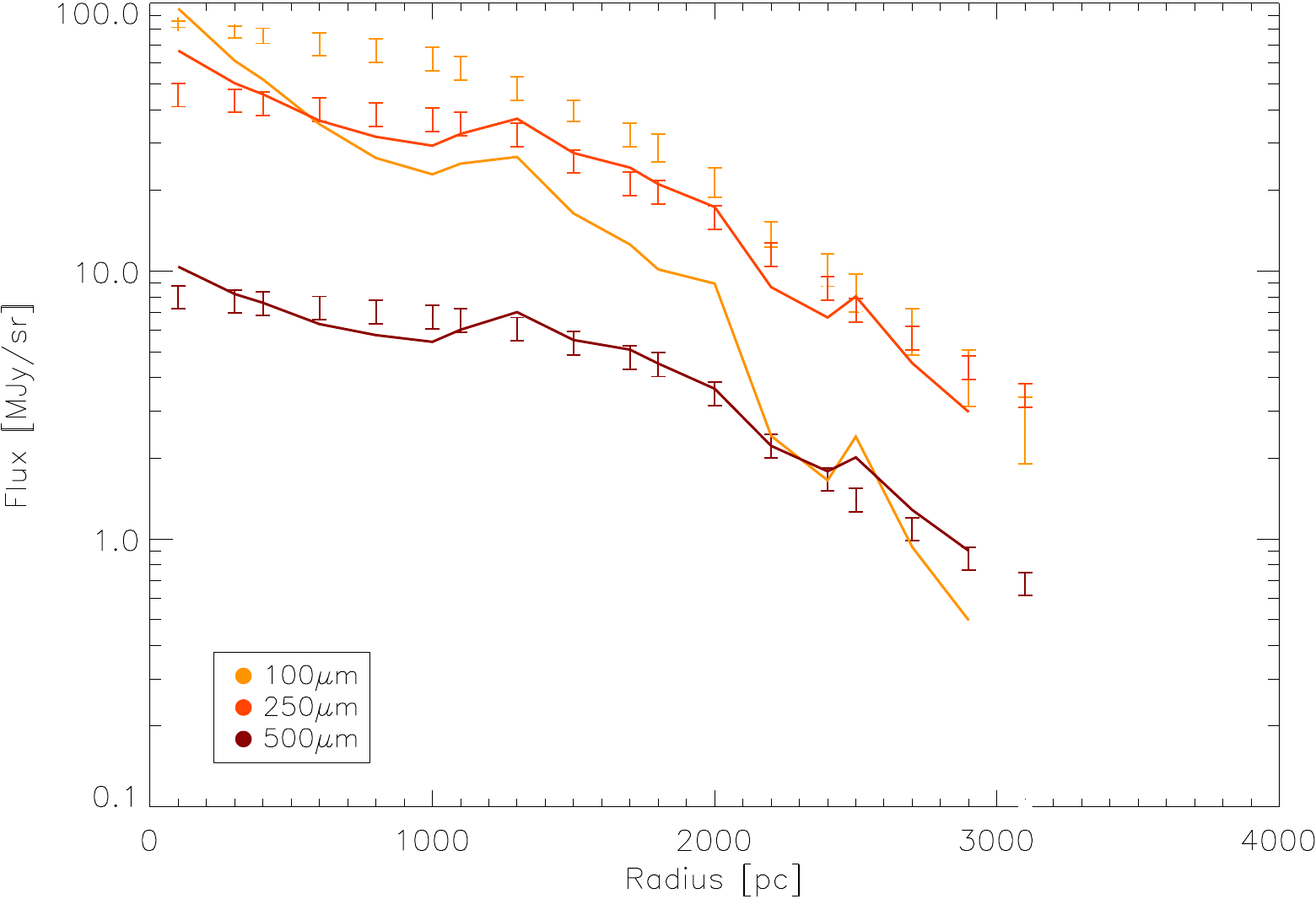}
            \end{subfigure}
            
            \begin{subfigure}{0.6\textwidth}
              \centering
              \caption{NGC~2976 main properties profiles}
              \includegraphics[width=0.8\linewidth]{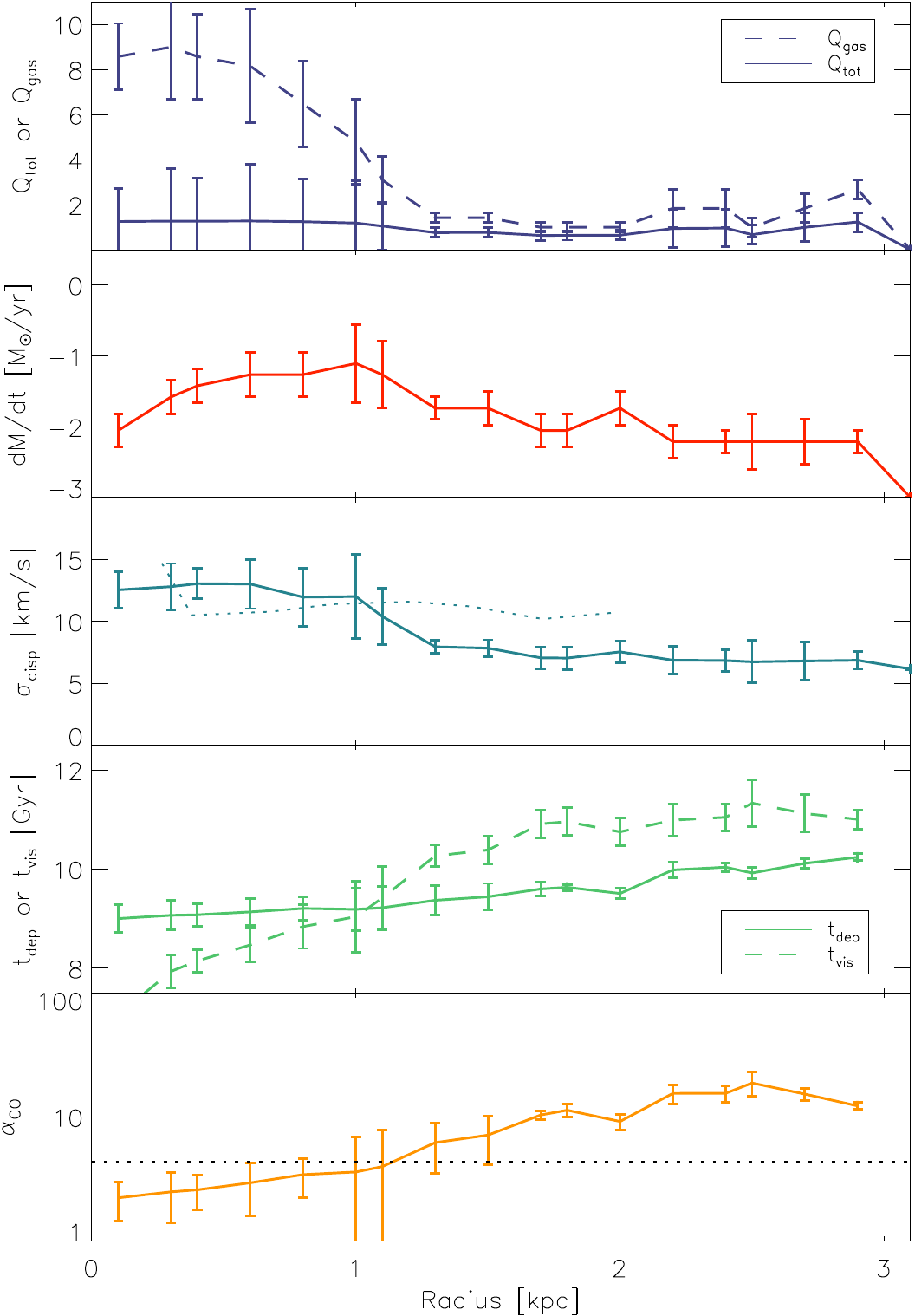}
            \end{subfigure}
             
            \caption{As Fig.~\ref{fig:n628} for NGC~2976.}
            \label{fig:2976}
            \end{figure*}

              \begin{figure*} 
                \center
                \begin{subfigure}{.45\textwidth}
                  \centering
                  \caption{NGC~4214 best model}
                  \includegraphics[width=1.\linewidth]{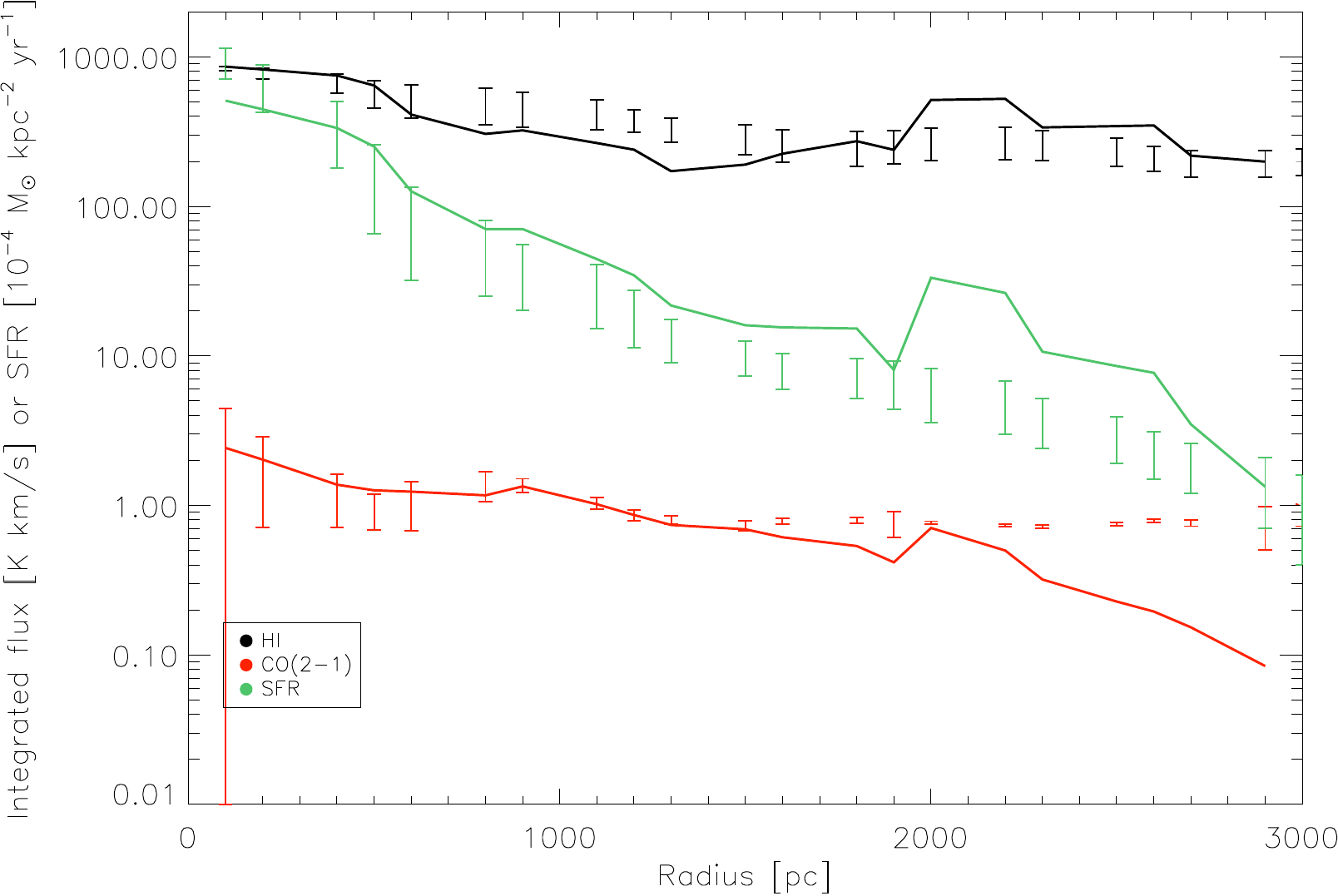}
                 
                \end{subfigure}
                \begin{subfigure}{.45\textwidth}
                  \centering
                  \caption{NGC~4214 infrared profiles}
                  \includegraphics[width=1.\linewidth]{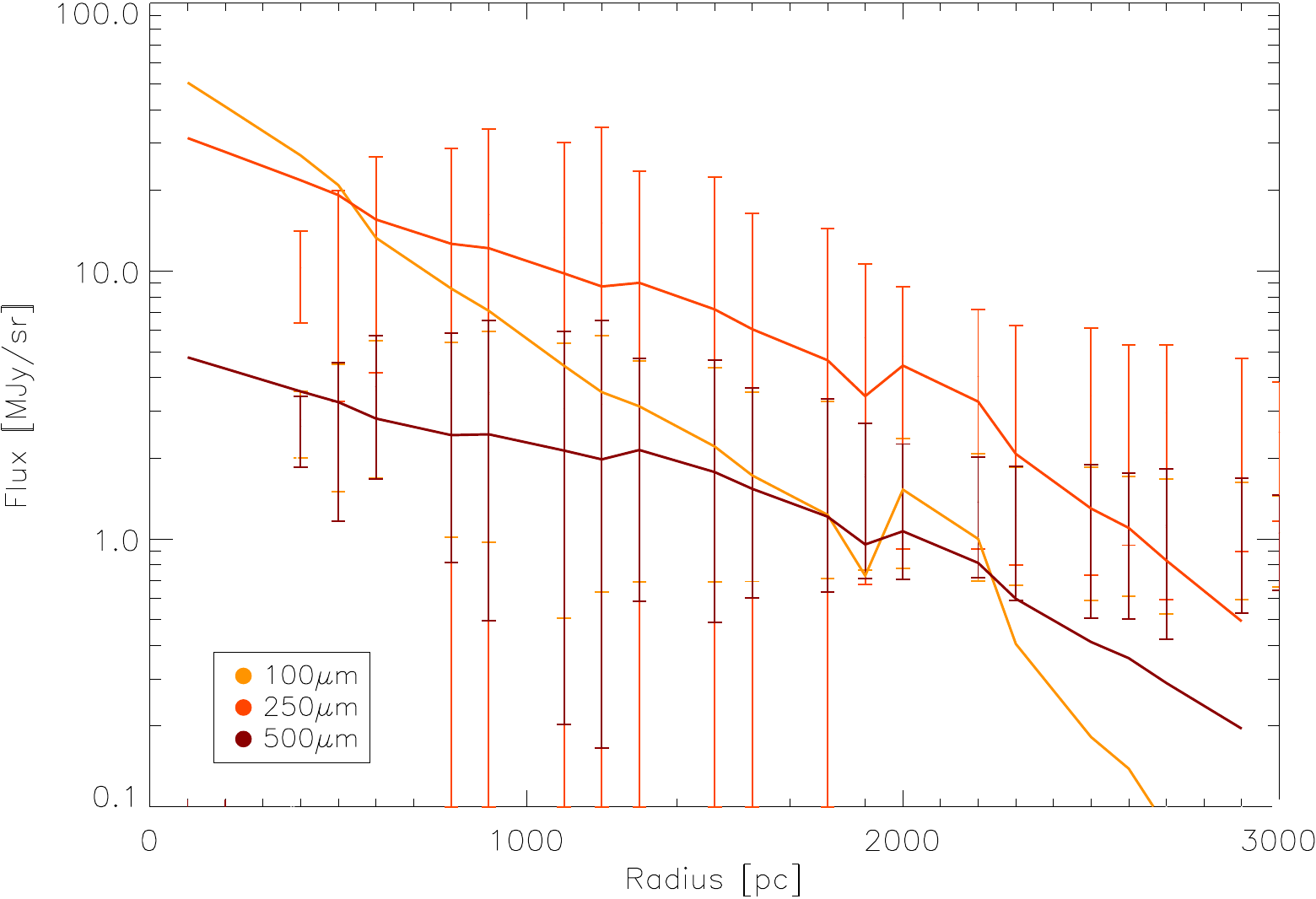}
                \end{subfigure}
                
                \begin{subfigure}{0.6\textwidth}
                  \centering
                  \caption{NGC~4214 main properties profiles}
                  \includegraphics[width=0.8\linewidth]{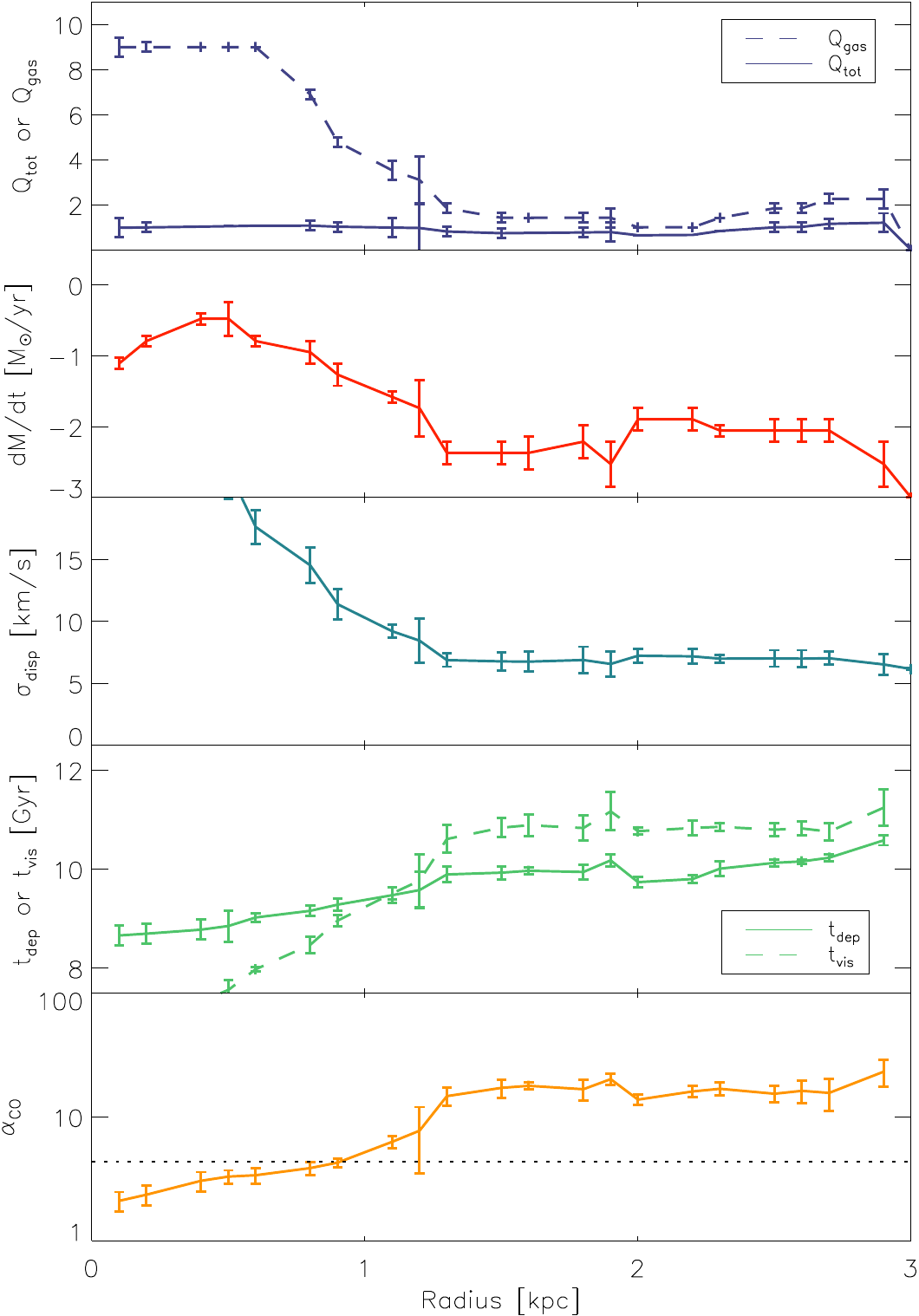}
                \end{subfigure}
                 
                \caption{As Fig.~\ref{fig:n628} for NGC~4214.}
                \label{fig:4214}
                \end{figure*}

              \begin{figure*} 
                \center
                \begin{subfigure}{.45\textwidth}
                  \centering
                  \caption{NGC~7793 best model}
                  \includegraphics[width=1.\linewidth]{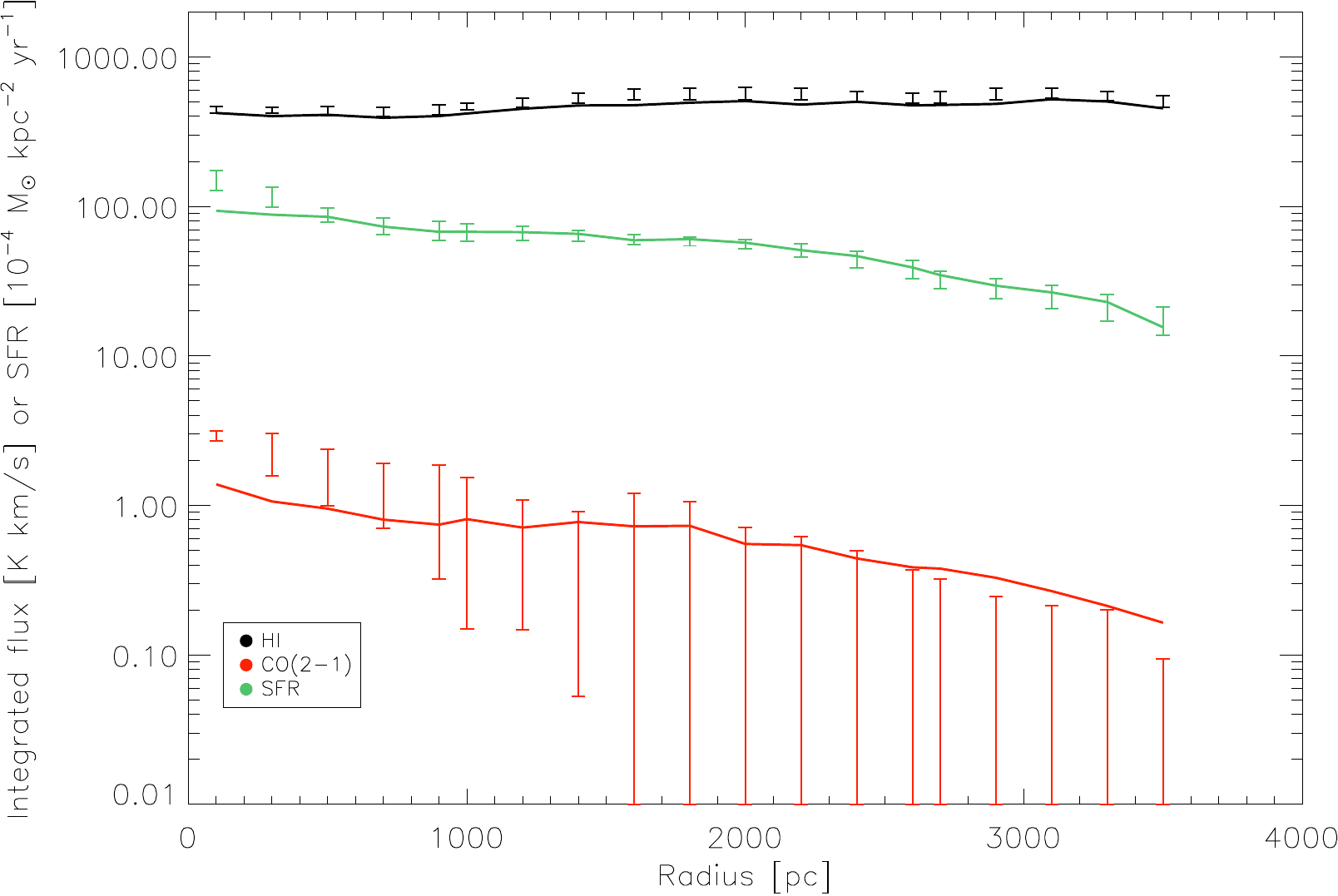}
                 
                \end{subfigure}
                \begin{subfigure}{.45\textwidth}
                  \centering
                  \caption{NGC~7793 infrared profiles}
                  \includegraphics[width=1.\linewidth]{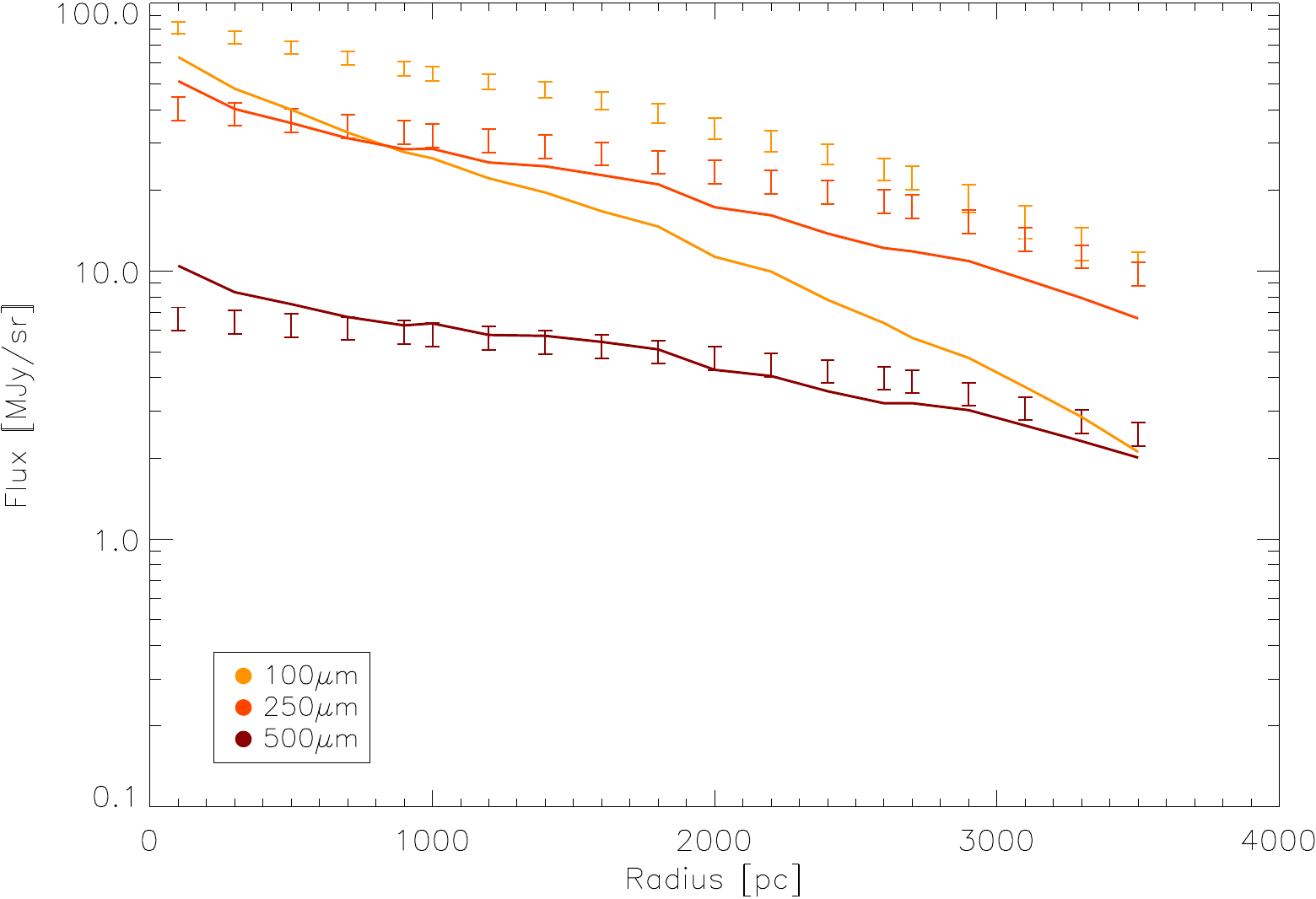}
                \end{subfigure}
                
                \begin{subfigure}{0.6\textwidth}
                  \centering
                  \caption{NGC~7793 main properties profiles}
                  \includegraphics[width=0.8\linewidth]{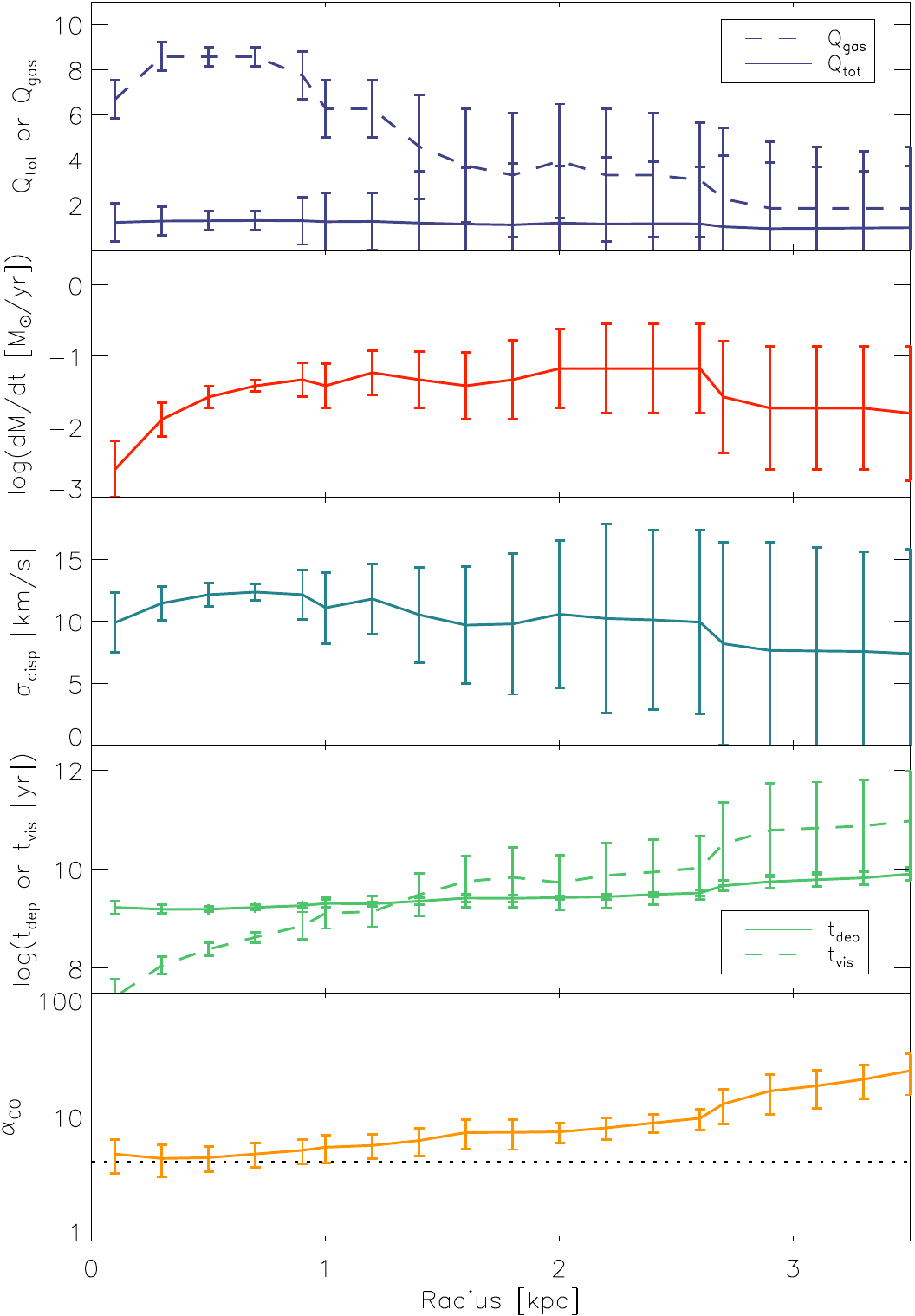}
                \end{subfigure}
                 
                \caption{As Fig.~\ref{fig:n628} for NGC~7793.}
                \label{fig:7793}
                \end{figure*}
             
\section{Emission from dense gas}

The fractions of total flux emitted by gas with densities higher than $n=10^{3}$, $10^{3.5}$, $10^4$, and $10^{4.5}$~cm$^{-3}$
are presented in Table~\ref{tab:fracint11}.
  
\begin{table*}[!ht]
  \caption{Fraction of the total flux.}
  \label{above}
  \begin{center}
   \renewcommand{\arraystretch}{1.2}
   \addtolength{\tabcolsep}{8.pt}
  \begin{tabular}{c|c|cccc}
     \hline
     \hline
      Galaxy & Molecule & $n > 10^{3}$ cm$^{-3}$ & $n > 10^{3.5}$ cm$^{-3}$ & $n > 10^{4}$ cm$^{-3}$ & $n > 10^{4.5}$ cm$^{-3}$ \\
      \hline

NGC628& CO &          20 \%&            9 \%&            3 \%&            2 \%\\
 & HCN &          62 \%&           50 \%&           28 \%&           17 \%\\
 & HCO &          48 \%&           31 \%&           14 \%&            8 \%\\
\hline
NGC3184& CO &          13 \%&            6 \%&            4 \%&            1 \%\\
 & HCN &          53 \%&           38 \%&           28 \%&           10 \%\\
 & HCO &          39 \%&           23 \%&           15 \%&            4 \%\\
\hline
NGC3627& CO &          19 \%&            7 \%&            4 \%&            1 \%\\
 & HCN &          59 \%&           41 \%&           31 \%&           12 \%\\
 & HCO &          46 \%&           24 \%&           16 \%&            6 \%\\
\hline
NGC5055& CO &          45 \%&           11 \%&            7 \%&            2 \%\\
 & HCN &          79 \%&           48 \%&           38 \%&           16 \%\\
 & HCO &          76 \%&           31 \%&           22 \%&            8 \%\\
\hline
NGC5194& CO &          39 \%&           10 \%&            6 \%&            2 \%\\
 & HCN &          80 \%&           42 \%&           34 \%&           15 \%\\
 & HCO &          69 \%&           26 \%&           20 \%&            8 \%\\
\hline
NGC6946& CO &          28 \%&            9 \%&            4 \%&            2 \%\\ 
 & HCN &          73 \%&           43 \%&           25 \%&           15 \%\\
 & HCO &          62 \%&           27 \%&           13 \%&            8 \%\\
      \hline
  \end{tabular}
     \end{center}
     \label{tab:fracint11}
\end{table*}   

\end{document}